\documentclass[12pt,twoside,a4paper]{report}

\usepackage{fancyhdr}
\usepackage{graphicx}
\usepackage{psfrag,amsmath,amssymb,amsfonts,
bbm,latexsym,color,dcolumn,epsf}
\usepackage{XTocinc} 
\renewcommand{\baselinestretch}{1.30}

\newlength{\captsize}		\let\captsize=\normalsize
\newlength{\captwidth}		
\newlength{\beforetableskip} 	
\newcommand{\capt}[1]{
	\renewcommand{\baselinestretch}{0.9}		
        \begin{minipage}{\captwidth}
	\let\normalsize=\captsize
	\caption[#1]{\sf #1}
	\end{minipage}\\ \vspace{\beforetableskip}}

\makeatletter
	\long\def\@makecaption#1#2{\vskip 10\p@
	\setbox\@tempboxa\hbox{{\bf #1:} #2} 
	\ifdim \wd\@tempboxa >\hsize
           {\bf #1:} #2\par
        \else
           \hbox to\hsize{\box\@tempboxa\hfill}
        \fi}
\makeatother     

\begin{document}


\newcommand\aap{Astron. \& Astrophys.{} }
\newcommand\apj{Astrophys. J.{} }
\newcommand\geo{Geophys. Astrophys. Fluid Dyn.{} }
\newcommand\mnras{Mon. Not. Roy. Astr. Soc.{} }
\newcommand\nat{Nature{} }
\newcommand\physrep{Phys. Rep.{} }
\newcommand\pra{Phys. Rev. A{} }
\newcommand\pre{Phys. Rev. E{} }
\newcommand\prl{Phys. Rev. Lett.{} }
\newcommand\sci{Science{} }
\newcommand\solphys{Solar Phys.{} }


\newcommand\fem{\mathcal{E}}
\newcommand\vA{\mbox{\boldmath $ A $}}
\newcommand\vB{\mbox{\boldmath $ B $}}
\newcommand\vE{\mbox{\boldmath $ E $}}
\newcommand\vF{\mbox{\boldmath $ F $}}
\newcommand\vG{\mbox{\boldmath $ G $}}
\newcommand\vJ{\mbox{\boldmath $ J $}}
\newcommand\vS{\mbox{\boldmath $ S $}}
\newcommand\vU{\mbox{\boldmath $ U $}}
\newcommand\vb{\mbox{\boldmath $ b $}}
\newcommand\vf{\mbox{\boldmath $ f $}}
\newcommand\vu{\mbox{\boldmath $ u $}}
\newcommand\vphi{\mbox{\boldmath $ \phi $}}
\newcommand\vomega{\mbox{\boldmath $ \omega $}}
\newcommand\oB{\overline{\vB}}
\newcommand\oF{\overline{\vF}}
\newcommand\oG{\overline{\vG}}
\newcommand\oU{\overline{\vU}}
\newcommand\of{\overline{\vf}}
\newcommand{\beq}{\begin{equation}}
\newcommand{\eeq}{\end{equation}}
\newcommand{\beqa}{\begin{eqnarray}}
\newcommand{\eeqa}{\end{eqnarray}}
\newcommand{\beqas}{\begin{eqnarray*}}
\newcommand{\eeqas}{\end{eqnarray*}}
\newcommand{\mbf}{\mathbf}
\newcommand{\itm}{\mathit}


\begin{titlepage}
\begin{center}

\begin{figure}
\centering
\epsfxsize=3.5cm
\epsfbox{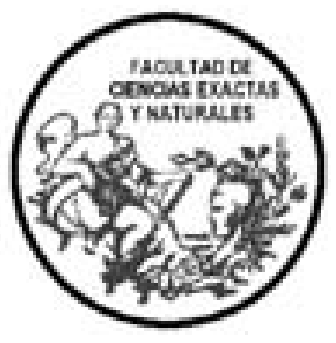}
\end{figure}

\vspace{0.3cm}
\Large{\textbf{UNIVERSIDAD DE BUENOS AIRES \\}}
\vspace{0.2cm}
\Large{Facultad de Ciencias Exactas y Naturales \\}
\vspace{0.2cm}
\Large{Departamento de F\'{\i}sica \\}
\vspace{2.5cm}

\LARGE{\textit{Sistemas cu\'anticos  bajo la influencia de condiciones
externas: fluctuaciones y decoherencia.\\}}
\vspace{0.5cm}
{\small por} {\textsc{Paula In\'{e}s Villar \\}}

\vspace{3cm}
\large{\centering{
Director de Tesis: Fernando C\'esar Lombardo \\ 
Lugar de Trabajo: Departamento de F\'{\i}sica, FCEN, UBA\\}}

\vspace{2.0cm}
\normalsize{
Trabajo de Tesis para optar por el t\'{\i}tulo 
de \\
Doctor de la Universidad de Buenos Aires
en el \'area de Ciencias F\'isicas.\\}

\vspace{0.2cm}
\normalsize{Diciembre 2007}  \\

\end{center}
\end{titlepage}

\newpage
\cleardoublepage
\thispagestyle{empty}
\vspace{1cm}
\hspace{11 cm}
{\Large{\textit{A mi familia}}}
\renewcommand{\abstractname}{Resumen}
\cleardoublepage

\abstract{ Un tema de mucho inter\'es en numerosas ramas de la F\'{\i}sica 
es el estudio del mecanismo  por  el  cual  ocurre  la  transici\'on
cu\'antico-cl\'asica.  Este  es  un tema largamente debatido en la historia
de  la  mec\'anica  cl\'asica. 

Actualmente, una  explicaci\'on  razonable es    que  los    sistemas
macrosc\'opicos se comportan cl\'asicamente debido a  su  interacci\'on con
el  entorno.  Son sistemas cu\'anticos abiertos que interact\'uan con otros 
sistemas (o partes de ellos). Esa interacci\'on produce en el sistema 
en estudio o relevante, efectos disipativos y/o m\'as sofisticados como ser 
la aparici\'on de una base privilegiada (entre los numerosos estados en su  
espacio de Hilbert) en la que el sistema presentar\'a 
aspectos cl\'asicos (base de punteros).

En esta Tesis estudiaremos el proceso de p\'erdida de coherencia, entrelazamiento
y excitaci\'on energ\'etica inducida en los sistemas originalmente cu\'anticos
debido a la presencia de un entorno.

En primer lugar, nos dedicaremos a analizar las fuctuaciones de
vac\'{\i}o del entorno como fuente
 del proceso de \textit{p\'erdida de coherencia} (o bien, decoherencia).
   El acoplamiento de un sistema cu\'{a}ntico  a un entorno generalmente
   produce fluctuaciones de la energ\'{\i}a de
 la part\'{\i}cula de prueba a\'un a temperatura cero. Luego, extenderemos
este an\'alisis an\'alitico y num\'erico a entornos m\'as generales
a temperatura extrictamente cero y en el l\'imite de alta temperatura.
Finalmente, mostraremos que una vez que el subsistema en cuesti\'on
perdi\'o coherencia y se ``hizo cl\'asico'',  tiene chances
de excitarse energ\'eticamente, a\'un en presencia de un  entorno cu\'antico
 a temperatura cero.

Por otro lado, estudiaremos el proceso de decoherencia en sistemas
compuestos y en 
el caso de part\'iculas cargadas en campos electromagn\'eticos.
En ese contexto, aplicaremos nuestras estimaciones anal\'iticas a
experimentos de interferencia  y pondremos condiciones
para la medici\'on de la fase geom\'etrica en el caso de estados
mixtos que evolucionan de manera no unitaria.

Finalmente, estudiaremos la p\'erdida de coherencia en teor\'ia de
campos durante transiciones de fase. En ese contexto, utilizaremos
el m\'etodo de la funcional de decoherencia, sin la necesidad de 
pasar por la ecuaci\'on maestra. Estudiamos configuraciones de campo
que difieren entre s\'i en su amplitud, pero tienen igual distribuci\'on
espacial, as\'i como tambi\'en historias de campo con diferente
localizaci\'on en el espacio.

{\small{{\it Palabras claves}: Movimiento Browniano C\'uantico-Teor\'ia de la 
Decoherencia-Fluctuaciones Cu\'anticas- Efecto T\'unel-Fases Geom\'etricas}}
\renewcommand{\abstractname}{Abstract}
\cleardoublepage

\abstract{
 Macroscopic quantum states are never isolated from their environments. They are not
closed quantum systems, and therefore they cannot behave according to the unitary quantum-mechanical rules. Consequently, these so often called ‘classical’ systems suffer a loss of quantum coherence that is absorbed by the environment. This decoherence destroys quantum interferences. For our everyday world, the time scale at which the quantum interferences are
destroyed is so small that, in the end, the observer is able to perceive only one outcome. As
far as we see, decoherence is the main process behind the quantum to classical transition.
Formally, it is the dynamic suppression of the interference terms induced on subsystems due
to the interaction with an environment.

Therefore, in this Thesis,  we shall analyze the effects that general environments, namely ohmic and non-ohmic, at zero and high temperature induce over a quantum Brownian particle. 
We state that the evolution of the system can be summarized in terms of two main environmental induced physical phenomena: decoherence
and energy activation. In this Thesis, we shall show that the latter is a post-decoherence phenomenon. As the energy is an observable, the excitation process can be consider
a direct indicator of the system-environment entanglement particularly useful at zero
 temperature.

From other point of view, we shall study different attempts to show the 
decoherence process in double-slit-like experiments both for 
charged particles (electrons) and for neutral particles
with permanent dipole moments. Interference will be studied when electrons or
atomic systems are coupled to classical or quantum electromagnetic fields.
In this context, we shall show that 
the interaction between the particles and time-dependent fields induces a time-varying Aharonov phase. In this context, we shall apply our results to a real matter wave interferometry experiment. We shall also show under which general conditions the geometry
phase of a quantum open system can be observed.

Finally, we shall study the decoherence process during a quantum phase transition.
In this framework, we shall show that it can be phrased easily in terms
of the decoherence functional, without having to use the master equation. To demonstrate this, we shall consider the decohering effects due to the
displacement of domain boundaries, with implications for the displacement of defects, in general. We shall see that decoherence arises so quickly in
this event, that it is negligible in comparison to decoherence due to field fluctuations in the way defined in  previous papers.

{\small{{\it Keywords}: Quantum Brownian Motion- Decoherence Theory-
Quantum Fluctuations-Tunel Effect-Geometric Phases}}
}



\cleardoublepage

\pagenumbering{arabic}
\fancyhead[LE]{\thepage}
\fancyhead[RE]{\sl Indice}
\fancyhead[RO]{\bf \thepage}
\fancyfoot[CE,CO]{}

\tableofcontents
\newpage
\thispagestyle{empty}
\cleardoublepage

\renewcommand{\chaptermark}[1]{\markboth{\sl #1}{}}
\renewcommand{\sectionmark}[1]{\markright{\sl \thesection . \hspace{0.05cm} 
#1}}

\fancyhead[LE,RO]{\bf \thepage}
\fancyhead[RE]{\leftmark}
\fancyhead[LO]{\rightmark}

\chapter*{Introducci\'on}
\addcontentsline{toc}{chapter}{Introducci\'on}
\markboth{Sistemas cu\'anticos bajo la influencia de condiciones
externas}
{Introducci\'on}

A comienzos de los a\~nos 1900, se formul\'o una teor\'ia para explicar
qu\'e era la materia, \mbox{asumiendo} que hab\'ia peque\~nas part\'iculas
cargadas  dentro de los \'atomos. Esta teor\'ia evolucion\'o \mbox{gradualmente}
hasta incluir un n\'ucleo pesado y electrones movi\'endose alrededor del
mismo. Primero se intent\'o explicar el movimiento de estos electrones
usando las leyes mec\'anicas de movimiento de manera an\'aloga a lo
que hab\'ia realizado Newton para explicar el movimiento de los planetas
alrededor del Sol. Pero \'esto result\'o un verdadero fracaso: todas las
predicciones eran err\'oneas. Incluso la teor\'ia del electromagnetismo
predec\'ia una vida media para el \'atomo de Hidr\'ogeno
40 \'ordenes de magnitud menor a la que en efecto ``viv\'ia".
Llev\'o mucho tiempo encontrar la
manera de explicar lo que ocurr\'ia a nivel at\'omico, ya que no resultaba
para nada intuitivo y hab\'ia que perder, en cierta forma, el {\it sentido com\'un}
para entender ese escenario. Finalmente, en 1926, una teor\'ia denominada
 Mec\'anica Cu\'antica, explic\'o el comportamiento de 
los electrones en la materia. La palabra ``cu\'antica" hac\'ia referencia a
 este aspecto peculiar
de la Naturaleza que iba en contra del sentido com\'un. Esta teor\'ia 
explicaba todo tipo de detalles, como por ejemplo, por qu\'e un \'atomo de Ox\'igeno
 se combinaba con dos \'atomos de Hidr\'ogeno para formar agua. 
La teor\'ia Cu\'antica fue un verdadero \'exito. A partir de entonces, surgieron
muchos aportes al desarrollo de esta teor\'ia, la
cual,  hoy d\'ia, es una de las teor\'{\i}as m\'as exitosas de la historia de
la f\'{\i}sica: todas sus predicciones concuerdan con los experimentos con gran precisi\'on
y su aplicaci\'on ha transformado el mundo tecnol\'ogico en diferentes \'areas. Se puede
aplicar para describir el comportamiento de los s\'olidos, la estructura y funcionamiento
del ADN y las propiedades de los superflu\'idos, entre otras tantas aplicaciones. Sin embargo, y
a pesar de todas sus virtudes, la Teor\'ia Cu\'antica es a\'un hoy una teor\'ia controversial.
Su descripci\'on de los fen\'omenos f\'isicos es, a menudo, enfrentada con nuestra percepci\'on
de la realidad. Peor a\'un, a veces, nos conduce a pensar en predicciones que son consideradas
parad\'ojicas. La ra\'iz de este sentimiento de desaz\'on se encuentra en la diferencia que notamos en
el principio de  superposici\'on que gobierna el mundo cu\'antico y el mundo cl\'asico
de todos los d\'ias, el cual pareciera
violar dicho principio.
Si bien la Mec\'anica Cu\'antica es imprescindible para una {\it descripci\'on
macrosc\'opica} de la Naturaleza, con la Mec\'anica Cl\'asica bastar\'ia, en principio,
para describir el comportamiento de sistemas a escalas macrosc\'opicas. Esta afirmaci\'on
est\'a basada en nuestro sentido com\'un: las cosas ``suceden'' o ``no suceden'' y los
objetos materiales siempre tienen propiedades bien definidas. Sin embargo, no es posible
identificar los sistemas macrosc\'opicos como sistemas cl\'asicos 
ya que hay muchos ejemplos, como las estrellas de neutrones o las junturas \mbox{Josephson}, donde
la Mec\'anica Cu\'antica es absolutamente necesaria a\'un a escala macrosc\'opica. La pregunta
 que nos surge luego, es si la ``clasicalidad'' de un sistema podr\'ia ser 
considerada una propiedad emergente de este \'ultimo.

El conflicto aparente entre la Mec\'anica Cu\'antica y nuestro sentido com\'un se 
basa en el hecho que nosotros no observamos efectos de interferencia entre estados
macrosc\'opicos distinguibles. Nosotros observamos una realidad con distintas
alternativas bien definidas en lugar de una superposici\'on coherente de ellas como
predecir\'ia la Mec\'anica Cu\'antica. En todos los casos, la interferencia cu\'antica
est\'a ausente y las probabilidades pueden sumarse como en la Mec\'anica Cl\'asica.

En general, la ausencia de interferencia cu\'antica entre estados macrosc\'opicamente
distinguibles se puede explicar como consecuencia de un proceso de 
``p\'erdida de coherencia'' o ``decoherencia''.  Este proceso considera como aspecto 
fundamental que los sistemas cu\'anticos macrosc\'opicos no est\'an aislados. Es decir,
siempre interact\'uan con un entorno formado de infinitos grados de libertad. Por lo mismo,
estos sistemas, no son sistemas cu\'anticos cerrados y no se comportan de acuerdo a las
leyes de la Mec\'anica Cu\'antica. Consecuentemente, estos sistemas, generalmente conocidos
como ``cl\'asicos'', sufren una p\'erdida de coherencia cu\'antica que es absorbida por el
entorno. La interacci\'on del sistema cu\'antico en cuesti\'on con el entorno es la que destruye
las interferencias cu\'anticas del sistema, y hace posible  una descripci\'on en t\'ermino
de variables cl\'asicas \cite{Zurek1,Zurek2,Zurek3}. Para nuestro mundo cotidiano, la
escala temporal en la cual se lleva a cabo este proceso es muy chica y es esa la raz\'on
por la cual nosotros precibimos una s\'ola realizaci\'on de los eventos. La decoherencia es
el principal proceso detr\'as de la transici\'on cu\'antico-cl\'asica. 

En este contexto,
la motivaci\'on fundamental de la presente Tesis es avanzar en la compresi\'on
del origen y de los mecanismos por los cuales la transici\'on cu\'antico-cl\'asica
tiene lugar. En particular, utilizando tanto el enfoque del formalismo de la funcional de influencia
de Feynman y Vernon \cite{Feynman} como el modelo de ``dephasing"
 de Stern, Aharonov e Imry
\cite{SternAhaImry}, estudiamos este proceso
en distintos sistemas cu\'anticos abiertos. 

El ejemplo paradigm\'atico de un sistema cu\'antico abierto es el llamado 
Movimiento Browniano Cu\'antico (MBC), en el cual una part\'\i cula cu\'antica 
(por ejemplo sujeta a un potencial de oscilador arm\'onico) interact\'ua con
un n\'umero (en principio infinito) de osciladores arm\'onicos que representan 
al entorno (o a los posibles entornos) con el cual la part\'\i cula de inter\'es 
se encuentra acoplada. T\'ecnicamente, el problema ha sido tratado a partir de la 
funcional de influencia de Feynman y Vernon \cite{GraIng,HuPazZhangI}.
 Integrando los grados de libertad 
asociados al entorno, uno obtiene una descripci\'on efectiva de la din\'amica 
del sistema donde la ``influencia" debida al entorno aparace en t\'erminos de 
la renormalizaci\'on de los par\'ametros del sistema, efectos de disipaci\'on y 
de difusi\'on. Se ha demostrado que la difusi\'on (o el ruido) est\'a directamente 
relacionada con los efectos de decoherencia y, a partir de ella, con la 
transici\'on cu\'antico-cl\'asica. Este ejemplo ser\'a resumido brevemente
en el Cap\'itulo \ref{c1} de esta Tesis.

Usualmente, el problema de la p\'erdida de coherencia en el MBC ha 
sido tratado para sistemas 
acoplados a entornos con una dada densidad espectral (\'ohmicos) y en estados 
t\'ermicos a  temperatura alta. Ejemplos m\'as generales de entornos no \'ohmicos 
a toda temperatura son m\'as escasos en la Literatura y, en general, est\'an 
circunscriptos a estudios num\'ericos solamente. La motivaci\'on original del
Cap\'itulo \ref{c2} de esta Tesis, consiste en entender el rol de las fluctuaciones
puramente cu\'anticas (de vac\'io) en el proceso de decoherencia.  
El acoplamiento de un sistema cu\'antico  a un entorno generalmente 
produce fluctuaciones de la energ\'ia de la part\'icula de prueba a\'un a temperatura cero
\cite{Buttiker}. Como las fases adquiridas por las componentes del 
sistema son integrales de la energ\'ia en el tiempo, las fluctuaciones cu\'anticas 
del vac\'io permiten que se produzca p\'erdida de coherencia a\'un a temperatura cero. 
Estas fluctuaciones son una consecuencia del acoplamiento finito 
entre la part\'icula de prueba (o sistema) y el ba\~no t\'ermico (o entorno), 
y del hecho que el Hamiltoniano del sistema aislado no conmuta con el 
Hamiltoniano de Interacci\'on. Por lo tanto, en este contexto, en el
Cap\'itulo \ref{c2} de esta Tesis, consideraremos el efecto 
que producen las fluctuaciones de vac\'io como \'unica
 fuente de decoherencia  en un sistema cu\'antico (o sea, no hay
 fluctuaciones t\'ermicas que induzcan la transici\'on cu\'antico-cl\'asico). 
 Estudiaremos los efectos difusivos 
en  una part\'{\i}cula Browniana cu\'antica acoplada a un
 entorno a temperatura cero,
realizando un estudio detallado del problema. 
Mostraremos el c\'alculo anal\'{\i}tico de los
coeficientes difusivos y del tiempo de decoherencia. Una
cantidad particularmente instructiva es la energ\'ia. 
Las fluctuaciones de la energ\'ia
a temperatura cero, son un buen indicador del entrelazamiento
existente entre el sistema y el entorno \cite{Jordan}. 
Por el contrario, si el sistema y el ba\~no no est\'an
entrelazados, el sistema se halla simplemente en su
estado de menor energ\'ia. Por todo \'esto, en el Cap\'itulo 
\ref{c2} de esta Tesis, tambi\'en analizaremos el proceso de activaci\'on
energ\'etica cuando la temperatura del entorno es cero.
En ese contexto, extenderemos el an\'alisis a entornos m\'as
generales, es decir supra\'ohmicos y sub\'ohmicos, para analizar
la p\'erdida de coherencia y la activaci\'on energ\'etica en estos entornos
m\'as generales. Este cap\'itulo est\'a 
principalmente basado en las Refs.\cite{PLA, PLA2}.

Una vez demostrada la existencia de estos fen\'omenos inducidos en 
el caso sencillo de la part\'icula Browniana, en el Cap\'itulo \ref{c3},
nos dedicaremos a aplicar estos conceptos a un caso m\'as complicado. 
Resolveremos num\'ericamente la din\'amica completa de una part\'icula en un
potencial pozo doble acoplada a un ba\~no t\'ermico. 
El potencial pozo doble resulta interesante, principalmente, por dos motivos.
Por un lado, es un modelo de ruptura de la simetr\'ia discreta, y por tanto,
nos permite estudiar el rol de las fluctuaciones de vac\'io en una transici\'on
de fase (efectos cu\'anticos en el punto cr\'itico). Por otro lado, permite estudiar
el rol de las fluctuaciones de vac\'io en la din\'amica
de superposiciones macrosc\'opicas de estados cu\'anticos, como por ejemplo
BECs \cite{Dounas}, los cuales ya han sido experimentalmente creados en 
un pozo doble \cite{Shin2}. En este contexto, en el Cap\'itulo \ref{c3}, veremos que
la din\'amica de este modelo est\'a dominada por tres procesos 
de naturaleza y escalas temporales distintas: p\'erdida de coherencia,
 efecto t\'unel y activaci\'on energ\'etica. Este cap\'itulo est\'a principalmente 
basado en
los trabajos publicados en \cite{dwPRE,dwJCS}.

Siguiendo nuestra motivaci\'on de estudiar el proceso de p\'erdida de
coherencia 
en distintas situaciones f\'isicas, en el Cap\'itulo \ref{c4} nos concentraremos
en el caso de un sistema  acoplado a un entorno compuesto. 
Es decir, en el modelo a estudiar
hay un subsistema $A$ acoplado a otro subsistema $B$, el cual a su vez
est\'a acoplado a un entorno compuesto por infinitos osciladores
arm\'onicos. En este caso, analizaremos la p\'erdida de coherencia en 
el subsistema $A$ debido a un entorno compuesto formado por  $B$ y 
el entorno de infinitos osciladores arm\'onicos. 
Veremos que, en todos los casos, ya sean $A$ y $B$ osciladores, u
 osciladores invertidos, hay p\'erdida de coherencia en el subsistema $A$. 
Sin embargo, lo m\'as notable es que la eficiencia de la difusi\'on depende
de los grados de libertad inestables de los subsistemas $A$ y $B$. Por lo tanto,
hacia el final del cap\'itulo, haremos un estudio minucioso de
 los grados de libertad  inestables presentes en el entorno ``efectivo''.
Este modelo ha tenido un gran impulso en el \'ultimo tiempo con
la aparici\'on de trabajos como \cite{ChouYuHu}, en donde 
utilizan 
este modelo, ligeramente modificado, 
para estudiar la transferencia de informaci\'on
entre los subsistemas con el entorno como mediador.
Este cap\'itulo est\'a basado en las 
Refs.\cite{compositePRA, compositeModern}.

En muchos casos, la interacci\'on con el entorno no puede ser
eliminada. Por ejemplo, para part\'{\i}culas cargadas o \'atomos
neutros con momento dipolar, la interacci\'on con el campo
electromagn\'etico resulta crucial. Esta interacci\'on induce una
reducci\'on en las franjas  que se observan en los
experimentos de interferencia. Por lo
tanto, las fluctuaciones de vac\'{\i}o del campo electromagn\'etico
han sido consideradas como factores ``decoherentes" en la Literatura. Los
experimentos de dos rendijas tambi\'en fueron estudiados en
presencia de conductores, los cuales cambian la estructura de
vac\'{\i}o y modifican as\'{\i} las predicciones acerca de la
decoherencia. En el Cap\'itulo \ref{c5}, basado en las Refs.
\cite{Casher, CasherJPA,fasesPRA, fringeModern,fringeJCS},
 veremos que en muchos
casos, la presencia del entorno se traduce en fases adicionales
en el sistema, que traen como consecuencia la clasicalizaci\'on
del mismo. 

Por \'ultimo, el estudio del proceso de p\'erdida de coherencia 
en Teor\'\i a Cu\'antica de 
Campos ha sido desarrollado recientemente y merece atenci\'on debido a su incumbencia 
en diferentes \'areas de la F\'\i sica actual. En 
particular, y gracias a resultados conocidos para la part\'\i cula Browniana 
cu\'antica, se extendi\'o el formalismo de la funcional de influencia de 
Feynman y Vernon para teor\'\i a de campos \cite{ferdiego}. 
El proceso de ``decoherencia'' ha sido 
estudiado para campos escalares en el espacio de Minkowski y para 
campos escalares acoplados a geometr\'\i as arbitrarias con el objeto de 
entender la transici\'on ``a lo cl\'asico'' de modelos de gravedad 
cu\'antica \cite{fer97, fer98,ferrusso,ferPhD}.
En estos casos, el estudio del mecanismo de p\'erdida de coherencia es visto como 
un primer paso hacia un entendimiento global del proceso de transici\'on 
cu\'antico-cl\'asica en teor\'\i a de campos.
El formalismo antes mencionado en teor\'\i a de campos fue aplicado al estudio 
del efecto de la decoherencia  
del par\'ametro de orden durante transiciones de fase continuas \cite{fermazziray,rayfermazzi-plb,fermazziray-npb}. 
Este an\'alisis implic\'o un estudio de la din\'amica del proceso de formaci\'on y evoluci\'on de los 
defectos topol\'ogicos que se generan durante la transici\'on de fase, ejemplo
que se discutir\'a en el cap\'itulo \ref{c6}, basado principalmente 
en la referencia \cite{PLB}.

Finalmente, en el \'ultimo cap\'itulo, desarrollamos nuestras conclusiones.
\newpage
\thispagestyle{empty}
\cleardoublepage


\chapter{Sistemas cu\'anticos abiertos}
\label{c1}
\markboth{Sistemas cu\'anticos abiertos}
{Cap\'itulo 1}

Los efectos de interferencia son la caracter\'istica m\'as notable
de la Mec\'anica Cu\'antica. El experimento de dos rendijas (o de 
Young) es generalmente usado como el punto de partida para 
hacer una descripci\'on cu\'antica de la Naturaleza. Por tanto, 
aqu\'i motivaremos este cap\'itulo con un paradigma de la transici\'on
cu\'antico-cl\'asica. Supongamos el siguiente experimento de 
interferencia: una fuente de luz F,
de un s\'olo color y tenue (un fot\'on a la vez), que ilumina un 
detector D. En el camino, es decir, entre la fuente y el detector, 
se coloca una pantalla con dos peque\~nos orificios, A y B, separados apenas
unos mil\'imetros de distancia. Si la fuente y el detector est\'an separados
100 cm, los orificios deben ser menores que 1 d\'ecimo de mil\'imetro.
Si cerramos el orificio B, obtenemos un cierto n\'umero de clicks en el
detector D, que representa el n\'umero de fotones que llegan pasando
por A (supongamos que el detector detecta 1 de cada 100 fotones que 
salen de F, es decir $1\%$). Si, por el contrario,  tapamos el orificio A,
y destapamos B, obtendremos aproximadamente el mismo n\'umero de
clicks.
Si destapamos ambos orificios, la respuesta es m\'as complicada.
Si \'estos est\'an separados cierta distancia, obtenemos m\'as clicks
que lo esperado: $4\%$ en vez de $2\%$ (este \'ultimo valor corresponde a 
la suma de las probabilidades cl\'asicas).
 Esto se debe a la
contribuci\'on de las interferencias cu\'anticas. Si, en cambio, est\'an separados
alguna otra distancia, no obtenemos ning\'un fot\'on en el detector.
Guiados por nuestra intuici\'on, hubi\'esemos esperado que la
cantidad de luz que llega al detector aumentara si ambos orificios
est\'an destapados. Pero no sucede as\'i; es decir, resulta incorrecto 
pensar que la luz ``va por un camino o el otro".

Ahora, supongamos una peque\~na variaci\'on a este sencillo
experimento de interferencia.  Le agregamos un detector a cada orificio, 
capaz de indicar cu\'ando un fot\'on pasa por \'el. De esta manera, podemos
saber por cu\'al orificio (A o B) pasa cada fot\'on cuando ambos orificios est\'an
destapados. Ya que la probabilidad de que un fot\'on viaje de F a D depende
de la distancia entre orificios, quiz\'as existe alguna forma extra\~na,  
para que ambos detectores, en A y B, suenen juntos de modo de indicar la contribuci\'on
de las interferencias. Sin embargo, lo que sucede es lo siguiente: los
detectores en A y B no suenan nunca juntos; es decir, suena A \'o suena B.
El fot\'on  no se divide en dos; es decir, elige un s\'olo camino. 
Ahora s\'i corresponde asumir que ``va por un camino o el otro". 
El experimento ha sido modificado
de manera no trivial y la cantidad de luz que llega a D es simplemente
la suma de las cantidades por separado, es decir $2\%$. Las interferencias
cu\'anticas desaparecen cuando colocamos los detectores en A y B, y
obtenemos un resultado cl\'asico.
As\'i es como funciona la Naturaleza: si colocamos instrumentos para determinar
el camino que recorre la luz, conocemos el orificio por el cual \'esta pas\'o, pero 
las interferencias cu\'anticas son destru\'idas
y el resultado obtenido es de naturaleza cl\'asica. Sin embargo, si no
los colocamos, no podemos saber cu\'al fue el camino recorrido, pero
el resultado es de naturaleza puramente cu\'antica, con las interferencias
jugando un rol importante en la cantidad de luz que alcanza al detector D.

Este experimento sencillo pone de manifiesto las diferentes evoluciones de
 los sistemas cu\'anticos cerrados y abiertos.
Cuando no colocamos detectores en los orificios, nuestro sistema cu\'antico
evoluciona libremente y de manera unitaria, aislado de cualquier influencia
del mundo exterior. El resultado es puramente cu\'antico y el principio
de superposici\'on tiene validez. Sin embargo,
cuando colocamos los detectores, nuestro sistema cu\'antico original est\'a
en interacci\'on con los detectores, y la evoluci\'on del sistema deja de ser
unitaria, ya que el entorno (los detectores en este caso) modifica
dicha evoluci\'on.  Las interferencias cu\'anticas son simplemente 
destru\'idas por la presencia del entorno.

El tipo de  experimentos de interferometr\'ia mencionado al comienzo, 
 tambi\'en se ha realizado con 
objetos materiales con el fin de comprobar la dualidad luz-materia 
que predice la Mec\'anica Cu\'antica. Esta propiedad 
ya ha sido demostrada para electrones, neutrones y \'atomos.
El avance de la tecnolog\'ia ha permitido incluso llegar m\'as lejos: se han
realizado experimentos de interferometr\'ia con sistemas mesosc\'opicos 
\cite{Facchi,Brezger}.
Estos objetos no son macrosc\'opicos ni microsc\'opicos. Generalmente,
son sistemas  formados por un n\'umero bastante grande de \'atomos, por  lo cual se
esperar\'ia que su comportamiento fuera cl\'asico. Sin embargo, 
 pueden ser descriptos por una funci\'on de onda e incluso, se han observado
interferencias cu\'anticas en experimentos con estas mol\'eculas \cite{Hornberger:2003}.
Los fulerenos
$C_{60}$ y $C_{70}$ son ejemplos conocidos de estos  sistemas.
En general, el problema que surge en interferometr\'ia con estas
mol\'eculas es que, a pesar que las franjas de interferencias est\'an y se
observan en la pantalla, muchas veces \'estas se ven atenuadas. Es decir, existe una 
p\'erdida de coherencia espacial que se ve reflejada en una falta de
contraste entre los m\'aximos y m\'inimos de interferencia del experimento.
De manera ilustrativa, una simulaci\'on de tal experimento se  presenta en la Fig.\ref{patrones}.
\begin{figure}[!ht]
\center
\includegraphics[width=14cm]{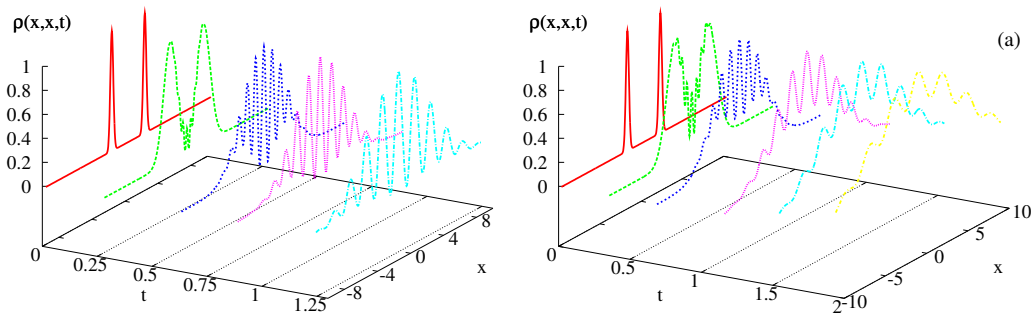}
\caption{Distintos patrones de interferencia para un experimento con
 part\'iculas. Las curvas demuestran el patr\'on para distintos tiempos.
La Figura de la izquierda, es el sistema cerrado. A la derecha, el sistema est\'a
abierto y las interferencias cu\'anticas se aten\'uan a los mismos tiempos de la 
evoluci\'on.}
 \label{patrones}
\end{figure}

En el gr\'afico de la izquierda, mostramos la evoluci\'on temporal del patr\'on de 
interferencia para un experimento con part\'iculas cuando el sistema est\'a
aislado. Inicialmente, tenemos  dos gaussianas que empiezan a evolucionar
y dan lugar a las interferencias cu\'anticas. Del lado derecho, el mismo sistema
cu\'antico  est\'a en interacci\'on con un entorno, y,
es esta interacci\'on la responsable de la atenuaci\'on o destrucci\'on de
 las franjas de interferencia. Es f\'acil ver que las franjas de 
interferencias se observan en la pantalla, pero la diferencia entre
los m\'aximos y m\'inimos no es tan grande como en el caso aislado. 
A medida
que transcurre el tiempo (o, equivalentemente, la pantalla est\'a m\'as lejos), 
el contraste entre las franjas es cada vez menor (en la Fig.\ref{patrones},
 la distancia es proporcional al tiempo). 
Resulta bastante intuitivo que
el patr\'on final depender\'a de la relaci\'on entre la intensidad del acoplamiento
sistema-entorno,
la separaci\'on de las rendijas y la distancia que 
las part\'iculas viajan hasta llegar a la 
pantalla \cite{fringeModern,fringeJCS}.

En general, esta ausencia de interferencia cu\'antica entre estados macrosc\'opicamente
distinguibles se puede explicar como consecuencia de un proceso de 
``p\'erdida de coherencia'' o ``decoherencia''.  Este proceso considera como aspecto 
fundamental que los sistemas cu\'anticos macrosc\'opicos no est\'an aislados. Es decir,
la ``clasicalidad'' de un sistema resulta  una propiedad emergente del mismo,
 debido a la interacci\'on con un entorno. Este proceso
de p\'erdida de coherencia es el factor principal en la transici\'on cu\'antico-cl\'asica.

En general, para que un sistema cu\'antico pueda ser considerado cl\'asico, deben satisfacerse
al menos dos condiciones. Por un lado, la funci\'on de onda debe predecir que las variables
can\'onicas est\'en fuertemente correlacionadas de acuerdo a las leyes cl\'asicas, o alguna
distribuci\'on constru\'ida a partir de ellas (como por ejemplo, la funci\'on de Wigner),
debe presentar un ``pico'' alrededor de una o un conjunto de configuraciones cl\'asicas. Por
otro lado, la interferencia entre las distintas configuraciones cl\'asicas debe ser despreciable,
de forma tal que sea posible predecir que un sistema est\'a en cierto estado definido,
entre los muchos estados posibles. Esto implica una ``p\'erdida de coherencia'', o bien
la destrucci\'on de los t\'erminos no diagonales de la matriz densidad, que representan las coherencias de dicha matriz.

 Por tanto, existe una necesidad de explicar te\'oricamente el efecto de la p\'erdida
de coherencia en la Naturaleza, tanto en estos ejemplos de interferometr\'ia como
en muchos otros sistemas f\'isicos relevantes donde se observa este proceso.
 Con nuestros ejemplos hemos mostrado los distintos 
comportamientos entre sistemas cerrados y abiertos. En lo que sigue, nos concentraremos 
en describir y cuantificar el proceso
de p\'erdida de coherencia en sistemas cu\'anticos abiertos. 
En particular, nos concentraremos en un caso  ampliamente
utilizado para modelar distintos sistemas f\'isicos. 
Con las herramientas te\'oricas que presentaremos
a continuci\'on, en un cap\'itulo posterior podremos tratar de darle una raz\'on
cuantitativa  al ejemplo de la Fig.\ref{patrones}.

\section{Sistemas cu\'anticos cerrados y abiertos}

A diferencia de lo que ocurre con los sistemas cu\'anticos cerrados, la
din\'amica de un sistema cu\'antico abierto no puede, en general, ser
descripta por una evoluci\'on unitaria (regida por la ecuaci\'on
de Schr\"{o}dinger). Frecuentemente, resulta mucho m\'as \'util
plantear la ecuaci\'on de movimiento del sistema cu\'antico abierto
en t\'erminos de su matriz densidad, es decir, escribir la ecuaci\'on
maestra cu\'antica del sistema. 

Comenzaremos por resumir brevemente las caracter\'{\i}sticas de la evoluci\'on 
din\'amica de sistemas cerrados y abiertos. Los procesos Markovianos cu\'anticos
son fundamentales (al menos a fines pr\'acticos) ya que representan el caso m\'as
sencillo para describir la din\'amica de estos sistemas no aislados. Este tipo
de proceso puede ser considerado una generalizaci\'on directa del concepto de
probabilidad cl\'asica de un semigrupo din\'amico de Mec\'anica Cl\'asica.
Es decir, an\'alogamente a la ecuaci\'on diferencial de Chapman-Kolmogorov
para la probabilidad cl\'asica, un semigrupo din\'amico cu\'antico da lugar,
a primer orden, a una ecuaci\'on diferencial para la matriz densidad reducida,
la cual resulta conocida con el nombre de ecuaci\'on maestra markoviana en
la forma de Lindblad.

De acuerdo con la Mec\'anica Cu\'antica, el vector estado $\vert \psi(t)\rangle$
evoluciona en el tiempo de acuerdo a la ecuaci\'on de Schr\"{o}dinger,
\begin{equation}
 i \frac{d}{dt} \vert \psi(t)\rangle = H(t) \vert \psi(t)\rangle,
\label{Schrodinger}
\end{equation}
donde $H(t)$ es el Hamiltoniano del sistema y la constante de Planck $\hbar$
ha sido fijada a uno. La soluci\'on de la ecuaci\'on de Schr\"{o}dinger
puede ser representada en t\'ermino de un operador de evoluci\'on unitario
$U(t,t_0)$, el cual transforma el estado $\vert \psi(t_0)\rangle$ a un tiempo
inicial $t_0$ en otro estado $\vert \psi(t)\rangle$ a un tiempo $t$,
\beq
\vert \psi(t)\rangle=U(t,t_0)\vert \psi(t_0)\rangle. \label{estadovector}
\eeq
Si reemplazamos la expresi\'on (\ref{estadovector}) en la ecuaci\'on 
(\ref{Schrodinger}),
obtenemos una ecuaci\'on para el operador evoluci\'on $U(t,t_0)$,
\beq
 i \frac{\partial}{\partial t} U(t,t_0)= H(t) U(t,t_0),
\label{evoU}
\eeq
 con la condici\'on inicial $U(t_0,t_0)=I$. Es f\'acil demostrar, con la
ayuda de las ecuaciones anteriores, que $U^{\dagger}(t,t_0)U(t_0,t)=
U(t_0,t)U^{\dagger}(t,t_0)\equiv I$ y, por lo tanto, $U(t_0,t)$ es un operador
unitario.

Para un sistema cu\'antico cerrado, el Hamiltoniano $H$ del sistema aislado
es \mbox{independiente} del tiempo y la ecuaci\'on (\ref{evoU}) es f\'acilmente
integrable en el tiempo obteni\'endose 
\begin{equation}
 U(t,t_0)= \exp(-iH(t-t_0)).
\end{equation}

Sin embargo, en distintas aplicaciones f\'{\i}sicas, uno se encuentra con
que el sistema en cuesti\'on, generalmente, es afectado por fuerzas externas;
por ejemplo, un campo electromagn\'etico externo. Si, en dicha situaci\'on, 
la din\'amica del sistema puede  ser formulada en t\'ermino de un
Hamiltoniano ``generador'' dependiente del tiempo $H(t)$, el sistema
ser\'a a\'un considerado cerrado.
Para un Hamiltoniano dependiente del tiempo, la ecuaci\'on (\ref{evoU}),
considerando la condici\'on inicial mencionada, puede ser representada
por una exponencial ordenada temporalmente,
\begin{equation}
 U(t,t_0)=T_{\leftarrow} \exp\big[ -i \int_{t_0}^{t} ds H(s) \big],
\end{equation}
 donde $T_{\leftarrow}$ denota el ordenamiento temporal, tal que el
argumento de los operadores aumenta temporalmente de derecha a izquierda
como indica la flecha.

Si, en cambio, el sistema en cuesti\'on se describe en funci\'on de  un estado mixto, el
correspondiente {\it ensemble} cu\'antico puede ser descripto
con la ayuda del operador estad\'istico $\rho$. Entonces, es f\'acil deducir
una ecuaci\'on de movimiento para la matriz densidad a partir de
la ecuaci\'on de Schr\"{o}dinger (\ref{Schrodinger}). Si, inicialmente,
el estado del sistema es 
\begin{equation}
 \rho(t_0)=\sum_{\alpha} \omega_{\alpha} \vert\psi_{\alpha}(t_0)\rangle
\langle\psi_{\alpha}(t_0)\vert,
\end{equation}
 donde $\omega_{\alpha}$ son los pesos estad\'{\i}sticos y 
$\vert\psi_{\alpha}(t_0)\rangle$ los estados vector normalizados  que evolucionan
de acuerdo a la ecuaci\'on  (\ref{estadovector}). El estado del sistema
a un tiempo posterior es,
\begin{equation}
 \rho(t)=\sum_{\alpha} \omega_{\alpha} U(t,t_0) \vert \psi_{\alpha}(t_0)\rangle
\langle\psi_{\alpha}(t_0)\vert U^{\dagger}(t,t_0),
\end{equation}
el cual puede ser escrito de manera m\'as sencilla como
\begin{equation}
 \rho(t)=U(t,t_0)\rho(t_0)U^{\dagger}(t,t_0).
\end{equation}

Si diferenciamos esta ecuaci\'on con respecto al tiempo, obtenemos una ecuaci\'on
de movimiento para la matriz densidad del sistema,
\begin{equation}
 \frac{d}{d t} \rho(t)=-i [H(t),\rho(t)], \label{vonN}
\end{equation}
la cual es conocida como ecuaci\'on de {\it von Neumann} o {\it 
Liouville- von Neumann}.

Para hacer la analog\'{\i}a de la ecuaci\'on (\ref{vonN}) con la correspondiente
ecuaci\'on para la probabilidad cl\'asica de Mec\'anica Estad\'istica,
la ecuaci\'on (\ref{vonN}) es generalmente escrita 
 \begin{equation}
 \frac{d}{d t} \rho(t)= {\cal L}(t) \rho(t),
\label{superLio}
\end{equation}
donde ${\cal L}(t)$ es el superoperador de Liouville que se define como
${\cal L}(t)\rho = -i [H(t),\rho]$.

\section{Din\'amica de los sistemas cu\'anticos abiertos}

Un sistema cu\'antico abierto es un sistema cu\'antico $S$ que est\'a acoplado a 
otro sistema  ${\cal E}$ llamado ``entorno''. Es por eso, que el sistema $S$
en realidad representa un {\it subsistema} del sistema total combinado $S+{\cal E}$,
generalmente considerado cerrado. Por consiguiente, un estado del sistema $S$
cambiar\'a como consecuencia de su din\'amica interna y de la interacci\'on con
su entorno. Esta interacci\'on genera ciertas correlaciones entre el sistema y
el entorno que har\'an que el estado de $S$ permanezca constante a partir de cierto
momento, y por lo tanto, $S$ responda a una evoluci\'on unitaria a partir de entonces.
La din\'amica del subsistema $S$ inducida por el Hamiltoniano del sistema total
es generalmente conocida como la {\it din\'amica del sistema reducido}, y $S$
es llamado el {\it sistema reducido}. 

Llamemos ${\cal H}_S$ al espacio de Hilbert del sistema $S$ y ${\cal H}_{\cal E}$
al espacio de Hilbert del entorno ${\cal E}$. El espacio de Hilbert del sistema
total $S+{\cal E}$ est\'a generado por el producto tensorial ${\cal H}={\cal H}_S
\otimes {\cal H}_{\cal E}$. El Hamiltoniano total del sistema se puede escribir
seg\'un
\begin{equation}
 H(t)=H_S \otimes I_{\cal E} + I_S \otimes H_{\cal E} + \hat{H}_I (t),
\end{equation}

donde $H_S$ es el Hamiltoniano del sistema $S$, $H_{\cal E}$ es el Hamiltoniano
libre del entorno ${\cal E}$ y $\hat{H}_I (t)$ describe la interacci\'on entre el
sistema y el entorno. Un diagrama esquem\'atico de una situaci\'on t\'ipica
se muestra en la Fig.\ref{figure1-1}.

\begin{figure}[ht]
\begin{picture}(400,150)(0,0)
\put(180,0){\resizebox{45mm}{!}{\includegraphics{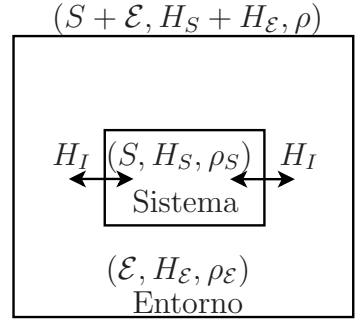}}}
\put(215,15){(${\cal E},H_{\cal E},\rho_{\cal E}$)}
\put(225,2){Entorno}
\put(215,58){($S, H_S,\rho_S$)}
\put(195,58){$H_I$} \put(280,58){$H_I$} 
\put(225,40){Sistema}
\put(195,110){($S+{\cal E},H_S+H_{\cal E},\rho$)}
\end{picture}
\caption{Esquema ilustrativo de un sistema cu\'antico abierto.}
\label{figure1-1}
\end{figure}

La motivaci\'on principal para estudiar los sistemas abiertos es que en
muchas situaciones f\'isicas importantes un modelo matem\'atico
completo  de la din\'amica del sistema (combinado con el entorno)
es muy complicado. El entorno puede estar representado por muchos grados
de libertad, con lo cual un tratamiento exacto del problema requiere resolver
un sistema de infinitas ecuaciones de movimiento acopladas. A\'un en el caso
en que la soluci\'on exista, uno se encuentra con la tediosa tarea de
aislar y determinar las cantidades f\'isicas importantes del sistema
a trav\'es de un promedio sobre todas las dem\'as cantidades y grados
de libertad irrelevantes. Por lo tanto, de aqu\'i en adelante,
utilizaremos una descripci\'on m\'as sencilla en t\'ermino de un espacio de estados
reducido, formado \'unicamente por un conjunto de variables f\'isicas relevantes
obtenido a partir del uso de m\'etodos anal\'iticos y de aproximaci\'on.
La matriz densidad reducida del subsistema $S$ a un tiempo $t$ 
se obtiene a partir de la matriz densidad del sistema total $\rho(t)$
haciendo  la traza parcial sobre los grados de 
libertad del entorno. La matriz densidad total implica un
comportamiento unitario, por lo cual podemos escribir: 
\begin{equation}
 \rho_S(t)={\rm tr}_{\cal E} \rho(t)={\rm tr}_{\cal E}\{U(t,t_0)\rho(t_0)
U^{\dagger}(t,t_0)\}. \label{rhos}
\end{equation}
De forma an\'aloga a lo anterior, se puede deducir una ecuaci\'on de 
movimiento para la matriz densidad reducida a partir de la ecuaci\'on
de Liouville-Von Neumann para el sistema total,
\begin{equation}
 \frac{d}{d t}\rho_S(t)=-i {\rm tr}_{\cal E} [H(t), \rho(t)].\label{ecrhos}
\end{equation}

En lo que resta de esta Tesis, veremos que 
la matriz densidad reducida del sistema, $\rho_S$, ser\'a de fundamental
inter\'es en el estudio de ls din\'amica de los sistemas cu\'anticos
abiertos.  Estudiaremos distintos modelos f\'isicos y, en todos ellos, escribiremos
la ecuaci\'on maestra cu\'antica para la matriz densidad reducida del sistema
en consideraci\'on. La ecuaci\'on maestra se obtendr\'a bajo distintas aproximaciones,
seg\'un corresponda.

 Por ejemplo, para un proceso de Markov,  uno puede asumir que 
las correlaciones temporales en 
las variables din\'amicas del entorno son cortas. Eso permite suponer
que los efectos de memoria son despreciables y formular la din\'amica
del sistema en funci\'on de un semigrupo din\'amico cu\'antico.
Dado este semigrupo y bajo ciertas condiciones matem\'aticas estrictas,
existe un mapa lineal ${\cal L}$ que resultar\'a ser el generador del semigrupo
y permite representar al mismo de forma exponencial  \cite{Breuer},
\begin{equation}
 V(t)=\exp({\cal L} t).
\end{equation}
Esta representaci\'on da lugar inmediatamente a una ecuaci\'on diferencial
de primer orden para la matriz densidad reducida,
\begin{equation}
 \frac{d}{d t}\rho_S(t)={\cal L} \rho_S(t),
\end{equation}
la cual es conocida por ser la ecuaci\'on maestra Markoviana cu\'antica.
El generador ${\cal L}$ del semigrupo representa a un super operador
y puede ser considerado una generalizaci\'on del super operador de Liouville
que mencionamos en la ecuaci\'on (\ref{superLio}), relacionado con el
Hamiltoniano del sistema.  En un caso as\'i, para conocer la din\'amica
del sistema reducido, es necesario encontrar la forma expl\'icita de dicho
generador ${\cal L}$.

\section{Movimiento Browniano Cu\'antico}
\label{QBM}

La relajaci\'on t\'ermica de sistemas en interacci\'on con entornos 
ha sido un tema de gran inter\'es en Mec\'anica Estad\'istica por 
mucho tiempo. En particular, es importante notar que la propia
descripci\'on de esta relajaci\'on debe ser analizada por medio de la 
Mec\'anica Cu\'antica \cite{Lindenberg}. La situaci\'on m\'as simple 
que podemos considerar en este contexto es aquella del 
movimiento Browniano de un oscilador arm\'onico cu\'antico en un 
 entorno de la misma naturaleza. Los modelos de Movimiento 
Browniano Cu\'antico (MBC) proveen un ejemplo t\'ipico de sistemas 
cu\'anticos abiertos, y han sido muy utilizados 
para el entendimiento de la teor\'ia de medici\'on en Mec\'anica 
Cu\'antica \cite{Zurek1}, \'Optica Cu\'antica \cite{Carmichel} y p\'erdida de coherencia
\cite{UnruhZurek}, por
citar s\'olo algunos de los intereses que presentan estos modelos.
 El objeto central de estudio es la
ecuaci\'on maestra para la matriz densidad reducida de la part\'icula 
Browniana, que se obtiene luego
de integrar los grados de libertad correspondientes al entorno. 
Una gran cantidad de trabajos en esta esta direcci\'on han sido 
realizados en el pasado \cite{GraIng,Lindenberg,Caldeira,Hakim,Haake}. 
La derivaci\'on m\'as general de
esta ecuaci\'on maestra es la efectuada por 
B.L. Hu, J.P. Paz y Y. Zhang \cite{HuPazZhangI}, 
donde utilizaron la
funcional de influencia de Feynman y Vernon \cite{Feynman} provista por t\'ecnicas 
de integrales de camino.

En el caso del MBC la din\'amica del sistema $S$  es 
mucho m\'as lenta que los tiempos de correlaci\'on del sistema. 
Bajo estas circunstancias, otro tipo de ecuaci\'on maestra
rige la evoluci\'on de la $\rho_S$ y su obtenci\'on es distinta
a la mencionada anteriormente.
Para obtener la ecuaci\'on maestra del MBC,
se presentar\'a y discutir\'a  el modelo de Caldeira-Legget \cite{Caldeira},
el cual es el prototipo del MBC y permite analizar los fen\'omenos
de disipaci\'on inducidos por el entorno. Para el caso
del l\'imite de  temperatura alta del entorno, este modelo da una
ecuaci\'on maestra markoviana (aunque no es de la forma de Lindblad).
Sin embargo, para  temperatura 
estrictamente cero, o en el caso de acoplamientos fuertes, la
matriz densidad reducida del sistema exhibe un comportamiento 
no markoviano y, entonces, se debe recurrir a distintas t\'ecnicas para
describir la din\'amica del sistema reducido. 

En esta secci\'on
presentaremos el modelo, discutiremos el Teorema de Fluctuaci\'on
Disipaci\'on y el m\'etodo de la funcional de influencia
de Feynman-Vernon. En particular, \mbox{nuestro} objetivo fundamental es introducir 
el formalismo que, luego, aplicaremos a distintos entornos y situaciones f\'isicas.
La ventaja de este modelo es que la ecuaci\'on maestra puede ser
 obtenida en forma exacta para todo tipo de entornos a 
considerarse \cite{HuPazZhangI,Paz}. Incluso acoplamientos m\'as generales 
$(x_n^k q_n)$ han sido  tambi\'en considerados en la Literatura, donde la ecuaci\'on
 maestra correspondiente a dichos acoplamientos ha sido obtenida perturbativamente
en \cite{HuPazZhangII}.

El modelo describe a una part\'icula Browniana de masa $m$ y 
coordenadas $x$ que se mueve bajo la acci\'on de un potencial
$V(x)$. El Hamiltoniano libre de la part\'icula es 
\begin{equation}
H_S=\frac{1}{2 m} p^2 + V(x),
\end{equation}
donde $p$ es el momento de la part\'icula. La part\'icula, adem\'as,
est\'a acoplada a un entorno, el cual consiste en un gran n\'umero 
de osciladores arm\'onicos de masas $m_n$ y frecuencias $\omega_n$,
descripto por el Hamiltoniano
\begin{equation}
 H_{\cal E}=\sum_n \hbar \omega_n \bigg(b_n^{\dagger} b_n +\frac{1}{2}\bigg)=
\sum_n \bigg( \frac{1}{2 m_n} p_n^2 + \frac{1}{2} m_n \omega_n^2 q_n^2
\bigg).
\end{equation}

En esta expresi\'on $b_n$ y $b_n^{\dagger}$ denotan los operadores
de creaci\'on y aniquilaci\'on de los modos del entorno, 
respectivamente, mientras que $q_n$ y $p_n$ son las correspondientes
coordenadas y momentos can\'onicos conjugados.
En el modelo considerado, la coordenada $x$ de la part\'icula Browniana
est\'a acoplada linealmente con la coordenada $q_n$ de los osciladores
del ba\~no. El Hamiltoniano de Interacci\'on $H_I$ est\'a definido por
\begin{equation}
 H_I=-x \sum_{n} \kappa_n q_n \equiv -x B,
\end{equation}
donde el operador del entorno es $B=\sum_n \kappa_n q_n =
\sum_n \kappa_n \sqrt{\frac{\hbar}{2 m_n \omega_n}}(b_n 
+b_n^{\dagger})$,  una suma pesada sobre las coordenadas
$q_n$ de los modos del ba\~no y las constantes de 
acoplamiento $\kappa_n$.

En el modelo MBC, la evoluci\'on 
del sistema combinado (sistema-entorno), 
puede \mbox{carac-} terizarse por cuatro escalas de tiempo 
diferentes: la primera est\'a asociada a
la frecuencia natural de la part\'icula aislada; 
la segunda est\'a representada por el tiempo de relajaci\'on
 (caracterizado por el acoplamiento entre la part\'icula y 
el entorno); la tercera corresponde
al ``tiempo de memoria" del entorno 
(en general asociado a la frecuencia m\'as alta presente en el
entorno) y, finalmente, la escala de tiempo asociada con la 
temperatura del entorno, que mide la
importancia relativa entre los efectos cu\'anticos y t\'ermicos.

 El efecto del entorno sobre la din\'amica del sistema est\'a 
caracterizado por los fen\'omenos de
fluctuaci\'on y disipaci\'on. Estos efectos pueden 
determinarse por una propiedad totalmente espec\'ifica del entorno: 
la densidad espectral $I(\omega)$. 
Esta densidad representa el n\'umero de osciladores  presentes
en el entorno con una  frecuencia dada y  que se acoplan al sistema
con una intensidad espec\'ifica. En el caso discreto de $N$
osciladores, esta funci\'on es 
\begin{equation}
 I(\omega)=\sum_n \delta(\omega-\omega_n) \frac{\kappa_n^2}
{2 m_n \omega_n}.
\end{equation}
De este modo, indicando la densidad $I(\omega)$ y el estado inicial 
del entorno, tanto la disipaci\'on como las
fluctuaciones quedan un\'ivocamente determinadas, 
como podremos ver a \mbox{continuaci\'on}.
Diferentes densidades espectrales $I(\omega)$ clasifican a los distintos
tipos de entornos; mientras que la constante de acoplamiento se
ajusta a la frecuencia del entorno como, por ejemplo,
$\kappa_n = m_n \omega_n^{\alpha}$ para cada modelo de entorno
\cite{GraIng}. Por razones f\'isicas, uno no espera que 
un entorno real contenga un n\'umero infinito de frecuencias,
y en general, se introduce una escala arbitraria, que llamaremos 
frecuencia de corte, que anule
la densidad espectral para aquellas frecuencias mayores 
que esta frecuencia de corte $\Lambda$; es
decir $I(\omega) \rightarrow 0$ cuando $\omega  > \Lambda$. 
Por  tanto, la escala temporal asociada a la memoria del
entorno, queda 
determinada por la inversa de esta frecuecia
de corte. El entorno se conoce usualmente como \'ohmico 
\cite{Caldeira} si la densidad espectral es tal que
 $I(\omega) \approx \omega$  ($\omega < \Lambda$); 
supra\'ohmico si $I(\omega) \approx \omega^{\alpha}$  ($\alpha >1$)
 o sub\'ohmico si $\alpha < 1$. Es sencillo ver que
el caso \'ohmico (en general el m\'as estudiado) 
corresponde a la situaci\'on f\'isica en la que el
entorno induce sobre el sistema una fuerza lineal con la velocidad.
En lo que sigue,  utilizaremos la 
siguiente expresi\'on para la densidad espectral 
\cite{HuPazZhangI}
\begin{equation}
I(\omega)= \frac{2}{\pi} M \gamma_0 \omega \bigg(\frac{\omega}
{\Lambda} \bigg)^{\alpha-1} e^{-\frac{\omega^2}{\Lambda^2}}
\end{equation}
donde con $\gamma_0$ representamos la constante de 
relajaci\'on del entorno que corresponde a la frecuencia
asociada al acoplamiento entre este \'ultimo y el sistema 
\cite{Caldeira}.

\subsection{Funcional de influencia de Feynman y Vernon}
\label{FVcap1}

El objetivo central de esta secci\'on es desarrollar
 un formalismo general que nos permita
encontrar todos los efectos  cu\'anticos inducidos por  un entorno 
sobre nuestro sistema de inter\'es. En este contexto,
 nos limitaremos a reproducir
la derivaci\'on de la ecuaci\'on maestra exacta para un entorno 
general  en funci\'on de una  representaci\'on
funcional del operador de evoluci\'on de la matriz 
densidad reducida, definida seg\'un
\begin{equation}
 \rho_S(x,x')=\int_{-\infty}^{\infty} dq_n 
\int_{-\infty}^{\infty} dq'_n ~\rho(x,q_n|x',q'_n)
\delta(q_n-q'_n).
\end{equation}

Para construir la representaci\'on de la integral de camino,
utilizamos  la representaci\'on de Schr\"{o}dinger
e introducimos el propagador $J$ definido por la siguiente
relaci\'on
\begin{equation}
 \rho_S(x_f,x'_f,t_f)= \int dx_i
\int dx'_i J(x_f,x'_f,t_f;x_i,x'_i,t_i) \rho_S(x_i,x'_i,t_i).
\end{equation}
Es f\'acil notar que el propagador es simplemente la funci\'on
de Green para la matriz densidad reducida en la representaci\'on
de posici\'on. Adem\'as, se puede ver que, en el caso
en que no hay acoplamiento entre el sistema y el entorno,
el propagador se reduce al producto de la funci\'on de Green
para la ecuaci\'on de Schr\"{o}dinger con Hamiltoniano $H_S$,
\begin{eqnarray}
 J(x_f,x'_f,t_f;x_i,x'_i,t_i)&=&\langle x_f t_f|x_i t_i \rangle
\langle x'_i t_i|x'_f t_f \rangle \nonumber \\
&=& G_S(x_f,t_f;x_i,t_i)G_S^*(x'_f,t_f;x'_i,t_i).
\end{eqnarray}

La  funcional
de influencia se obtiene escribiendo la funci\'on $G_S$ en t\'ermino
de las integrales de camino
\begin{equation}
 G_S(x_f,t_f;x_i,t_i)= \langle x_f |\exp(-i H_S (t_f-t_i)/\hbar)|
x_i \rangle = \int {\cal D}x~\exp\bigg(\frac{i}{\hbar} S_0 [x]
\bigg),
\end{equation}
donde 
\begin{equation}
 S_0[x]=\int_{t_i}^{t_f} dt~\bigg(\frac{1}{2} m \dot{x}^2 - V(x)
\bigg)
\end{equation}
es la acci\'on funcional cl\'asica de la part\'icula Browniana
libre. La integral de camino es una integral sobre todos los 
posibles caminos $x(t)$ que cumplen las condiciones
de contorno $x(t_i)=x_i$, y $x(t_f)=x_f$.

Considerando como condici\'on inicial que el sistema y el entorno
no est\'an correlacionados \cite{HuPazZhangI}, es decir $\hat{\rho}(t_0)=
\hat{\rho}_S(t_0) \otimes \hat{\rho}_{\cal E}(t_0)$, el operador de
evoluci\'on toma la forma
\begin{eqnarray}
 J_r(x_f,x'_f,t;x_i,x'_i,t_0)&=& \int_{x_i}^{x_f} {\cal D}x
\int_{x'_i}^{x'_f} {\cal D}x' \exp\bigg(\frac{i}{\hbar}(S[x]
- S[x'])\bigg) F[x,x'] \nonumber \\
&=& \int_{x_i}^{x_f} {\cal D}x
\int_{x'_i}^{x'_f} {\cal D}x'\exp\bigg(\frac{i}{\hbar} {\cal A}
[x,x']\bigg). \label{propagador}
\end{eqnarray}
${\cal A}[x,x']$ es la llamada ``acci\'on efectiva'' para el
sistema cu\'antico abierto y $F[x,x']$ es la {\it funcional de
influencia} de Feynman y Vernon, definida explic\'itamente como
\cite{Feynman}
\begin{eqnarray}
 F[x,x']&=& \int_{-\infty}^{\infty} dq_f \int_{-\infty}^{\infty} dq_i
 \int_{-\infty}^{\infty} dq'_i \int_{q_i}^{q_f} {\cal D}q 
\int_{q'_i}^{q_f} {\cal D}q'\exp\bigg(\frac{i}{\hbar}(S_{\cal E}[q]
+ S_{\rm int}[x,q])\bigg) \nonumber \\ 
&\times& \exp\bigg(\frac{i}{\hbar}(S_{\cal E}[q']
+ S_{\rm int}[x',q'])\bigg) \rho_B(q_i,q'_i,t_0) \nonumber \\ 
&=& \exp\bigg(\frac{i}{\hbar} {\cal A}
[x,x']\bigg),
\end{eqnarray}
donde llamamos $\delta {\cal A}[x,x']$ a la {\it acci\'on de 
influencia}. En consecuencia, la acci\'on efectiva para el 
sistema cu\'antico abierto es ${\cal A}[x,x']=S[x]-S[x']
+ \delta {\cal A} [x,x']$.

Gracias a la condici\'on inicial en la que no existen 
correlaciones, la funcional de \mbox{influencia} s\'olo
depende del estado inicial del entorno. En este ejemplo 
mostramos la funcional $F[x,x']$ para un
entorno que inicialmente est\'a en equilibrio 
termodin\'amico a temperatura $T=(\beta^{-1})$\footnote{ Otras condiciones 
m\'as generales acerca de la condici\'on de equilibrio 
entre el sistema y el entorno pueden encontrarse en
la Ref.\cite{Hakim}; acerca de condiciones iniciales con 
correlaciones se pueden consultar las Refs.\cite{GraIng,jppdavila}.}.
Para las presentes condiciones iniciales, la funcional de influencia 
se calcula  exactamente \cite{Feynman,HuPazZhangI,Caldeira}.

El resultado es
\begin{eqnarray}
 F[x,x']&=& \exp \bigg[ -\frac{i}{\hbar} \int_0^t ds_1 
\int_0^{s_1} ds_2 [x(s_1)-x'(s_1)] \eta(s_1-s_2)
[x(s_2)+x'(s_2)] \nonumber \\
&-& \int_0^t ds_1 
\int_0^{s_1} ds_2 [x(s_1)-x'(s_1)] \nu(s_1-s_2)
[x(s_2)-x'(s_2)] \bigg].
\label{Ffinal}
\end{eqnarray}

Los n\'ucleos $\nu$ y $\eta$, llamados de  ruido y disipaci\'on, 
respectivamente, son en general no-locales en el tiempo
y est\'an definidos como 
\begin{eqnarray}
 \nu (s)&=& \int_0^{\infty} d\omega I(\omega)
\coth(\frac{\beta \hbar \omega}{2}) \cos (\omega s),  \label{nucleonu}\\
\eta (s) &=& \frac{d}{d s} \gamma(s), \label{nucleoeta}
\end{eqnarray}
donde $\gamma(s)= \int_0^{\infty} d\omega \frac{I(\omega)}
{\omega} \cos(\omega s)$. Dada la expresi\'on de la funcional 
de influencia en t\'erminos
de los n\'ucleos $\nu$ y $\eta$, podemos escribir la acci\'on de
influencia como
\begin{eqnarray}
 \delta {\cal A}[x,x'] &=& -2 \int_0^{t} ds_1 \int_0^{s_1} ds_2
\Delta(s_1) \eta(s_1-s_2) \Sigma (s_1) \nonumber \\
&+& i \int_0^{t} ds_1 \int_0^{s_1} ds_2 \Delta(s_2)
\nu(s_1-s_2) \Sigma(s_2),
\end{eqnarray}
donde se ha introducido un cambio de variables $\Delta=x-x'$
y $\Sigma=1/2(x+x')$.

Las partes real e imaginaria de la acci\'on efectiva ${\cal A}
[x,x']$ pueden interpretarse como responsables de la disipaci\'on
y el ruido. Los n\'ucleos de ruido y disipaci\'on est\'an 
siempre relacionados por una ecuaci\'on integral conocida
como la relaci\'on de fluctuaci\'on-disipaci\'on \footnote{El teorema 
de fluctuaci\'on-disipaci\'on vale para
un sistema general en equilibrio t\'ermico y un observable $z(t)$
en la representaci\'on de Heisenberg.
La funci\'on $\chi$ es la respuesta del sistema 
ante una fuerza externa $F(t)$ que se aplica para $t>0$ y 
se representa por la
perturbaci\'on temporal de la forma $V(t)=-z F(t)$ en 
el Hamiltoniano del sistema total. 
La transformada de Fourier de la funci\'on de respuesta del
sistema es 
\begin{equation}
 \tilde{\chi}(\omega)=\int_{-\infty}^{\infty} d\tau
e^{i \omega \tau} \chi(\tau)=\int_{0}^{\infty} d\tau
e^{i \omega \tau} \chi(\tau) \equiv \tilde{\chi'(\omega)}
+ i \tilde{\chi''(\omega)},
\end{equation}
El teorema de fluctuaci\'on-disipaci\'on relaciona la respuesta
lineal del sistema a una fuerza externa y las 
fluctuaciones de equilibrio \cite{Breuer}.}. Para el caso 
del MBC, 
esta relaci\'on puede escribirse como
\begin{equation}
 \upsilon(s)=\int_{-\infty}^{\infty} ds' K(s-s')\gamma(s'),
\end{equation}
donde $K(s)$ est\'a definido por 
\begin{equation}
 K(s)=\int_{0}^{\infty} d\omega \frac{\omega}{\pi}
\coth(\frac{\beta \hbar \omega}{2}) \cos(\omega s).
\end{equation}

En el l\'imite de temperatura alta, el n\'ucleo $K$ es proporcional
a una funci\'on delta de Dirac $K(s)=2 k_B T \delta(s)$ y 
la relaci\'on de fluctuaci\'on-disipaci\'on no es m\'as que la
relaci\'on de Einstein usual. 

La ecuaci\'on de movimiento para la part\'icula Browniana puede
deducirse a partir de la acci\'on efectiva ${\cal A}[x,x']$ mediante
el c\'alculo $\frac{\partial {\cal A}[x,x']}{\partial x} \vert_{x=x'}=0$.
Si efectuamos esta variaci\'on, el n\'ucleo de disipaci\'on $\eta$
aparecer\'a en la ecuaci\'on de movimiento  de manera expl\'icita;
mientras que dicha ecuaci\'on no contendr\'a contribuciones provenientes
del n\'ucleo de ruido $\nu$, debido a la forma funcional de la parte 
imaginaria de la acci\'on efectiva ${\cal A}[x,x']$. Por otro lado, podemos
notar en la Ec.(\ref{Ffinal}), por ejemplo, 
que dicha contribuci\'on imaginaria act\'ua
``pesando" las distintas trayectorias que contribuyen a la integral
funcional. De este modo, la
acci\'on efectiva ${\cal A}[x,x']$ puede identificarse como la resultante de
un promedio efectuado sobre las distintas realizaciones de una 
``fuerza estoc\'astica" \cite{Caldeira}.

Para hacer evidente 
esto \'ultimo, utilizamos la siguiente identidad matem\'atica:
\begin{eqnarray}
F[y(t)] &\equiv& \exp \bigg\{ -\frac{i}{\hbar} \int_0^t dt \int_0^s ds_1 
y(s) \nu(s-s_1) y(s_1) \bigg\} \nonumber \\
& =& \int {\cal D} \xi(t) P[\xi(t)] \exp \bigg\{ -\frac{i}{\hbar} \int_0^s ds
y(s)\xi(s) \bigg\}
\end{eqnarray}
con
\begin{equation}
P[\xi(t)]=N_{\xi} \exp \bigg\{ -\frac{1}{2} \int_0^t  ds_1 \int_0^{s_1}
 ds_2 \xi(s_1) \nu(s_1-s_2)^{-1} \xi(s_2) \bigg\},
\end{equation}
donde $N_{\xi}$ es una constante de normalizaci\'on. Con todo esto, la parte 
 imaginaria de la acci\'on efectiva puede escribirse en funci\'on de la
fuente de ruido estoc\'astico como
\begin{equation}
\exp \bigg\{ \frac{i}{\hbar} {\cal A}[x,x'] \bigg\}= \int {\cal D} \xi(t) P[\xi(t)]
\exp \bigg\{ \frac{i}{\hbar} S_{\rm ef} [x,x',\xi ]\bigg\},
\end{equation}
con 
\begin{equation}
S_{\rm ef} [x,x',\xi ] = S[x]-S[x'] + {\rm Re} \{ \delta {\cal A} [x,x']\} -\int_0^t
ds ~y(s) \xi(s),
\end{equation}
donde $\xi(s)$ puede interpretarse como una fuerza estoc\'astica de ruido
y definida por la probabilidad gaussiana $P[\xi(t)]$.

Las propiedades del ruido pueden ser obtenidas directamente
a partir de la funcional $F[y(t)]$, 
\begin{eqnarray}
\langle \xi(t) \rangle_P &=& -\frac{\hbar}{i} \frac{\delta F[y]}{\delta y(t)} \vert_{y=0}
=0, \nonumber \\
\langle \xi(t_1) \xi(t_2) \rangle_P &=& \bigg(-\frac{\hbar}{i}\bigg) \frac{\delta F[y]}{\delta y(t_1) \delta y(y_2)} \vert_{y=0}
= \hbar \nu (t_1-t_2),
\end{eqnarray}
donde $\langle \xi(t_1) \xi(t_2) \rangle_P $ es la funci\'on de correlaci\'on
del entorno. Con todo esto,   podemos
obtener una  ecuaci\'on de movimiento expl\'icitando 
la presencia de un t\'ermino de ruido.  
Esta es la ecuaci\'on para la part\'icula browniana
asociada de Langevin, donde la din\'amica de la part\'icula cl\'asica
est\'a afectada por la presencia de una fuente de ruido 
\cite{Feynman,GraIng,UnruhZurek,Caldeira}:
\begin{equation}
\ddot x(t) + \Omega^2 x(t) + 2 \int_0^t ds \eta(s-t) x(s)=\xi(t).
\end{equation}
En esta ecuaci\'on se pone de manifiesto que las trayectorias 
cl\'asicas de la part\'icula \mbox{browniana} adquieren, debido al entorno,
un comportamiento estoc\'astico, conocido como efecto de ruido.
Adem\'as, si uno considera la contribuci\'on del n\'ucleo $\eta$, aparecen
efectos disipativos y de renormalizaci\'on.

\subsection{Ecuaci\'on maestra}

En general, por consideraciones f\'isicas, el conocimiento 
de la ecuaci\'on maestra para la evoluci\'on
de la matriz densidad reducida es m\'as \'util que la 
evoluci\'on exacta de la matriz densidad total
misma. En realidad, a partir de la ecuaci\'on maestra podemos 
extraer muchos aspectos cualitativos
acerca del comportamiento del sistema, los cuales 
son independientes de las condiciones iniciales.

 El modelo presentado en este cap\'itulo tiene un 
acoplamiento lineal y puede resolverse exactamente. 
Las integrales de camino de la Ec.(\ref{propagador}) 
pueden evaluarse debido a que son integrales
gaussianas \cite{HuPazZhangI,Paz}. 
Por este motivo, la deducci\'on 
de la ecuaci\'on maestra est\'a basada en la 
representaci\'on  integral funcional para el 
operador de evoluci\'on de la matriz densidad reducida
(Ec.(\ref{propagador})) \cite{leshouches}. 
La no-localidad de los n\'ucleos presentes 
en la funcional de influencia es la \'unica complicaci\'on; 
por  tanto, obtener formalmente la ecuaci\'on maestra 
es conceptualmente equivalente
a derivar la ecuaci\'on de Schr\"{o}dinger a partir de la 
representaci\'on funcional del propagador en
Mec\'anica Cu\'antica.

Los resultados basados en el procedimiento funcional 
seguido por B.L. Hu, J.P. Paz y Y.
Zhang en la Ref.\cite{HuPazZhangI} pueden obtenerse 
de  manera extremadamente sencilla, a segundo orden en la
constante de acoplamiento,  mediante
 un m\'etodo perturbativo.

Para hallar la ecuaci\'on maestra debemos evaluar la derivada
temporal del operador de evoluci\'on reducido. Como la funcional
de influencia es no local, no podemos calcular esta derivada simplemente
expandiendo en $dt$ al propagador $J_r(t+dt,t)$ y rest\'andole $J_r(t,t)$;
ya que el propagador $J_r(t+dt,t)$ depende del estado del sistema
a tiempo $t$. Esto \'unicamente podr\'ia hacerse en el caso de temperatura
muy alta porque en ese l\'imite la funcional de influencia se vuelve local
en el tiempo \cite{Caldeira}.

En la representaci\'on de interacci\'on, la matriz densidad total
evoluciona seg\'un la ecuaci\'on
\begin{equation}
i \hbar \dot{\rho} = [V(t),\rho],
\label{ecm1}
\end{equation}
donde el potencial de interacci\'on $V(t)$, en la representaci\'on de interacci\'on
es, simplemente,
\begin{equation}
V(t)= \exp \bigg[ \frac{i}{\hbar} (H_S + H_{\cal E})t \bigg] V \exp
 \bigg[- \frac{i}{\hbar} (H_S + H_{\cal E})t \bigg],
\end{equation}
y la matriz densidad $\rho$,
\begin{equation}
\rho(t)= \exp \bigg[ \frac{i}{\hbar} (H_S + H_{\cal E})t  \bigg] \rho 
\exp \bigg[- \frac{i}{\hbar} (H_S + H_{\cal E})t \bigg].
\end{equation}

La soluci\'on a la Ec.(\ref{ecm1}) puede ser obtenida de manera perturbativa
utilizando la serie de Dyson:
\begin{equation}
\rho(t)= \sum_{n \geq 0} \int_0^{t} dt_1 \int_0^{t_1} dt_2 ... \int_0^{t_n}
\bigg( \frac{1}{i \hbar}\bigg)^n [V(t_1),[V(t_2),[...,[V(t_n),\rho(0)]...]].
\end{equation}
A partir de esta serie, podemos obtener una ecuaci\'on perturbativa,
a segundo orden en la constante de acoplamiento con el entorno, para 
la matriz densidad reducida $\rho_r={\rm Tr}_{\cal E} \rho$, 
en la representaci\'on de interacci\'on
\begin{eqnarray}
\rho_r(t) &\approx& \rho_r(0) + \frac{1}{i \hbar} \int_0^t dt_1 {\rm Tr}_{\cal E}
([V(t_1),\rho(0)])  \nonumber \\
&-& \frac{1}{\hbar^2}  \int_0^t dt_1 \int_0^{t_1} dt_2
 {\rm Tr}_{\cal E} ([V(t_1),[V(t),\rho(0)]]).
\label{ecpert}
\end{eqnarray}

Para obtener la ecuaci\'on maestra, debemos hacer la derivada
temporal de  esta ecuaci\'on, de modo de obtener
\begin{equation}
\dot{\rho}_r= \frac{1}{i \hbar} \rm{Tr}_{\cal E} [V(t),\rho(0)] -
\frac{1}{\hbar^2} \int_0^t dt_1 {\rm Tr}_{\cal E} [V(t), [V(t_1),\rho(0)]].
\end{equation}
Si  asumimos que en el instante inicial, el sistema
y el entorno no est\'an correlacionados, es decir $\rho(0)=\rho_r(0)
\otimes \rho_{\cal E}(0)$, la ecuaci\'on para la matriz densidad reducida
queda
\begin{equation}
\dot{\rho}_r=\frac{1}{i \hbar} {\rm Tr}_{\cal E} [V(t), \rho_r(0) \otimes
\rho_{\cal E}(0)] - \frac{1}{\hbar^2} \int_0^t \int dt_1 {\rm Tr}_{\cal E}
[V(t),[V(t_1),\rho_r(0) \otimes \rho_{\cal E}(0)]].
\end{equation}
Al observar el lado derecho de esta ecuaci\'on, vemos que podr\'iamos reemplazar el
estado inicial del sistema $\rho_r(0)$ en funci\'on de $\rho_r(t)$, despejando
el primero  de la Ec.(\ref{ecpert}). De esta manera, nos independizamos
de la matriz densidad reducida a tiempo $t=0$, y \'unicamente obtenemos una
dependencia 
del tiempo $t$. Bajo estas suposiciones, la ecuaci\'on maestra que se obtiene es:
\begin{eqnarray}
\dot{\rho}_r &=& \frac{1}{i \hbar} {\rm Tr}_{\cal E} [V(t), \rho_r(t) \otimes \rho_{\cal E}(0)] - \frac{1}{\hbar^2} \int_0^t dt_1  {\rm Tr}_{\cal E}[V(t), [V(t_1),\rho_r(t) \otimes \rho_{\cal E}(0)]] \nonumber \\
&+&  \frac{1}{\hbar^2} \int_0^{t} dt_1  {\rm Tr}_{\cal E}([V(t), {\rm Tr}_{\cal E}([V(t_1), \rho_r(t) \otimes \rho_{\cal E}(0)])\otimes \rho_{\cal E}(0)]).
\end{eqnarray}

En el caso que nos interesa, $V=\sum_n \lambda_n q_n x$. 
Si el entorno est\'a en equilibrio t\'ermico a temperatura $T=1/k_B\beta$,
la ecuaci\'on maestra se reduce a
\begin{equation}
\dot{\rho}_r = - \frac{1}{\hbar^2} \int_0^t dt_1  {\rm Tr}_{\cal E}
[V(t), [V(t_1),\rho_r(t) \otimes \rho_{\cal E}(0)]],
\end{equation}
y el t\'ermino dentro de la integral temporal se puede escribir como
\begin{eqnarray}
 {\rm Tr}_{\cal E}[V(t), [V(t_1),\rho_r(t) \otimes \rho_{\cal E}(0)]] 
&=& \frac{1}{2} \sum_n \lambda_n^2 (\langle \{q_n(t),q_n(t_1)\} 
\rangle [x(t),[x(t_1),\rho_r]] \nonumber \\
&+& \langle [q_n(t),q_n(t_1)] \rangle [x(t),\{x(t_1),\rho_r\}]),
\end{eqnarray}
donde $\langle ... \rangle$ significa el promedio sobre el estado 
inicial del entorno. Si volvemos a la representaci\'on de Schr\"odinger, 
la ecuaci\'on maestra se escribe
\begin{equation}
 \dot{\rho}_r = \frac{1}{i \hbar} [H_S,\rho_r] - \int_0^t dt_1
\bigg(\nu(t-t_1) [x,[x(t_1-t),\rho]]-i\eta(t-t_1)[x,\{x(t_1-t),\rho\}] \bigg),
\end{equation}
en funci\'on de los n\'ucleos de ruido y disipaci\'on, definidos respectivamente como:
\begin{eqnarray}
 \nu(t) &=& \frac{1}{2 \hbar^2} \sum_n \lambda_n^2 \langle \{q_n(t),q_n(0)\} \rangle = \sum_n \frac{\lambda_n^2}{2 m_n \hbar \omega_n} \cos(\omega_n t) (1+2 N_n) \nonumber \\
\eta(t) &=& \frac{i}{2 \hbar^2} \sum_n \lambda_n^2 \langle [q_n(t),q_n(0)] \rangle =\sum_n \frac{\lambda_n^2}{2 m_n \hbar \omega_n} \sin(\omega_n t)
\end{eqnarray}
Usando que $1+2 N_n=\coth(\frac{\beta \hbar \omega}{2})$, 
la expresi\'on para la densidad espectral y la relaci\'on
\begin{equation}
 \sum_{n=1}^N \frac{\lambda_n^2}{2 m_n \omega_n} f(\omega_n)=
\int_0^{\infty} d\omega I(\omega) f(\omega),
\end{equation}
obtenemos
\begin{eqnarray}
 \nu(t)&=& \int_0^{\infty} d\omega I(\omega)\cos(\omega t)
(1+2N(\omega)), \nonumber \\
\eta(t)&=& \int_0^{\infty} d\omega I(\omega) \sin(\omega t).
\label{kernelcap2}
\end{eqnarray}

Si resolvemos las ecuaciones de Heisenberg para 
el sistema reducido, el operador de posici\'on est\'a dado por
\begin{equation}
 x(t)= x\cos(\Omega t) + \frac{1}{M \Omega} p \sin(\Omega t).
\end{equation}
De esta forma, podemos escribir la expresi\'on final de la ecuaci\'on maestra, 
para un entorno general y a cualquier temperatura:
\begin{equation}
 \dot{\rho}_r= \frac{i}{\hbar} [H_S+\frac{1}{2} M \delta \Omega^2(t) x^2, \rho_r] + 2 i \gamma(t) [x,\{p,\rho_r\}]-
{\cal D}(t)[x,[x,\rho_r]]-f(t) [x,[p,\rho_r]],
\label{ecmaestracap1}
\end{equation}
donde los coeficientes se definen
\begin{eqnarray}
 \delta \Omega^2(t)&=&-\frac{2 \hbar}{M} \int_0^t dt'
\cos(\Omega t') \eta(t'), \nonumber \\
\gamma(t)&=& -\frac{1}{2 M \Omega} \int_0^t dt'
\sin(\Omega t') \eta(t'), \nonumber \\
D(t)&=& \int_0^t dt' \cos(\Omega t') \nu(t'),  \label{coefdef} \\
f(t)&=& -\frac{1}{M \Omega} \int_0^t dt'
\sin(\Omega t') \nu(t').\nonumber 
\end{eqnarray}
Estos coeficientes, expl\'icitamente dependientes del tiempo, est\'an definidos
a segundo orden en la constante de acoplamiento. $\tilde \Omega^2= 
\Omega^2 +\delta \Omega^2(t)$ es la frecuencia natural \mbox{renormalizada}
del entorno; $\gamma(t)$ es la tasa de relajaci\'on; $D(t)$ y
$f(t)$ son los coeficientes de difusi\'on (responsables de los
efectos de p\'erdida de coherencia en el sistema).
En el l\'imite de  temperatura alta estos coeficientes toman
formas m\'as sencillas y se vuelven constantes en el tiempo:
$\gamma(t)=\gamma_0$, ${\cal D}(t)=2 \gamma_0 k_B T$ y $f(t) \sim 1/k_BT$,
por lo cual, este coeficiente es despreciable.

La ecuaci\'on maestra Ec.(\ref{ecmaestracap1}), 
obtenida a partir de un desarrollo perturbativo, 
es la misma que se deduce de manera exacta para el
MBC con acoplamiento lineal \cite{HuPazZhangI,Paz}. Sin embargo,
en este \'ultimo caso, los coeficientes son formalmente distintos,
ya que su formulaci\'on es exacta.
\section{ El proceso de p\'erdida de coherencia}
\label{decoherencia}

La interacci\'on entre un subsistema cu\'antico abierto y su entorno
crea correlaciones entre los estados del subsistema y aquellos del entorno.
El entorno se ``lleva'' informaci\'on del subsistema a trav\'es
de estas correlaciones. Despu\'es de trazar sobre los grados de
libertad del entorno, un determinado conjunto de estados exhiben una
fuerte estabilidad frente a la interacci\'on con el entorno, mientras
que superposiciones lineales de estos estados son destru\'idas 
en la evoluci\'on din\'amica  del sistema, a veces de forma
muy r\'apida o, incluso, de manera instant\'anea. Esta destrucci\'on
din\'amica de las coherencias cu\'anticas, inducida puramente por 
la presencia de un entorno, se llama {\it decoherencia}. Este
proceso implica una selecci\'on din\'amica de un conjunto privilegiado 
de estados del subsistema denominados estados punteros o ``pointer states'', los
cuales resultan invariantes a la acci\'on del Hamiltoniano de
Interacci\'on que describe la din\'amica entre el subsistema y el entorno.
En general, el punto central en el estudio de decoherencia es
estimar el tiempo en el cual se lleva a cabo la destrucci\'on
de las coherencias. Este tiempo define una nueva escala en el
problema ($t_D$), llamada tiempo de decoherencia, y resulta extremadamente
\'util si uno quiere medir efectos cu\'anticos en un experimento, pues,
una vez destru\'idas estas coherencias, el sistema exhibir\'a un
comportamiento ``cl\'asico''.

Para poder hacer una estimaci\'on de este tiempo, uno debe identificar
el coeficiente de difusi\'on de la ecuaci\'on maestra para la
matriz densidad reducida. Siguiendo con el ejemplo de una part\'icula
Browniana cu\'antica, en el l\'imite de  temperatura alta, el
coeficiente de difusi\'on es ${\cal D}=2 \gamma_0 k_B T$. Como est\'a
ampliamente mostrado en la Literatura \cite{HuPazZhangI}, la
tasa de p\'erdida de coherencia $\Gamma_{D}=1/t_D$, est\'a definida por:
\begin{equation}
 \Gamma_{D}= 4 L_0^2 {\cal D} \approx 8 L_0^2 \gamma_0 k_B T.
\end{equation}
$L_0$ implica alguna longitud caracter\'istica del problema en
consideraci\'on (por ejemplo, una superposici\'on lineal inicial
de estados separados una distancia $L_0$). Por lo tanto, los t\'erminos 
fuera de la diagonal decaen como $\Gamma_D$, y las 
coherencias cu\'anticas desaparecen exponencialmente
en una escala que podemos identificar como
\begin{equation}
t_D=\gamma_0^{-1} \bigg(\frac{\lambda_T}{L_0}\bigg)^2,
\end{equation}
donde $\lambda_T=\hbar/\sqrt{2 m k_B T}$ es la longitud de
onda de de Broglie. As\'i, para un objeto 
macrosc\'opico, el tiempo de p\'erdida de coherencia $t_D$ es, 
tip\'icamente, varios \'ordenes de magnitud
menor que el tiempo de relajaci\'on 
$t_R = \gamma_0^{-1}$. Por ejemplo, para un sistema en 
una habitaci\'on a
una temperatura $T = 300K$, con una masa de $m = 1$g y 
con una separaci\'on $L_0 = 1$cm, el cociente
$t_D /t_R = 10^{-40}$. En consecuencia, aunque el tiempo 
de relajaci\'on sea del orden de la edad del
Universo, $t_R \sim 10^{17}$ segs, la coherencia 
cu\'antica ser\'a destru\'ida en $t_D \sim  10^{-23}$ segs 
\cite{Zurek3}. 
Una diferencia
tan grande puede obtenerse s\'olo para objetos 
macrosc\'opicos, y puede aceptarse  cuando las
condiciones por las cuales se deriv\'o la ecuaci\'on 
maestra se satisfacen. No obstante, es simple de 
entender, en este contexto, por qu\'e la p\'erdida de 
coherencia entre trayectorias macrosc\'opicamente
distinguibles es casi instant\'anea, a\'un para 
sistemas casi aislados. 

En este contexto, a lo largo de esta Tesis, nos centraremos
en el estudio del proceso de p\'erdida de coherencia 
que ocurre en distintos sistemas f\'isicos relevantes, y 
en los casos en que nos sea posible, estimaremos el
 tiempo de p\'erdida de coherencia para dichos sistemas.            
\newpage
\thispagestyle{empty}
\cleardoublepage

\chapter{Efectos difusivos inducidos  por fluctuaciones 
cu\'anticas}
\label{c2}
\markboth{Efectos difusivos inducidos por fluctuaciones 
cu\'anticas}
{Cap\'itulo 2}


En este cap\'itulo, analizaremos el efecto de las fluctuaciones de
vac\'io de un entorno a temperatura cero como fuente del
proceso de p\'erdida de coherencia.
Las fluctuaciones de vac\'io tienen muchos efectos visibles.
El corrimiento de la frecuencia de Lamb o el efecto Casimir
son claras manifestaciones de su existencia.  En estos ejemplos,
el efecto del vac\'io simplemente se traduce en la renormalizaci\'on de los
par\'ametros originales del sistema. Sin embargo, las fluctuaciones
que trataremos en adelante, no  inducen \'unicamente una renormalizaci\'on
de los par\'ametros de la part\'icula de prueba. Por el contrario,
representan una fuente de ruido y disipaci\'on para el sistema acoplado al
entorno a temperatura cero. En primer lugar, estudiaremos
 anal\'iticamente el proceso de decoherencia y estimaremos
 la escala temporal en la cual este proceso se lleva a cabo
en el caso de entornos \'ohmicos. Luego, extenderemos este estudio al
caso de entornos m\'as generales, es decir supra\'ohmicos 
y sub\'ohmicos, tambi\'en a temperatura estrictamente cero. En todos los  casos, 
completaremos el desarrollo anal\'itico con simulaciones
num\'ericas. En la segunda parte de este cap\'itulo, nos concentraremos
en estudiar otro tipo de efecto difusivo, inducido tambi\'en por 
las fluctuaciones de vac\'io del entorno, cuando \'este
est\'a a temperatura cero: la excitaci\'on energ\'etica. En particular,
a trav\'es de un an\'alisis num\'erico, estudiaremos su existencia para entornos
generales a temperatura cero y, mostraremos que este fen\'omeno
es posterior al proceso de decoherencia que se lleva a cabo en el
sistema cu\'antico original. 

La presencia de un entorno cu\'antico a temperatura cero y su
injerencia en el fen\'omeno de interferencia ha sido estudiado
en los \'ultimos a\~nos por varios autores \cite{ford,imry,sinha}. 
En muchos trabajos, particularmente donde se estudia 
el fen\'omeno de p\'erdida
de coherencia en el Movimiento Browniano Cu\'antico (MBC), la
mayor\'ia de las conclusiones acerca del proceso difusivo  a temperatura 
cero han sido en base a resultados num\'ericos. 
Por el contrario, el l\'imite de  temperatura alta del MBC
\cite{jpphabzurek} ha sido ampliamente estudiado. Este
mismo proceso, pero en el caso
de baja temperatura (aunque no estrictamente cero), tambi\'en ha
sido discutido  num\'ericamente (a trav\'es de los coeficientes
de difusi\'on normal y an\'omalo) para entornos generales
\cite{HuPazZhangI,jppdavila,leshouches}. Sin embargo, 
poco se ha estudiado el caso de un entorno a temperatura cero. 
En lo que sigue, nos ocuparemos del proceso de p\'erdida de coherencia
en entornos generales a temperatura cero.

\section{El modelo}
\label{modelocap2}

 El modelo que consideraremos consiste en una 
part\'icula de prueba (de masa $M$
y frecuencia caracter\'istica $\Omega$) 
acoplada a un entorno representado
por un conjunto de infinitos osciladores arm\'onicos.
La acci\'on total, correspondiente al sistema m\'as su entorno, 
resulta 
(fijando $\hbar=1$)
\begin{eqnarray}S[x,q_n] &=& S[x] + S[q_n] + S_{\rm int}[x,q_n]\nonumber \\
&=& \int_0^t ds \left[\frac{1}{2} M ({\dot x}^2 - \Omega^2 x^2) +
\sum_n \frac{1}{2} m_n ({\dot q}_n^2 - \omega_n^2 q_n^2)\right] -
\sum_n \lambda_n x q_n,\end{eqnarray}
donde $x$ y $q_n$ son las coordenadas de la part\'icula y de los osciladores,
respectivamente. La part\'icula est\'a acoplada a cada oscilador de manera
lineal a trav\'es de una  constante de acoplamiento $\lambda_n$.
Los objetos que resultan relevantes para analizar la transici\'on cu\'antico-
cl\'asica en este modelo son la matriz densidad reducida y 
la funci\'on de Wigner asociada definidas, respectivamente, como
\begin{eqnarray}
\rho_{\rm r} (x,x',t)&=& \int d{\bar q} \,\, 
\rho (x,{\bar q},x',{\bar q},t)\nonumber\\
W_{\rm r} (x,p,t)&=& \frac{1}{2\pi} \int_{-\infty}^{+\infty} dy~
e^{ipy} ~ \rho_{\rm r}(x+ \frac{y}{2}, x-\frac{y}{2},t).
\end{eqnarray}
La matriz densidad reducida satisface una ecuaci\'on maestra 
(Ec.(\ref{ecmaestracap1})) mencionada en el Cap\'itulo \ref{c1},
la cual fue derivada de manera exacta por  Hu, Paz y Zhang 
 para el MBC \cite{HuPazZhangI}. La ecuaci\'on 
maestra, se puede escribir alternativamente como
\begin{eqnarray}
i\frac{\partial}{\partial t} \rho_{\rm r}(x,x',t) &=&
\Bigg[-\frac{1}{2 M^2} \bigg( \frac{\partial ^2}{\partial x^2} -
\frac{\partial ^2}{\partial x'^2} \bigg)\Bigg] \rho_{\rm
r}(x,x',t)
+ \frac{1}{2}M \Omega^2 (x^2-x'^2)  \rho_{\rm r}(x,x',t)
\nonumber \\
&+& \frac{1}{2}M {\delta \Omega}^2(t) (x^2-x'^2) \rho_{\rm
r}(x,x',t)
- i \gamma(t) (x-x') \bigg(\frac{\partial }{\partial x}
-\frac{\partial }{\partial x'} \bigg) \rho_{\rm
r}(x,x',t)\nonumber \\
&-&i M {\cal D}(t)(x-x')^2 \rho_{\rm r}(x,x',t)
-f(t)(x-x')\bigg(\frac{\partial }{\partial x} + \frac{\partial
}{\partial x'} \bigg) \rho_{\rm r}(x,x',t),
\label{master}\end{eqnarray} 
donde resulta evidente que el coeficiente de difusi\'on ${\cal D}(t)$
est\'a relacionado con el decaimiento de los t\'erminos no 
diagonales de la matriz, ya que es proporcional a $(x-x')^2$.
Los coeficientes de esta ecuaci\'on son aquellos definidos 
en la Ec.(\ref{coefdef}) del Cap\'itulo \ref{c1}. 
 Estos coeficientes, expl\'icitamente 
dependientes del tiempo
(salvo en situaciones particulares), est\'an definidos a 
segundo orden en la constante de acoplamiento.
Inicialmente, es decir a tiempo $t=0$, 
todos los coeficientes son cero si suponemos que
el sistema y el entorno no est\'an correlacionados.
 Los coeficientes de renormalizaci\'on de la frecuencia
y de disipaci\'on se deducen a partir del n\'ucleo 
hom\'onimo a este \'ultimo coeficiente; 
mientras que los coeficientes difusivos, lo
hacen a partir del n\'ucleo de ruido, ambos definidos en la Ec.(\ref{kernelcap2}).
En lo que sigue, nos dedicaremos a escribir expl\'icitamente
estos coeficientes para los distintos tipos de entorno, es decir
 \'ohmico, sub\'ohmico y supra\'ohmico, seg\'un corresponda.

\section{Entornos \'ohmicos}
\label{ohmcap2}
Para analizar c\'omo es el proceso de
 p\'erdida de coherencia cuando la part\'icula
Browniana est\'a acoplada a un entorno \'ohmico a temperatura
cero, debemos definir la densidad espectral de dicho
entorno. En este caso particular, usaremos $I(\omega ) =
 (2/\pi) M\gamma_0 \Lambda^2
\omega/(\omega^2 + \Lambda^2)$ donde $\Lambda$ es una
frecuencia f\'isica de corte, que representa la frecuencia
m\'as alta presente en el entorno.
En el l\'imite de  temperatura alta ($\hbar \Lambda \ll 
k_B T$), los coeficientes de la Ec.(\ref{coefdef})
resultan constantes. En particular, el coeficiente de disipaci\'on
es $\gamma_0$ y, el coeficiente de difusi\'on
 ${\cal D}=2 M\gamma_0 k_B T$. El coeficiente de 
difusi\'on an\'omalo puede ser despreciado
ya que resulta ser inversamente proporcional a la temperatura
del ba\~no, la cual es muy alta en este caso ($f \propto 1/T$). 
Por lo tanto, el t\'ermino relevante de la ecuaci\'on maestra
 a la hora de evaluar la p\'erdida de coherencia del sistema,
es \'unicamente  aquel 
proporcional a ${\cal D}$.

Nuestro objetivo primario es analizar este fen\'omeno 
cuando el ba\~no carece de temperatura, es decir, el entorno 
tiene temperatura estrictamente cero.
Esta condici\'on se traduce en fijar 
$\coth (\beta \hbar \omega/2)=1$
en la definici\'on del n\'ucleo de ruido 
Ec.(\ref{kernelcap2}), y por lo tanto,
en los coeficientes Ec.(\ref{coefdef}).
De esta manera, podemos evaluar cada uno de los coeficientes de
la ecuaci\'on maestra. En particular, estamos interesados en escalas
temporales mayores que el tiempo de memoria del entorno, es 
decir $t > 1/\Lambda$. Encontraremos, tambi\'en, 
el comportamiento para tiempos largos
 ($\Lambda t \gg 1$) de cada coeficiente.  
En la Figura \ref{figure1-2}, 
mostramos
el comportamiento de los coeficientes calculados para distintos casos.

Empezaremos calculando  el corrimiento de la frecuencia 
(Fig.\ref{figure1-2}(a)),
\begin{equation}
{\delta\Omega}^2 (t) = \frac{4 M \gamma_0}{\pi}\Lambda^2 \int_0^\infty
d\omega \int_0^t ds \frac{\omega}{\omega^2 + \Lambda^2} \sin\omega
s \cos\Omega s,
\label{Omegaohm}\end{equation} el cual, tras realizar las integrales, es
\begin{equation}
{\delta\Omega}^2 (t) = 2 M \gamma_0 \frac{\Lambda^3}{\Lambda^2 +
\Omega^2} \left[1 - e^{-\Lambda t} \left(\cos\Omega t -
\frac{\Omega}{\Lambda} \sin\Omega t\right)\right].
\label{Omega2ohm}\end{equation}
Si, en particular, queremos conocer el comportamiento a tiempos largos,
tales que $\Lambda t \gg1$, entonces el cambio en la frecuencia natural 
del sistema se cuantifica seg\'un
\begin{equation}
{\delta\Omega}^2  = 2 M \gamma_0 \frac{\Lambda^3}{\Lambda^2 +
\Omega^2}.\label{Omegalargeohm}
\end{equation}

Por su parte, el coeficiente de disipaci\'on est\'a graficado
en la Fig.\ref{figure1-2}(b) y se calcula a partir de
\begin{equation}
\gamma (t) = -\frac{2 M \gamma_0}{\pi\Omega}\Lambda^2 \int_0^\infty
d\omega \int_0^t ds \frac{\omega}{\omega^2 + \Lambda^2} \sin\omega
s \sin\Omega s.\label{gammaohm}
\end{equation}
Luego de realizar estas integrales, se obtiene
\begin{equation}
\gamma (t) =  - M \gamma_0 \frac{\Lambda^2}{\Lambda^2 + \Omega^2}
\left[1 - e^{-\Lambda t} \left(\cos\Omega t + \frac{\Lambda}{\Omega}
\sin\Omega t\right)\right], \label{gamma2ohm}\end{equation}
que tiene el siguiente comportamiento para tiempos largos
\begin{equation}
\gamma = - M \gamma_0 \frac{\Lambda^2}{\Lambda^2 + \Omega^2}.
\label{gammalargeohm}\end{equation}

El coeficiente de difusi\'on normal se define seg\'un la expresi\'on
\begin{equation}
D(t)= \frac{2M\gamma_0}{\pi}\Lambda^2 \int_0^\infty d\omega \int_0^t ds
\frac{\omega}{\omega^2 + \Lambda^2} \cos\omega s \cos\Omega s,
\label{Dohm}\end{equation}
la cual, al ser expl\'icitamente calculada, resulta
\begin{figure}[!ht]
\centering
\includegraphics[width=12cm]{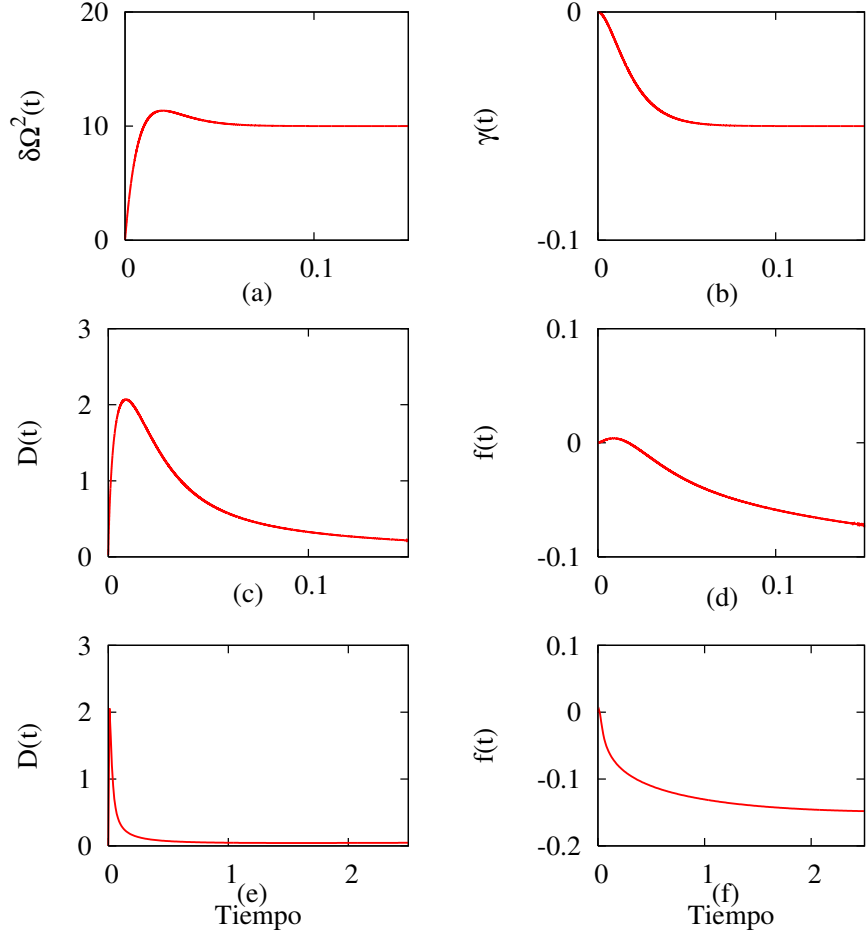}
\caption{ Evoluci\'on temporal de los coeficientes de la ecuaci\'on maestra
para un entorno a temperatura cero. En los gr\'aficos superiores mostramos
la renormalizaci\'on de la frecuencia en (a) y la disipaci\'on en (b).
Los gr\'aficos en la fila del medio corresponden a la difusi\'on normal 
${\cal D}(t)$ (c) y  an\'omalo $f(t)$ (d) para tiempos cortos con la
intenci\'on de mostrar el transitorio inicial.
Los valores para tiempos m\'as largos de los coeficientes
difusivos se muestran en los gr\'aficos inferiores (e) y (f).
En todos los casos, los par\'ametros usados son: 
$\gamma_0 = 0.05$, $\Lambda = 100$, $\Omega =1$.}
 \label{figure1-2}
\end{figure}

\begin{eqnarray}
D(t)&=&\frac{2M\gamma_0}{\pi} 
\frac{\Lambda^2 \Omega}{\Omega^2 + \Lambda^2}
 \left[{\rm Shi}(\Lambda t) 
\left(\frac{\Lambda}{\Omega}\cos\Omega t
\cosh\Lambda t +
\sin\Omega t \sinh\Lambda t\right)\right. 
\nonumber \\
&-& \left.
{\rm Chi}(\Lambda t) \left(\frac{\Lambda}
{\Omega}\cos\Omega t \sinh\Lambda t +
\sin\Omega t \cosh\Lambda t\right) \right. 
+ \left. {\rm Si}(\Omega t) \right],
\label{D2ohm}\end{eqnarray}
donde $\rm Shi(x)$ y $\rm Chi(x)$ son el 
SinIntegral y CosIntegral hiperb\'olico,
respectivamente; y $\rm Si(x)$ es el 
SinIntegral. Esta expresi\'on puede ser
simplificada, en el caso que $\Lambda t \gg1$, de manera
 de hacer evidente 
el comportamiento  de la forma
\begin{equation}
D(t) = \frac{2M\gamma_0}{\pi} \frac{\Lambda^2 
\Omega}{\Omega^2 + \Lambda^2}
\rm {Si}(\Omega t).
\label{Dlargeohm}\end{equation}
Este coeficiente vale para cualquier valor de la 
frecuencia natural del sistema 
($\Omega$). Su comportamiento resulta oscilatorio
en el tiempo. De hecho, tiene un comportamiento \mbox{constante} s\'olo en el
caso en que $\Omega t \gg 1$, ya que $\rm Si(x) \rightarrow \pi/2$,
$D_\infty \sim M \gamma_0 \Lambda^2 \Omega/(\Lambda^2 
+ \Omega^2)$, como se puede observar en la Fig \ref{figure1-2} (e). 
En cualquier otro caso, el coeficiente tiene un transitorio inicial
y se aproxima al valor asint\'otico $D_{\infty}$, como muestra la
Fig.\ref{figure1-2}(c).
Por el contrario, cuando $\Omega \ll 1$,
la difusi\'on normal crece linealmente con el tiempo, seg\'un
$D \sim 2M(\gamma_0/\pi)
\Lambda^2\Omega^2 t/(\Lambda^2 + \Omega^2)$, de forma similar a lo
obtenido en la Ref.\cite{sinha}.

Finalmente, el coeficiente de difusi\'on an\'omalo se puede calcular 
seg\'un
\begin{equation}
f(t)= -\frac{
2\gamma_0}{\pi\Omega}\Lambda^2 \int_0^\infty d\omega \int_0^t ds
\frac{\omega}{\omega^2 + \Lambda^2} \cos\omega s \sin\Omega s,
\label{fohm}\end{equation} que, al hacer las integrales,
resulta
\begin{eqnarray}
f(t)&=& -2\gamma_0 \frac{\Lambda^2}{\Omega^2 + \Lambda^2}
\left[{\rm Shi}(\Lambda t) \left(\frac{\Lambda}{\Omega}\sin\Omega t
\cosh\Lambda t -
\cos\Omega t \sinh\Lambda t\right)\right. \nonumber \\
&+& \left.
{\rm Chi}(\Lambda t) \left(- \frac{\Lambda}{\Omega}\sin\Omega t \sinh\Lambda t +
\cos\Omega t \cosh\Lambda t\right) \right.
- \left.{ \rm Ci}(\Omega t) - \log\frac{\Lambda}{\Omega} \right].
\label{f2ohm}\end{eqnarray}
Nuevamente, si queremos conocer el comportamiento a tiempos
largos, $\Lambda t \gg 1$, el coeficiente es
\begin{equation}
f(t) = 2\gamma_0 \frac{\Lambda^2}{\Omega^2 + \Lambda^2}
\left({\rm Ci}(\Omega t) + \log\frac{\Lambda}{\Omega}\right).
\label{flargeohm}\end{equation}
Este coeficiente tambi\'en toma un valor constante cuando
$\Omega t \gg 1 $ siendo $f_\infty \sim 2\gamma_0 
(\Lambda^2/(\Lambda^2 + \Omega^2))
\log\Lambda/\Omega$ (Fig.\ref{figure1-2}(f)); y si, por el contrario,
 $\Omega t \ll 1$, entonces,
$f(t) \sim  2 \gamma_0 (\log\Lambda t + \Gamma_{\rm Euler} )$, con
$\Gamma_{\rm Euler}$ el n\'umero de Euler Gamma.
Vale destacar que el comportamiento  asint\'otico de estos coeficientes
fue recientemente estudiado en la Ref.\cite{Hu2}. Esto fue desarrollado
de manera exacta y coincide con nuestros resultados en el caso de
$\gamma_0 \ll 1$ y frecuencias naturales y de corte chicas.

\subsection{P\'erdida de coherencia a temperatura cero}
\label{decoohm}

Analizaremos el proceso de p\'erdida de coherencia
en un caso sencillo. Nuestro estado inicial est\'a formado
por 
una superposici\'on lineal de dos estados deslocalizados,
tanto en posici\'on como en momento. Para ello, asumiremos
que tenemos dos paquetes gaussianos localizados
sim\'etricamente en el espacio de fases, es decir $x_0=\pm L_0$,
de la misma forma a lo realizado por los autores en 
 \cite{jpphabzurek}: $\Psi(x,t=0) = 
\Psi_1(x) + \Psi_2(x)$, tal que
\begin{equation}
\Psi_{1,2} = N \exp\left(-\frac{(x\mp L_0)^2}{2\delta^2}\right)
\exp(\pm i P_0x),\end{equation}
\begin{equation} N^2 = \frac{{\tilde N}^2}{\pi\delta^2}=
\frac{1}{2\pi\delta^2}\left[1 + \exp\left(-\frac{L_0^2}{\delta^2}
- \delta^2 P_0^2\right)\right]^{-1},\end{equation}
donde $N$ es la normalizaci\'on y $\delta$ es el ancho inicial
del paquete de ondas. En t\'ermino de la funci\'on de Wigner,
el estado a un tiempo dado $t$ es 
$W(x,p,t) = W_1(x,p,t) + W_2(x,p,t) + W_{\rm int}(x,p,t)$,
donde 
\begin{equation}W_{1,2}= \frac{{\tilde
N}^2}{\pi}\frac{\delta_1}{\delta_2}\exp\left(-\frac{(x\mp
x_c)^2}{\delta_1^2}\right)\exp\left(-\delta_2^2(p\mp p_c - \beta
(x \mp x_c))^2\right),
\end{equation}
y
\begin{equation}W_{\rm int}=\frac{2{\tilde
N}^2}{\pi}\frac{\delta_1}{\delta_2}\delta_2^2 (p - \beta x)^2
\cos(2k_p p + 2 (k_x - \beta k_p)x).\end{equation}

Todos los par\'ametros que aparecen en estas ecuaciones son funciones
del tiempo, determinados por la evoluci\'on del propagador de la matriz
densidad reducida y de la condici\'on a tiempo cero \cite{jpphabzurek}. El estado 
inicial ser\'a tal que 
$\delta_1^2=\delta_2^2=\delta^2$, $k_x = P_0=p_c$, $k_p=L_0=x_c$.
Los par\'ametros $k_p$ y $k_x$, indican la evoluci\'on de las franjas de interferencia
en las coordenadas de momento y posici\'on del espacio de fases, 
respectivamente.

Como ha sido definido en la Literatura, por ejemplo en \cite{leshouches},
el efecto de la p\'erdida de coherencia se cuantifica a trav\'es de un factor 
exponencial $\Gamma(t)$ o factor de ``decoherencia", 
definido seg\'un
\begin{equation}
\Gamma(t)=\exp(-A_{\rm int}) = \frac{1}{2}\frac{W_{\rm int}(x,p)|_{\rm
peak}}{\left[W_1(x,p)|_{\rm peak} W_2(x,p)|_{\rm
peak}\right]^{\frac{1}{2}}}.
\label{visibilityaint}
\end{equation}
Inicialmente el ``factor de visibilidad" de las 
franjas de interferencia $A_{\rm int}$ 
es nulo y siempre est\'a acotado por el valor m\'aximo
$A_{\rm int} \leq L_0^2/\delta^2 + \delta^2 P_0^2 = A_{\rm int}|_{\rm max}$.
Este factor de visibilidad evoluciona en el tiempo seg\'un 
${\dot A}_{\rm int}= 4 {\cal D}(t) k_p^2 - 4 f(t) k_p (k_x - \beta
k_p)$. En el caso en que el entorno est\'a a temperatura muy alta,
y vale  $\hbar \omega
\ll k_B T$, el coeficiente de difusi\'on an\'omalo es despreciable
y se obtiene una tasa de p\'erdida de coherencia proporcional a la 
temperatura del entorno, derivada \'unicamente del t\'ermino de
difusi\'on normal. En nuestro caso particular, en donde el
entorno est\'a a temperatura cero, ambos coeficientes ${\cal D}(t)$
y $f(t)$ contribuyen al factor de visibilidad $A_{\rm int}$. La
estimaci\'on de este factor es, entonces, m\'as complicada. 
Una simplificaci\'on posible es asumir 
que las franjas de interferencia
se mantienen aproximadamente en las posiciones
iniciales  y, as\'i, fijar $k_p=L_0$ y $k_x=1/(2L_0)$. Si miramos
tiempos tales que $\Lambda t \gg 1$ de manera de ignorar
el transitorio inicial, se pueden usar las expresiones
para tiempos largos de los coeficientes difusivos 
(Ecs.(\ref{Dlargeohm}) y (\ref{flargeohm})) para evaluar
el coeficiente $A_{\rm int}$.  De modo de obtener la
forma m\'as sencilla para la ecuaci\'on de evoluci\'on de
$A_{\rm int}$, tomaremos $\beta \sim 0$, equivalente a 
quedarnos con $\beta$ a tiempos cortos \cite{jpphabzurek}. De 
este modo, estamos haciendo una elecci\'on conservativa de este
par\'ametro, la cual nos permitir\'a obtener una cota
superior para el tiempo de decoherencia $t_{\cal D}$.
La ecuaci\'on de evoluci\'on para el coeficiente
de visibilidad resulta
\begin{equation}
\dot{A}_{\rm int}\approx 4 L_0^2 {\cal D}(t) - 2 f(t) 
.\label{aintcap2}
\end{equation}

Para conocer el tiempo de p\'erdida de coherencia $t_D$, debemos
resolver  $1 \approx A_{\rm int}(t = t_D)$. Si pedimos esta
condici\'on, se puede observar en la Ec.(\ref{aintcap2}) que no es
posible despejar anal\'iticamente una escala global de p\'erdida
de coherencia para el caso de un entorno a temperatura cero. De todas
maneras, podemos encontrar las correspondientes escalas para situaciones
particulares.
Por ejemplo, en el caso de un sistema con frecuencia natural $\Omega$,
tal que $\Omega \sim \Lambda$ ($\Omega t \gg 1$), se puede obtener
f\'acilmente
\begin{equation} A_{\rm int} \sim 2 L_0^2 M \gamma_0 \Lambda t +
 4 \gamma_0 \left(t ~ {\rm Ci}(\Lambda t) - 
\frac{\sin \Lambda t}{\Lambda}\right),
\end{equation}
que implica un escala de p\'erdida de coherencia corta 
\begin{equation}
t_D \sim \frac{1}{2 M L_0^2 \gamma_0 \Lambda}.
\label{td1}\end{equation} Este resultado es v\'alido mientras 
sea cierto que el producto $M L_0^2 \gamma_0 \leq 1$, de 
manera de poder despreciar el transitorio inicial. Esta escala
coincide con el tiempo de p\'erdida de coherencia evaluado
directamente a partir de ${\cal D}_{\infty}$, como por 
ejemplo, hacen los autores en la Ref.\cite{leshouches}. En 
este l\'imite particular, el coeficiente an\'omalo no juega un papel
importante como se puede verificar usando $f_{\infty}$ en la
Ec.(\ref{aintcap2}).

En el caso contrario, cuando $\Omega t \ll 1$ (para tiempos
$\frac{1}{\Lambda} < t < \frac{1}{\Omega} < \frac{1}{\gamma_0}$), podemos
evaluar el coeficiente $A_{\rm int}$ usando el 
l\'imite asint\'otico de $\rm Si(x)$ y $\rm Ci(x)$. De esa forma, obtenemos
\begin{equation}
A_{\rm int} \approx \frac{8\Lambda^2}{\Lambda^2 + \Omega^2} \gamma_0 \left[
\frac{M L_0^2}{2\pi} (\Omega t)^2 + t ~ (\log\Lambda t + \Gamma - 1)\right]  ,
\label{a2}\end{equation}
que resulta  una cota para el tiempo de p\'erdida de coherencia
$t_{\cal D} \leq \frac{1}{8  \gamma_0}$, el cual puede ser muy grande
para sistemas subamortiguados. En este caso particular, la
correcci\'on logar\'itmica se debe al t\'ermino de difusi\'on an\'omalo
$f(t)$, a diferencia de lo que hicieran los autores en la Ref.\cite{sinha}
despreciando este efecto a\'un a temperatura cero. Esta escala
es m\'as larga que la escala temporal $t_{\cal D}$ correspondiente
al caso de  temperatura alta, incluso en el caso que el sistema de prueba
tiene una frecuencia natural alta y el entorno una temperatura ``baja"
 (pero sigue valiendo que $\hbar \omega \ll k_B T$). Resulta
necesario remarcar que, nuestra estimaci\'on es menor que el
tiempo de saturaci\'on del sistema $t_{\rm sat} 
= \gamma_0^{-1}$, tiempo en el cual $A_{\rm int}$
alcanza su m\'aximo valor.
Finalmente, podemos ver que, en el caso que sea posible despreciar
el segundo t\'ermino de la Ec.(\ref{a2}), por ejemplo cuando
las trayectorias son macrosc\'opicas ($M L_0 \gg 1$), obtenemos
\begin{equation}
t_{\cal D} \approx \frac{1}{2L_0\Omega}\sqrt{\frac{\pi}{M\gamma_0}}.
\end{equation}

Con el fin de complementar nuestras estimaciones anal\'iticas
sobre los tiempos en los cuales el proceso de p\'erdida 
de coherencia se lleva a cabo para los distintos casos,
hemos resuelto num\'ericamente la ecuaci\'on maestra.
 Para ello, utilizamos un estado inicial formado por dos estados
deslocalizados como se explica en detalle en el Ap\'endice A.
En la Fig.\ref{figexpohmcap2} mostramos la evoluci\'on temporal
del factor de p\'erdida de coherencia $\Gamma(t)$ 
(definido en la Ec.(\ref{visibilityaint}))
para distintos valores de la frecuencia de corte $\Lambda$ y la constante
de acoplamiento $\gamma_0$.  Podemos afirmar que el sistema
pierde coherencia para tiempos $\Lambda t >1$. Esto ocurre
proporcionalmente con la frecuencia de corte y la constante
de acoplamiento, es decir, cuando m\'as grandes son estos par\'ametros,
m\'as r\'apido desaparecen las interferencias. Adem\'as,
en el recuadro inferior de la  figura, 
resulta evidente que el sistema alcanza un punto 
de ``\mbox{saturaci\'on}"
 de su Entrop\'ia Lineal (definida en la 
Secci\'on \ref{Aentropia} del Ap\'endice \ref{dospaquetes})
 para tiempos $t \leq t_{\rm sat}$.  Esta situaci\'on
se alcanza m\'as r\'apido para valores m\'as grandes de $\gamma_0$
pero valores m\'as chicos de $\Lambda$. La raz\'on es, por un lado, 
el acoplamiento al entorno es m\'as grande y, por el otro, el
espacio de Hilbert del entorno es menor, y por  tanto, es m\'as
f\'acil alcanzar el valor de saturaci\'on.
\begin{figure}[!ht]
\centering
\includegraphics[width=10cm]{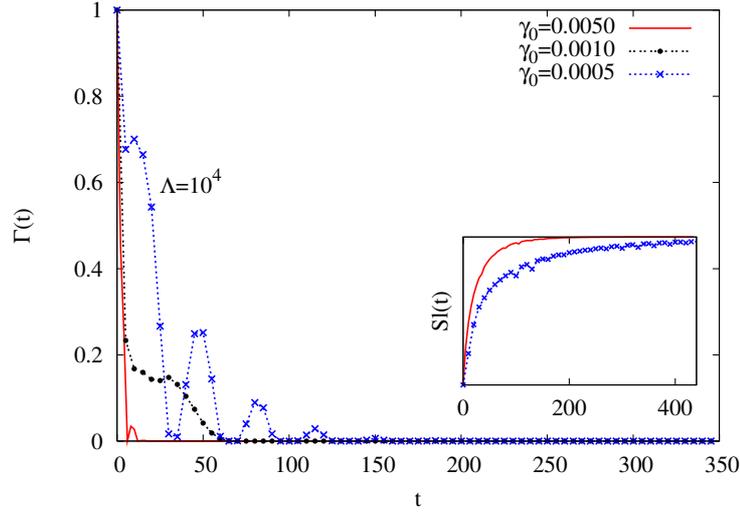}
\caption{El factor de p\'erdida de coherencia 
$\Gamma(t)=\exp(-4 L_0 (A(t)-C(t)))$, calculado con el m\'etodo num\'erico
detallado en el Ap\'endice A, para un entorno \'ohmico a temperatura cero,
para distintos valores de la frecuencia de corte $\Lambda$ y la
constante de acoplamiento $\gamma_0$. Los par\'ametros est\'an
medidos en unidades de la frecuencia natural $\Omega$.
Consideramos los casos $\Omega=15$, $L_0=0.5$, $\Lambda=10^3$ 
( y $\Lambda=10^4$ cuando es indicado en el gr\'afico) para distintos
valores de $\gamma_0$ en el gr\'afico principal. En el recuadro inferior,
presentamos la evoluci\'on temporal de la Entrop\'ia Lineal para dos
casos ($\gamma_0=0.005$ y $\gamma_0=0.0005$) y 
$\Lambda=10^3$. Ambos gr\'aficos muestran que el 
entorno a temperatura cero efectivamente induce p\'erdida de
coherencia ($\Gamma(t) \rightarrow 0$) en el sistema.}
\label{figexpohmcap2}
\end{figure}

La \'unica diferencia que hemos notado hasta el momento, entre
el entorno \'ohmico a temperatura cero y el mismo a temperatura alta,
es la escala temporal en la cual el sistema pierde coherencia $t_D$. En el \'ultimo
caso, se espera que este proceso se lleve a cabo en un
tiempo aproximado del orden ${\cal O}(1/2M\gamma_0k_BT)$
mientras que, en el primero, se estima que lo hace en
tiempos m\'as largos, menores que tiempos del orden 
${\cal O}(1/\gamma_0)$ \cite{PLA}.

En s\'intesis, en esta secci\'on, hemos deducido expresiones anal\'iticas
para la tasa de p\'erdida de coherencia en un sistema acoplado
a un entorno \'ohmico a temperatura cero para distintos casos 
y l\'imites particulares. 
 Reobtuvimos los resultados conocidos de las Refs. \cite{leshouches,sinha}
y presentamos resultados num\'ericos que coinciden con lo
estimado anal\'iticamente.

\section{Entornos no \'ohmicos}
\label{nonohmcap2}
En esta secci\'on, continuaremos analizando los efectos que induce
la presencia de un entorno a temperatura cero 
en una part\'icula Browniana, extendiendo
el desarrollo de la Secci\'on \ref{ohmcap2} al caso de entornos generales.
 Los entornos que
estudiaremos en adelante ser\'an llamados supra\'ohmicos o sub\'ohmicos
dependiendo de la forma de su densidad espectral, la cual definiremos
oportunamente.  La importancia de estos entornos a temperatura cero radica en 
su utilidad para modelar varias situaciones f\'isicas. 
El entorno \'ohmico es el m\'as estudiado en 
la Literatura y produce una fuerza disipativa que, en el l\'imite que la frecuencia de
corte $\Lambda$ tiende a cero, es proporcional a la velocidad. Por su parte,
el entorno supra\'ohmico, es usado generalmente para modelar la interacci\'on
entre defectos y fonones en metales \cite{legget}, o tambi\'en, 
la interacci\'on entre una carga y su propio campo electromagn\'etico
\cite{sonnentag}. En particular, en el caso del entorno supra\'ohmico
nos permitir\'a encontrar una relaci\'on directa con el proceso de p\'erdida
de coherencia en la Teor\'ia Cu\'antica de Campos \cite{ferdiego}. El caso
sub\'ohmico, generalmente usado en el contexto del modelo
spin-bos\'on de Legget {\it et. al} \cite{legget}, es utilizado, por ejemplo, 
para estudiar las transiciones de fase cu\'anticas en un 
anillo met\'alico mesosc\'opico  \cite{tongvojta}.

En este contexto, calcularemos los coeficientes dependientes
del tiempo Ec.(\ref{coefdef}) en cada caso y haremos estimaciones
anal\'iticas del coeficiente de visibilidad $A_{\rm int}$. Cuando
sea posible, deduciremos las escalas temporales asociadas a este
fen\'omeno. En todos los casos, completaremos el estudio anal\'itico,
resolviendo num\'ericamente la ecuaci\'on maestra Ec.(\ref{master}) con 
el m\'etodo num\'erico detallado en el Ap\'endice A.

Cuando tratemos con los entornos generales, usaremos  la siguiente
 densidad espectral \begin{equation}
I(\omega ) = \frac{2}{\pi} M \gamma_0 \omega 
(\frac{\omega}{\Lambda}) ^{n-1}
e^{-\omega^2/\Lambda^2},
\label{densidadnoohm}
\end{equation} la cual nos facilitar\'a la resoluci\'on de algunas integrales
relacionadas con los coeficientes de la ecuaci\'on maestra.
En la Ec.(\ref{densidadnoohm}), el par\'ametro $\Lambda$
representa nuevamente la frecuencia m\'axima presente
en el entorno.

Para el caso de una part\'icula Browniana acoplada a un entorno
supra\'ohmico, por ejemplo $n=3$ en la Ec.(\ref{densidadnoohm}),
 los coeficientes de la ecuaci\'on maestra
 a temperatura  cero, son:
\begin{eqnarray}
\delta \Omega^2(t)_{\rm n=3}&=& \frac{ 2 M \gamma_0 }{ \pi} \bigg\{2
\Lambda + \frac{\Omega^2}{\Lambda} - \frac{1}{\Lambda (
1 + \Lambda^2 t^2)^3} \bigg[ (2 \Lambda^2
- 6 t^2 \Lambda^4 + (\Omega + \Lambda ^2 t^2 \Omega)^2 )
\cos(\Omega t) \nonumber \\
&+& 2 \Lambda^2 t (1+ \Lambda^2 t^2) \Omega \sin(\Omega t) \bigg]
+ \frac{\Omega^3}{2 \Lambda^2}\bigg[\sinh(\frac{\Omega}{\Lambda})
\bigg(\rm{Ci}(\frac{-i \Omega}{\Lambda}) + \rm{Ci}(\frac{i \Omega}{\Lambda})
 \nonumber \\
&-& \rm{Ci}(\Omega(t -\frac{i\Omega}{\Lambda})) - \rm{Ci}(\Omega(t
+\frac{i\Omega}{\Lambda}))\bigg) \nonumber \\
&+& \frac{\Omega^3}{2 \Lambda^2} \cosh(\frac{\Omega}{\Lambda})
\bigg(\rm{Si} (\Omega(t-\frac{i \Omega}{\Lambda}))-\rm{Si}
(\Omega(t+\frac{i \Omega}{\Lambda})) 
- \rm{Shi}(\frac{\Omega}{\Lambda}) \bigg) \bigg] \bigg\}, \nonumber
\end{eqnarray}

\begin{eqnarray}
\gamma(t)_{\rm n=3}&=&-\frac{ 2 M \gamma_0 }{ \pi ( 1 + \Lambda^2
t^2)^3} \bigg\{2 \Omega \Lambda t (1+ \Lambda^2 t^2 )
\cos(\Omega t)
-\frac{1}{\Lambda} [2 \Lambda^2 - 6 \Lambda^4 t ^2 + (\Omega \nonumber \\
&+& \Omega \Lambda^2 t^2)^2] \sin(\Omega t) 
- \frac{\Omega^3}{2 \Lambda^2}( 1 + \Lambda^2 t^2)^3
\bigg[\cosh(\frac{\Omega}{\Lambda}) \bigg(i~\rm{Ci}(\frac{-i
\Omega}{\Lambda}) \nonumber \\
&-& i~\rm{Ci}(\frac{i
\Omega}{\Lambda}) 
- i~\rm{Ci}(\Omega (t-\frac{i}{\Lambda}))
+ i~\rm{Ci}(\Omega (t
+\frac{i}{\Lambda}))\bigg) 
+ \sinh(\frac{\Omega}{\Lambda})\bigg(\rm{Si}(\Omega (t
+\frac{i}{\Lambda}))  \nonumber \\
&-& \rm{Si}(\Omega (t
-\frac{i}{\Lambda}))\bigg)\bigg] \bigg\}, \nonumber 
\end{eqnarray}

\begin{eqnarray}
{\cal D}(t)_{\rm
n=3}&=& \frac{ M \gamma_0 }{ \pi \Lambda^2 ( 1 + \Lambda^2 t^2)^3}
\bigg\{ 2 \Lambda^2
        \bigg[( 6 t \Lambda^2 - 2 t^3 \Lambda^4
+ t ( \Omega  + t^2 \Lambda^2 \Omega)^2)
    \cos (\Omega t )
+  ( -1 \nonumber \\
&+& \Lambda^4 t^4) \Omega \sin (\Omega t )\bigg]
 +  ( \Omega  +  \Lambda^2 t^2 \Omega)^3
 \bigg[ 2 \cosh (\frac{\Omega }{\Lambda})
  \bigg(\rm{Si}(\Omega (t-\frac{i}{\Lambda}))
+ \rm{Si}(\Omega (t+\frac{i}{\Lambda})) \bigg) \nonumber \\
&+& \rm{i} \sinh({\frac{\Omega}{\Lambda}})\bigg(\rm{Ci}(\Omega
(t-\frac{i}{\Lambda})) 
 -  \rm{Ci}(\Omega
(t+\frac{i}{\Lambda})) - \log( \frac{-i \Omega}{\Lambda}) +  \log(
\frac{i \Omega}{\Lambda}) \bigg)  \bigg]\bigg\} 
 \label{Dsupohm}  
\end{eqnarray}
y
\begin{eqnarray}
f(t)_{\rm n=3}&=& \frac{2 M \gamma_0 }{ \pi ( 1 +
\Lambda^2 t^2)^3} \bigg\{ \frac{ \Omega^3}{2 \Lambda^2}
( 1 + \Lambda^2 t^2)^3
\cosh(\frac{\Omega}{\Lambda})
\bigg[\rm{Chi}(\frac{-i\Omega}{\Lambda})
+ \rm{Chi}(\frac{i\Omega}{\Lambda}) \bigg]
-\Omega( 1 + \Lambda^2 t^2)^3\nonumber \\
&+& (\Omega - \Lambda^4 t^4 \Omega) \cos(\Omega t)
+ t(6 \Lambda^2 - 2 t^2 \Lambda^4 + (\Omega + t^2 \Lambda^2
\Omega)^2 )\sin(\Omega t)\nonumber \\
&+& \frac{\Omega^3}{\Lambda^2}\sinh(\frac{\Omega}{\Lambda})( 1 +
\Lambda^2 t^2)^3 \bigg[ \rm{Shi}(\frac{\Omega}{\Lambda}) 
+ \rm{i}~ \rm{Si}(\Omega (t- \frac{i}{\Lambda})) - \rm{i}~
\rm{Si}(\Omega (t+ \frac{i}{\Lambda}))\bigg] \bigg\}.
\label{fsupohm}
\end{eqnarray}
Es importante destacar que, a pesar de su apariencia
abrumadora, estos coeficientes son todos reales. 
En la Figura \ref{figure3-2} presentamos la evoluci\'on 
temporal de cada uno de ellos.\\
\begin{figure}[!ht]
\centering
\includegraphics[width=10cm]{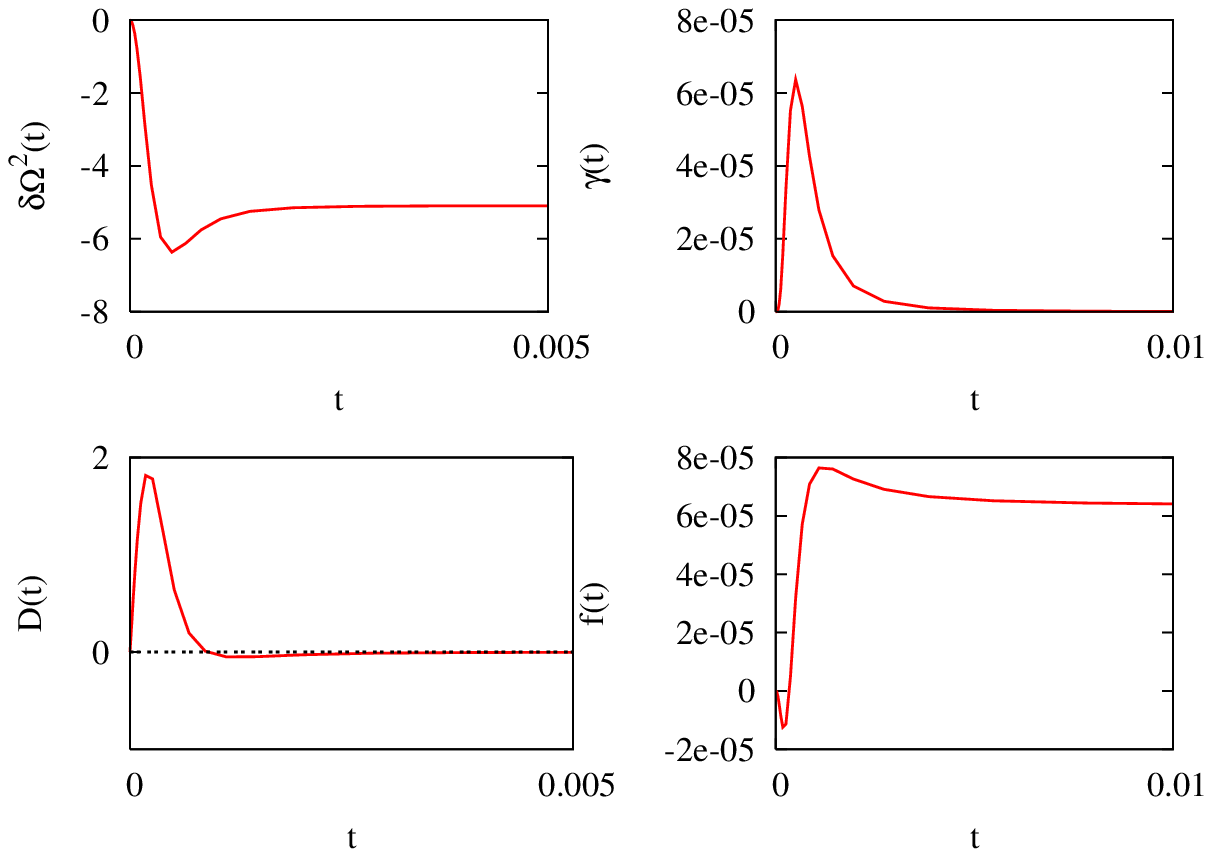}
\caption{Evoluci\'on temporal de los coeficientes
 de la ecuaci\'on maestra para un entorno supra\'ohmico a temperatura
cero. En los gr\'aficos superiores mostramos el coeficiente
de renormalizaci\'on $\delta \Omega^2 (t)$ y el de disipaci\'on 
$\gamma(t)$. Los gr\'aficos en la parte inferior muestran
la evoluci\'on temporal de los coeficientes de difusi\'on normal
${\cal D}(t)$ y an\'omalo $f(t)$. Los par\'ametros usados en 
todos los casos son:  $\gamma_0=0.001$, $\Lambda=2000$, 
y $\Omega=0.1$; y est\'an medidos en unidades de 
$\Omega$.} 
\label{figure3-2}
\end{figure}

Para el caso sub\'ohmico, es decir $n=1/2$ en la Ec.(\ref{densidadnoohm}),
los coeficientes de la ecuaci\'on maestra pueden ser estimados, si se verifica
que $\Lambda/\Omega \gg 1$,
\begin{eqnarray}
\delta \Omega^2_{\rm n=1/2}(t)&\approx&\frac{ 4 M \gamma_0
\Lambda}{2 \pi} \bigg\{ 2 \rm{Ci}(\Omega t)-
 \rm{Ci}[(\Lambda-\Omega)t]
- \rm{Ci}[(\Lambda+\Omega)t] + \log(\Lambda-\Omega) 
\nonumber \\
&+& \log(\Lambda + \Omega) 
-2 \log(\Omega) \bigg\}, \nonumber \\
\gamma(t)_{\rm n=1/2}&\approx&- \frac{2 M \gamma_0 \Lambda}{ \pi}
\rm{Si}(\Omega t), \nonumber \\
{\cal D}(t)_{\rm n=1/2}&\approx&\frac{2 M \gamma_0 \Lambda}{3 \pi}
\bigg\{\rm{Si}[(\Lambda-\Omega)t]+ \rm{Si}[(\Lambda+\Omega)t]
\bigg\}~~~~~{\rm y}  \label{coefsub} \\
f(t)_{\rm n=1/2}&\approx&-\frac{2 M \gamma_0 \Lambda}{3
\pi}\bigg\{\Gamma_{\rm Euler} - \rm{Ci}(2 \Lambda t) + \log(2 \Lambda t)
\bigg\}. \nonumber
\end{eqnarray}
\begin{figure}[!ht]
\centering
\includegraphics[width=10cm]{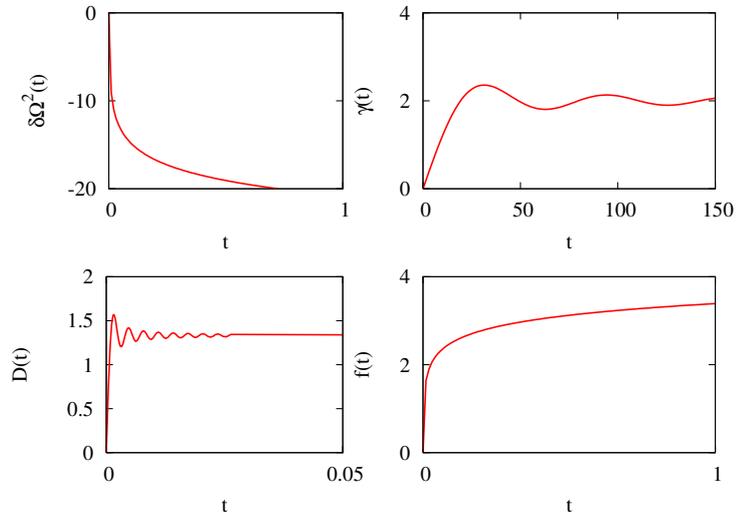}
\caption{Evoluci\'on temporal de los coeficientes
 de la ecuaci\'on maestra para un entorno sub\'ohmico a temperatura
cero. En los gr\'aficos superiores mostramos el coeficiente
de renormalizaci\'on $\delta \Omega^2 (t)$ y el de disipaci\'on 
$\gamma(t)$. Los gr\'aficos en la parte inferior muestran
la evoluci\'on temporal de los coeficientes de difusi\'on normal
${\cal D}(t)$ y an\'omalo $f(t)$. Los par\'ametros usados en 
todos los casos son:  $\gamma_0=0.001$, $\Lambda=2000$, 
y $\Omega=0.1$; y est\'an medidos en unidades de 
$\Omega$.} 
\label{figure4-2}
\end{figure}

Podemos ver que, en el caso
supra\'ohmico, el coeficiente disipativo tiende a cero para tiempos largos
a pesar de su  importante crecimiento inicial, producto
de la interacci\'on entre el sistema y el entorno (que a tiempo $t=0$
es cero), que se registra en la escala temporal de la frecuencia de
corte $t_{\Lambda} \sim 1/\Lambda$. Este coeficiente tiene un comportamiento
bastante diferente en el caso sub\'ohmico, ya que tambi\'en presenta un 
crecimiento considerable en la en la escala temporal $t_{\Lambda}$ pero,
luego, contin\'ua creciendo hasta que logra un valor constante
en tiempos $ \Lambda t > 1$. El coeficiente correspondiente
al caso \'ohmico, tiene un comportamiento similar a \'este \'ultimo,
pero logra un valor constante en una escala temporal m\'as corta.
En la Fig.\ref{figure5-2} est\'an graficados todos los coeficientes
de disipaci\'on para poder apreciar sus semejanzas y diferencias
de forma m\'as sencilla, de manera similar a lo realizado por
 Hu, Paz y Zhang en \cite{HuPazZhangI} para otras temperaturas
diferentes a cero.
\begin{figure}[!ht]
\centering
\includegraphics[width=10cm]{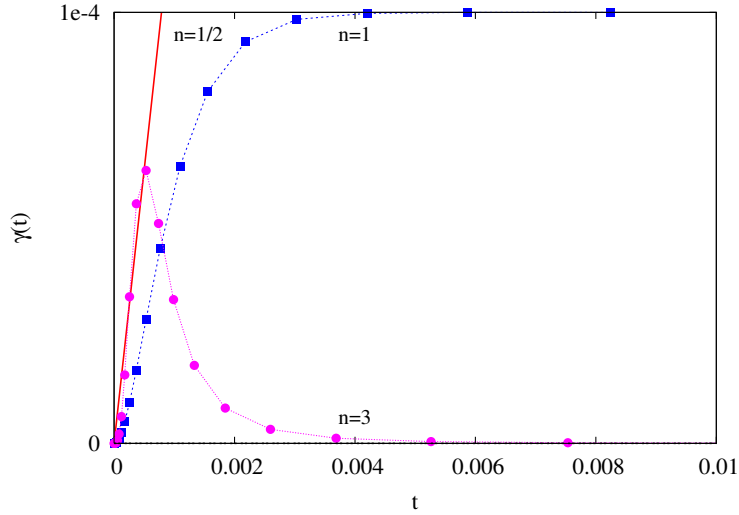}
\caption{El coeficiente de disipaci\'on para el entorno \'ohmico
 ($n=1$), sub\'ohmico ($n=1/2$) y supra\'ohmico ($n=3$). 
Los par\'ametros est\'an medidos en unidades de la frecuencia natural
del sistema  $\Omega$. Consideramos los casos  
$\gamma_0=0.001$, $\Lambda=2000$, y
$\Omega=0.1$. El caso supra\'ohmico no resulta disipativo
despu\'es del crecimiento inicial (cuando se enciende la interacci\'on).}
 \label{figure5-2}
\end{figure}

El comportamiento del coeficiente $\delta \Omega^2(t)$ no difiere 
mucho entre los tres entornos, sin embargo, resulta considerablemente
mayor en el caso sub\'ohmico (Fig.\ref{figure6-2}). 
La importancia de este coeficiente, radica en
 que su valor asint\'otico es aquel que fija
el valor de la frecuencia normalizada del sistema en cada caso.
\begin{figure}[!ht]
\centering
\includegraphics[width=10cm]{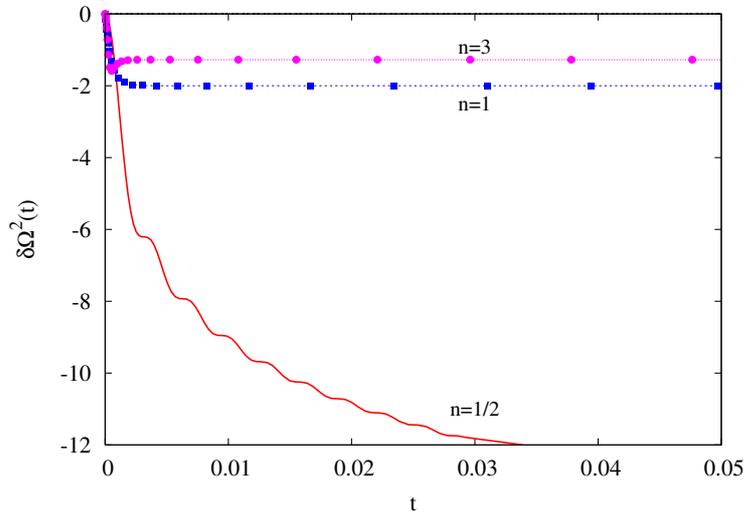}
\caption{El coeficiente de renormalizaci\'on de la frecuencia
 para el entorno \'ohmico
 ($n=1$), sub\'ohmico ($n=1/2$) y supra\'ohmico ($n=3$). 
Los par\'ametros est\'an medidos en unidades de la frecuencia natural
del sistema  $\Omega$. Consideramos los casos  
$\gamma_0=0.001$, $\Lambda=2000$, y
$\Omega=0.1$.}
\label{figure6-2}
\end{figure}
\begin{figure}[!ht]
\centering
\includegraphics[width=10cm]{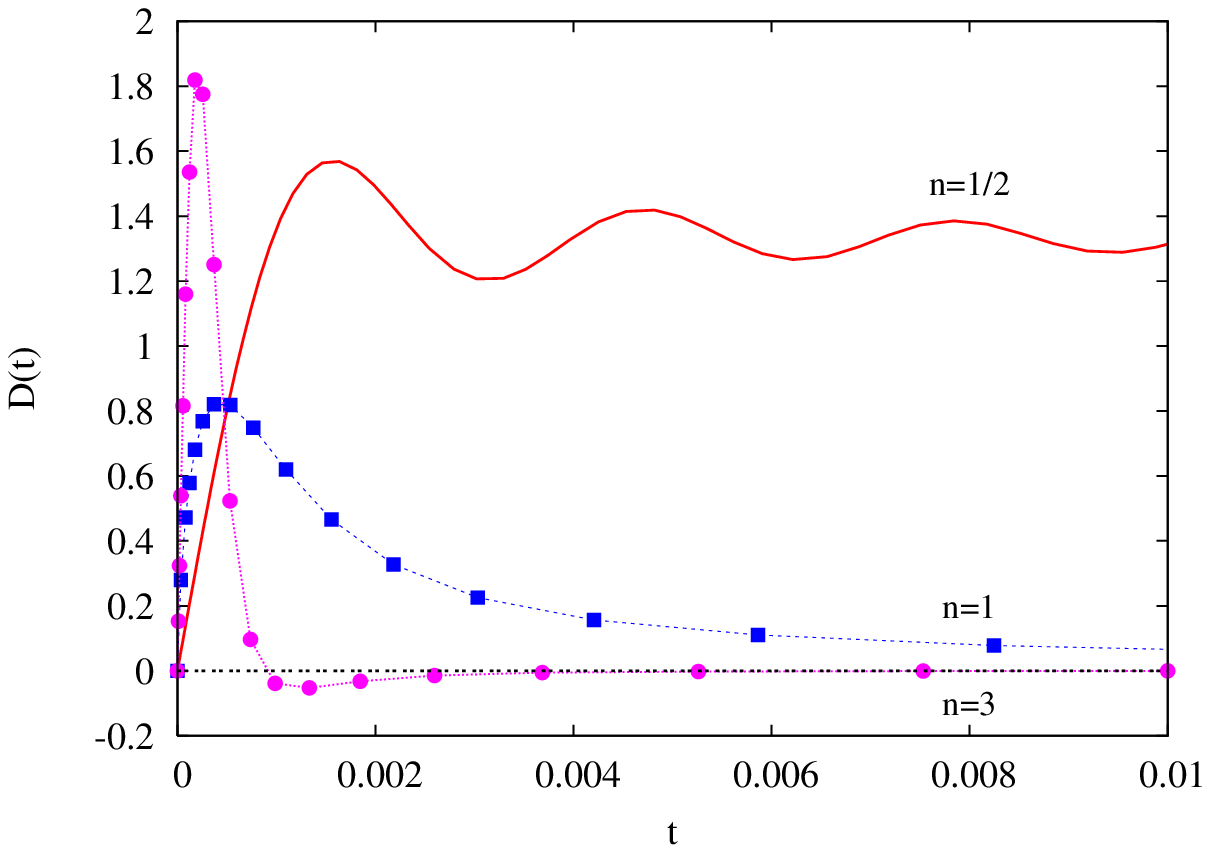}
\caption{El coeficiente de difusi\'on normal 
 para el entorno \'ohmico
 ($n=1$), sub\'ohmico ($n=1/2$) y supra\'ohmico ($n=3$). 
Los par\'ametros est\'an medidos en unidades de la frecuencia natural
del sistema  $\Omega$. Consideramos los casos  
$\gamma_0=0.001$, $\Lambda=2000$, y
$\Omega=0.1$. De nuevo vemos que el coeficiente tiende
a cero en el entorno supra\'ohmico.}
\label{figure7-2}
\end{figure}

En cuanto a los efectos difusivos inducidos por el entorno, las cantidades relevantes para analizar son los
coeficientes ${\cal D}(t)$ y $f(t)$ \cite{dwPRE}. En la Fig.\ref{figure7-2}
se puede observar que todos los coeficientes de difusi\'on normal
tienen un abrupto crecimiento inicial en la escala temporal $t_{\Lambda}$,
similar al que se reportara en la Ref.\cite{ferdiego} para un escenario 
de Teor\'ia de Campos. Sin embargo, luego del crecimiento inicial, todos ellos
alcanzan valores asint\'oticos distintos. Por ejemplo, para el caso supra\'ohmico, este valor
es cero, mientras que en el caso sub\'ohmico, el valor asint\'otico 
 es significativamente mayor que cero y se obtiene en una escala
temporal mas tard\'ia (la misma en la cual desarrolla un comportamiento
 muy similar a aquel del coeficiente disipativo \cite{HuPazZhangI}).
Es evidente que, para tiempos $t > t_{\Lambda}$, el entorno
supra\'ohmico no resulta difusivo. Este hecho puede ser asociado
con la baja intensidad de la densidad espectral supra\'ohmica en
el sector infrarrojo del espectro de frecuencias. Para el caso \'ohmico,
el valor asint\'otico es ${\cal D}_{\infty}=M \gamma_0 \Lambda^2
\Omega/(\Omega^2 + \Lambda^2)$ como fue mencionado en
la Secci\'on \ref{ohmcap2} y en la Ref.\cite{PLA}; mientras que,
en el caso sub\'ohmico, es ${\cal D}_{\infty}= 2/3 M \gamma_0 \Lambda$.
Ambos casos l\'imites coinciden con el gr\'afico de la Fig.\ref{figure7-2}.

Finalmente, los coeficientes de difusi\'on an\'omalo $f(t)$
 de los tres entornos est\'an 
graficados en la Fig.\ref{figure8-2}. All\'i se observa que dicho
coeficiente resulta mucho m\'as grande en el caso sub\'ohmico. Este hecho
puede resultar importante cuando analicemos el proceso de excitaci\'on
energ\'etica en este tipo de entornos en la Secci\'on \ref{excitacion}.
\begin{figure}[!ht]
\centering
\includegraphics[width=10cm]{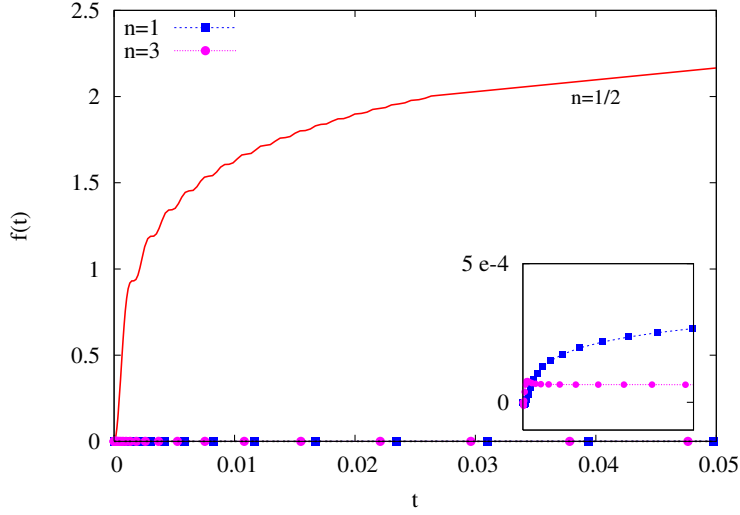}
\caption{El coeficiente de difusi\'on an\'omalo
 para el entorno \'ohmico
 ($n=1$), sub\'ohmico ($n=1/2$) y supra\'ohmico ($n=3$). 
Los par\'ametros est\'an medidos en unidades de la frecuencia natural
del sistema  $\Omega$. Consideramos los casos  
$\gamma_0=0.001$, $\Lambda=2000$, y
$\Omega=0.1$. El entorno sub\'ohmico tiene un valor asint\'otico bastante
mayor que en los otros dos casos.} 
\label{figure8-2}
\end{figure}

\subsection{P\'erdida de coherencia a temperatura cero}
\label{deconoohm}

En esta secci\'on, extenderemos lo realizado en la Secci\'on
\ref{decoohm} para un entorno \'ohmico, es decir, analizaremos
la p\'erdida de coherencia de una superposici\'on inicial de 
dos paquetes gaussianos deslocalizados (separados una
distancia $2 L_0$ en posici\'on), en el caso
que la part\'icula est\'a acoplada a un entorno no \'ohmico a
temperatura cero. Como dijimos \mbox{anteriormente}, el coeficiente 
de visibilidad $A_{\rm int}$ resulta un buen indicador de 
este proceso. Este coeficiente cumple con la ecuaci\'on de
evoluci\'on temporal (\ref{aintcap2}), donde ${\cal D}(t)$ y $f(t)$
son, en este caso,  los coeficientes difusivos no \'ohmicos
correspondientes.

Comenzaremos por el entorno supra\'ohmico. La din\'amica de este
tipo de entorno es bastante peculiar, ya que todo lo que ocurre en
el sistema browniano se debe al impulso inicial en tiempos 
del orden $t \leq t_{\Lambda}$  \cite{HuPazZhangI}. 
El coeficiente de visibilidad $A_{\rm int}$
puede ser estimado anal\'iticamente para distintas situaciones 
f\'isicas. Por ejemplo, si calculamos los coeficientes de difusi\'on
  a partir de las Ecs.(\ref{Dsupohm}) y (\ref{fsupohm}) 
para tiempos cortos $\Omega t \ll 1$ 
(pero  $t > t_{\Lambda}$) se verifica f\'acilmente que,
\begin{equation}
D(t)_{n=3} \sim (2M\gamma_0)/(\pi\Lambda^2)\Omega^4 t
~~~~~{\rm y}~~~~~
f(t)_{n=3} \sim - (2\gamma_0)/(\pi)\Omega t.
\end{equation}
Reemplazando estos coeficientes en la Ec.(\ref{aintcap2}), obtenemos
la evoluci\'on temporal de ${A}_{\rm int}$
\begin{equation}
\dot {\rm A}_{\rm int} \sim 4 \gamma_0\Omega t 
(1 + 2 L_0^2 M\Omega^3/\Lambda^2).
\end{equation}
De esta forma, podemos estimar el factor de visibilidad de las
franjas como
\begin{equation}
{\rm A}_{\rm int} \sim 2\frac{\gamma_0}{\Omega}\left(1
 + 2L_0^2M \frac{\Omega^3}{\Lambda^2}\right)\Omega^2t^2,
\end{equation}
el cual es menor que uno, en particular  si se verifica que 
$\Lambda \gg \Omega$. Esto implica que el factor de
p\'erdida de coherencia $\Gamma(t)=\exp(-A_{\rm int})$
no es una funci\'on exponencial decreciente y la p\'erdida
de coherencia no ser\'a efectiva en este caso.
Por otro lado, si pedimos tanto $\Lambda t \gg 1$ como
 $\Omega t \geq 1$, ambos coeficientes difusivos se
anulan r\'apidamente, ya que
\begin{equation}
D(t)_{n=3} \sim 2M\gamma_0\Lambda 
\cos (\Lambda t)/\Lambda t, ~~~~~{\rm y}~~~~~ f(t)_{n=3} 
\sim \gamma_0,
\end{equation}
con $\gamma_0 \ll 1$ porque estamos en el r\'egimen subamortiguado.
De esta forma, $\dot {\rm A}_{\rm int} \rightarrow 0$ y
el coeficiente de visibilidad permanece constante en el tiempo. 
Podr\'iamos estimar su valor aproxim\'andolo por el valor
que alcanza a tiempos largos de forma de asegurar la continuidad
de este coeficiente, es decir 
$\rm A_{\rm int} \approx 2 M L_0^2 \gamma_0$. Resulta evidente, entonces,
que no hay p\'erdida de coherencia en este caso, salvo para 
valores de $\gamma_0$ no v\'alidos en este modelo  (valores que no
cumplen con la condici\'on del r\'egimen subamortiguado). Para soluciones
 m\'as generales se puede mirar una referencia muy 
reciente  \cite{Hu2}. Por lo dicho anteriormente, el factor de 
p\'erdida de coherencia resulta constante ($\Gamma \sim
e^{-2ML_0^2\gamma_0}$) para todo tiempo. En el r\'egimen subamortiguado
el exponente nunca ser\'a de orden uno y $\Gamma$ no se har\'a cero. 
Los efectos de p\'erdida de coherencia ser\'an \'unicamente
relevantes cuando se trate de trayectorias macrosc\'opicamente distinguibles, 
es decir $ML_0^2 \geq 1/\gamma_0$. Estos resultados son  llamativamente
diferentes al caso supra\'ohmico a  temperatura alta, donde la p\'erdida
de coherencia ocurre en una escala temporal corta $
t_D ^{n=3,HT}\sim (\Lambda M k_B T L_0^2\gamma_0)^{-1/2}$,
de manera muy similar a un entorno \'ohmico a la misma temperatura (para 
m\'as detalles del entorno supra\'ohmico a  temperatura alta ver el Ap\'endice B).
\begin{figure}[!ht]
\centering
\includegraphics[width=10cm]{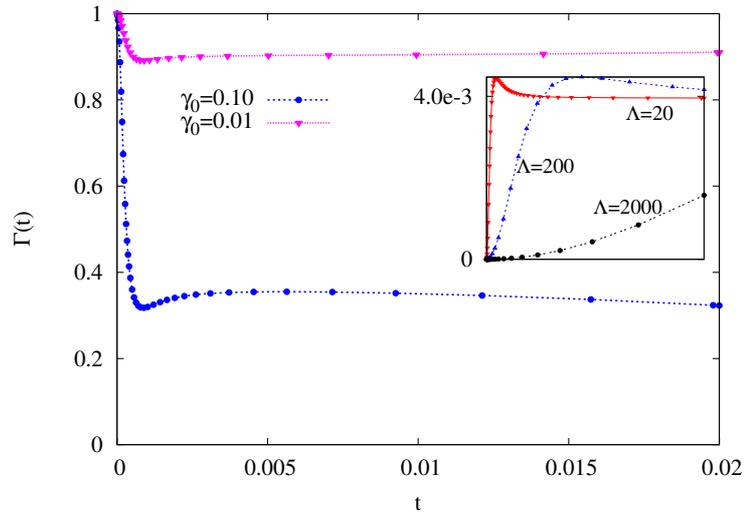}
\caption{Factor de p\'erdida de coherencia $\Gamma(t)$
para un entorno supra\'ohmico ($n=3$)  a temperatura cero.
Los par\'ametros est\'an medidos en unidades de la frecuencia
natural del sistema $\Omega$. Consideramos los casos
$\Lambda=2000$ ($\Lambda=200$ si es indicado en la Figura), 
$\Omega= 0.1$, $L_0=2$ para distintos valores de $\gamma_0$.
En el recuadro, mostramos la evoluci\'on temporal
de la Entrop\'ia Lineal $Sl(t)$ para distintos valores de la frecuencia
de corte $\Lambda$ cuando $\gamma_0= 0.5$. No hay p\'erdida
de coherencia en el r\'egimen subamortiguado para
el entorno supra\'ohmico a temperatura cero.}
\label{decosupcap2}
\end{figure}

En la Fig.\ref{decosupcap2}, mostramos el comportamiento del 
factor de p\'erdida de coherencia $\Gamma(t)$ para 
valores distintos de la constante de acoplamiento en el caso
de un entorno supra\'ohmico a temperatura cero. Como es
esperable, a mayor acoplamiento, mayor decaimiento de la funci\'on 
$\Gamma(t)$. Sin embargo,  este factor nunca se hace estrictamente
cero. Por ejemplo, para $\gamma_0=0.01$, se observa que
$\Gamma(t)  \approx 0.9$, valor que alcanza durante el transitorio
inicial y mantiene despu\'es durante su evoluci\'on posterior. Este 
comportamiento se explica considerando que, cuanto mayor
es el acoplamiento con el entorno, m\'as grande es
la ca\'ida inicial que sufre el factor de decoherencia 
del sistema y, m\'as importante
resulta la atenuaci\'on inicial de  las
franjas de interferencia. Sin embargo,  nunca llegan a suprimirse 
completamente
como en el caso \'ohmico que ya estudiamos. Del mismo modo,
podemos observar en el recuadro superior, que la Entrop\'ia
Lineal nunca llega a saturar (ni se aproxima al valor m\'aximo
posible para un estado mixto) como en el caso \'ohmico.

El c\'aracter no disipativo del entorno supra\'ohmico se debe a una combinaci\'on
de dos factores; por un lado, de la forma de la densidad espectral en el
sector infrarrojo, y por el otro, de la dependencia con $\Lambda$,
ya que es m\'as sensible al sector ultravioleta del mismo. El entorno
supra\'ohmico puede ser visto como un modelo de juguete 
de un escenario t\'ipico de la Teor\'ia Cu\'antica de Campos (TCC).
En la Ref.\cite{ferdiego}, se demostr\'o las condiciones
que deb\'ian cumplirse para que, efectivamente, hubiera
p\'erdida de coherencia a $T=0$ en el caso de un campo
con interacciones no lineales. El caso supra\'ohmico
es d\'ebilmente difusivo ya que su coeficiente hom\'onimo
tiende a cero despu\'es del transitorio inicial. En este caso,
la p\'erdida de coherencia depende fuertemente del valor de 
la constante de acoplamiento con el entorno. En TCC, los
efectos difusivos se deben a la creaci\'on  de
part\'iculas en el entorno, debido a la interacci\'on con el
sistema. Cuando hay una frecuencia de corte en el entorno, s\'olo
los modos del sistema que tienen una frecuencia parecida o
del orden de \'esta, tienen posibilidades de crear part\'iculas
y por lo tanto, perder coherencia. Esta es la raz\'on por la cual, en TCC,
el coeficiente de difusi\'on es no nulo s\'olo para algunos valores
particulares de los par\'ametros. Este resultado es bastante similar
a los resultados que hemos obtenido a lo largo de esta secci\'on
para un entorno supra\'ohmico. En el modelo que estamos estudiando
aqu\'i, la relaci\'on entre los par\'ametros $\Omega$, $\Lambda$ y 
$\gamma_0$ es crucial para obtener efectos difusivos en el sistema.
En particular, si el entorno est\'a a temperatura cero, y 
 adem\'as $\Omega \ll \Lambda$,
el sistema es incapaz de excitarlo 
 para ``crear part\'iculas'' \cite{ferdiego}. 

Como conclusi\'on, podemos decir que mientras el entorno supra\'ohmico 
a temperatura alta resulta muy efectivo induciendo p\'erdida
de coherencia en el sistema bajo ciertas condiciones \cite{HuPazZhangI,PLA},
a temperatura cero \'esto no es as\'i, ya que, para que realmente haya efectos
difusivos en el sistema, se debe cumplir una condici\'on muy fuerte
sobre $\gamma_0$ \cite{PLA2}. Finalmente, en la Fig.\ref{comohmsupcap2},
mostramos  los coeficientes de p\'erdida
de coherencia para un entorno \'ohmico y otro supra\'ohmico,
tanto en el caso que  \'estos se hallen  a temperatura alta como cero. 
La p\'erdida de coherencia resulta tan r\'apida en el caso
supra\'ohmico a temperatura alta como en el \'ohmico a la
misma temperatura. Sin embargo, cuando el entorno 
 est\'a a temperatura cero, no hay p\'erdida de coherencia
en el caso supra\'ohmico. El entorno \'ohmico a \mbox{temperatura}
cero logra  finalmente supramir las franjas de interferencia 
en una escala temporal m\'as larga, como se puede corroborar
en el recuadro de la Fig.\ref{comohmsupcap2}.
\begin{figure}[!ht]
\centering
\includegraphics[width=10cm]{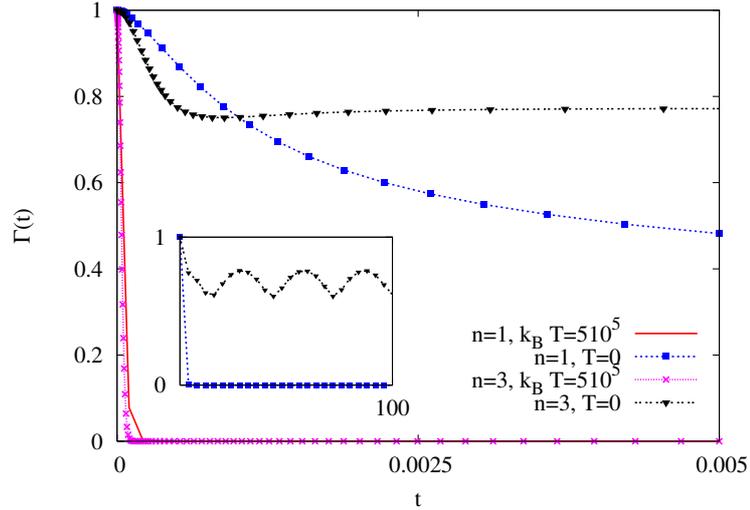}
\caption{Comparaci\'on entre las tasas de p\'erdida de 
coherencia para el entorno \'ohmico y supra\'ohmico en
los casos de temperatura alta y cero. La p\'erdida de coherencia
ocurre en el entorno supra\'ohmico  tan r\'apido como 
en el caso \'ohmico en el l\'imite de  temperatura alta ($ \hbar \Lambda \ll
k_B T$); mientras que para esa escala temporal, a\'un no hay evidencia
de este proceso cuando la temperatura del entorno es cero.
En el recuadro, mostramos el comportamiento del coeficiente 
$\Gamma(t)$ para una escala temporal m\'as larga.
En el caso supra\'ohmico no hay p\'erdida de coherencia
mientras que en el \'ohmico el entorno finalmente logra
destruir las franjas de interferencia del sistema de prueba.
Los par\'ametros usados: $\Lambda=2000$,
$\gamma_0= 0.1$, $L_0=1$ y est\'an medidos en unidades de
 $\Omega$.}
\label{comohmsupcap2}
\end{figure}\\

En cuanto al caso de un entorno sub\'ohmico, podr\'iamos repetir el mismo
procedimiento que realizamos para los otros dos casos, y calcular
el coeficiente de visibilidad $A_{\rm int}$. Por un lado, si se verifica que  
$\Lambda t \gg 1$ y $\Omega \geq 1$, de la Ec.(\ref{aintcap2})  se obtiene 
la siguiente  ecuaci\'on para la variaci\'on temporal de $A_{\rm int}$:
\begin{equation}
\dot {\rm A}_{\rm int} \sim \gamma_0 \Lambda \left(2 M L_0^2+
\frac{\Gamma_{\rm Euler}}{\Omega}+ \frac{\log(2 \Lambda t)}{\Omega}\right).
\end{equation}
Integrando esta ecuaci\'on, podemos estimar
el valor del coeficiente de visibilidad $A_{\rm int}$ como $\rm A_{\rm int} \sim$
$ \gamma_0 \Lambda t/\Omega  \log(2 \Lambda t)$. A partir de esta
expresi\'on, es f\'acil obtener una cota para el tiempo de p\'erdida de coherencia 
$ t_D \leq \Omega/(\gamma_0\Lambda)$.
Del mismo modo, si se cumple  $\Omega t \ll 1$, obtenemos la misma
escala temporal, ya que los coeficientes que hemos podido
estimar en el caso sub\'ohmico, dependen d\'ebilmente del cociente
$\Omega /\Lambda$. Es importante remarcar que, para poder
despreciar el transitorio inicial, uno debe verificar que se cumpla
la relaci\'on 
$\Omega /\gamma_0 > 1$.
\begin{figure}[!ht]
\centering
\includegraphics[width=10cm]{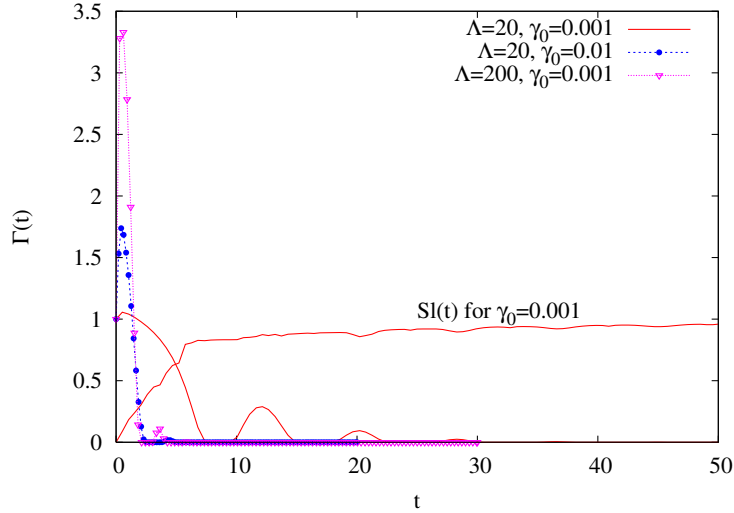}
\caption{Coeficiente $\Gamma(t)$ para el caso de una part\'icula
de prueba acoplada a un entorno sub\'ohmico ($n=1/2$)  a temperatura cero.
Los par\'ametros est\'an medidos en unidades de la frecuencia natural
$\Omega$. Consideramos los casos $\Omega=0.1$, $L_0=2$
para distintos valores de $\gamma_0$ y de la frecuencia de corte 
$\Lambda$. Adem\'as se muestra la evoluci\'on temporal de la Entrop\'ia
Lineal $Sl(t)$ para un caso.} \label{decosubcap2}
\end{figure}

Podemos verificar nuestras estimaciones num\'ericas con la ayuda de
la Fig.\ref{decosubcap2}, donde presentamos la evoluci\'on del
coeficiente $\Gamma(t)$ para un sistema Browniano acoplado
a un entorno sub\'ohmico a temperatura cero. Este coeficiente tiene
un rasgo peculiar en este caso, ya que inicialmente crece (a diferencia
de los otros casos), pero luego, inmediatamente comienza a decrecer
hasta que se hace cero \cite{HuPazZhangI}. Podemos notar en 
dicha figura, que su dependencia con la frecuencia de corte y la constante
de acoplamiento es similar a la observada en los otros dos entornos
(cuanto m\'as grandes $\Lambda$ y $\gamma_0$, m\'as r\'apido 
$\Gamma(t)$ tiende a cero). En la figura, adem\'as, presentamos
una curva para la evoluci\'on de la Entrop\'ia Lineal, donde 
se observa claramente que esta cantidad alcanza su valor de saturaci\'on.
Este hecho coincide, pr\'acticamente, con el tiempo en el cual la curva 
 de $\Gamma(t)$ correspondiente, disminuye considerablente, para
luego hacerse cero.

Finalmente,  en la 
Fig.\ref{comohmsubcap2}, mostramos una comparaci\'on
entre los coeficientes $\Gamma(t)$ para un entorno \'ohmico y uno
sub\'ohmico, en el caso de temperatura alta y cero.
All\'i, se puede observar que el entorno sub\'ohmico es muy efectivo induciendo
p\'erdida de coherencia en el sistema, tanto a temperatura alta como en el
caso que su temperatura sea cero.
\begin{figure}[!ht]
\centering
\includegraphics[width=10cm]{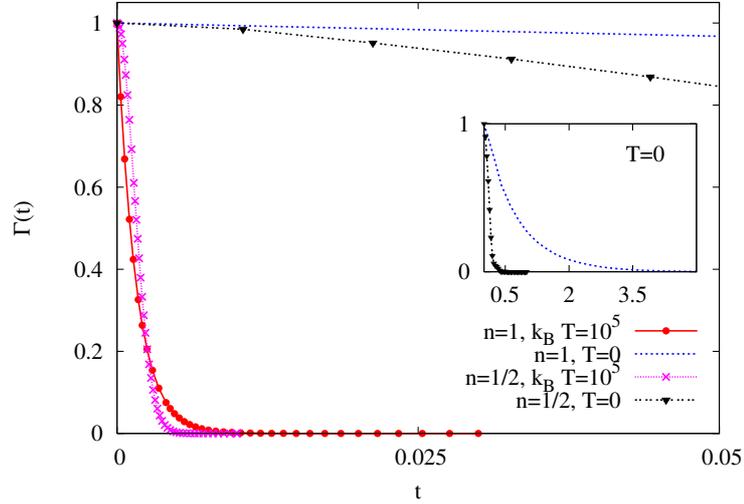}
\caption{Comparaci\'on de los coeficientes de p\'erdida de
coherencia entre un entorno \'ohmico y uno sub\'ohmico,
tanto a temperatura alta como cero.  La p\'erdida de coherencia
es tan r\'apida en  el entorno sub\'ohmico como
en el \'ohmico a temperatura alta.  En el recuadro,
mostramos el comportamiento de los coeficientes
para ambos entornos a temperatura cero en una escala
temporal donde es posible observar que se anulan. Los
par\'ametros usados: $\Lambda=200$,
$\gamma_0=0.01$, $L_0=0.1$ y est\'an medidos en 
unidades de $\Omega$.} \label{comohmsubcap2}
\end{figure}

\section{Excitaci\'on energ\'etica debido a los efectos de ruido del entorno}
\label{excitacion}

El oscilador arm\'onico aislado, en su estado fundamental, obedece dos 
principios muy  \mbox{importantes}: por un lado, el de m\'inima incerteza y, por el
otro, el de equipartici\'on de la energ\'ia. Cuando estudiamos la din\'amica
de sistemas cu\'anticos abiertos, los efectos del entorno sobre el sistema
se manifiestan,  justamente, a trav\'es de la violaci\'on de dichos principios.
Como la energ\'ia del sistema es un observable, \'esta cantidad
ejemplifica la diferencia entre estados separables y estados
entrelazados.

En esta secci\'on, estudiaremos la existencia del fen\'omeno
de ``excitaci\'on energ\'etica"  inducido por el ruido, como
otra manifestaci\'on de los efectos difusivos  inducidos en el sistema 
debido a la interacci\'on con una ba\~no t\'ermico. 
Mostraremos que es un fen\'omeno 
posterior a la p\'erdida de coherencia en el sistema y, que
aparece a\'un cuando el entorno est\'a a 
temperatura cero. En el caso de un entorno a  temperatura alta, este
fen\'omeno est\'a asociado a la  ``activaci\'on t\'ermica", pero
aqu\'i hemos decidido llamarlo de otra forma, para hacer evidente
que se induce en el sistema  debido al ruido o fluctuaciones en 
el entorno y  
no depende, exclusivamente, de la temperatura  del mismo.

El fen\'omeno de ``activaci\'on t\'ermica" es un fen\'omeno conocido
 y ampliamente estudiado. 
Sin embargo, no ha sido estudiado
en el caso de entornos no \'ohmicos, ni tampoco se ha 
relacionado la escala en la cual el fen\'omeno se lleva
a cabo $t_{\rm act}$ con la escala de p\'erdida de coherencia $t_D$.
Por lo tanto, en esta secci\'on, analizaremos los dos r\'egimenes t\'ermicos
que resultan interesantes: el l\'imite de temperatura alta y el caso
de temperatura estrictamente cero. En cada uno de ellos, estudiaremos
si el sistema, una vez que perdi\'o sus franjas de interferencias y se volvi\'o
``cl\'asico", puede aumentar su energ\'ia media a costa de la interacci\'on
con el ba\~no de osciladores arm\'onicos.

Para todos los casos, evaluaremos num\'ericamente la cantidad 
\begin{equation}
\langle E(t)\rangle = \frac{1}{2M}\langle p^2\rangle (t) +
\frac{M{\tilde \Omega}^2(t)}{2}\langle x^2\rangle (t) ,\nonumber
\end{equation} 
donde  $\langle x^2\rangle = {\rm Tr}[\rho_{\rm r}(t) x^2]$
y $\langle p^2\rangle = {\rm Tr}[\rho_{\rm r}(t) p^2]$ se calculan 
usando la soluci\'on $\rho_{\rm r}(x,x',t)$ (que representa una
superposici\'on lineal de dos paquetes gaussianos
localizados inicialmente en $x_0=\pm L_0$) de la ecuaci\'on maestra
Ec.(\ref{master}) (ver Ap\'endice A).

\subsection{L\'imite de temperatura alta}
\label{excitacionHT}

Empezaremos estudiando el proceso de ``activaci\'on energ\'etica"
 en entornos \mbox{generales}, ya sean \'ohmicos o no \'ohmicos,
en el caso del l\'imite de temperatura alta ($\hbar \Lambda \ll k_B T$).
Para cada caso, daremos argumentos anal\'iticos y mostraremos
evidencia num\'erica de la existencia de este fen\'omeno. Ser\'a evidente,
adem\'as, que en todos los sistemas donde es observable este fen\'omeno,
las interferencias fueron efectivamente suprimidas en una escala temporal
anterior.

\subsubsection{Entornos \'ohmicos}
\label{ohmHT}

La tasa de activaci\'on para un sistema cl\'asico
puede ser deducida a partir de la ecuaci\'on de Fokker-Planck para la funci\'on
de Wigner, 
\begin{eqnarray}
\dot{W}=\{H_{\rm sys},W\}_{\rm PB} +2 \gamma_0 \partial_p(pW) +
{\cal D}
\partial^2_{pp}W, \label{fp}
\end{eqnarray}
la cual es el an\'alogo cl\'asico de la ecuaci\'on maestra cu\'antica.
La evoluci\'on cl\'asica media para cualquier observable f\'isico $A(x,p)$,
en este r\'egimen, se calcula seg\'un
\begin{equation}
\partial_t \langle A \rangle = -\langle\{H_{\rm sys},A\}_{\rm PB} \rangle +
 {\cal D} \langle \partial^2_p A \rangle - 2 \gamma_0 \langle p
\partial_p A\rangle. \label{Em}
\end{equation}
Si consideramos que $A(x,p)$ puede ser, por ejemplo, el Hamiltoniano
de la part\'icula de prueba, obtenemos $\partial_t \langle H \rangle 
= 2 \gamma_0 (k_BT - \langle p^2 \rangle)$ (con ${\cal D}= 
2 M \gamma_0 k_B T$ en este caso).  Esta expresi\'on puede 
ser  simplificada a\'un un poco m\'as, asumiendo que la temperatura $k_B T$
es mucho m\'as importante que las escalas relevantes de la energ\'ia,
es decir $ k_B T \gg \langle p^2 \rangle$, al menos durante los primeros
tiempos de la evoluci\'on. Como resultado, obtenemos
la dependencia temporal de la energ\'ia media del sistema seg\'un,
\begin{equation}
\partial_t \langle H \rangle =
2 \gamma_0 k_B T \,\,\,\rightarrow\,\,\, E= 2 \gamma_0 k_BT t + E_0~,
\label{EmediaHT}
\end{equation}
donde $E_0$ es la energ\'ia inicial del mismo. De esta forma, podemos
estimar el tiempo de activaci\'on $t_{\rm act}$, para este
caso sencillo, como
\begin{equation}
t_{\rm act}= \frac{E-E_0}{2 \gamma_0 k_BT}.
 \label{t_th}
\end{equation}

En la Fig.\ref{EohmHT} graficamos la evoluci\'on temporal de la energ\'ia media del
sistema (part\'icula Browniana) para un entorno \'ohmico. En dicha figura,
se observa que el comportamiento es proporcional a la temperatura 
del ba\~no como indica la Ec.(\ref{EmediaHT}). Inicialmente, el sistema sufre
un impulso inicial (al igual que hab\'iamos notado en las Secciones \ref{decoohm}
y \ref{deconoohm} para el factor p\'erdida de coherencia) debido a
que la interacci\'on con el entorno a tiempo cero es nula. El crecimiento inicial
es apenas un transitorio de escala temporal corta ($\sim 1/\Lambda$).
La energ\'ia crece proporcionalmente a $\gamma_0$ y $k_B T$  y no 
depende de la frecuencia de corte $\Lambda$. Es importante recordar
que nosotros estamos estudiando la  evoluci\'on din\'amica del sistema
para tiempos comprendidos entre $1/\Lambda \ll
t \ll t_{\rm sat} \sim 1/\gamma_0$.  Como ya se mencion\'o, la escala temporal
de p\'erdida de coherencia en este caso es 
$t_D\sim 1/(2 M \gamma_0 k_B T L_0^2)$ \cite{HuPazZhangI}, la cual
resulta una escala muy chica para los tiempos de la figura \ref{EohmHT}(a).
En el recuadro, podemos observar el tiempo en el cual
la Entrop\'ia Lineal del sistema alcanza su valor m\'aximo. Para este
tiempo, las interferencias ya han sido suprimidas. Adem\'as, all\'i mismo 
se indica
la energ\'ia media del sistema aislado, la cual siempre es menor que en
el resto de los casos donde el sistema es abierto. En la Fig.\ref{EohmHT}(b)
resulta evidente la dependencia de la energ\'ia media del sistema con
la constante de acoplamiento $\gamma_0$ para una temperatura y 
frecuencia de corte fijas. La evoluci\'on a tiempos largos se encuentra
en \cite{dwPRE,dwJCS}, donde se puede observar
que el sistema alcanza un estado de equilibrio (el costo num\'erico 
para ir a tiempos $t \sim 1/\gamma_0$ es muy grande).

\begin{figure}[!ht]
\centering
\includegraphics[width=17cm]{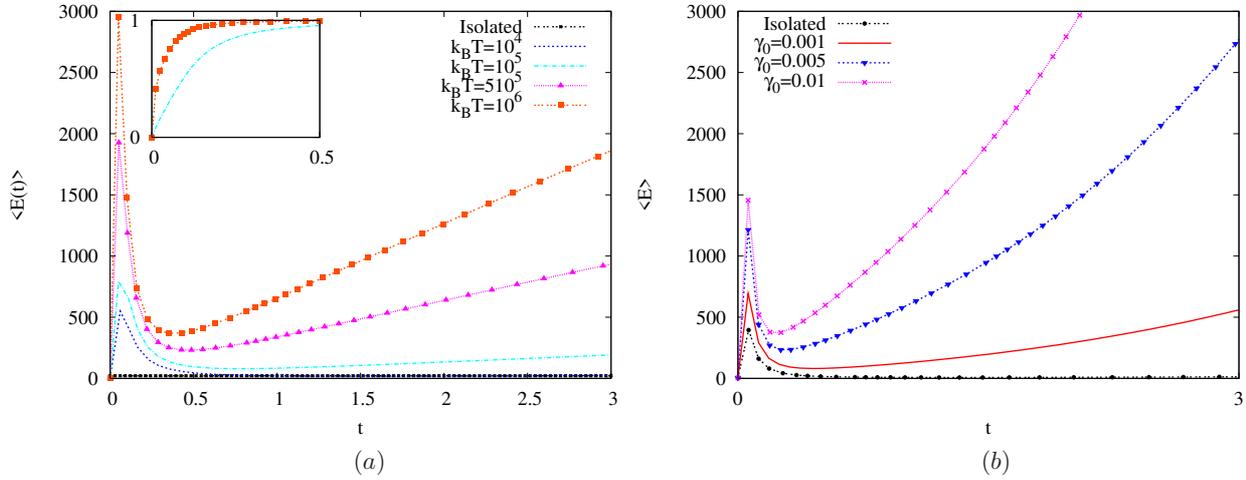}
\caption{(a) Evoluci\'on temporal de la energ\'ia media del sistema
acoplado a un entorno \'ohmico ($n=1$) en el l\'imite de 
temperatura alta. Consideramos los casos 
$\gamma_0=0.001$, $\Lambda=2000$, $\Omega=0.1$, $L_0=2$
para distintas temperaturas del entorno.  En el recuadro, mostramos
la evoluci\'on temporal de la Entrop\'ia Lineal $Sl(t)$ para dos casos
distintos: $k_B T=10^5$ y $k_B T=10^6$ (los mismos colores
que en el gr\'afico principal).
(b) Evoluci\'on temporal de la energ\'ia media del sistema
acoplado a un entorno \'ohmico ($n=1$) en el l\'imite de alta
temperatura. Consideramos los casos $k_BT=10^5$,
$\Lambda=2000$, $\Omega=0.1$, $L_0=2$  para distintos valores
de la constante de acoplamiento $\gamma_0$ y la frecuencia de corte
$\Lambda$.  El sistema se 
``activa" antes cuanto m\'as grande es el valor de $\gamma_0$.
Los par\'ametros son medidos en unidades de la frecuencia natural
del sistema $\Omega$ en todos los casos.}
\label{EohmHT}
\end{figure}

\subsubsection{Entornos no \'ohmicos}
\label{noohmHT}

El an\'alisis previo, correspondiente a entornos \'ohmicos, tambi\'en 
puede ser realizado para aquellos entornos que no lo son. Sin embargo,
las expresiones anal\'iticas no son tan sencillas ya que, en los entornos
m\'as generales, no se cumple que ${\cal D}(t)$ sea constante ni que $f(t)$
sea despreciable. Por lo tanto, nos restringiremos a un an\'alisis num\'erico,
\'unicamente basado en los estudios del proceso de p\'erdida
de coherencia que existen en la Literatura (en el caso de  temperatura alta) 
\cite{GraIng,HuPazZhangI} y las estimaciones anal\'iticas que nosotros hemos
realizado y se presentan en el Ap\'endice B de esta Tesis.

Comenzaremos por el entorno supra\'ohmico. En las Fig.\ref{fig4c}(a)
and Fig.\ref{fig4c}(b), presentamos la evoluci\'on temporal  para la energ\'ia 
media del sistema para diferentes valores de la constante de acoplamiento
y la temperatura del entorno, respectivamente. En ambas figuras podemos
ver que el sistema aumenta m\'as su energ\'ia media a medida que
 $\gamma_0$ es m\'as grande y el entorno m\'as caliente.
\begin{figure}[!ht]
\centering
\includegraphics[width=16cm]{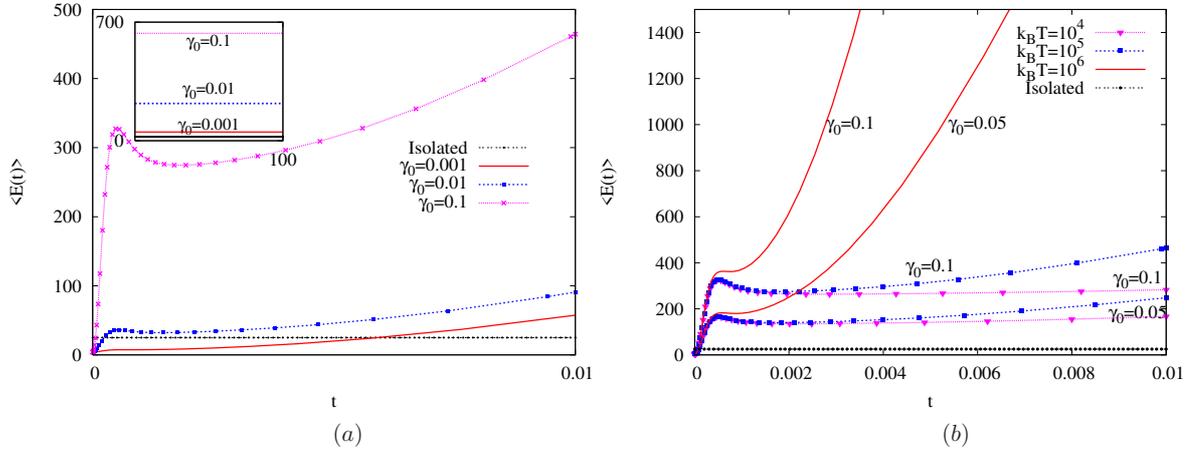}
\caption{(a) Evoluci\'on temporal de la energ\'ia media del sistema
acoplado a un entorno supra\'ohmico ($n=3$) en el l\'imite de tenperatura alta.
Consideramos los casos $k_BT=10^5$, $\Lambda=2000$,
$\Omega=0.1$, $L_0=2$ para diferentes valores de $\gamma_0$.
En el recuadro, graficamos la energ\'ia media para tiempos m\'as 
largos. (b) Evoluci\'on temporal de la energ\'ia media del sistema
acoplado a un entorno supra\'ohmico ($n=3$) en el l\'imite de tenperatura alta.
Consideramos los casos $\gamma_0=0.001$, $\Lambda=2000$,
$L_0=2$  para distintas temperaturas del entorno y valores de 
$\gamma_0$.} \label{fig4c}
\end{figure}

La ``intensidad" de un entorno est\'a dada por la relaci\'on entre los tres
par\'ametros del mismo: $\gamma_0$, $k_BT$ y $\Lambda$.
Para un entorno supra\'ohmico ``fuerte" 
($2Mk_BTL_0^2\gamma_0 \gg \Lambda$), la p\'erdida de coherencia
sucede en escalas temporales
 $t_{\cal D}\sim (\Lambda M \gamma_0 k_B T L_0^2)^{-1/2}$  
(como se muestra en el Ap\'endice B). En esos casos, se puede
observar que la energ\'ia media del sistema crece considerablemente.
Para entornos supra\'ohmicos ``d\'ebiles", es decir 
$Mk_BTL_0^2\gamma_0<\Lambda$, como por ejemplo, 
$\gamma_0=0.001$ en la  Fig.\ref{fig4c}(a), la energ\'ia media del
sistema no presenta un crecimiento considerable. Esto se debe
a que las interferencias del sistema no han sido efectivamente suprimidas
 (Fig.\ref{figsupapb} del Ap\'endice B), ya que los
efectos difusivos inducidos en el sistema no resultan importantes. 
Por esa misma raz\'on, el intercambio de energ\'ia entre el sistema y el entorno no est\'a
completamente destinado a la excitaci\'on energ\'etica del sistema 
(el entorno a\'un
intenta suprimir las franjas de interferencia del sistema). Este caso 
difiere cualitativamente
del caso \'ohmico. En particular, alcanza un r\'egimen asint\'otico que, el
\'ultimo, s\'olo logra a escalas del tiempo de saturaci\'on del entorno. En el
recuadro superior de la Fig.\ref{fig4c}(a), mostramos la evoluci\'on de la energ\'ia media del
sistema a tiempos m\'as largos, donde se puede observar claramente
que el sistema entra en un r\'egimen asint\'otico y su energ\'ia media se
mantiene constante.

\begin{figure}[!ht]
\centering
\includegraphics[width=9cm]{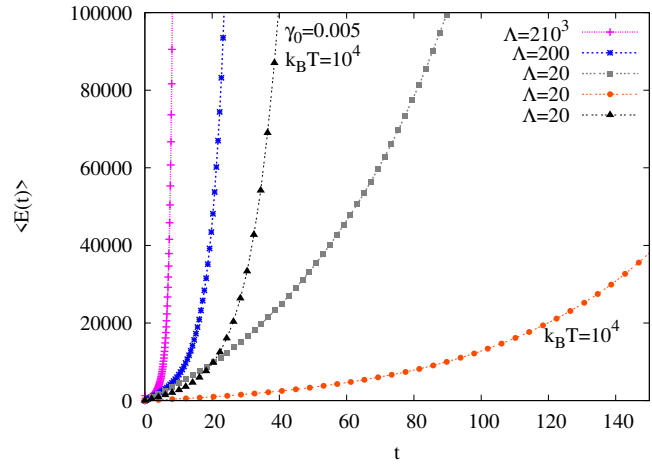}
\caption{Evoluci\'on temporal de la energ\'ia media del sistema
acoplado a un entorno sub\'ohmico ($n=1/2$) en el l\'imite de tenperatura alta.
Consideramos los casos 
$\Omega=0.1$, $L_0=2$, $\gamma_0=0.001$ (y $\gamma_0=0.005$ 
cuando es indicado) and $k_B T=10^5$ ($k_B T=10^4$ cuando es indicado)
para distintos valores de la frecuencia de corte $\Lambda$.  
Los par\'ametros est\'an medidos en unidades de la frecuencia
natural del sistema $\Omega$ en todos los casos..} \label{fig4cc}
\end{figure}
Finalmente, la activaci\'on t\'ermica o energ\'etica es muy clara en el 
entorno sub\'ohmico. En la Fig.\ref{fig4cc} graficamos la energ\'ia
media del sistema para diferentes valores de la constante de acoplamiento
($\gamma_0=0.001$ y $\gamma_0=0.005$), diferentes temperaturas
($k_BT=10^5$ y $k_BT=10^4$) y diferentes frecuencias de corte
en el l\'imite de temperatura alta. En este caso, es f\'acil ver que la energ\'ia
crece m\'as r\'apido a medida que $\gamma_0$, $\Lambda$ y $k_B T$
aumentan. Si recordamos el comportamiento del factor de 
p\'erdida de coherencia $\Gamma(t)$ (Fig.\ref{figsubapb} del Ap\'endice
B), podemos verificar, nuevamente, que la activaci\'on energ\'etica se lleva a cabo 
a tiempos posteriores al tiempo de p\'erdida de coherencia 
$t_{\cal D}\sim (M \gamma_0 L_0^2 k_B T)^{-1}$.

\subsection{Temperatura cero}
\label{ET0}

Como ya hemos mencionado, a\'un hoy en d\'ia, en algunas ramas de la
f\'isica, existe la idea err\'onea de que la p\'erdida de coherencia
tiende a cero como funci\'on de la temperatura, y por lo tanto, no
existe tal fen\'omeno en el caso de un entorno a temperatura extrictamente
cero. Si esto fuera cierto, la f\'isica deber\'ia ser cualitativamente 
diferente en el caso de ausencia de temperatura en el entorno.
Muchas preguntas surgir\'ian en ese caso. \textquestiondown Qu\'e esperamos
encontrar si un sistema cu\'antico est\'a acoplado a un entorno
con temperatura cero? \textquestiondown Es posible que un sistema se ``active",
como en el caso de temperatura alta, si el entorno carece de temperatura?
\textquestiondown C\'omo se justificar\'ia semejante fen\'omeno?

En la Secci\'on \ref{ohmcap2}, ya hemos demostrado que efectivamente hay
p\'erdida de coherencia a\'un en el caso en que la temperatura del
entorno es estrictamente cero. Incluso, hemos estimado las escalas
temporales a las cuales ocurre dicho proceso y hemos mostrado dicho
efecto tambi\'en en forma num\'erica. Nuestra pr\'oxima afirmaci\'on
es que, efectivamente, tambi\'en hay activaci\'on energ\'etica en este caso.
El sistema se ``activa" energ\'eticamente, ya no inducido por la temperatura
del entorno sino por sus fluctuaciones cu\'anticas.  M\'as a\'un, mostraremos
que existe una relaci\'on muy estrecha entre la p\'erdida de coherencia
y la activaci\'on de la energ\'ia, ambos procesos inducidos por la interacci\'on
entre el sistema y el entorno. Los sistemas que sufren una mayor o total
p\'erdida de coherencia, son aquellos cuya energ\'ia tiene un 
crecimiento m\'as evidente.

Cuando se intenta interpretar el comportamiento posterior a la p\'erdida
de coherencia del sistema, varios factores de la din\'amica deben ser
tenidos en cuenta. En primer lugar, hay que destacar que el estado
fundamental del sistema no corresponde al estado fundamental del sistema total
formado por la part\'icula y su entorno. Tan
pronto como la interacci\'on entre el sistema y el entorno se enciende
a tiempo $t=0$, el sistema se encontrar\'a en un estado 
excitado del mismo.  El entorno tendr\'a, entonces, una cantidad no
nula de energ\'ia en relaci\'on al nuevo estado inicial. Desde el punto
de vista netamente cl\'asico, esta energ\'ia no puede ser responsable
de la excitaci\'on de la part\'icula de prueba a niveles energ\'eticos
m\'as altos. Este argumento puede ser  explicado de forma m\'as cuantitativa
de la siguiente forma. El potencial total, es decir de la part\'icula m\'as el
entorno, es 
\begin{equation}
V(x,q_n) = V_{\rm sys}(x) + V_{\rm env}(q_n) + V_{\rm int}(x,q_n),
\end{equation}
con $V_{\rm sys}(x) = -\frac{1}{2} \Omega^2 x^2$, 
$V_{\rm env}(q_n)= \sum_n \frac{1}{2} \omega^2 m_n^2 q_n^2$
y $V_{\rm int}(x,q_n)= \sum_n C_n x q_n$.
Cl\'asicamente, la condici\'on inicial es $x=0$, y, como adem\'as el entorno
est\'a a temperatura $T=0$, resulta $q_n=0$. Luego, para la acci\'on total,
los t\'erminos de energ\'ia de la condici\'on inicial son: $V_{\rm sys} =0$
(valor m\'inimo de $V_{\rm sys}$), $V_{\rm env}=0$, y  $V_{\rm
int}=0$. Por lo tanto, la condici\'on inicial de la energ\'ia  potencial del 
sistema total es $V=0$. Resulta importante que, desde el punto de vista
cl\'asico, la energ\'ia total del sistema total es igual al valor de la energ\'ia
del sistema aislado, a\'un cuando la interacci\'on se ``enciende" a $t=0$.
Esta es una consecuencia importante de que la temperatura del entorno es cero.

Las fluctuaciones cu\'anticas del entorno, presentes en el estado inicial
del mismo, claramente deben jugar alg\'un papel relevante en la activaci\'on.
Estas fluctuaciones ya no son fluctuaciones de vac\'io del sistema total.
Igualmente,  resulta sorprendente que tengan tanta injerencia en la evoluci\'on del
sistema. La naturaleza puramente cu\'antica del entorno, la cual puede
ser correctamente despreciada en el caso de temperatura alta, da lugar
a efectos importantes en el caso del entorno a temperatura cero.

En t\'erminos de la ecuaci\'on maestra, las fluctuaciones cu\'anticas del
entorno generan t\'erminos difusivos no nulos, tanto normales ${\cal D}(t)$
como an\'omalos $f(t)$. Esto es particularmente cierto, en el caso del
coeficiente  $f(t)$. En el caso \'ohmico, hemos demostrado  en la
Secci\'on \ref{ohmcap2} que dicho 
coeficiente depende logar\'itmicamente de la frecuencia de corte $\Lambda$,
con lo cual puede ser de gran magnitud \cite{PLA}. Los efectos difusivos,
inducidos por las fluctuaciones cu\'anticas del entorno, son la mayor
diferencia con el caso de alta temperatura, y resultan responsables, 
a $T=0$ de la excitaci\'on energ\'etica de la part\'icula. A pesar que
este proceso es muy diferente a la ``activaci\'on t\'ermica", creemos
que a\'un as\'i puede ser interpretado en funci\'on de un escenario
cl\'asico modificado. La mayor dificultad para encontrarlo, es lograr 
que un ba\~no t\'ermico
cl\'asico simule las propiedades cu\'anticas de un entorno a $T=0$.
Considerando las versiones cu\'anticas y cl\'asicas del n\'ucleo de
ruido $\nu(s)$, es posible mostrar que un conjunto de osciladores
cl\'asicos con una temperatura dependiente de la frecuencia, es decir 
$T(\omega)=\hbar \omega/2$, reproduce los efectos del estado
inicial cu\'antico. En particular, para esta elecci\'on de ba\~no cl\'asico,
se obtienen los mismos t\'erminos difusivos ${\cal D}(t)$ y $f(t)$ que 
en el caso de $T=0$. Es decir, una vez que se
lleva a cabo el proceso de p\'erdida de coherencia en el sistema,
un entorno puramente cu\'antico a $T=0$ deber\'ia influenciar a la 
part\'icula (ahora cl\'asica) de la misma forma que 
un ba\~no cl\'asico cuyos osciladores est\'an excitados de forma
particular (tal que simulan las fluctuaciones del correspondiente 
entorno cu\'antico). 

En lo que sigue, mostraremos ejemplos del
 fen\'omeno de activaci\'on energ\'etica 
en diferentes entornos a temperatura estrictamente
cero.

\subsubsection{Entornos \'ohmicos}
\label{EohmT0}

En la Fig.\ref{fig6c} mostramos la evoluci\'on temporal de la energ\'ia media
del sistema para distintos valores de $\Lambda$ y $\gamma_0$ en el
caso que la part\'icula est\'a acoplada a un entorno \'ohmico.
\begin{figure}[!ht]
\centering
\includegraphics[width=16cm]{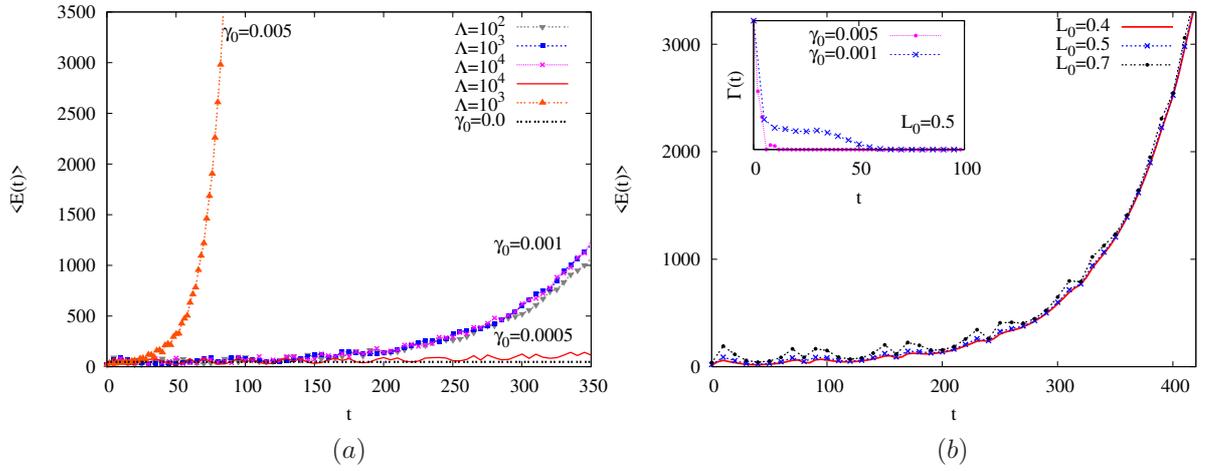}
\caption{(a)  Evoluci\'on temporal de la energ\'ia media
para un sistema cu\'antico acoplado a un entorno \'ohmico
a temperatura cero. Se ve claramente la dependencia de la
energ\'ia con la constante de acoplamiento $\gamma_0$ pero
no con la frecuencia de corte $\Lambda$. (b) Evoluci\'on 
temporal de la energ\'ia media
para un sistema cu\'antico acoplado a un entorno \'ohmico
a temperatura cero. Consideramos los casos $\gamma_0=0.001$,
$\Omega=15$, $\Lambda=1000$ para distintos valores de $L_0$.
En el recuadro superior, mostramos la evoluci\'on 
temporal del factor de p\'erdida de coherencia $\Gamma(t)$
para distintos valores de $\gamma_0$. La excitaci\'on energ\'etica 
empieza una vez que la p\'erdida de coherencia se hizo efectiva.
Los par\'ametros est\'an medidos en unidades de la frecuencia 
natural $\Omega$ y fueron elegidos en base a una conveniencia
num\'erica.}
\label{fig6c}
\end{figure}
\begin{figure}[!ht]
\centering
\includegraphics[width=16cm]{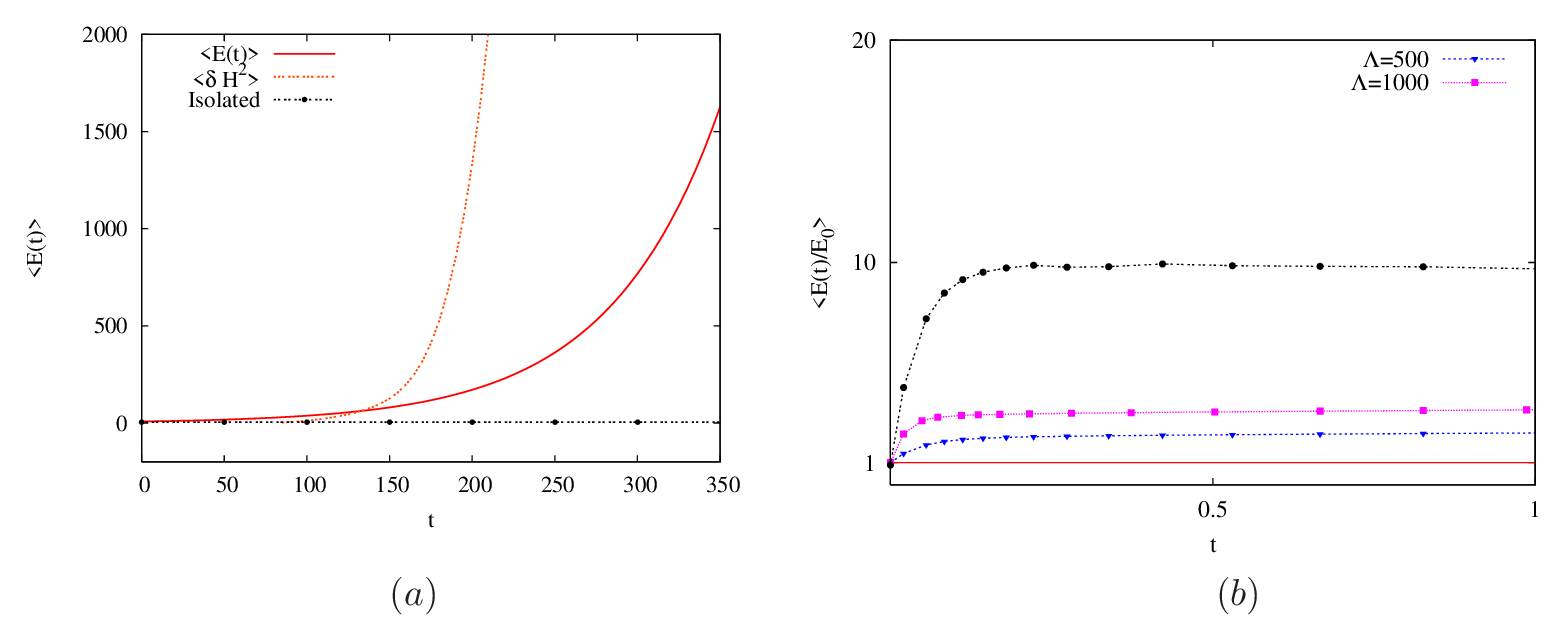}
\caption{(a) Evoluci\'on temporal de la energ\'ia media del sistema
y de la dispersi\'on media de la energ\'ia para un entorno \'ohmico
a temperatura cero. Adem\'as, presentamos la evoluci\'on temporal
de la energ\'ia media del sistema aislado. Los par\'ametros usados son:
$\Lambda=10^3$, $\Omega=15$, $L_0=0.5$ y
$\gamma_0=0.001$. Las fluctuaciones de la energ\'ia resultan 
fundamentales en el proceso de activaci\'on de la energ\'ia. Los par\'ametros 
est\'an medidos en unidades de la frecuencia natural del sistema 
$\Omega$. (b) Comportamiento de la energ\'ia media del sistema
a tiempos largos para el caso de un entorno \'ohmico a temperatura cero.
En todos los casos, la energ\'ia media del sistema es mayor  que la energ\'ia
media del sistema aislado (l\'inea s\'olida).
Los par\'ametros 
est\'an medidos en unidades de $\gamma_0$ para poder incluir varias
corridas en el mismo gr\'afico. Consideramos los casos: 
$\Omega=1$ y $\gamma_0=0.1$ y para distintos
valores de $L_0$: $L_0=0$, que representa una \'unica gaussiana (cuadrados
y tri\'angulos) y   $L_0=0.5$ (c\'irculos).  $E_0$ es la energ\'ia del sistema
aislado. Los par\'ametros fueron deliberadamente seleccionados
en funci\'on del costo num\'erico de las corridas a escalas temporales muy grandes.}
 \label{fig7c}
\end{figure}

All\'i se observa que, cuanto m\'as grande es el valor
de $\gamma_0$, m\'as r\'apido se inicia el proceso de activaci\'on de
la energ\'ia. Inicialmente, la energ\'ia media del sistema es menor
que la energ\'ia media del sistema aislado. Sin embargo, a un tiempo
$t \geq t_{\cal D}$, la energ\'ia media del sistema comienza a crecer
considerablemente debido a la interacci\'on con el entorno. El 
sistema gana energ\'ia a expensas del entorno, el cual le sirve
como  fuente de energ\'ia. En la Fig.\ref{fig6c}(b),  graficamos la
evoluci\'on temporal de la energ\'ia media del sistema
para distintos valores de la separaci\'on inicial de los paquetes
gaussianos ($L_0$). Como era de esperar, la energ\'ia no
depende considerablemnte de este par\'ametro:  cuanto m\'as alejados
est\'an los paquetes de ondas ($L_0$ m\'as grande), la  p\'erdida
de coherencia se hace efectiva antes, y, por lo tanto, 
la activaci\'on comienza m\'as temprano
 (el tiempo $t_{\cal D}$ es proporcional a $L_0^{-2}$).  En el recuadro
de la Fig.\ref{fig6c}(b), podemos observar el comportamiento del
factor de p\'erdida de coherencia, el cual da una medida de la escala
temporal en la cual las interferencias del sistema son efectivamente
suprimidas. Notablemente, esta escala coincide con el comienzo
del proceso de activaci\'on de la energ\'ia.

 \begin{figure}[!ht]
\centering
\includegraphics[width=16cm]{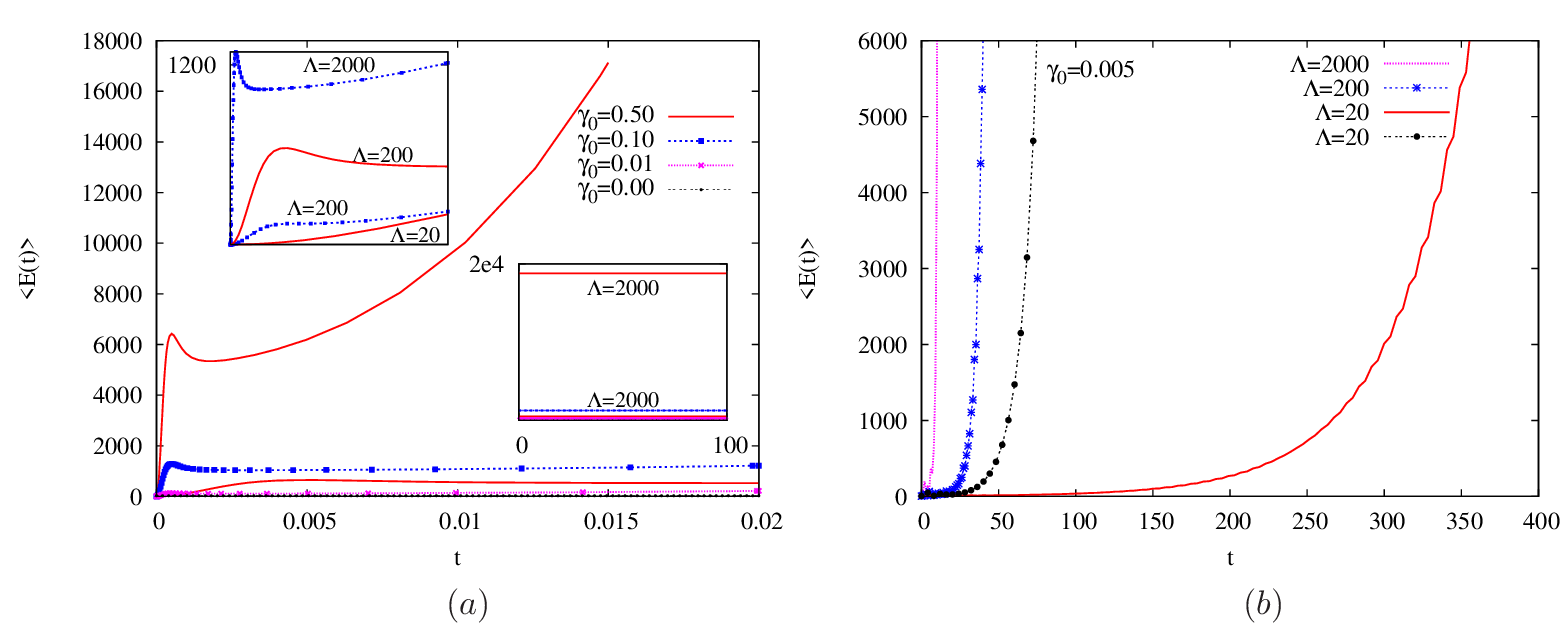}
\caption{(a) Evoluci\'on temporal de la energ\'ia media
para un sistema cu\'antico acoplado a un entorno supra\'ohmico
a temperatura cero. Los par\'ametros usados son:
$\Lambda=2000$, $\Omega=0.1$, $L_0=2$  para diferentes valores
de la frecuencia de corte $\Lambda$ y la constante de acoplamiento
$\gamma_0$. En el recuadro superior, mostramos la energ\'ia media en
una escala temporal m\'as corta para ver claramente la dependencia con
$\Lambda$. En el recuadro inferior, se muestra
la energ\'ia media para tiempos muy
largos de la evoluci\'on para algunos valores del gr\'afico principal.
(b) Evoluci\'on temporal de la energ\'ia media
para un sistema cu\'antico acoplado a un entorno sub\'ohmico
a temperatura cero. Los par\'ametros usados son:
$\Omega=0.1$, $L_0=2$,
$\gamma_0=0.001$ para diferentes valores $\Lambda$.
Adem\'as graficamos el caso $\gamma_0=0.005$ y $\Lambda
=20$ para comparar.  Los par\'ametros est\'an medidos en
unidades de la frecuencia natural del sistema $\Omega$ 
en todos los casos.}
\label{EnoohmT0}
\end{figure}
En la Fig.\ref{fig7c}(a), presentamos la evoluci\'on temporal 
de la energ\'ia media de
un sistema cerrado y uno abierto, junto con la dispersi\'on media de la
energ\'ia de un sistema abierto cuando la temperatura del entorno es cero.
Por supuesto, la energ\'ia media del sistema cerrado permanece constante.
En el caso de un sistema abierto, observamos que, inicialmente, la energ\'ia media
es menor que la correspondiente al caso cerrado, pero inmediatamente despu\'es,
comienza a crecer. La dispersi\'on media de la energ\'ia muestra que
las fluctuaciones resultan ser una funci\'on
uniformemente creciente del tiempo.  Esto es un indicador de que las mismas
son muy importantes y responsables del aumento de la
energ\'ia media del sistema.
Finalmente, en la Fig.\ref{fig7c}(b), mostramos
el comportamiento de la energ\'ia media del sistema a tiempos
largos del orden del tiempo de saturaci\'on $t_{\rm sat} \sim 1/\gamma_0$.
Para ello, resolvimos num\'ericamente la ecuaci\'on maestra para tiempos
de este orden. Como nosotros trabajamos en el r\'egimen subamortiguado,
estas corridas de tiempos muy largos resultan num\'ericamente muy costosas
y pueden acarrar errores num\'ericos importantes en algunos casos. Sin embargo, 
presentamos aqu\'i dicho comportamiento para un conjunto
apropiado de los par\'ametros f\'isicos del modelo.

\subsubsection{Entornos no \'ohmicos}
\label{secEnoohmT0}

Extenderemos el an\'alisis anterior al caso m\'as general de entornos
no \'ohmicos. 
Empezaremos con el supra\'ohmico, cuya evoluci\'on 
temporal de la energ\'ia media se observa en la Fig.\ref{EnoohmT0}(a). 
\begin{figure}[!ht]
\centering
\includegraphics[width=11cm]{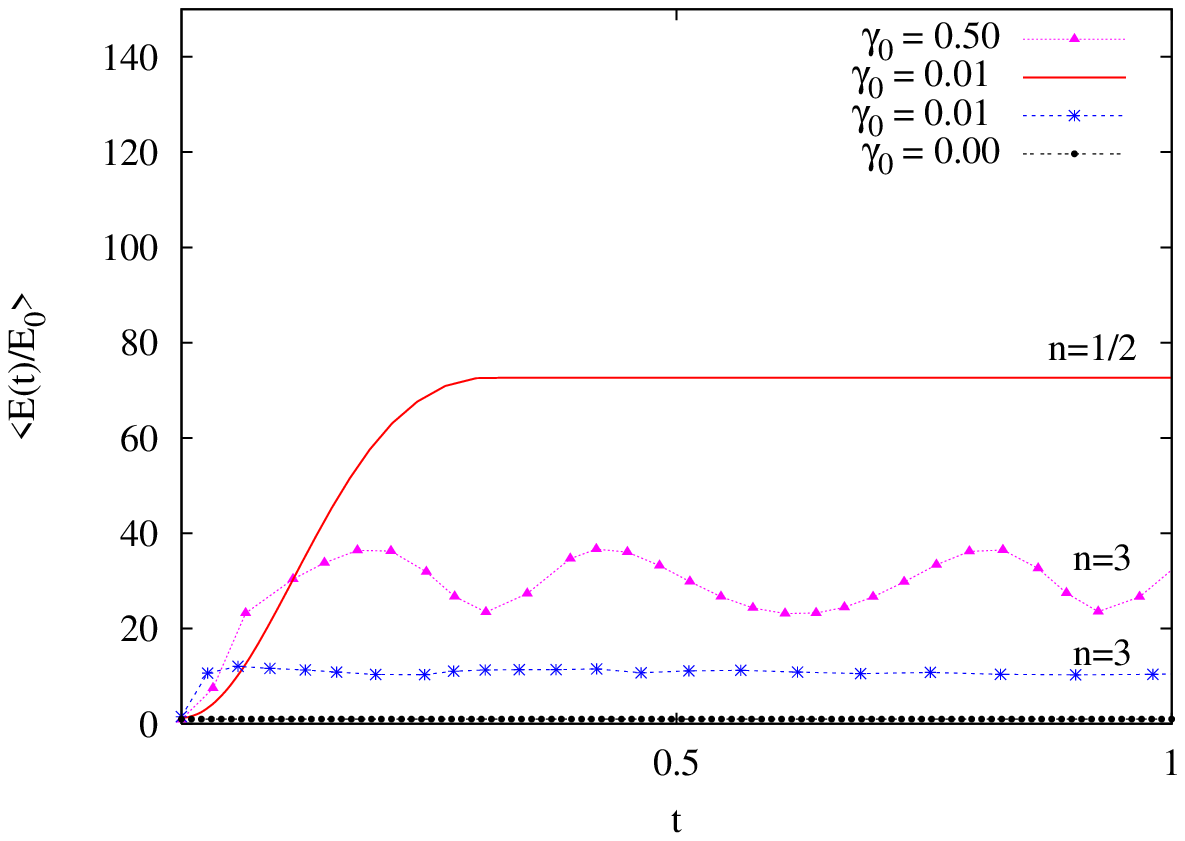}
\caption{ Evoluci\'on de la energ\'ia media del sistema a tiempos largos, cuando
la part\'icula est\'a acoplada a distintos entornos no \'ohmicos, todos a temperatura
cero. Consideramos el caso supra\'ohmico $n=3$ y el sub\'ohmico $n=1/2$, y
los casos $\Omega=1$ y $L_0=2$  para distintos valores de 
 $\gamma_0$  para poder distinguir entre entornos supra\'ohmicos
``fuertes" y ``d\'ebiles".
 $E_0$ es la energ\'ia media del sistema aislado. 
Los par\'ametros usados est\'an medidos en unidades de $\gamma_0$, para
poder incluir distintas corridas (y escalas temporales) en el mismo gr\'afico.}
\label{EnoohmT0largos}
\end{figure}

En dicha
figura, podemos observar que la energ\'ia media presenta 
un \mbox{crecimiento} importante a tiempos muy cortos, 
y, luego, se mantiene constante
como indica el recuadro inferior derecho. En la Secci\'on
\ref{deconoohm} hemos mencionado que, este tipo de entorno,
 no resulta muy efectivo
induciendo efectos difusivos en el sistema  y, por lo mismo, ahora notamos que
no  hay activaci\'on energ\'etica. 
Es decir, sin p\'erdida de coherencia no 
hay excitaci\'on de la energ\'ia. En la Fig.\ref{EnoohmT0}(a), podemos observar
un valor de $\gamma_0$ para el cual la activaci\'on es evidente. Sin embargo,
y como ya mencionamos,
este valor de la constante de acoplamiento ($\gamma_0=0.5$) no resulta
un valor apropiado en el caso del r\'egimen subamortiguado. Todos los otros
valores de $\gamma_0$ de esa figura corresponden a entornos ``d\'ebiles"
que no son capaces de suprimir las interferencias (Fig.\ref{decosupcap2}) ni
activar energ\'eticamente el sistema.
En cuanto al entorno sub\'ohmico, podemos notar en la Fig.\ref{EnoohmT0}(b)
 un comportamiento similar al observado en el caso 
de temperatura alta, pero ocurriendo en 
escalas temporales m\'as largas. En este caso, tambi\'en
 es posible verificar que la escala
de activaci\'on energ\'etica $t_{\rm act}$ es posterior a la de p\'erdida
de coherencia  $t_D \sim \Omega/(\Lambda \gamma_0)$. 

Finalmente, en la Fig.\ref{EnoohmT0largos} presentamos la evoluci\'on temporal 
de la energ\'ia media del sistema acoplado a entornos no \'ohmicos
a temperatura cero para diferentes valores de los par\'ametros
del modelo. En todos los casos, podemos observar, un comportamiento
asint\'otico correcto y esperado.

\newpage
\thispagestyle{empty}
\cleardoublepage
\chapter{Efecto t\'unel, p\'erdida de coherencia y activaci\'on energ\'etica}
\label{c3}
\markboth{Efecto t\'unel, p\'erdida de coherencia y activaci\'on energ\'etica}
{Cap\'itulo 3}

La observaci\'on del efecto t\'unel, un proceso de caracter\'isticas
puramenta cu\'anticas,  en escalas macrosc\'opicas
resulta un aspecto muy intrigante de la Mec\'anica Cu\'antica 
 \cite{caleg,gral}. Los sistemas
macrosc\'opicos, por lo general, son sistemas  abiertos que interact\'uan
con un entorno externo. En este contexto, el efecto t\'unel resulta
cualitativamente diferente a su an\'alogo microsc\'opico experimentalmente
verificado \cite{hanggi}.

Un sistema cu\'antico cerrado, descripto por ejemplo por un 
estado localizado en uno de los m\'inimos de un potencial
de pozo doble,  podr\'ia ser encontrado
del otro lado de la barrera de potencial, es decir, localizado en el otro m\'inimo,
en una escala temporal definida. Este tiempo, llamado tiempo de
``tuneleo" ($\tau$), puede ser estimado a trav\'es del uso de t\'ecnicas
conocidas como, por ejemplo, el m\'etodo del instant\'on  \cite{Coleman}.
Sin embargo, cuando el sistema es abierto, la posibilidad de observar
dicho efecto, depende de factores adicionales relacionados con el entorno.
Como hemos visto a lo largo del cap\'itulo anterior, 
 la din\'amica de los sistemas cu\'anticos abiertos,
presenta rasgos y escalas temporales intr\'insicas, originadas en 
la interacci\'on entre el sistema y el entorno. Los efectos difusivos
inducidos por el entorno en el sistema, generan p\'erdida de coherencia
en este \'ultimo. El sistema presenta un comportamiento cl\'asico
tan pronto como sus interferencias cu\'anticas son destru\'idas por el
ruido externo \cite{jpphabzurek}.  Esta transici\'on cu\'antico-cl\'asica
se lleva a cabo en una escala temporal caracter\'istica, la cual hemos denominado
tiempo de p\'erdida de coherencia ($t_{\cal D}$). Esta cantidad depende
de las propiedades del sistema cu\'antico, del entorno y de la constante
de acoplamiento entre ellos. Si el tiempo de p\'erdida de coherencia es
significativamente menor que el tiempo de ``tuneleo", uno esperar\'ia,
siguiendo con el ejemplo del pozo doble,
que el estado del sistema se mantuviera confinado al m\'inimo inicial.  Es decir,
si $t_{\cal D} \ll \tau$, entonces, la supresi\'on de las interferencias 
y la clasicalizaci\'on del sistema cu\'antico se llevan a cabo antes, por lo 
cual, cualquier posibilidad de encontrar al estado final del otro lado de 
la barrera de potencial, es nula.

Sin embargo, hemos estudiado que los efectos difusivos tambi\'en
generan una excitaci\'on energ\'etica en el sistema, una vez que
\'este ya se ha vuelto cl\'asico. Por lo tanto, el sistema, a\'un luego
de perder sus interferencias cu\'anticas, podr\'ia ser encontrado
en el otro m\'inimo de potencial, si ganase la suficiente energ\'ia
como para ``saltar" dicha barrera. La naturaleza de este proceso
es completamente distinta al efecto t\'unel cu\'antico, ya que es de
naturaleza puramente cl\'asica.  Como hemos visto en el cap\'itulo
anterior, este proceso de activaci\'on es efectivo tanto en el caso de
un entorno a temperatura alta como a  temperatura cero.

La influencia de un entorno en el efecto t\'unel fue estudiada por
Caldeira y Legget en  \cite{caleg,legget} para un modelo
esp\'in-bos\'on. Los autores mostraron
que la disipaci\'on inhibe el efecto t\'unel. Desde ese entonces, 
numerosos trabajos se han realizado estudiando varios aspectos
distintos del mismo fen\'omeno, y por lo general, arrivaron a 
 conclusiones similares \cite{gral2}. Lo que resulta m\'as llamativo
de estos trabajos, es que la mayor\'ia de ellos basan sus conclusiones
en t\'ecnicas anal\'iticas, ya sea usando 
m\'etodos funcionales o generalizaciones del m\'etodo del 
instant\'on.  Estos estudios est\'an principalmente fundamentados
en conceptos de equilibrio y, en algunos casos, \'esto podr\'ia
omitir aspectos importantes de la evoluci\'on din\'amica del sistema. En
particular, resulta muy d\'ificil estudiar simult\'aneamente el efecto
t\'unel y la activaci\'on energ\'etica, y discernir entre ambas contribuciones
individuales al final de la evoluci\'on. Llevando esta afirmaci\'on
al plano de nuestro ejemplo del pozo doble, si dada la evoluci\'on din\'amica
de nuestro sistema cu\'antico abierto, observamos que el estado final
del sistema se halla del otro lado de la barrera de potencial,
 \textquestiondown c\'omo
podr\'iamos explicar dicha observaci\'on? En funci\'on de la
din\'amica de los sistemas cu\'anticos abiertos, el estado 
podr\'ia tanto haber pasado por debajo de la barrera (efecto
t\'unel) como haber saltado la misma (activaci\'on). Resulta
evidente entonces, que el proceso de p\'erdida de coherencia
tiene un rol fundamental para determinar la respuesta.

Desde aquellos primeros trabajos sobre sistemas abiertos, la maquinaria
computacional y las t\'ecnicas asociadas a ella, han evolucionado
de manera asombrosa, al punto que resulta posible, hoy d\'ia,
corroborar y extender resultados anal\'iticos conocidos. En el
contexto de simulaciones num\'ericas en sistemas cu\'anticos
abiertos, el principal objeto de estudio han sido los sistemas
``forzados", donde el efecto t\'unel se observa entre 
islas regulares y no regulares \cite{tunel,diana_jpp}. En estos modelos
resulta interesante el juego entre la disipaci\'on inducida por el entorno
y el caos cl\'asico intr\'insico al sistema, ya que, por ejemplo, el comportamiento
ca\'otico del sistema elimina las interferencias cu\'anticas, las cuales,
a su vez, son restitu\'idas por la disipaci\'on inducida en el sistema.

En este cap\'itulo, nos concentraremos en estudiar la 
din\'amica completa de una part\'icula, localizada inicialmente
en un m\'inimo de un potencial de  pozo doble, acoplada a un
entorno externo. Estudiaremos tanto el caso de temperatura alta
como el de ausencia de \mbox{temperatura} en el entorno.
En funci\'on del estudio detallado acerca
del proceso de p\'erdida de coherencia
y activaci\'on energ\'etica, ambos fen\'omenos inducidos por el
ruido del entorno, que hemos desarrollado en Cap\'itulo \ref{c2} de esta
Tesis, estimaremos las escalas temporales asociadas
a los tres procesos f\'isicos que gobiernan la din\'amica del sistema.
Estos son: p\'erdida de coherencia, efecto t\'unel y activaci\'on
energ\'etica. Usando simulaciones num\'ericas de gran escala, 
obtendremos una descripci\'on completa de la din\'amica del
sistema y verificaremos las estimaciones anal\'iticas realizadas.
Pero lo que resulta fundamental, es que tendremos acceso
al estado del sistema a cualquier tiempo, y eso nos permitir\'a
distinguir entre los efectos de los distintos procesos mencionados
anteriormente para poder, as\'i, encontrar la respuesta a la pregunta
anteriormente formulada.

\section{El modelo}
\label{modeldw}

En lo que sigue, consideraremos un oscilador cu\'antico, en
un potencial no lineal 
$V(x)=-\frac{1}{4}\Omega^2 x^2 + \lambda x^4$, acoplado a un
entorno formado por  infinitos osciladores arm\'onicos. 
El potencial $V(x)$ tiene dos m\'inimos absolutos 
$x_0=\pm \Omega/\sqrt{8\lambda}$
separados por una barrera de potencial de altura 
$V_0=\Omega^4/(64\lambda)$. La acci\'on total del sistema
est\'a dada por
\begin{eqnarray}S[x,q_n] &=& S_{\rm sistema}[x] + S_{\rm entorno}[q_n] +
 S_{\rm interaccion}[x,q_n]  \\
&=& \int_0^t ds \left[ \frac{1}{2}{\dot x}^2 + \frac{1}{4}\Omega^2 x^2 -
\lambda x^4 \right. 
+ \left. \sum_n \frac{1}{2} m_n ({\dot q}_n^2 - \omega_n^2 q_n^2)\right] - 
\sum_n \lambda_n x q_n,\nonumber
\label{actiondw} \end{eqnarray}  
donde $q_n$, $m_n$ y $\omega_n$ son las coordenadas, masas y frecuencias
de los osciladores del entorno, respectivamente. La masa del oscilador
principal (sistema) fue fijada a uno.  El sistema est\'a acoplado linealmente a
cada uno de los osciladores del entorno, a trav\'es de la constante de
acoplamiento $\lambda_n$.

La din\'amica del oscilador no lineal puede ser estudiada a trav\'es de
la traza sobre los grados de libertad del entorno y, 
la ecuaci\'on maestra que resulta para la matriz densidad reducida
$\rho_{\rm r}(t)$.  Asumiremos, nuevamente, que
el sistema y el entorno est\'an inicialmente no correlacionados, y que,
este \'ultimo, est\'a en un estado de equilibrio t\'ermino a temperatura
$T$ (la cual puede ser cero inclusive). El estado inicial es, entonces, el
producto del estado inicial del sistema (un estado localizado en el
m\'inimo izquierdo del potencial) y del estado inicial t\'ermico del entorno.
S\'olo cuando la interacci\'on entre ambos es ``encendida'', el sistema
comienza a evolucionar bajo la influencia del ba\~no t\'ermico. 
Es importante recordar que
la condici\'on inicial no es una condici\'on de equilibrio del sistema total.
Bajo estas suposiciones y, considerando que el acoplamiento entre el sistema
y el entorno es muy peque\~no (r\'egimen subamortiguado), la matriz
densidad reducida satisface la ecuaci\'on maestra de la Ec.(\ref{master}).

Como ya mencionamos, la ecuaci\'on
maestra que resulta es perturbativa. Por tanto, nosotros
trabajaremos con la matriz densidad reducida obtenida a segundo
orden en la constante de acoplamiento entre el sistema y el entorno.
Al considerar el r\'egimen subamortiguado, aseguramos la validez
de este tratamiento perturbativo para las escalas temporales que queremos
estudiar \cite{HuPazZhangI,diana_jpp}. Por esta raz\'on, 
trabajaremos con un estorno \'ohmico y asumiremos que $\gamma_0 \ll 
\hbar=1$,
lo cual fija el dominio temporal de las soluciones perturbativas que
encontraremos.

El estado del sistema est\'a inicialmente localizado en un m\'inimo
del potencial. Lo describiremos con una
funci\'on de onda gaussiana centrada en el m\'inimo izquierdo del mismo,
es decir en $x_0=-\Omega/\sqrt{8\lambda}$,
\begin{equation}
\Psi_0(x)=\frac{1}{(2\pi\sigma_x^2)^{1/4}} 
            \exp{\left[ -\frac{(x-x_0)^2}{4 \sigma_x^2}\right]}.
\label{psi0}
\end{equation} 
El ancho de esta gaussiana es $\sigma_x=1/\sqrt{2\Omega}$, que corresponde
al estado de vac\'io de un oscilador arm\'onico de frecuencia $\Omega$.
Debido a la presencia del entorno, el oscilador 
sufre un cambio en su frecuencia natural $\tilde \Omega^2=\Omega^2+
\delta \Omega^2(t)$; de modo que eligiremos los par\'ametros de forma
tal que el coeficiente de renormalizaci\'on $\delta \Omega^2(t)$
pueda ser despreciado a todo tiempo. La frecuencia natural del
sistema se puede obtener expandiendo el potencial $V(x)$ alrededor
del m\'inimo de potencial $x_0$ con la ayuda de $\Psi_0(x)$; de manera
de describir  la part\'icula que est\'a {\it localmente} en el {\it vac\'io}.

Para el sistema cerrado, esperamos que la part\'icula pase al otro
m\'inimo en un tiempo de ``tuneleo'' $\tau$. La funci\'on de onda, a 
ese tiempo, deber\'ia corresponder a una gaussiana similar al estado
inicial, pero localizada en el m\'inimo derecho del potencial.
El estado inicial puede ser bien aproximado por una
combinaci\'on lineal de los dos primeros autoestados de energ\'ia
del potencial $V(x)$. Si llamamos $E_0$ y $E_1$ a las autoenerg\'ias
sim\'etricas y antisim\'etricas, respectivamente, esperamos que
el tiempo en que el efecto t\'unel se lleva a cabo, sea del
orden $\tau \simeq 1/(E_1-E_0)$. Sin embargo, como la condici\'on 
inicial no es exactamente la combinaci\'on lineal de dos autoestados,
sino de una cantidad mayor de los mismos, este n\'umero se ver\'a
apenas corregido en nuestras simulaciones num\'ericas 
($\tau=3./(E_1-E_0)$).
Usando el m\'etodo del instant\'on, hemos calculado 
 el tiempo de ``tuneleo'' anal\'iticamente 
\cite{Coleman},
\begin{equation}
\tau=\frac{3.}{E_1-E_0}=
\frac{3}{8}\sqrt{\frac{\pi}{2}\frac{\Omega}{V_0}}
\frac{1}{\Omega} \exp{\left[\frac{16}{3}\frac{V_0}{\Omega}\right]},
\label{initial}
\end{equation}
donde la expresi\'on dentro de la exponencial es la acci\'on cl\'asica
para el instant\'on, $S_0=(16/3) V_0/\Omega$.

\section{Efecto t\'unel y  p\'erdida de coherencia 
en el l\'imite de temperatura alta}
\label{tunelydecoHT}

Cuando el ba\~no t\'ermico est\'a a temperatura muy alta, la ecuaci\'on 
maestra (Ec.(\ref{master})), se puede expresar de manera mucho m\'as
sencilla por medio de la distribuci\'on de Wigner reducida en el espacio de
fase, $W=W(x,p;t)$ \cite{leshouches,jpphabzurek}:
\begin{eqnarray}\dot W = \{H_{\rm sys},W\}_{\rm PB}
- \frac{\lambda}{4} x 
\partial^3_{ppp}W
&+&2\gamma (t) \partial_p(pW) + D(t) \partial^2_{pp}W 
- f(t) \partial^2_{px}W,
\label{wigner_eq}
\end{eqnarray}
donde $\gamma(t)$, ${\cal D}(t)$ y $f(t)$ son los coeficientes de
disipaci\'on, difusi\'on normal y an\'omalo, respectivamente, definidos
en las Ecs.(\ref{coefdef}). El primer t\'ermino del lado derecho de esta
expresi\'on, corresponde a la evoluci\'on cl\'asica del sistema, mientras
que el segundo incluye las correcciones cu\'anticas a la din\'amica 
del mismo.  Los \'ultimos tres t\'erminos de esta expresi\'on, describen
los efectos de disipaci\'on y ruido debido al acoplamiento del sistema
con entorno. Para simplificar el problema, asumiremos, en adelante, que
el entorno es \'ohmico.

Como hemos discutido anteriormente, 
el ba\~no t\'ermico genera dos efectos distintos
sobre la evoluci\'on de la part\'icula de prueba. En el r\'egimen donde
el acoplamiento (d\'ebil) con el entorno es suficientemente intenso,
la difusi\'on es la responsable de la supresi\'on de las interferencias
cu\'anticas del sistema, y por ende, de  su  p\'erdida de coherencia. 
El resultado de este proceso es la ``clasicalizaci\'on'' del sistema,
ya que, luego de un tiempo $t_{\cal D}$, el comportamiento cu\'antico
es inhibido, anulando tambi\'en la posibilidad de observar el efecto
t\'unel del estado inicial. Dado que la energ\'ia inicial del sistema es
menor que la barrera de potencial (de altura $V_0$), se espera que la 
part\'icula permanezca confinada en la posici\'on del m\'inimo inicial.
Sin embargo, la part\'icula sigue en contacto con el ba\~no t\'ermico, el
cual en este caso en particular, est\'a a una temperatura muy alta. Por
esto, la part\'icula se ``calentar\'a'', y en alg\'un tiempo, $t_{\rm act}$,
su energ\'ia aumentar\'a. En ese momento, existir\'a cierta posibilidad
de  que la part\'icula {\it salte} la barrera de potencial, via activaci\'on
energ\'etica inducida por el ruido del entorno, y sea encontrada del
otro lado de la misma. A tiempos muy largos de la evoluci\'on,
el sistema deber\'ia alcanzar un estado de equilibrio t\'ermico, en donde
encontrar a la part\'icula en cada uno de los m\'inimos
sea equiprobable.

En esta secci\'on, estimaremos las escalas temporales que juegan un
rol importante en este modelo. En particular, nos interesa conocer la relaci\'on
entre $t_{\cal D}$ y $t_{\rm act}$, ya que es \'esta la responsable que
la part\'icula pueda o no pasar la barrera de potencial durante las
distintas etapas de su evoluci\'on.

Por un lado, el tiempo de p\'erdida de coherencia, 
en el l\'imite de temperatura muy alta,
es inversamente proporcional al coeficiente de difusi\'on ${\cal D}(t)$ y el
cuadrado del tama\~no t\'ipico inicial de la funci\'on de onda  $L_0$. En este
caso, elegiremos por conveniencia, que a tiempos cortos de la
evoluci\'on,  $L_0$ sea el ancho
inicial de la funci\'on de onda gaussiana, es decir $L_0=
2\sigma_x=2/\sqrt{2\Omega}$. Usando que ${\cal D}=2 \gamma_0 T$, obtenemos
el tiempo estimado de p\'erdida de coherencia,
\begin{equation}
t_{\cal D}=\frac{\Omega}{4 \gamma_0 T}
\label{tddw}
\end{equation}
en unidades de $\hbar=1=k_B$ \cite{jpphabzurek}.
Esta estimaci\'on no resulta un tiempo exacto, sino una sobreestimaci\'on 
(debido a la elecci\'on de $L_0$) de la escala temporal 
 en la cual los efectos de p\'erdida de coherencia en el sistema resultan visibles. 

Por otro lado, la tasa de activaci\'on t\'ermica para un sistema cl\'asico puede ser
obtenida trabajando con el an\'alogo cl\'asico de la Ec.(\ref{wigner_eq}), es decir
la ecuaci\'on de Fokker-Planck:
\begin{eqnarray}
\dot{W}=\{H_{\rm sys},W\}_{\rm PB}
+2 \gamma_0 \partial_p(pW) + D \partial^2_{pp}W\,.
\label{fpcap3}
\end{eqnarray}
Se puede notar que, posteriormente a la p\'erdida de coherencia  en 
el sistema en cuesti\'on, los t\'erminos cu\'anticos se vuelven irrelevantes y la 
Ec.(\ref{wigner_eq}) se reduce a esta \'ultima ecuaci\'on (Ec.(\ref{fpcap3})).
En el cap\'itulo anterior, hemos demostrado que que el tiempo de
activaci\'on puede ser estimado a partir de las Ecs.(\ref{Em}) y (\ref{EmediaHT}).
Por completitud, volvemos a escribir el tiempo de activaci\'on t\'ermica,
\begin{equation}
t_{\rm act}=\frac{V_0-E_0}{2 \gamma_0 T}. \nonumber
\end{equation}
Esta estimaci\'on nos demuestra que existe una regi\'on muy grande de 
posibles valores de los par\'ametros f\'isicos ($V_0, \Omega,T, \gamma_0$)
 para los cuales es posible
obtener primero p\'erdida de coherencia en el sistema (antes que se
observe el efecto t\'unel) y retrasar, considerablemente, la activaci\'on
energ\'etica en el mismo. Si este es el caso, mientras  $t<t_{\rm act}$,
la part\'icula debe
permanecer confinada  a la regi\'on del lado de la barrera donde
el estado del sistema estaba ubicado inicialmente. La situaci\'on ideal ser\'ia
lograr que las tres escalas temporales en juego estuvieran lo m\'as
separadas posible para poder, as\'i,  distinguir entre los procesos
mencionados. A fines pr\'acticos, esto equivale a
$t_{\cal D} \ll \tau \ll t_{\rm act}$, por lo cual pedimos,
\begin{equation}
a \, t_{\cal D}  =  \tau  =  b \,t_{\rm act}\,.
\label{tDlltaulltth}
\end{equation}
De la primera y segunda condici\'on, obtenemos la siguiente restricci\'on
en los par\'ametros del potencial:
\begin{equation}
\frac{V_0}{\Omega} = \frac{1}{2} \left( \frac{a}{b} + 1 \right )\,.
\label{V0Omega}
\end{equation}
Esta expresi\'on, junto a la elecci\'on del tiempo de ``tuneleo" $\tau$,
fijan el valor del potencial del sistema. Los par\'ametros del entorno,
pueden ser fijados utilizando la primer parte de la Ec.(\ref{tDlltaulltth}),
\begin{equation}
\gamma_0 T = \frac{a \Omega}{4 \tau}\,.
\label{gamma0T}
\end{equation}
Finalmente, eligiendo $a>>1$ y $b<<1$, nos permitir\'a obtener
la evoluci\'on deseada en el sistema, manteniendo  la part\'icula
confinada de un lado de la barrera de potencial para un tiempo
arbitrariamente largo.

\subsection{La p\'erdida de coherencia inhibe el efecto t\'unel}
\label{subsub1}

Con ayuda de las simulaciones num\'ericas, intentaremos ilustrar el
mecanismo de inhibici\'on del efecto t\'unel detallado arriba.  Para ello,
resolvimos la ecuaci\'on maestra  (a trav\'es de la evoluci\'on de los
autoestados de energ\'ia) para valores de los par\'ametros elegidos
convenientemente $V_0=100$ y $\Omega=5$. De la ecuaci\'on 
(\ref{V0Omega}), se ve que esta situaci\'on corresponde a un valor
$n=V_0/\Omega=20$, n\'umero que representa 
la estimaci\'on semicl\'asica
de la cantidad de estados atrapados en el pozo de potencial. Para este
conjunto de par\'ametros, $\tau = 4.63155403 \, 10^{10}$. Hemos elegido,
adem\'as, $a=24.5$ y $b = 0.6282$ de modo que
$\gamma_0 T = 3.9 \, 10^{-11}$, y se verifique la condici\'on de r\'egimen
subamortiguado. Con esta elecci\'on, buscamos lograr que el sistema 
se caliente de a poco, retrasando la activaci\'on energ\'etica hasta bastante
despu\'es del tiempo de ``tuneleo". De esta forma, obtenemos la relaci\'on
entre las tres escalas temporales, siendo $t_{\cal D}
 \sim 0.0408 \;\tau$ y $t_{\rm act} \sim  1.6326 \;\tau$, en unidades del tiempo
$\tau$.

Como el estado inicial resulta bien representado por la superposici\'on lineal
de 10 autoestados de $H_{\rm sis}$, hemos elegido un espacio de 
Hilbert de dimensi\'on $N=40$, el cual es el valor m\'as grande de dicho
espacio que podemos manejar num\'ericamente. La frecuencia m\'as grande
presente en el entorno es $\Lambda = 10 \times \Delta_{40,0} = 10 
\times 102.237307$ ($\Delta_{40,0} $ es la diferencia de energ\'ia entre la
autofrecuencia 40 y la fundamental).

Tambi\'en resolvimos la evoluci\'on del sistema para el caso aislado, 
verificando que el tiempo de ``tuneleo" de nuestro estado inicial Ec.(\ref{psi0}) 
es, efectivamente, muy cercano al estimado anal\'iticamente en la Ec.(\ref{initial}). 
Esto est\'a ilustrado en la 
Fig.\ref{figure1dw}, donde mostramos la evoluci\'on temporal de la
probabilidad de encontrar a la part\'icula en el pozo inicial, tanto en el
caso aislado como en el abierto (sistema en interacci\'on con el ba\~no).
\begin{figure}
\centering
\epsfxsize=10cm
\epsfbox{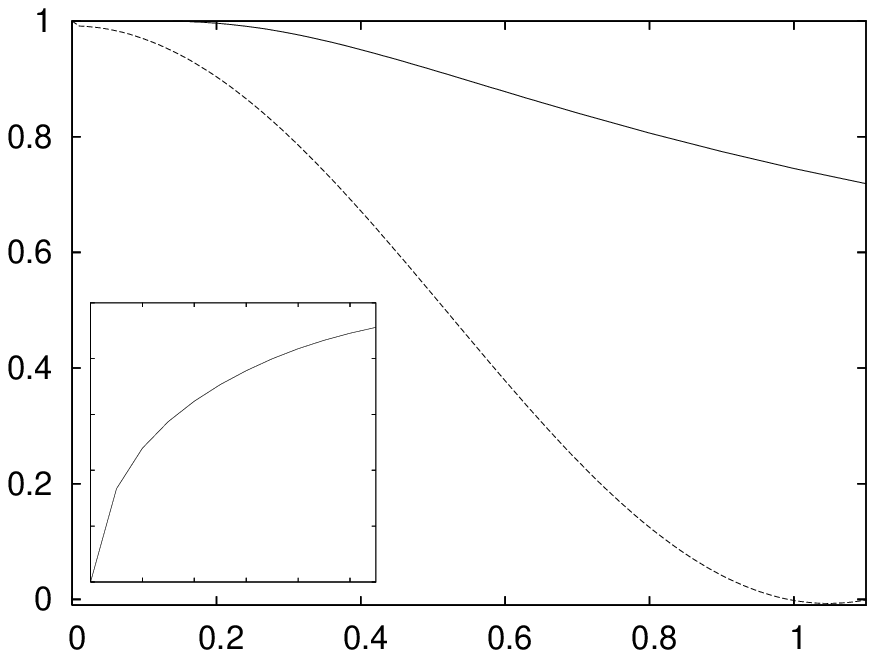}
\caption{Evoluci\'on temporal de la probabilidad 
de encontrar a la part\'icula en el pozo inicial del sistema abierto (l\'inea de
traza entero) y del sistema cerrado (l\'inea punteada). El tiempo est\'a medido
en unidades del tiempo de ``tuneleo" $\tau$. En el recuadro inferior, se muestra
la evoluci\'on temporal de ls Entrop\'ia lineal  del sistema abierto, $S_L/\ln N$,
donde $N=40$ es el tama\~no del espacio de Hilbert elegido. Para
$t\sim\tau$, la Entrop\'ia lineal alcanza un valor cercano al de saturaci\'on,
es decir $S_L/\ln N = 1$.
\label{figure1dw}}
\end{figure}
Esta probabilidad comienza en uno a $t=0$ y decrece a medida que
la part\'icula va pasando por debajo de la barrera de potencial, llegando
a cero cuando $t \sim \tau$. Para tiempos m\'as largos, aunque no se
muestra en esa figura, se puede ver que la part\'icula pasa de un lado
a otro de la barrera con un per\'iodo de oscilaci\'on de $2 \tau$. 

El comportamiento del sistema abierto es claramente diferente al 
correspondiente del sistema cerrado. La probabilidad de encontrar a la
part\'icula en el pozo de potencial original tambi\'en decrece inicialmente,
pero lo hace con una tasa m\'as lenta. Como demostraremos m\'as adelante,
y, en parte debido a nuestra elecci\'on de los par\'ametros, esta disminuci\'on se debe a
la activaci\'on t\'ermica m\'as que al  efecto t\'unel (es suprimido a tiempos
muy cortos). La probabilidad nunca se hace cero y tampoco se observan
oscilaciones; sino que decrece mon\'otonamente, y a tiempos muy largos,
esperamos que alcance el valor $0.5$. Es decir, el estado final deber\'ia
ser un estado de equilibrio t\'ermico con el entorno, con igual probabilidad
de encontrar a la part\'icula en cada uno de los m\'inimos de potencial. En 
la Fig.~\ref{figure1dw}, tambi\'en mostramos la evoluci\'on temporal de la 
Entrop\'ia Lineal $S_L$ del sistema abierto, definida como
$S_L/\ln N=-\ln\left[ {\rm Tr}\rho^2\right]/\ln N$ (donde $\ln N$ es el 
valor m\'aximo que puede alcanzar para el espacio de Hilbert elegido).  
Despu\'es de un tiempo, esta cantidad
alcanza su valor de saturaci\'on. Esto indica que la dimensi\'on del espacio 
de Hilbert ($N=40$) resulta chica para la simulaci\'on num\'erica y los estados 
de energ\'ia m\'as altos ya se est\'an poblando. Como consecuencia, los
resultados num\'ericos son menos confiables para tiempos mayores
que el  tiempo $\tau$. De todas formas, resulta evidente  que la
p\'erdida  de coherencia en el sistema inhibe el efecto t\'unel
para tiempos bastante anteriores a dicha cota temporal num\'erica.

\begin{figure}
\centering
\epsfxsize=14cm
\epsfbox{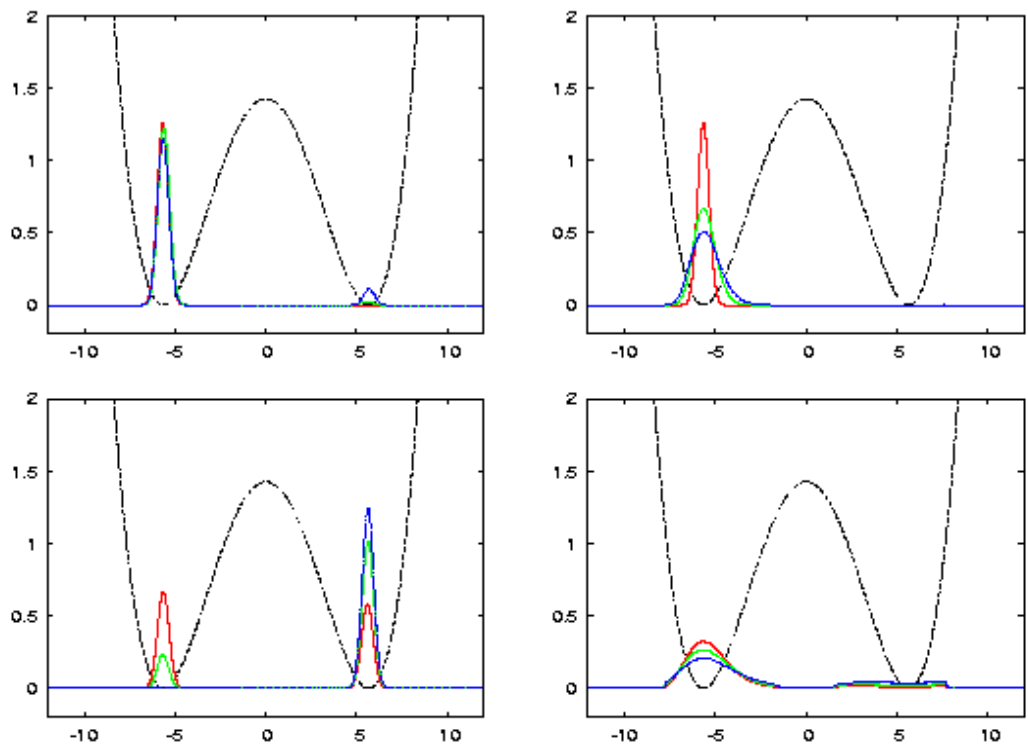}
\caption{Distribuci\'on de probabilidad $\sigma(x,x)$ para el sistema aislado
(izquierda) y abierto (derecha) para $t=0$; $t=0.1 \tau$ y $t=0.2\tau$ (superior);
 $t=0$; $t=0.5 \tau$ y $t=\tau$ (inferior).  Esta graficado, adem\'as, el potencial
rescalado $V(x)$ para usarlo de referencia.
\label{figure2dw}}
\end{figure}
Las caracter\'isticas particulares de la evoluci\'on, tanto del sistema aislado
como del abierto, se pueden observar en las Figs.~\ref{figure2dw} y
 \ref{figure3dw}, en las cuales graficamos la distribuci\'on de probabilidad
$\sigma(x,x) = \langle x | \rho_{\rm r} | x\rangle$ y la funci\'on de 
Wigner $W(x,p)$,
respectivamente, para ciertos tiempos. De nuevo, los diferentes
comportamientos entre el caso aislado y el abierto resultan evidentes.
Para tiempos cortos ($t \sim 0.2 \tau$), en el caso aislado, 
 la probabilidad de encontrar a la part\'icula del lado original 
 disminuye, mientras que del otro lado de la barrera  aumenta, 
sugiriendo que el estado comienza a pasar por debajo de ella.
\begin{figure}
\centering
\epsfxsize=13cm
\epsfbox{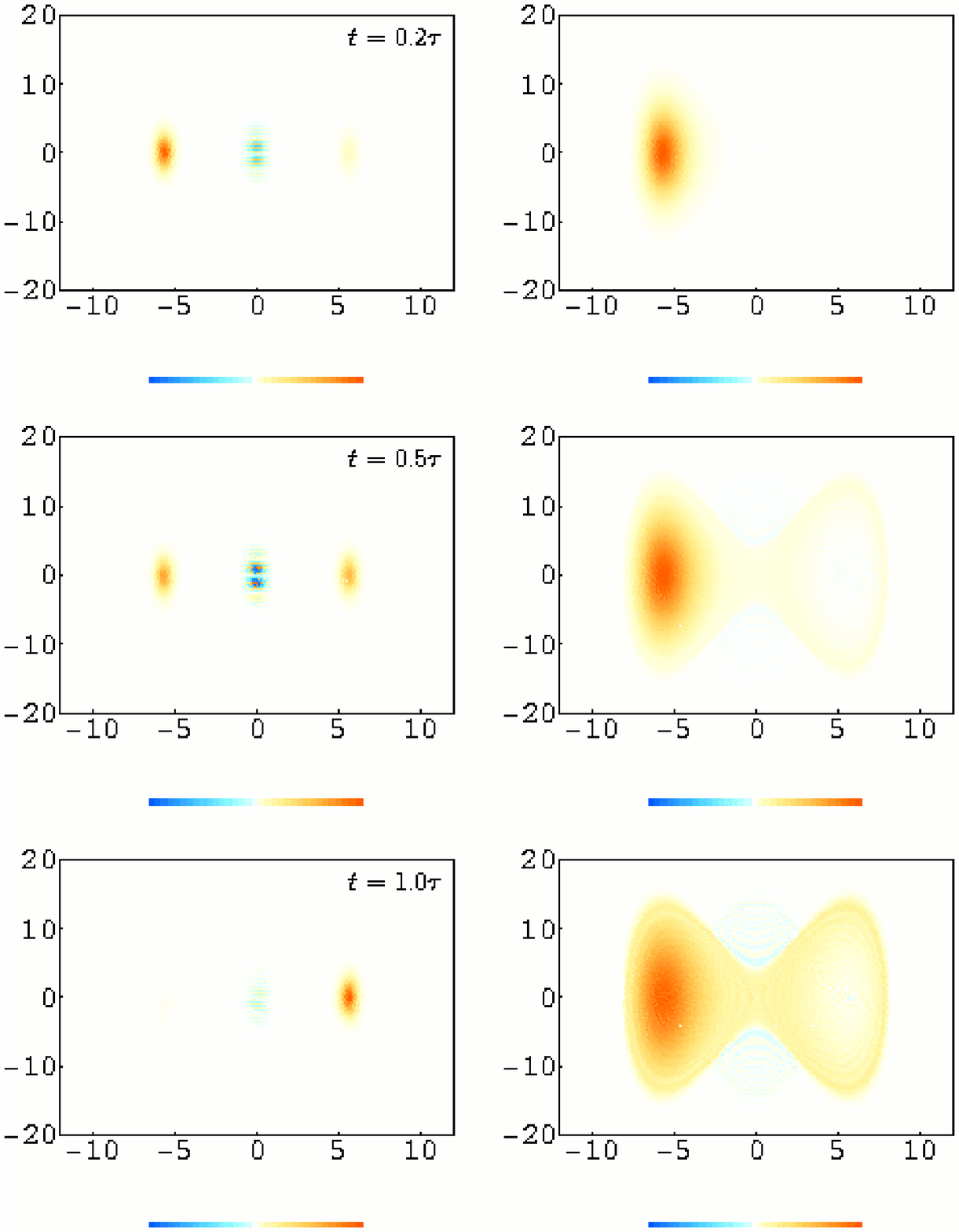}
\caption{Funci\'on de Wigner para el caso del sistema aislado (izquierda) 
y abierto (derecha), para los tiempos indicados. El eje horizontal corresponde
a $x$, mientras que el vertical a $p$. El fondo blanco corresponde a 
valores nulos de $W(x,p)$. Colores del amarillo al rojo, representan valores
positivos de esta funci\'on, mientras que los colores en la gama azul, 
corresponden a valores negativos.}
\label{figure3dw}
\end{figure}
El mismo efecto puede ser observado para el correspondiente tiempo
de la funci\'on de Wigner de la Fig.~\ref{figure3dw}.  All\'i, los efectos
cu\'anticos se notan en los valores negativos (interferencias) que
aparecen en el centro del espacio de fases. Para esos mismos tiempos,
en el sistema abierto, no observamos evidencia alguna de que el
sistema est\'e pasando a trav\'es de la barrera de potencial. Por
el contrario, vemos que la part\'icula est\'a estrictamente confinada
al pozo de potencial inicial. En este caso, observamos que tanto
$\sigma(x,x)$ como  $W(x,p)$ aumentan como consecuencia de la
difusi\'on inducida por la presencia del entorno. Como el tiempo
de p\'erdida de coherencia es muy corto para nuestra elecci\'on 
de los par\'ametros f\'isicos del modelo, la destrucci\'on de las
interferencias cu\'anticas ya se hizo efectiva a este tiempo ($t=0.2 \tau$) y, 
consecuentemente, la funci\'on de Wigner es positiva en todas las regiones
del espacio de fases. A un tiempo posterior, por ejemplo $t=0.5 \tau$,
tanto la distribuci\'on de probabilidad como la funci\'on de Wigner, 
son sim\'etricas para el sistema aislado. Sin embargo, en el sistema
abierto, la funci\'on de onda se sigue propagando en el espacio de
fases. Adem\'as, podemos notar los primeros signos de la activaci\'on
t\'ermica; es decir, comienza a haber distribuci\'on de probabilidad no
nula del lado derecho de la barrera de potencial.  Cuando se alcanza
el tiempo de ``tuneleo", a pesar que el sistema est\'a mayormente
localizado del lado izquierdo (inicial) de la barrera de potencial, 
el sistema comienza a calentarse y su funci\'on de Wigner 
 explora regiones m\'as grandes del espacio de fases motivada
por la activaci\'on t\'ermica.  Podemos notar peque\~nas franjas de 
interferencia de valor negativo en la funci\'on de Wigner ya que 
\'esta, ahora, ocupa \'areas de mayor no linealidad del potencial. Este
comportamiento transitorio es una consecuencia conocida de la
aparici\'on de efectos no lineales en el sistema y no tiene relaci\'on
con el efecto t\'unel \cite{nunoferdiana}.  Mientras, a tiempo $t=\tau$,
el sistema aislado se encuentra completamente localizado del otro
lado de la barrera de potencial, tras haber ``atravesado" dicha
barrera por efecto t\'unel.

\subsection{Activaci\'on t\'ermica en el l\'imite cl\'asico}

En esta secci\'on, presentaremos un ejemplo num\'erico sencillo
de activaci\'on t\'ermica en el l\'imite cl\'asico de temperatura muy alta.
Nuestra intenci\'on es mostrar la evoluci\'on del sistema en presencia de
un entorno cl\'asico a temperatura alta.
Un sistema cl\'asico estad\'istico se puede describir por la Ec.(\ref{fpcap3}).
Aqu\'i, en lugar de resolver dicha ecuaci\'on para obtener 
$W(x,p)$, nos limitaremos a hacer
evolucionar un conjunto muy grande de trayectorias de part\'iculas
cl\'asicas que interact\'uan con un ba\~no t\'ermico  v\'ia t\'erminos de
difusi\'on y disipaci\'on. La ecuaci\'on de movimiento para cada 
una est\'a dada por
\begin{equation}
\ddot{x}(t)=-2 \gamma_0 \dot{x}(t) -V'(x(t)) + \xi(t),
\label{langevindw}
\end{equation}
donde $\xi$ un ruido gaussiano no correlacionado temporalmente con
una varianza definida seg\'un 
$\langle\xi(t)\xi(t')\rangle=\gamma_0 T \delta (t - t')$
 (ruido blanco). Resulta f\'acil
demostrar que un conjunto de part\'iculas que evolucionan de
acuerdo a la ecuaci\'on de Langevin Ec.(\ref{langevindw}),  tambi\'en
satisfacen la ecuaci\'on maestra Ec.(\ref{fpcap3}). Sin embargo,
resulta menos costoso num\'ericamente encontrar la soluci\'on
para un n\'umero muy grande de part\'iculas que cumplen con la
primer ecuaci\'on.  De este modo,
las condiciones iniciales son elegidas de forma, que tanto $x$ como $p$,
sean variables gaussianas  aleatorias distribu\'idas seg\'un el an\'alogo
cl\'asico de la funci\'on de onda de la Ec.(\ref{psi0}),
\begin{equation}
W_0(x,p)=\frac{1}{\pi} \exp\left[ -\frac{(x-x_0)^2}{2\sigma_x^2}
-2 \sigma_x^2 p^2\right].
\end{equation}
Luego, a un tiempo dado $t$, podemos obtener valores de expectaci\'on
de las magnitudes f\'isicas promediando sobre el conjunto de 
part\'iculas. De este modo, podemos determinar la funci\'on de Wigner,
evaluando la cantidad de part\'iculas dentro del conjunto que tienen
un valor de posici\'on y momento en el intervalo $(x,x+dx)\times(p,p+dp)$. 
\begin{figure}
\centering
\epsfxsize=10cm
\epsfbox{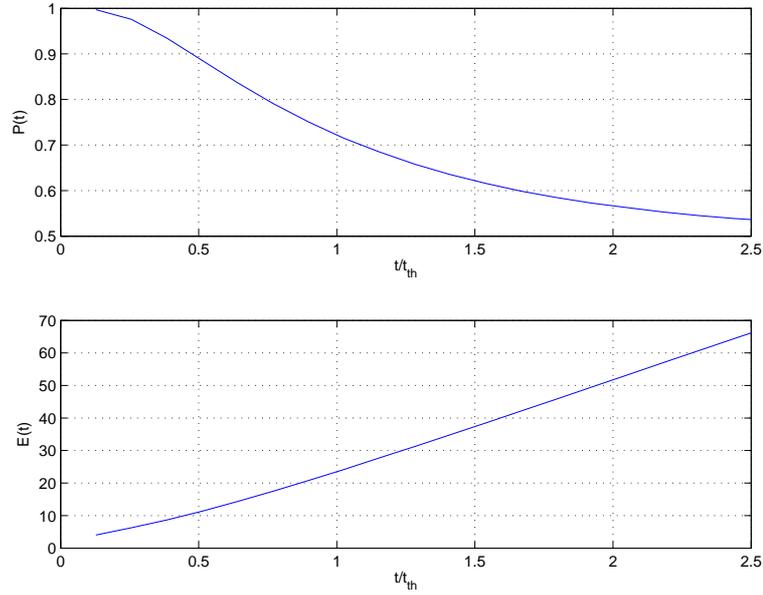}
\caption{Evoluci\'on temporal de la probabilidad de permanecer en el
pozo de potencial original (superior) y de la energ\'ia media del sistema 
(inferior). El tiempo est\'a expresado en unidades de la escala temporal 
de Eq.~(\ref{t_th}).}
\label{class}
\end{figure}
En las Figs.~\ref{class} y \ref{wigner_class}, mostramos los resultados
de la simulaci\'on para los valores de los par\'ametros 
$\Omega^2=12$, $V_0=23$, $T=10^7$, y $\gamma_0=2.5\times 10^{-9}$ \footnote{
Si hubi\'eramos usado los valores de la secci\'on
anterior, hubi\'esemos obtenido soluciones temporales poco pr\'acticas.
Sin embargo, a pesar de las distintas escalas temporales,
los aspectos cualitativos de estas simulaciones deber\'ian coincidir
con aquellos de dicha secci\'on, ya que la intenci\'on de nuestra simulaci\'on
cl\'asica es ilustrar las propiedades gen\'ericas del proceso de activaci\'on
t\'ermica.}.

El tiempo de activaci\'on t\'ermica estimado, para estos valores de los par\'ametros,
es $t_{\rm act}=390$.  En la Fig.~\ref{class}, mostramos la probabilidad de
encontrar a la part\'icula del lado izquierdo de la barrera de potencial $P(t)$ (gr\'afico
superior) y  la energ\'ia media del sistema (gr\'afico inferior). Como esper\'abamos,
cuando $t\simeq t_{\rm act}$, la energ\'ia media de la part\'icula es del mismo
orden que la altura de la barrera de potencial $V_0$. La probabilidad  $P(t)$ en ese
tiempo es $P\sim0.7$. Es importante destacar que hemos realizado distintas simulaciones 
num\'ericas para un amplio rango de par\'ametros y encontramos que la Ec.(\ref{t_th})
se cumple bien en todos los casos. En particular, notamos que la probabilidad
de ``permanencia" del lado original de la barrera, a un tiempo $t =t_{\rm act}$,
siempre estuvo alrededor de $P\simeq 0.65-0.75$, rango en el cual podemos
ubicar el valor de la simulaci\'on cu\'antica de la secci\'on anterior. Igualmente,
ese valor deber\'ia ser considerado s\'olo de modo cualitativo,
ya que fue obtenido despu\'es que la Entrop\'ia Lineal alcanzara su valor de
saturaci\'on, lo cual, como ya indicamos, hace que los resultados num\'ericos
sean menos confiables. La evoluci\'on de $P(t)$ en el caso cl\'asico, en general,
es muy similar a la correspondiente al caso cu\'antico para tiempos posteriores
a $t_{\cal D}$, ya que es una probabilidad que decrece mon\'otonamente y 
se acerca $0.5$ para valores temporales largos.
\begin{figure}
\centering
\epsfxsize=13cm
\epsfbox{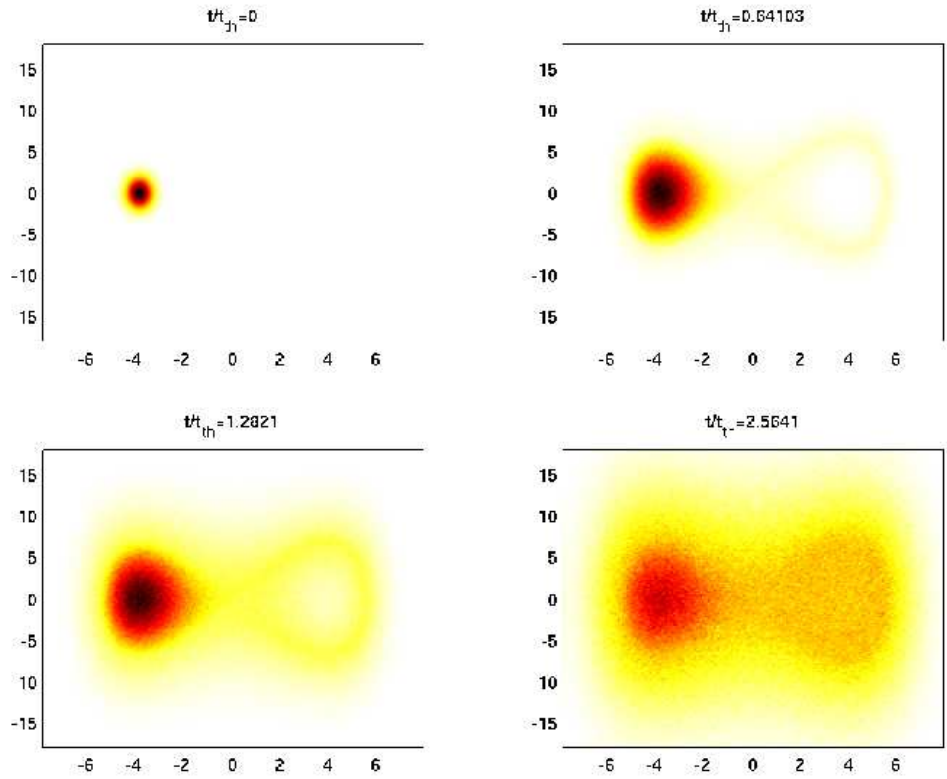}
\caption{Funci\'on de distribuci\'on cl\'asica para el sistema en 
los tiempos indicados. El eje horizontal corresponde a  $x$ y
el vertical a $p$.   El color blanco de fondo, corresponde a valores
nulos de la funci\'on de Wigner, mientras que el colores del amarillo al rojo,
a valores positivos de $W(x,p)$ y, colores en la gama del azul a 
valores negativos de la misma.}
\label{wigner_class}
\end{figure}
En la Fig.~\ref{wigner_class}, presentamos  la funci\'on de Wigner cl\'asica, 
es decir, la distribuci\'on
de probabilidad en el espacio de fases, para ciertos tiempos. Para toda la
evoluci\'on, y como esper\'abamos, $W(x,p)>0$, ya que la ecuaci\'on
de Fokker-Planck conserva la positividad de la distribuci\'on. A medida que
el tiempo transcurre, el paquete inicial gaussiano se ensancha. Esto genera
que su energ\'ia decrezca y permite, a una mayor cantidad de part\'iculas,
explorar el espacio de fases. Para $t=t_{\rm act}$, cuando la energ\'ia media
de la part\'icula iguala la altura de la barrera de potencial, la activaci\'on 
t\'ermica empieza a jugar un rol m\'as significativo en la evoluci\'on
del sistema. Resulta interesante destacar que, para este tiempo de la
evoluci\'on, la separatriz del espacio de fases muestra una gran densidad
de part\'iculas del lado derecho de la barrera de potencial. Esto confirma que
las part\'iculas cruzan dicha barrera cuando su energ\'ia media es del orden
de la altura de esta \'ultima (correspondiente a la energ\'ia de la separatriz).
Esta caracter\'istica de la activaci\'on t\'ermica tambi\'en puede observarse 
en la Fig.~\ref{figure3dw} del sistema cu\'antico abierto. En el caso del sistema
cu\'antico cerrado, por el contrario, observamos que la funci\'on de Wigner
tiene valores nulos en la regi\'on de la separatriz  a lo largo de toda la evoluci\'on.
En este caso, el efecto t\'unel es f\'acilmente reconocible debido a la presencia
de grandes interferencias cu\'anticas negativas en el origen del espacio de fases.
Si volvemos al caso cl\'asico que nos concierne en esta secci\'on, podemos
observar que para tiempos largos, los efectos difusivos y disipativos
se combinan para poblar las regiones centrales del m\'inimo derecho 
de potencial. Finalmente, el aspecto general de la funci\'on de Wigner
se vuelve sumamente sim\'etrico, indicando que el sistema converge
asint\'oticamente a un estado de equilibrio t\'ermico con el entorno.

\section{Efecto t\'unel y  p\'erdida de coherencia 
a \mbox{temperatura} estrictamente cero}
\label{tunelydecoT0}

En esta secci\'on, repetiremos todo el an\'alisis anterior para el caso en
que el entorno tiene temperatura cero.  Asumiremos, de nuevo, un entorno
\'ohmico y usaremos los tiempos de p\'erdida de coherencia estimados en la
Secci\'on \ref{decoohm}. Es necesario mencionar que las escalas
temporales obtenidas en dicha secci\'on fueron derivadas
 para un oscilador arm\'onico. Por tanto, s\'olo nos resultar\'an \'utiles
cuando empecemos con un estado inicial muy angosto localizado
en uno de los m\'inimos de potencial. Para tiempos cortos, el sistema
evolucionar\'a como si estuviera bajo la influencia de un potencial arm\'onico.
Luego de un tiempo corto, las no linealidades del potencial se volver\'an
importantes generando interferencias din\'amicamente \cite{nunoferdiana}.
De esta manera, esperamos que los tiempos all\'i estimados nos 
resulten \'utiles si, efectivamente, se cumple que la p\'erdida de coherencia
ocurre muy temprano para el sistema.

Para resolver la ecuaci\'on maestra utilizamos una rutina de paso variable 
para distintos par\'ametros del sistema y del entorno. Todos los resultados
obtenidos son estables ante cambios en los par\'ametros del m\'etodo
de integraci\'on Runge Kutta utilizado. Como ejemplo, mostraremos los resultados
obtenidos cuando $\Omega=100$ y $V_0 = 200$, para los cuales el tiempo de 
``tuneleo" resulta $\tau\approx 158.27$. De nuevo, queremos que la p\'erdida
de coherencia en el sistema sea anterior al efecto t\'unel, por tanto debemos
fijar los par\'ametros del entorno consecuentemente a $a t_{\rm D} = \tau$. Eligiremos
$a=10$. Usando la expresi\'on del tiempo de p\'erdida de coherencia cuando 
$\Omega t \ll 1$, es decir $t_{\cal D} \leq 1/(8 \gamma_0)$, 
obtenemos  $\gamma_0 = a/(8 \tau) \approx
0.007897$. Fijando la frecuencia de corte en $\Lambda =10 V_0= 2000$, y
considerando la Ec.(\ref{Omegaohm}), vemos que la renormalizaci\'on 
de la frecuencia $\delta
\Omega^2(t)$ puede ser despreciada a todo tiempo. De hecho, es f\'acil
verificar que, para estos valores, ${\tilde \Omega}^2$ es $0.32\%$  del valor 
de $\Omega^2$.  De esta forma, podemos  asumir tranquilamente
que el estado inicial est\'a dado por el vac\'io de un oscilador arm\'onico
de frecuencia $\Omega^2$, en lugar de ${\tilde \Omega^2}$.
\begin{figure}
\centering
\epsfxsize=10cm
\epsfbox{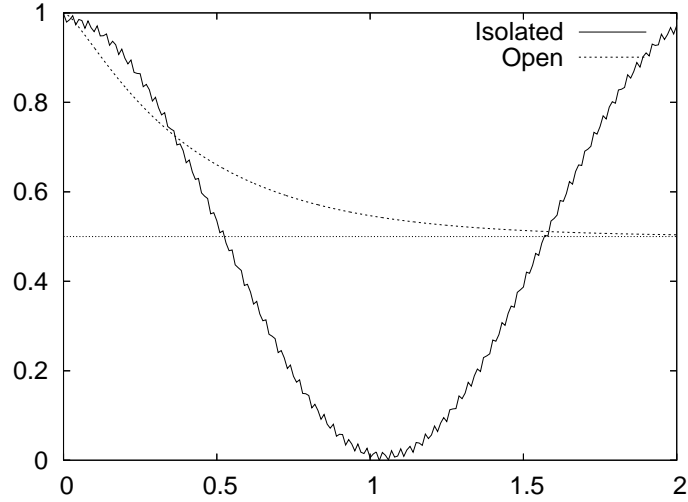}
\caption{Evoluci\'on temporal de la probabilidad de permanencia del
lado original de la barrera de potencial. El eje horizontal es el tiempo
medido en unidades del tiempo de ``tuneleo" $\tau$, mientra que el
vertical es la probabilidad $P(t)$.}
\label{fig1T=0dw}
\end{figure}
En la Fig.\ref{fig1T=0dw}, mostramos la probabilidad de encontrar a la part\'icula
del lado original de la barrera en funci\'on del tiempo (medido en unidades
del tiempo de ``tuneleo" estimado $\tau$), mientras que en las 
Figs. \ref{fig2T=0dw} y \ref{fig3T=0dw}, se grafica la distribuci\'on de probabilidad 
$\sigma(x,x)$ y la funci\'on de Wigner
del sistema, respectivamente.
\begin{figure}
\centering
\epsfxsize=14cm
\epsfbox{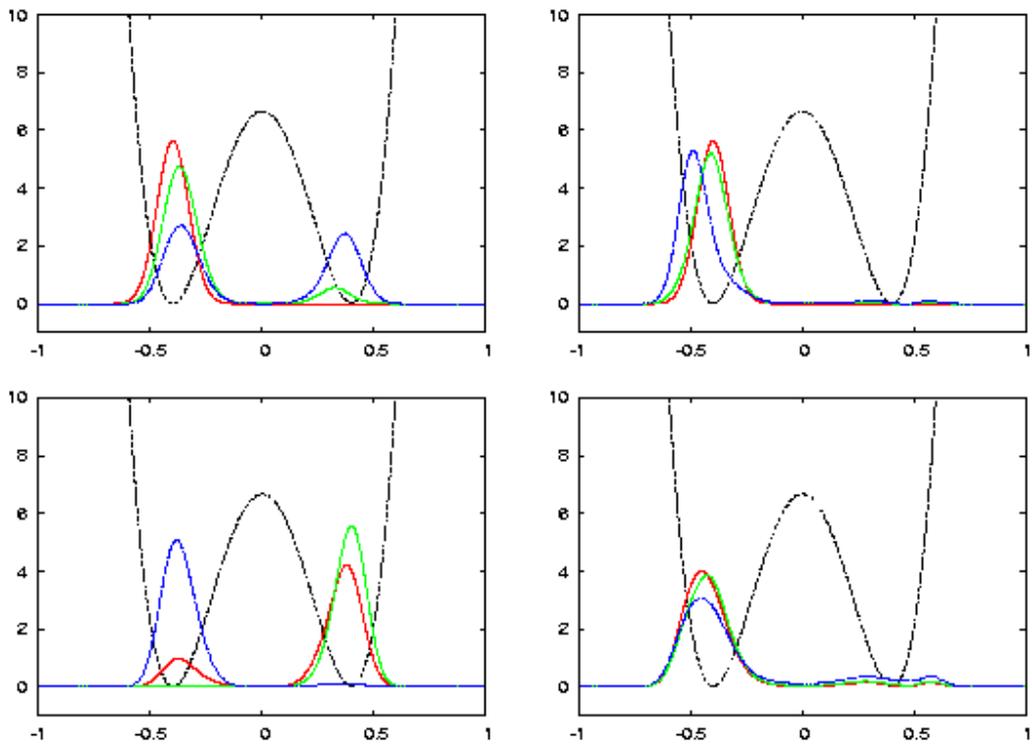}
\caption{Distribuci\'on de probabilidad $\sigma(x,x)$ (eje y) para el sistema 
aislado (izquierda) y el sistema abierto (derecha) para los tiempos
$t=0$; $t=0.1\;\tau$ y $t=0.2\;\tau$ (en los gr\'aficos superiores);
$t=0.5\;\tau$ y $t=\tau$ (en los inferiores).  El potencial
$V(x)$ rescalado aparece como referencia.
\label{fig2T=0dw}}
\end{figure}
\begin{figure}
\centering
\epsfxsize=12.5cm
\epsfbox{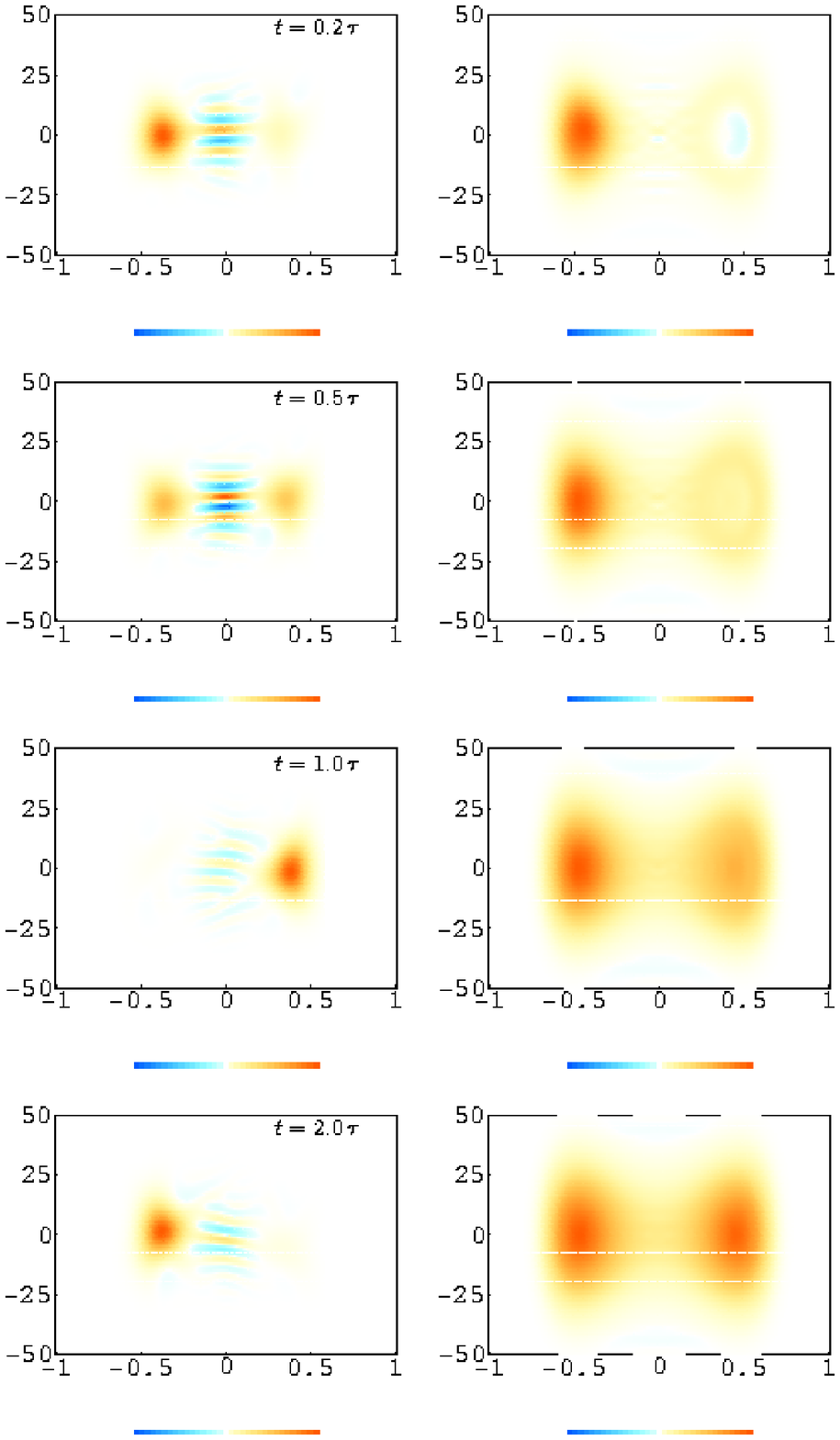}
\caption{La distribuci\'on de Wigner para el sistema aislado (izquierda)
y el sistema abierto (derecha) para los tiempos indicados. Los colores
son similares a aquellos definidos en la Fig.\ref{figure3dw}.}\label{fig3T=0dw}
\end{figure}
La evoluci\'on temporal del sistema abierto cuando el entorno est\'a a
temperatura cero, presenta tanto similitudes como diferencias significativas con
su an\'alogo a temperatura alta. Para tiempos muy cortos, la probabilidad de 
permanencia decrece r\'apidamente, acerc\'andose a $0.5$ cuando $t\sim 2\tau$.
Como esper\'abamos,  no hay signos de oscilaci\'on en el sistema ya que
 $t_{\cal D} < \tau$. Incluso, el comportamiento
asint\'otico tampoco presenta oscilaciones que impliquen que el efecto
t\'unel tiene un rol importante en la din\'amica a tiempos largos. Por el contrario,
observamos que el sistema evoluciona hacia un estado de ``equilibrio",
donde existe igual probabilidad de encontrar a la part\'icula en cada m\'inimo
del potencial. Tanto la distribuci\'on de probabilidad como la funci\'on de Wigner
corroboran este resultado. Para tiempos muy cortos, las regiones negativas
de la funci\'on de Wigner son considerablemente suprimidas si comparamos
con el escenario del sistema cerrado al mismo tiempo. Para $t>\tau/2$,  
 $W(x,p)$ se vuelve positiva y el efecto t\'unel es definitivamente suprimido.
Como ocurr\'ia en el caso de temperatura alta, la separatriz se puebla
considerablemente cuando la activaci\'on {\it t\'ermica} se vuelve relevante. Tanto $\sigma(x,x)$ como $W(x,p)$ se vuelven sim\'etricas
alrededor de $x=0$, lo cual sugiere que la din\'amica del sistema abierto deber\'ia
ser posible de describirse en t\'erminos de un proceso semejante a la activaci\'on 
t\'ermica. 
\begin{figure}
\centering
\epsfxsize=10cm
\epsfbox{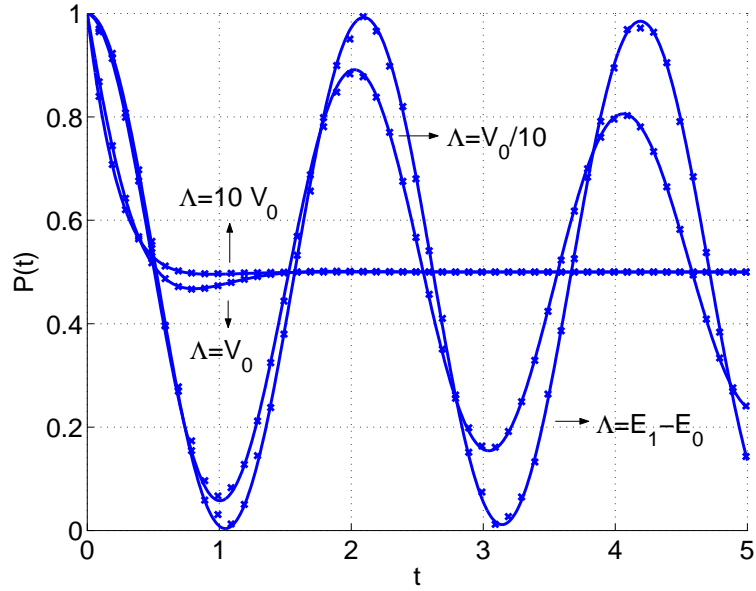}
\caption{Probabilidad de permanecer en el pozo original para
diferentes valores de $\Lambda$ (los otros par\'ametros son iguales a los gr\'aficos anteriores). Las cruces corresponden
a un grupo de los valores de la simulaci\'on num\'erica (no est\'an todos los puntos de la simulaci\'on para que la curva sea visible). La l\'inea de trazo s\'olido corresponde a un
ajuste no lineal de $\chi^2$ de los valores num\'ericos a la
expresi\'on de la Ec.(\ref{fitdw}). El tiempo est\'a medido
en unidades del tiempo $\tau$ del sistema cerrado.}
\label{prob-L2dw}
\end{figure}
Esto fue precisamente lo que estudiamos en la Secci\'on \ref{ET0}
para entornos generales y denominamos excitaci\'on energ\'etica inducida por
el ruido del entorno. En este caso, adem\'as del an\'alisis del rol de las fluctuaciones de vac\'io y el rol importante que juega la difusi\'on an\'omala, podemos mencionar
la relaci\'on particular que existe entre la p\'erdida de coherencia y la excitaci\'on 
de energ\'ia a $T=0$. En ambos procesos, la frecuencia de corte $\Lambda$ es
de gran importancia, ya que los afecta en la misma direcci\'on.
En las Figs. \ref{prob-L2dw} y  \ref{energia-L2dw}  se muestran
la probabilidad de permanecer en el m\'inimo original y la energ\'ia media del 
sistema, respectivamente, para distintos valores de la frecuencia de corte. 
$\Lambda$ var\'ia desde la frecuencia m\'as chica presente en el entorno
a la m\'axima $\Lambda = 10 V_0$.  Adem\'as, mostramos los valores intermedios
$\Lambda = V_0/10$ y $\Lambda = V_0$. 

Vemos que, si disminuimos el valor de $\Lambda$, el tiempo de activaci\'on
se ve retrasado, pero el de p\'erdida de coherencia tambi\'en. En esta 
situaci\'on, la part\'icula puede oscilar entre los m\'inimos de potencial, atravesando
la barrera por efecto t\'unel, por un per\'iodo m\'as largo. Valores m\'as
grandes de la frecuencia, conducen a que ambos procesos tengan lugar m\'as
temprano en la evoluci\'on del sistema. Por tanto, no es posible
localizar la part\'icula en uno de los m\'inimos de potencial, suprimiendo
simult\'aneamente el efecto t\'unel y la activaci\'on en el sistema, para esos
valores de $\Lambda$.

La dependencia de la activaci\'on energ\'etica con la frecuencia de corte
puede ser obtenida de manera m\'as cuantitativa realizando un
ajuste num\'erico de la probabilidad de permanencia de la part\'icula con una
simple expresi\'on evolutiva. En la Fig.\ref{prob-L2dw}, una selecci\'on
de puntos de la simulaci\'on num\'erica (cruces) son mostrados junto al
resultado del ajuste (l\'inea de traza s\'olido) de la forma,
  \begin{equation}
     P(t)=\frac{1}{2}+\frac{1}{2} \cos(\pi t/\tau) \exp(-t/t_{\rm
     act}).
  \label{fitdw}
  \end{equation}
La expresi\'on anal\'itica se ajusta extremadamente bien a los datos
de la simulaci\'on num\'erica, lo cual nos permite determinar las dos
escalas relevantes $\tau$ y $t_{\rm act}$ para cada elecci\'on de
$\Lambda$.
\begin{figure}
\centering
\epsfxsize=10cm
\epsfbox{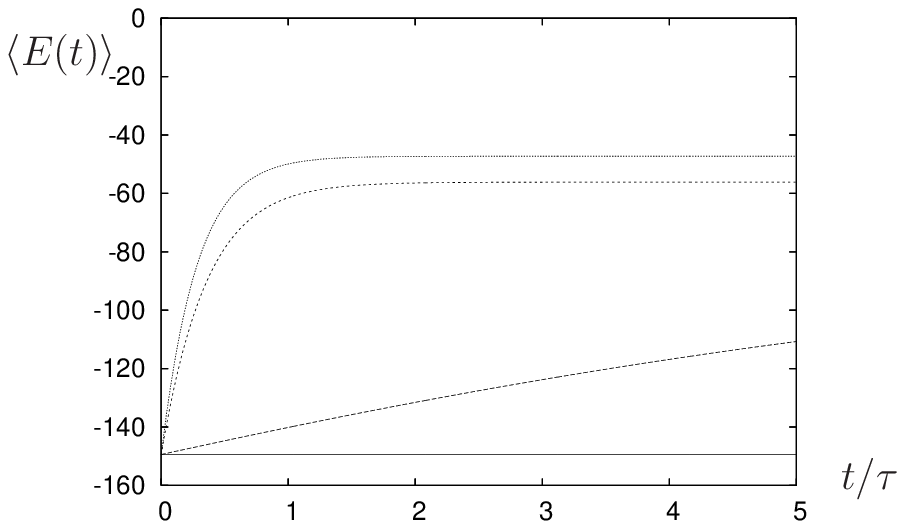}
\caption{Evoluci\'on temporal de la energ\'ia media del sistema
principal para el mismo conjunto de par\'ametros que en la Fig.\ref{prob-L2dw}. El tiempo est\'a medido en unidades del tiempo $\tau$. La l\'inea de trazo s\'olido corresponde al valor
m\'as chico de $\Lambda$; la l\'inea de trazo cortado al valor 
$\Lambda = V_0/10$. La l\'inea punteada es para los valores m\'as grandes de la frecuencia de corte: $\Lambda = 10 V_0$ arriba y $\Lambda = V_0$ abajo.}
\label{energia-L2dw}
\end{figure}
En la Fig.\ref{t_actdw}, mostramos el tiempo de  
activaci\'on calculado de esta forma, en funci\'on de la frecuencia
de corte. La curva incluye un n\'umero mayor de curvas, de modo de
cubrir una regi\'on m\'as amplia de posibles valores de $\Lambda$.
Podemos observar que, el tiempo de activaci\'on inicialmente decrece
a medida que $\Lambda$ aumenta.  Sin embargo, cuando $\Lambda$
es del orden de la altura de la barrera de potencial $V_0$, se nota un cambio
de r\'egimen. Para valores $\Lambda<V_0$, el efecto t\'unel es a\'un
observable en el sistema. Incluso, el valor deducido del ajuste de la
Ec.(\ref{fitdw}) coincide con un error menor al $5\%$ de aquel estimado
para el sistema aislado. Por el contrario, para valores de $V_0$ m\'as grandes que 
$\Lambda$, este efecto est\'a completamente inhibido. Esto implica
que el t\'ermino oscilatorio de la Ec.(\ref{fitdw}) se vuelve completamente
irrelevante para el ajuste realizado. Con todo esto, podemos considerar
que $\Lambda\simeq V_0$ es una suerte de umbral, a partir del cual
las fluctuaciones de vac\'io tienen un rol principal en la evoluci\'on
din\'amica del sistema.

En todos los casos, el comportamiento asint\'otico para la probabilidad
$P(t)$ fue 0.5 con gran precisi\'on. A modo de revisi\'on, realizamos
nuevos ajustes, pero dejando el valor asint\'otico de $P(t)$  como otro
par\'ametro libre del modelo. En todo el rango de valores de $\Lambda$
considerado, siempre obtuvimos resultados que se diferenciaban de 
0.5 en menos de un $0.8\%$.
\begin{figure}
\centering
\epsfxsize=9cm
\epsfbox{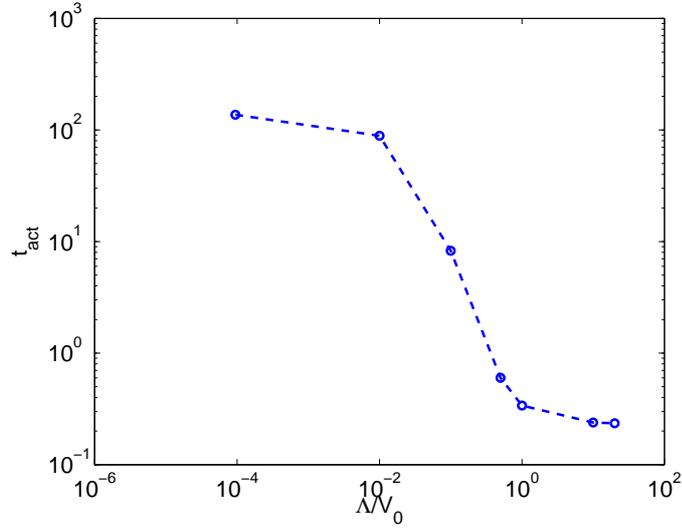}
\caption{Gr\'afico logar\'itmico del tiempo de activaci\'on
$t_{\rm act}$ como funci\'on de $\Lambda/V_0$.}
\label{t_actdw}
\end{figure}
Finalmente, podemos estudiar c\'omo afecta el patr\'on de la evoluci\'on
el valor de $\gamma_0$. En el caso particular de $\Lambda = 10 V_0$,
encontramos que, cuando $\gamma_0$ disminuye, el efecto t\'unel reaparece
y el tiempo de  activaci\'on aumenta. Estos resultados se presentan
en la Fig~\ref{t_act_gammadw}. 
 \begin{figure}
\centering
\epsfxsize=9cm
\epsfbox{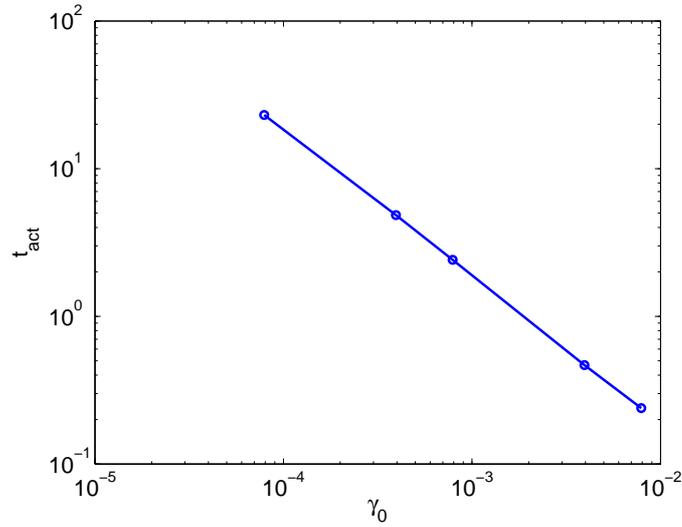}
\caption{Gr\'afico logar\'itmico del tiempo de activaci\'on
$t_{\rm act}$ como funci\'on de $\gamma_0$ para el valor $\Lambda=10 V_0$.}
\label{t_act_gammadw}
\end{figure}
Notablemente, el valor del tiempo estimado de ``tuneleo" $\tau$
var\'ia con $\gamma_0$. Para valores muy chicos del acoplamiento,
obtenemos $\tau\sim 1$ (en unidades de $\tau$). Este valor aumenta
a medida que aumenta $\gamma_0$ hasta un $50\%$ respecto del valor
original de este tiempo; en ese caso, el efecto t\'unel se ve suprimido.
Esto indica que, cuando la interacci\'on con el entorno se
hace m\'as grande, probablemente los efectos de renormalizaci\'on
de la frecuencia del sistema ya no sean despreciables. Adem\'as,
debemos notar que, dentro de la regi\'on de los par\'ametros
considerados,  el tiempo de activaci\'on var\'ia inversamente proporcional
a $\gamma_0$. Esto guarda semejanza con los tiempos de p\'erdida
de coherencia estimados en las Ecs.(\ref{tddw}) y Ec.(\ref{td1}); es decir,
que  ambos procesos dependen de igual forma de $\gamma_0$,
y por lo tanto, est\'an estrechamente relacionados.

En s\'intesis, hemos descripto la din\'amica del sistema abierto en funci\'on
de tres fen\'omenos principales: p\'erdida de coherencia, efecto t\'unel
y excitaci\'on  energ\'etica. Tanto el primero como el \'ultimo proceso son
inducidos por la presencia del entorno, mientras que el segundo es de
naturaleza puramente cu\'antica. Hemos demostrado que el entorno 
puede ser modelado de modo de que $t_{\cal D} \ll \tau$, o de hecho, 
cualquier permutaci\'on posible de los tres procesos que describen
la din\'amica del sistema. Esto se motiva en la comparaci\'on de las 
escalas temporales estimadas
en cada caso. 
Para el caso del sistema cerrado, las simulaciones num\'ericas reprodujeron
todas las propiedades conocidas del efecto t\'unel. El an\'alisis entre el caso
abierto y el cerrado nos result\'o fundamental; por un lado, para
 entender que, cuando el
entorno est\'a a temperatura cero, las fluctuaciones cu\'anticas del entorno
tienen un rol importante en la evoluci\'on y, por el otro, 
 confirmar nuestro estudio de la Secci\'on \ref{ET0}: que a\'un en contacto con un 
entorno a $T=0$, el sistema se excita energ\'eticamente para tiempos
posteriores al tiempo de p\'erdida de coherencia.

El efecto  t\'unel macrosc\'opico cu\'antico puede ser relacionado con el
efecto t\'unel de una funci\'on de onda de muchos cuerpos a trav\'es de
una barrera de potencial. De esta forma, provee una prueba fehaciente
de la v\'alidez de la Mec\'anica Cu\'antica, mucho m\'as concluyente que
el caso de este mismo efecto para una \'unica part\'icula. Un ejemplo concreto
donde el efecto t\'unel macrosc\'opico resulta experimentalmente accesible
es en el caso de un condensado de Bose-Einstein (BEC) en una trampa \'optica.
Estos sistemas tienen un n\'umero  finito de \'atomos (manejable) y, por tanto, se
extienden a lo largo de la frontera entre lo microsc\'opico y macrosc\'opico, y
por que no, entre los sistemas cu\'anticos y cl\'asicos.

Las propuestas experimentales apuntan al uso de \'atomos confinados \cite{Shin1}.
En general, consisten en la separaci\'on y fusi\'on de dos barreras de potencial,
de modo que los paquetes de onda at\'omicos se separen y recombinen, correspondientemente \cite{Anderson}. Por lo pronto, BECs en un pozo doble de potencial han sido
creados en un experimento \cite{Shin2}. Las interacciones \'atomos-\'atomos a distancias
finitas tienden
a localizar las part\'iculas en uno de los m\'inimos de potencial y reducen la
coherencia en los procesos de separaci\'on y recombinaci\'on \cite{Menotti}, mientras
que el efecto t\'unel entre los m\'inimos de potencial, sirve para ``delocalizar"
los paquetes de onda at\'omicos y  asegurarse una fase relativa bien definida
entre los pozos de potencial. La interacci\'on entre los \'atomos condensados
y aquellos que no lo est\'an favorecen la decoherencia en el sistema \cite{Pitaevskii}.

En este contexto, resulta muy interesente la extensi\'on del estudio de la din\'amica
del sistema  al
caso de superposiciones macrosc\'opicas de estados cu\'anticos, como por ejemplo
BECs en un pozo doble de potencial \cite{Dounas}. En \cite{dwJCS}, hemos extendido
nuestras consideraciones acerca  del efecto t\'unel, la decoherencia y la activaci\'on
al caso en el que la condici\'on
inicial en el sistema es una superposici\'on coherente de dos paquetes gaussianos.
En los experimentos de BECs de gases at\'omicos dilu\'idos alkalinos-met\'alicos, los
\'atomos atrapados son enfr\'iados por evaporaci\'on e intercambian part\'iculas 
con el entorno. La coherencia de BECs macrosc\'opicos cu\'anticos resulta en el
efecto t\'unel de \'atomos entre  modos, de manera an\'aloga al efecto t\'unel
en pares de Cooper de una juntura Josephson. Por lo tanto, nuestra motivaci\'on
en \cite{dwJCS}, fue estudiar cu\'an robusta es la coherencia de los BECs
en presencia de un entorno a temperatura cero, y de qu\'e manera, el
efecto t\'unel se ve afectado.            
\newpage
\thispagestyle{empty}
\cleardoublepage

\chapter{P\'erdida de coherencia inducida por entornos compuestos}
\label{c4}
\markboth{P\'erdida de coherencia inducida por  entornos compuestos}
{Cap\'itulo 4}


Generalmente, un sistema ``extenso", es decir compuesto
por dos o m\'as subsistemas, se encuentra en interacci\'on con
un entorno externo, posiblemente modelado por un ba\~no
t\'ermico con un n\'umero muy grande de grados de libertad. 
Este tipo de sistemas es usualmente conocido como sistema compuesto.
La transferencia de electrones en soluciones \cite{prezhdo}, 
una mol\'ecula biol\'ogica grande, la relajaci\'on vibracional 
de las mol\'eculas en soluciones, los excitones en semiconductores 
acoplados a modos ac\'usticos u \'opticos de los fonones, 
pueden ser ejemplos v\'alidos de este tipo de sistemas en la Naturaleza.

En este cap\'itulo analizaremos el efecto inducido en un sistema 
cu\'antico debido a la presencia del entorno. El modelo
considera al sistema cu\'antico compuesto formado por
un subsistema $A$ acoplado a un subsistema $B$, el cual, a su vez,
est\'a en interacci\'on con el entorno $\cal{E}$. El acoplamiento de
$A$ al entorno es s\'olo a trav\'es del subsistema $B$, por lo cual, 
consideramos el subsistema $A$ en interacci\'on con un entorno 
compuesto formado por $B + {\cal E}$. Para estudiar
este problema, consideraremos un modelo sencillo en el cual
el subsistema $A$ est\'a representado por un  oscilador 
arm\'onico y  el subsistema $B$ por un oscilador invertido. La principal
motivaci\'on para estudiar este modelo es doble. Por un lado,
buscamos ahondar en el estudio de la p\'erdida de coherencia en
sistemas cu\'anticos debido a la presencia de entornos ca\'oticos.
El oscilador invertido ha sido  usado recientemente para modelar
esta situaci\'on \cite{robin}. A pesar que esa consideraci\'on es una
simplificaci\'on muy grande para modelar  un entorno ca\'otico, el modelo
permite mostrar la sensibilidad exponencial del sistema en cuesti\'on
a las perturbaciones, lo cual resulta fundamental para analizar las
evoluciones ca\'oticas. Por otro lado, se busca enfatizar la idea que 
aislar un sistema cu\'antico de un entorno ca\'otico resulta muy
d\'ificil, como ya ha sido notado en \cite{robin}. M\'as a\'un, resulta
mucho m\'as complicado aislar el sistema cu\'antico de un entorno
ca\'otico que de los muchos osciladores arm\'onicos que modelan 
el entorno en el Movimiento Browniano Cu\'antico (MBC).
En este contexto, consideraremos dos situaciones diferentes. 
Primero, el caso en el cual el grado de libertad ca\'otico est\'a en
el entorno, es decir, $B$ est\'a representado por un oscilador invertido
y directamente acoplado, por un lado,  al entorno externo ${\cal E}$,
y por el otro, al subsistema $A$. Segundo, el caso en el cual, el grado
de libertad inestable est\'a en el sistema, es decir $A$ es un oscilador
arm\'onico acoplado a un oscilador invertido $B$, a su vez en
interacci\'on con el entorno ${\cal E}$. Estas son las extensiones
naturales de los estudios realizados en \cite{robin} y 
\cite{fermazzidiana,nunoferdiana,elze}
para el primer y segundo caso, respectivamente. En ambas situaciones,
estimaremos el tiempo de p\'erdida de coherencia, el cual  resulta
distinto para cada caso.
Finalmente, el an\'alisis del modelo se completar\'a con la inclusi\'on de las
otras dos posibilidades para el sistema, es decir,  el subsistema $A$ y el subsistema $B$, 
ambos representados por osciladores arm\'onicos en un caso, y ambos
osciladores invertidos en el otro. Por supuesto, y como en los otros dos
casos mencionados con anterioridad, el susbsistema $B$ est\'a a la vez
acoplado al entorno ${\cal E}$, formando un entorno compuesto en todos los casos.
Resumiendo, este cap\'itulo contar\'a con cuatro casos para analizar.
En todos ellos, estudiaremos la din\'amica del subsistema $A$, 
inducida por la presencia del entorno cmpuesto o ``efectivo''.  
Veremos que, todos los casos estudiados tienen distintas escalas de p\'erdida de
coherencia asociadas, las cuales dependen no s\'olo de los 
par\'ametros del entorno ${\cal E}$ (como su temperatura por ejemplo), sino adem\'as,
de los par\'ametros del subsistema $B$. En todos los casos, 
la p\'erdida de coherencia es mayor cuando $A$ es un grado de libertad 
inestable.
En trabajos previos, un sistema compuesto diferente ha sido analizado.
Por ejemplo, en \cite{kapral1}, el subsistema $A$ est\'a representado por
un sistema de dos niveles, el cual est\'a acoplado bilinealmente a un 
oscilador arm\'onico (subsistema $B$), el cual a su vez est\'a acoplado
a un entorno \'ohmico modelado por infinitos osciladores arm\'onicos. Los autores
han mostrado que $B$ pierde coherencia m\'as r\'apidamente que $A$. 
Este sistema de dos niveles tambi\'en fue estudiado en \cite{anu}, un trabajo
en el cual buscaban encontrar la soluci\'on exacta para la matriz densidad
reducida del sistema compuesto $AB$. Sin embargo, a pesar de que 
el sistema compuesto presentado en este cap\'itulo pueda resultar muy
similar a aquel de los trabajos citados \cite{kapral1} y \cite{anu}, nuestro
modelo tiene una din\'amica completamente diferente y merece ser
estudiado de forma completa y separada. 

\section{El modelo}
\label{modelocomp}

\subsection {Formulaci\'on general}
En adelante, consideraremos el sistema cu\'antico $AB{\cal E}$
formado de tres subsistemas interactuantes: un subsistema $A$
acoplado a un subsistema $B$, el cual a la vez est\'a en directa
interacci\'on con un entorno ${\cal E}$. La acci\'on cl\'asica 
total de $AB{\cal E}$ es:
\begin{equation}
 S[x,q,Q] = S_{\rm A}[x] + S_{\rm B}[q] + S_{\rm {\cal E}}[Q] +
S_{\rm AB}^{\rm int}[x,q] + S_{B \cal E}^{\rm int}[q,Q].
\end{equation}

Siguiendo el ejemplo del MBC, el entorno se modela con
un conjunto de N osciladores independientes de frecuencias 
$\tilde{\omega }_n$, masas $m_n$ y coordenadas y momentos
conjugados $(\hat{Q},\hat{P})=(\hat{Q}_1,...,\hat{Q}_N,
\hat{P}_1,...,\hat{P}_N)$, de forma tal que la acci\'on cl\'asica del
mismo es
\begin{equation}
  S_{\rm {\cal E}}[Q] = \int_0^t ds \sum_n \frac{m_m}{2} ({\dot Q}_n^2 -
\tilde{\omega}_{n}^2 Q_n^2).
\end{equation}

El subsistema $B$ consiste de un \'unico oscilador (ya sea \'armonico o
invertido, dependiendo del caso en cuesti\'on) con masa $M_B$, frecuencia
$\Omega$ y operador de posici\'on $\hat{q}$,
\begin{equation}
 S_{\rm B}(x) = \int_0^t ds \frac{M_{\rm B}}{2} (\dot q^2 \pm
\Omega^2 q^2).
\end{equation}
La interacci\'on entre el subsistema $B$ y el entorno es considerada como
bilineal y se modela seg\'un:
\begin{equation} S_{\rm B {\cal E}}^{\rm int} = \int_0^t ds \sum_n c_n q(s)
Q_n(s), \end{equation}
donde $c_n$ es la constante de acoplamiento con el oscilador en\'esimo.
El entorno est\'a caracterizado por la densidad espectral 
$I_{\cal E} \equiv  \pi \sum_n \frac{c_n^2}{2 m_n
\tilde{\omega}_n} \delta (\tilde{\omega}-\tilde{\omega}_n)$. 
Asumiremos
que el entorno es \'ohmico, aunque las generalizaciones son directas.
 De esta manera, la densidad espectral se
simplifica $I_{\cal E}(\omega ) = 2 M \gamma_0 \tilde{\omega}
e^{{-\frac{\tilde{\omega}^2}{\Lambda^2}}}$, donde $\Lambda$ es una
frecuencia f\'isica de corte, relacionada a la m\'axima frecuencia
presente en el entorno. 
Por \'ultimo, consideraremos al subsistema $A$ formado por un
oscilador (ya sea, arm\'onico o invertido), con operador de posici\'on
$\hat{x}$, cuya acci\'on cl\'asica es
 \begin{equation}
 S_{\rm A}[x] = \int_0^t ds \frac{M_{\rm A}}{2} (\dot x^2 \pm
\omega^2 x^2).
\end{equation}

Supondremos, adem\'as, que el subsistema $A$ est\'a acoplado
bilinealmente al subsistema $B$ a trav\'es del siguiente t\'ermino
\begin{equation}
 S_{\rm AB}^{\rm int} = - \lambda \int_0^t ds
x(s) q(s).\label{intAB} 
\end{equation}

Las propiedades din\'amicas de inter\'es pueden ser le\'idas
de la matriz densidad reducida del sistema $A$ a un tiempo dado $t$. 
La matriz densidad total se escribe de forma integral en funci\'on del propagador
(tomando $t_0=0$)
\begin{equation}
 \hat{K}(x,q,Q;t|x_0,q_0,Q_0;0) \equiv \hat{K}(t|0) 
= <x q Q\vert
\exp(-i \hat{H}t / \hbar)\vert x_0 q_0 Q_0>. \nonumber
\end{equation}
De esta forma, 
\begin{equation}
\hat{\rho}(x,q,Q,x\prime,q\prime,Q\prime) =\int
dx_0 dx_0\prime dq_0 dq_0\prime dQ_0 dQ_0\prime \hat{K}(t|0)
\hat{\rho}(0)
\hat{K}^\ast(t|0).\end{equation}

En primer lugar, nos ocuparemos de la din\'amica del sistema compuesto
$AB$ bajo la influencia del entorno $\cal E$, para, m\'as tarde, estudiar 
\'unicamente la din\'amica de $A$ debida al entorno efectivo. 
Por lo tanto, el objeto de estudio
primero es la matriz densidad reducida $\hat{\rho}_{\rm r}$, 
que se obtiene integrando
los grados de libertad irrelevantes del entorno $\cal E$. Esta reducci\'on de
grados de libertad es correcta si el tiempo caracter\'istico del
entorno (esencialmente del orden de $\sim 1/\Lambda$) es mucho menor
que aquellos correspondientes al subsistema $A$ y al subsistema $B$.
Como es usual en estos problemas, supondremos que la condici\'on inicial
del sistema compuesto $AB$ y del entorno $\cal E$
es factorizable, es decir,
\begin{equation} \hat{\rho}(x_0,x_0\prime,q_0,q_0
\prime, Q_0,Q_0\prime;0) = \hat{\rho}_{\rm AB}(x_0,x_0
\prime,q_0,q_0\prime;0) \hat{\rho}_{\cal
E}(Q_0,Q_0\prime;0), \end{equation}
y, adem\'as, que el ba\~no est\'a inicialmente en equilibrio
t\'ermico a temperatura $T$.

De esta forma, escribimos la forma integral de la matriz densidad reducida
a un tiempo dado $t$ como
\begin{equation}
\hat{\rho}_{\rm r}(x,x\prime,q,q\prime,t)= \int
dx_0 dx_0 \prime dq_0 dq_0\prime \hat{J}_{\rm r}(x,x\prime,q,q\prime;t
\vert x_0,x_0\prime,q_0,q_0\prime;0) 
\hat{\rho}_{\rm AB}(x_0,x_0 \prime,q_0,q_0\prime;0), \end{equation}
donde el operador de evoluci\'on reducido  $\hat{J}_{\rm r}$ es
\begin{eqnarray}
\hat{J}_{\rm r}(x,x\prime,q,q\prime;t \vert
x_0,x_0\prime,q_0,q_0\prime;0) &=& \int
dQ_0dQ_0\prime \hat{K}(x,q,Q;t\vert x_0,q_0,Q_0;0) \rho_{\cal
E}(Q_0,Q_0\prime,0)\nonumber \\
&\times & \hat{K^{\ast}}(x\prime,q\prime,Q\prime;t\vert
x_0\prime,q_0\prime,Q_0\prime;0).\end{eqnarray}


\subsection{M\'etodo de la funcional de influencia}
\label{influececomp}

La formulaci\'on de la matriz densidad reducida en t\'erminos de la funcional
de influencia ha sido largamente discutida en la Literatura
\cite{Feynman,GraIng,HuPazZhangI}. En la Secci\'on \ref{FVcap1}, hemos
utilizado el m\'etodo de la funcional de Feynman y Vernon en el marco
del movimiento Browniano cu\'antico. 
En esta secci\'on, aplicaremos
el mismo m\'etodo funcional 
al sistema cu\'antico compuesto $AB$, formado por
dos osciladores, ya sean arm\'onicos o invertidos.
En el caso m\'as general, el operador de evoluci\'on  $\hat{J}$  para
la matriz densidad $\hat\rho$ es $\hat{\rho}(t)=\hat{J}(t,0)\hat{\rho}(0)$,
donde
\begin{eqnarray}
\hat{J}(x_{\rm f},q_{\rm f},Q_{\rm f},x_{\rm f}\prime,q_{\rm f}\prime,Q_{\rm f}'
\vert x_0,q_0,Q_0,x_0',q_0',Q_0')
&=&  \int_{x_0}^{x_{\rm f}}{\cal D}x\int_{q_0}^{q_{\rm f}}{\cal
D}q\int_{Q_0}^{Q_{\rm f}}{\cal D}Q e^{\frac{i}{\hbar}S(x,q,Q)} \nonumber \\
&\times & \int_{x_0'}^{x_{\rm f}'}{\cal D}x'\int_{q_0'}^{q_{\rm f}'}{\cal D}q'
\int_{Q_0'}^{Q_{\rm f}'}{\cal D}Q' e^{-\frac{i}{\hbar}S(x',q',Q')} \nonumber
.\end{eqnarray} 
Las integrales de camino son sobre todas las posibles historias compatibles
con las condiciones de contorno. Como ya se mencion\'o, nuestro 
inter\'es primario es el efecto del ba\~no t\'ermico en el sistema compuesto
$AB$, definido por
\begin{equation}
\rho_{\rm r}(x,x',q,q') 
=  \int_{-\infty}^{+\infty}dQ \int_{-\infty}^{+\infty}dQ' \rho
(x,q,Q\vert x',q',Q') \delta (Q - Q'), \end{equation}
y el operador de evoluci\'on en el tiempo est\'a dado por el
operador de evoluci\'on reducido $\hat{J}_{\rm r}$
\begin{equation}
\rho_{\rm r}(x,x',q,q';t)=\int_{-\infty}^{+\infty}
\int_{-\infty}^{+\infty}dx_0 dx_0' dq_0 dq_0' ~ J_{\rm
r}(t\vert 0) 
\rho_{\rm AB}(x_0,x_0',q_0,q_0';0), \end{equation}
donde escribimos $J_{\rm r}(t\vert 0)=
J_{\rm r}(x_{\rm f},x_{\rm f}',q_{\rm f},q_{\rm f}';t
\vert x_0,x_0',q_0,q_0';0)$ para simplificar la notaci\'on.
Suponiendo condiciones iniciales separables, el propagador
reducido queda 
\begin{eqnarray}
J_{\rm r}(x_{\rm f},x_{\rm f}',q_{\rm f},q_{\rm f}';t
\vert x_0,x_0',q_0,q_0';0) 
&=&  \int_{x_0}^{x_{\rm f}}{\cal D}x \int_{x_0'}^{x_{\rm f}'}{\cal
D}x' \int_{q_0}^{q_{\rm f}}{\cal D}q \int_{q_0'}^{q_{\rm f}'}{\cal D}q'
e^{\frac{i}{\hbar}(S_{\rm A}(x) - S_{\rm A}(x'))} \nonumber \\ &\times &
e^{\frac{i}{\hbar}( S_{\rm B}(q) - S_{\rm B}(q'))} e^{\frac{i}{\hbar}( S_{\rm AB}(x,q)
- S_{\rm AB}(x',q'))} ~ F^{(AB)}(x,x',q,q'), \nonumber 
\end{eqnarray}
donde $F^{(AB)}(x,x',q,q')$ es la funcional de influencia de Feynman-Vernon
\cite{Feynman} (para el sistema compuesto) definida por
\begin{eqnarray}
F^{(AB)}(x,x',q,q') 
&=&\int_{-\infty}^{+\infty} dQ_{\rm f}
\int_{-\infty}^{+\infty}dQ_0\int_{-\infty}^{+\infty} dQ'_0
\int_{Q_0}^{Q_{\rm f}}{\cal D}Q\int_{Q_0'}^{Q_{\rm f}'}{\cal D}Q'  \nonumber \\
&\times &e^{\frac{i}{\hbar}(S_{{\cal E}}(Q) + S_{\rm B{\cal E}}(q,Q) -
S_{{\cal E}}(Q') - S_{\rm B{\cal E}}(q',Q'))} \ \rho_{{\cal E}}(Q_0,Q_0')  \nonumber \\
&\equiv & e^{\frac{i}{\hbar} \delta A^{(AB)}(x,x',q,q')},\end{eqnarray}
donde $\delta A^{(AB)}(x,x',q,q')$ es la acci\'on de influencia del sistema compuesto
$AB$.  Por lo tanto, se puede definir $A^{(AB)}(x,x',q,q')$ como la acci\'on efectiva
de granulado grueso: $A^{(AB)}= S_{\rm A}(x) - S_{\rm A}(x') + S_{\rm B}(q)
- S_{\rm B}(q')+ S_{\rm AB}(x,q)-S_{\rm AB}(x',q') + \delta A^{(AB)}(x,x',q,q')$.
Es importante destacar que en nuestro modelo, el subsistema $A$ no est\'a
acoplado directamente al entorno $\cal E$. Consecuentemente,
la funcional de influencia es la misma que para un ba\~no de osciladores
arm\'onicos \cite{HuPazZhangI} (y es \'unicamente funci\'on de $q$ y $q'$)
\begin{equation} \delta A^{(AB)}(q,q') =  -2
\int_0^t ds \int_0^s ds' \Delta q(s) ~\eta (s - s') ~\Sigma q(s')
+ i \int_0^t ds \int_0^s ds' \Delta q(s) ~\nu (s -
s') ~\Delta q(s') \end{equation}
con 
\begin{equation}  \Delta q(s) = q(s) - q'(s) ~;~
 \Sigma q(s) = \frac{1}{2} (q(s) +
q'(s)). \label{Deltaq} \end{equation} 

Los n\'ucleos $\eta$ and $\nu$, son los correspondientes
a la  disipaci\'on y el
ruido, respectivamente  y, 
fueron definidos en las Ecs.(\ref{nucleonu}) y (\ref{nucleoeta}) 
del Cap\'itulo \ref{c1}. Como hemos visto, 
en el l\'{\i}mite de temperatura alta, estos n\'ucleos
adoptan una forma m\'as sencilla, siendo 
$\nu \sim 2 M \gamma_0k_B T \delta (s)/\hbar$ y
$\eta \sim  M \gamma_0 \dot{\delta}(s)$
\cite{HuPazZhangI,jpphabzurek}.
Como nosotros trabajaremos en el l\'{\i}mite subamortiguado
y de temperatura 
alta\footnote{Es decir, $\gamma_0 << \Omega$  y $\hbar \omega \ll k_B T$ , pero
sin restricciones en $\gamma_0k_BT$.}, podemos usar estas expresiones
para los n\'ucleos. Por lo tanto, si evaluamos la expresi\'on
de $\delta A^{(AB)}(q,q')$, obtenemos
\begin{equation} \delta A^{(AB)}
(q,q\prime) \simeq - 2 M_{\rm B} \gamma_0 \int_0^t ds \Delta q(s)
\dot{\Sigma} q(s)+ i \frac{2M_{\rm B}\gamma_0k_B
T}{\hbar} \int_0^t (\Delta q(s))^2 ds. \label{deltaA} 
\end{equation}

Despu\'es de integrar los grados de libertad del ba\~no t\'ermico, podemos
escribir la funcional de influencia $F^{(AB)}(q,q\prime)$ en el l\'imite
de  temperatura alta como
\beq F^{(AB)}(q,q\prime) = e^{-\frac{i2 M_{\rm B} \gamma_0}{\hbar}
\int_0^t ds \Delta q(s)
\dot {\Sigma} q(s) } e^{\frac{-2M_{\rm B}\gamma_0k_B T}{\hbar^2} 
\int_0^t (\Delta q(s))^2 ds}, \eeq
y obtener, luego, la matriz densidad reducida,
\beqa
\rho_{\rm r}(x,x',q,q') 
& = & \int_{-\infty}^{\infty} dx_0 dx_0\prime
\int_{-\infty}^{\infty}dq_0 dq_0\prime \int_{-\infty}^{\infty}dq_{\rm f} dq_{\rm f}\prime 
\int_{q_0}^{q_{\rm f}} {\cal D} q \int_{q_0\prime}^{q_{\rm f}\prime} {\cal D}
q\prime \ e^{\frac{i}{\hbar}(S_{\rm B}(q)-S_{\rm B}(q\prime))}\nonumber \\ &\times &
\int_{x_0}^{x_{\rm f}} {\cal D} x
 \int_{x_0\prime}^{x_{\rm f}\prime} {\cal D} x\prime \
e^{\frac{i}{\hbar}(S_{\rm A}(x)-S_{\rm A}(x\prime))}
e^{\frac{i}{\hbar}\big(S_{\rm AB}(x,q)-S_{\rm AB}(x\prime,q\prime)+ \delta
A^{(AB)}(q,q\prime)\big)}. \eeqa

A esta altura, ya tenemos toda la informaci\'on necesaria
para estimar el efecto inducido por la presencia del ba\~no
en el sistema compuesto $AB$. Sin embargo, si buscamos conocer
c\'omo es la p\'erdida de coherencia en el subsistema $A$, debemos
trazar sobre los grados libertad del nuevo entorno. Si miramos las
expresiones obtenidas hasta el momento, veremos que podemos
pensar al problema como un subsistema $A$ acoplado a un subsistema
$B$ a trav\'es de alguna interacci\'on ``efectiva"  
 $S_{\rm eff}^{\rm int}(x,q,x',q')$ definida como
 \begin{eqnarray}
S_{\rm eff}^{\rm int}(x,q,x',q')&=& S_{\rm AB}(x,q)- S_{\rm AB}(x',q')
- 2 M_{\rm B} \gamma_0
\int_0^t ds \Delta q(s) \dot{\Sigma} q(s) \nonumber \\
&+& i\frac{2M_{\rm B}\gamma_0k_B T}{\hbar} \int_0^t ds
(\Delta q(s))^2.\label{Sinteff}
\end{eqnarray}

\subsection{La funcional de influencia aplicada al subsistema A}

Si queremos analizar el efecto producido en el subsistema $A$
debido a la interacci\'on con el subsistema $B$ y el entorno $\cal E$
(a trav\'es de $B$), debemos buscar la matriz densidad reducida
para el subsistema $A$ \'unicamente. Esto es, debemos calcular
\beq \rho^{(A)}_{\rm r}(x,x\prime)=\int_{-\infty}^{\infty}dq
\int_{-\infty}^{\infty}dq\prime \rho(x,q \vert x\prime,q \prime)
\delta(q-q\prime), \nonumber \eeq 
la cual es propagada en el tiempo por el operador evoluci\'on
reducido  $\hat{{\cal J}}^{(A)}_{\rm r}(x,x\prime)$
\beq
\rho_{\rm r}^{(A)}(x,x\prime;t) = \int_{-\infty}^{\infty}dx_0
\int_{-\infty}^{\infty}dx_0\prime {\cal J}^{(A)}_{\rm r}(x,x\prime;t\vert
x_0,x_0\prime;0) 
~ \rho_{\rm r}(x_0,x_0\prime;0).\label{finalrho}\eeq

Por simplicidad, asumiremos que a tiempo $t=0$, el subsistema $A$
y el nuevo entorno ``efectivo'' no est\'an correlacionados,
esto es $\hat{\rho}_{\rm AB}(t=0) = \hat{\rho}_{\rm A}(t=0) 
\otimes \hat{\rho}_{\rm B}(t=0)$. Asumiremos tambi\'en que inicialmente
el subsistema $B$ es un paquete gaussiano de la forma
 $e^{-((q_0-q_0\prime)^2)/2 \sigma}$. Esta elecci\'on 
resulta conveniente ya que este tipo de estados forma un
set cerrado al evolucionar linealmente \cite{fermazzidiana,robin}.
De esta forma, el operador evoluci\'on no depende
del estado inicial y puede ser escrito, como en \cite{HuPazZhangI},
\begin{eqnarray} 
{\cal J}_{\rm r}(x_{\rm f},x_{\rm f}\prime;t \vert x_0,x_0\prime;0)_{(A)}
&=&  \int_{x_0}^{x_{\rm f}} {\cal D}x \int_{x_0\prime}^{x_{\rm
f}\prime} {\cal D}x\prime  e^{\frac{i}{\hbar} (S_{\rm A}(x)-S_{\rm
A}(x\prime))} {\cal F}^{(A)}(x,x\prime) \nonumber \\
& \equiv & 
\int_{x_0}^{x_{\rm f}} {\cal D}x \int_{x_0\prime}^{x_{\rm f}\prime}
{\cal D}x\prime ~ \exp\left\{{\frac{i}{\hbar} {\cal
A}^{(A)}(x,x\prime)}\right\}, \label{calprop}
\end{eqnarray}

donde hemos definido ${\cal F}^{(A)}(x,x\prime) = 
e^{\frac{i}{\hbar} \delta {\cal
A}^{(A)}(x,x\prime)}$ y ${\cal A}^{(A)}(x,x\prime)=
S_{\rm A}(x)-S_{\rm
A}(x\prime)+\delta {\cal A}^{(A)}(x,x\prime)$ 
como la funcional de influencia y la acci\'on de influencia
para el subsistema $A$, respectivamente. 
Para evaluar
$ \delta {\cal A}^{(A)}(x,x\prime)$, debemos realizar las siguientes
integrales,
\beq \delta {\cal
A}^{(A)}(x,x\prime) = \int_{-\infty}^{\infty} dq_0 
\int_{-\infty}^{\infty} dq_0\prime
\int_{-\infty}^{\infty} dq_{\rm f} \int_{q_0}^{q_{\rm f}} 
{\cal D} q 
\int_{q_0'}^{q_{\rm f}'} {\cal D} q'
e^{\frac{i}{\hbar}(S_{\rm B}(q)-S_{\rm B}(q\prime))}
e^{\frac{i}{\hbar}S_{\rm eff}^{\rm int}(x,q,x',q')} .
\label{deltaA2}\eeq

Las integrales funcionales pueden ser calculadas despu\'es de
resolver la ecuaci\'on cl\'asica de movimiento para el subsistema $B$,
dada por
\beq \ddot q(s)
\pm \Omega^2 q(s)=\frac{\lambda}{M_{\rm B}} x(s).\label{ec.mov.}\eeq
En esta \'ultima expresi\'on, hemos despreciado el t\'ermino relacionado 
con la disipaci\'on que surge a ra\'iz de la presencia del entorno $\cal E$.
Esta suposici\'on es v\'alida ya que estamos trabajando en el
r\'egimen subamortiguado ($\gamma_0 \ll 1$).

A esta altura, debemos remarcar nuevamente que nuestro objetivo es
estudiar todas las posibles combinaciones de subsistema $A$ y $B$.
En algunos casos, el
subsistema $B$ ser\'a un oscilador arm\'onico (signo positivo 
en Ec.(\ref{ec.mov.})) y en otros casos, ser\'a un oscilador
invertido (signo menos en la Ec.(\ref{ec.mov.})).  Por lo pronto,
nosotros escribiremos expl\'icitamente la soluci\'on para un s\'olo
caso, ya que es posible obtener las dem\'as soluciones
realizando la correspondiente modificaci\'on ($\Omega$ por $i \Omega$)
en la soluci\'on que obtendremos. De todos modos, los detalles
del resto de los casos se inducir\'an brevemente.

En adelante, entonces, supondremos que el subsistema $B$ es un oscilador
invertido que obedece la siguiente ecuaci\'on de movimiento 
$\ddot q(s) - \Omega^2 q(s)=\frac{\lambda}{M_{\rm B}} x(s)$. 
El subsistema $A$ ser\'a un oscilador arm\'onico 
y este caso ser\'a denominado el caso (a).  Para encontrar
la soluci\'on a dicha ecuaci\'on, debemos imponer condiciones
iniciales y finales, como por ejemplo, $q(s=0)=q_0$ 
y $q(s=t)=q_{\rm f}$. En ese caso, la soluci\'on es
\beqa
q_{\rm cl}(s) &=& q_0\frac{\rm {\sinh}(\Omega(t-s))}{\rm
{\sinh}(\Omega t)}+ q_{\rm f} \frac{\rm {\sinh}(\Omega s)}{\rm
{\sinh}(\Omega t)} -\bigg(\frac{\lambda}{M_{\rm B} \Omega}\bigg)
\frac{\rm {\sinh}(\Omega s)}{\rm {\sinh}(\Omega t)} \int_0^t x(u)
\rm{\sinh}(\Omega (s-u)) du \nonumber
\\ & + & \bigg(\frac{\lambda}{M_{\rm B} \Omega}\bigg)\int_0^s x(u) \rm{\sinh}(\Omega
(s-u)) du. \label{classol}\eeqa
Una vez conocida la soluci\'on cl\'asica para 
el oscilador invertido, se puede escribir, de manera
expl\'icita, la acci\'on de influencia
a partir de la integraci\'on de los grados de libertad del entorno
$\cal E$ (Ec.(\ref{deltaA})) y, por ende, el t\'ermino
de la acci\'on de interacci\'on efectiva Ec.(\ref{Sinteff}). 
En particular, podemos despreciar nuevamente
el t\'ermino de disipaci\'on\footnote{Es decir,
nos quedamos a orden $\gamma_0$.}, pero, esta vez, aquel correspondiente a la
acci\'on de influencia del sistema compuesto $AB$, $\delta 
A^{(AB)}(q,q')$, ya que el coeficiente de disipaci\'on
$\gamma_0$ es b\'asicamente el cuadrado de la constante de 
acoplamiento con el entorno $\gamma_0 \sim c_n^2$, obteniendo
la siguiente expresi\'on para la acci\'on de influencia
\beq \delta A^{(AB)}(q,q\prime)=
 i 2 M_{\rm B} \gamma_0 K T \int_0^t ds(\Delta q(s))^2. 
\label{deltaapend}\eeq

En esta \'ultima expresi\'on, haciendo
un poco de algebra a partir de la soluci\'on cl\'asica $q_{\rm cl}
(s)$, $\Delta q(s)$ est\'a definido seg\'un
\beq \Delta q_{\rm cl}(s)= (q_0
-q_0\prime)^2 \bigg(\frac{\sinh(\Omega(t-s))}{\sinh(\Omega
t)}\bigg)^2 + 2(q_0-q_0\prime)
\frac{\sinh(\Omega(t-s))}{\sinh(\Omega t)} g(s,t)+ g(s,t)^2,
 \eeq
con
\beq g(s,t)= \frac{\lambda}{M_{\rm B} \Omega}
\bigg(-\int_0^s du \Delta x(u) \sinh(\Omega(s-u)) 
+ \frac{\sinh(\Omega s)}{\sinh(\Omega t)} \int_0^t du \Delta
x(u) \sinh(\Omega(t-u))\bigg)\nonumber \eeq y
$\Delta x(u)=x(u)-x\prime(u)$. Como el \'ultimo t\'ermino
en la expresi\'on de $\Delta q_{\rm cl}(s)$ 
no depende de las condiciones iniciales, ser\'a
transparente a las integrales de la acci\'on de influencia 
del subsistema $A$, $\delta {\cal A}^{(A)}(x,x')$ de la 
Ec.(\ref{deltaA2}). Juntando estas expresiones y la
condici\'on inicial para el subsistema $B$, podemos escribir
la expresi\'on para la funcional de influencia del
subsistema $A$ como
\beqa
{\cal F}^{(A)}(x,x') &=&  \exp\left\{{\frac{i}{\hbar}
\delta {\cal A}^{(A)}(x,x')}\right\} \nonumber \\
&=& 
\int_{-\infty}^{+\infty}dq_{\rm f}
\int_{-\infty}^{+\infty}dq_0\int_{-\infty}^{+\infty} dq_{\rm f}'
\int_{q_0}^{q_{\rm f}}{\cal D}q\int_{q_0'}^{q_{\rm f}'}{\cal D}q'
e^{\frac{i}{\hbar} \delta A^{(AB)}(q_{\rm cl},q_{\rm
cl}\prime)} \ \rho_{\rm r}^{\rm B}(q_0,q_0') \nonumber \\
&\times & e^{\frac{i}{\hbar}
(S_{\rm B}(q_{\rm cl}) + S_{\rm AB}(x,q_{\rm cl}) -
S_{\rm B}(q'_{\rm cl}) - S_{\rm AB}(x',q'_{\rm cl}))}.\eeqa
Estas integrales
pueden ser f\'acilmente realizadas (en el l\'imite que el
sistema compuesto $AB$ est\'a debilmente acoplado) y dan
como resultado
\beq
\delta {\cal A}^{(A)}(x,x\prime)= 2 \int_0^t ds_1 \int_0^{s_1}
ds_2 ~y(s_1) \tilde{\eta}(s_1-s_2) r(s_2) + i \int_0^t ds_1
  \int_0^{s_1} ds_2 ~y(s_1) \tilde{\nu}(s_1-s_2) y(s_2), 
\label{newdeltaA}
\eeq 
con $y(s)=x(s)-x\prime(s)$ y $r(s)=(x(s)+x\prime(s))/2$. 
Las cantidades $\tilde{\eta}$ y $\tilde{\nu}$ 
son los nuevos n\'ucleos de disipaci\'on
y ruido, respectivamente, dados por
\beqa \tilde{\eta}(s_1-s_2)&=&\bigg(\frac{\lambda^2}{2 M_{\rm B} \Omega}\bigg)
\rm{\sinh}(\Omega(s_1-s_2)),\nonumber \\
\tilde{\nu}(s_1-s_2)&=&\bigg(\frac{\lambda^2\sigma}{32 \hbar}\bigg)
\rm{\cosh}(\Omega(s_1-s_2)).\label{nuevanu}\eeqa

Para evaluar esta nueva funcional de influencia, usaremos
el m\'etodo de ``fase estacionaria". Las integrales est\'an dominadas
por la soluci\'on cl\'asica de la ecuaci\'on de movimiento
del oscilador libre (subsistema $A$) \cite{kapral1}: 
$\ddot x(s) + \omega^2 x(s)=0$.
Imponiendo, nuevamente, condiciones iniciales y finales 
$x(s=0)=x_0$ y $x(s=t)=x_{\rm f}$,
es posible obtener la siguiente soluci\'on cl\'asica
\beq
x_{\rm cl}(s)=x_0 \frac{\rm {\sin}(\omega(t-s))}{\rm {\sin}
(\omega t)} +
x_{\rm f} \frac{\rm{\sin}(\omega s)}{\rm {\sin}(\omega t)}.
\eeq

En este caso, el operador de evoluci\'on reducido
para el subsistema $A$ resulta
\beq 
{\cal J}^{(A)}_{\rm r}(x_{\rm f},x_{\rm f}\prime;t
\vert x_0,x_0\prime;0)
=  e^{\frac{i}{\hbar}(S_{\rm A}(x_{\rm cl})
-S_{\rm A}(x\prime_{\rm cl}))}
e^{-g^2(s,t)}
e^{\frac{i}{\hbar} \delta {\cal A}^{(A)}(x,x\prime)}
 \equiv ~e^{\frac{i}{\hbar}U(t)} e^{-D(t)}
\label{Jr2}.\eeq \\
con 
$U$ y $D$, relacionados con la evoluci\'on unitaria y el 
proceso de p\'erdida de coherencia, respectivamente, definidos
seg\'un
\beq
U=(x_0-x_0\prime)\frac{\sin (\omega(t-s))}{\sin (\omega t)}
+(x_{\rm f}-x_{\rm f}\prime)\frac{\sin (\omega s)}{
\sin (\omega t)} - 2\gamma_0 \int_0^t ds_1
\int_0^{s_1}ds_2 ~ y(s_1)\tilde{\eta}(s_1-s_2)r(s_2),\eeq 
y 
\begin{equation} 
D =\frac{2 \gamma_0 k_B T}{\hbar \Omega^2}
\lambda^2\int_0^t ds (\Delta q_{\rm cl}(s))^2 +
\frac{\lambda^2\sigma}{32 \hbar}\int_0^t ds_1\int_0^{s_1}ds_2
y(s_1)\tilde{\nu}(s_1-s_2)y(s_2).\label{propagator}
\end{equation}

A partir de esta \'ultima ecuaci\'on, podemos notar
dos contribuciones diferentes al coeficiente de difusi\'on.
La primera, proporcional a la temperatura del entorno $\cal E$,
derivada del acoplamiento del subsistema $B$ al ba\~no t\'ermico.
La segunda, proporcional a $\lambda^2$, es la acci\'on del
subsistema $B$ sobre el subsistema $A$. A pesar de que estamos
trabajando en el r\'egimen subamortiguado, es v\'alido recalcar
que ambas contribuciones pueden ser del mismo orden. Por lo
tanto, tendremos en cuenta ambos t\'erminos a la hora de estudiar
los efectos de p\'erdida de coherencia en el subsistema $A$.

\section{Coeficiente de difusi\'on de la ecuaci\'on maestra}
\label{difcoefcomp}

En esta secci\'on, nos dedicaremos a obtener el coeficiente de
difusi\'on de la ecuaci\'on maestra, el cual nos dar\'a
una medida cuantitativa de la p\'erdida de coherencia que sufre
el subsistema $A$ en todos los casos analizados.
Una forma muy com\'un de analizar la p\'erdida de coherencia
es examinando c\'omo evolucionan los elementos 
fuera de la diagonal de la matriz densidad reducida de acuerdo
con la ecuaci\'on maestra.
Utilizando las mismas t\'ecnicas que se usan en el MBC
\cite{HuPazZhangI} para obtener la ecuaci\'on maestra, podemos
calcular la derivada temporal del propagador ${\cal J}_{\rm r}$,
y eliminar la dependencia de las condiciones iniciales 
 $x_0$, $x_0\prime$ que aparecen tras utilizar la soluci\'on
cl\'asica  $x_{\rm cl}(s)$. Este procedimiento es relativamente
f\'acil si se usa la propiedad de la soluci\'on \cite{HuPazZhangII}:
\begin{equation}
\Delta_0 J_{\rm r}^{(A)} (t,0)=\left[\cos(\omega(t-s))\Delta_{\rm f} 
+ \frac{\sin(\omega(t-s))}
{\omega} i \hbar \frac{\partial}{\partial \Sigma_{\rm f}} 
\right] J_{\rm r}^{(A)}(t,0),
\end{equation}
donde $\Delta_0=(x_0-x_0\prime)$, $\Delta_{\rm f}=(x_{\rm f}-x_{\rm f}
\prime)$ y $\Sigma_{\rm f}=
 (x_{\rm f} + x_{\rm f}\prime)$.

La ecuaci\'on maestra es muy complicada en este caso y, como
en el MBC, depende del acoplamiento del subsistema y el entorno.
En este caso particular, adem\'as depende del acoplamiento
entre los subsistemas $A$ y $B$, es decir $\lambda$. Como nosotros
estamos interesados en estudiar s\'olo el proceso de p\'erdida
de coherencia en $A$, s\'olo nos concentraremos en calcular
la correcci\'on a la evoluci\'on unitaria  que se
induce del n\'ucleo de ruido \'unicamente (es decir, de la
parte imaginaria de la acci\'on de influencia). El
resultado obtenido ser\'a proporcional a
\beqa
\dot{\rho}_{\rm r}(x_{\rm f},x_{\rm f}';t) &\sim &  
- i [H_{\rm ren},\rho_{\rm r}] - 
\frac{\partial}{\partial t} \bigg( \frac{2 \gamma_0 k_B
T}{\hbar}\frac{\lambda^2}{\Omega^2}  \int_0^t ds ~g(s)^2 
\nonumber \\
&-&  \int_0^t \int_0^s ds ds' \Delta x_{\rm cl}(s) 
\tilde{\nu}(s,s')
\Delta x_{\rm cl}(s')\bigg) \rho_{\rm r} + ...
 \nonumber \\ &= & - i [H_{\rm ren},\rho_{\rm r}] 
- \bigg( \frac{2
\gamma_0 K T}{\hbar}\frac{\lambda^2}{\Omega^2} 
\int_0^t 2
\dot{g}(s) g(s) ds  \nonumber \\
&+& \frac{\lambda^2\sigma}
{32 \hbar} (x_{\rm f}-x_{\rm f}')\int_0^t ds 
\cosh(\Omega(t-s))\Delta
x_{\rm cl}(s)\bigg)\rho_{\rm r} \nonumber \\
&+& ...  ,\nonumber \eeqa
donde los puntos suspensivos indican t\'erminos que no 
contribuyen a la difusi\'on.
Esta expresi\'on resulta equivalente a escribir
 \beq \dot{\rho_{\rm r}} \approx  
- i [H_{\rm ren},\rho_{\rm r}] 
-(x_{\rm f}-x_{\rm f}')^2
{\cal D}(t) \rho_{\rm r}, \eeq
con ${\cal D}(t)$ el coeficiente de difusi\'on que aparece
en la ecuaci\'on maestra Ec.(\ref{master}). De esta forma,
el efecto de la difusi\'on puede ser estudiado
a partir de la soluci\'on aproximada para la matriz densidad
reducida (soluci\'on de la ecuaci\'on maestra)
\beq \rho_{\rm r}(x_{\rm f},x_{\rm f}';t)
\approx \rho_{\rm r}^{\rm u}(x_{\rm f},x_{\rm f}';t) ~ 
e^{-(x_{\rm f} - x_{\rm f}')^2\int_0^t{\cal D}(s)ds},
\label{decofactor}\eeq
donde $\rho_{\rm r}^{\rm u}$ es la soluci\'on de la parte
unitaria de la ecuaci\'on maestra, es decir, la evoluci\'on sin tener en cuenta
el entorno (sistema cerrado).

En el resto de la secci\'on, mostraremos el resultado
exacto del coeficiente de difusi\'on para cada uno de los
casos estudiados, a saber:

\begin{itemize}
\item {\bf Caso ({\mbox a}): 
 Oscilador arm\'onico +  Oscilador invertido + ${\cal E}$}.
Este es el  ejemplo que hemos  desarrollado hasta
ahora, donde el subsistema $A$ est\'a representado por
un oscilador arm\'onico y el $B$ por un oscilador invertido. 
El coeficiente de difusi\'on 
para este caso es ${\cal D}_{\rm a}$, obtenido a partir
del desarrollo anterior
\beqa {\cal D}_{\rm a}
&=&  \frac{\Omega^2}{(\omega^2+\Omega^2)^2}\Big\{\frac{2 \gamma_0 k_B
T}{\hbar \Omega^2} \lambda^2 \int_0^t ds 
\Big[\frac{
\sinh(\Omega s)}{\sinh(\Omega t)}(\cosh(\Omega t)\cos(\omega
t)-1)\nonumber \\
&-& \cosh(\Omega s)\cos(\omega t)
+\cos(\omega(t-s))\Big]\nonumber \\
&\times & \Big[\Omega\Big(\frac{
\sinh(\Omega s)}{\sinh(\Omega t)^2}\cosh(\Omega t) 
(1- \cosh(\Omega t)\cos(\omega t))+\sinh(\Omega s)\cosh(\omega t)\Big)
\nonumber \\
&+&  \omega \Big(-\frac{ \sinh(\Omega s)}{\sinh(\Omega
t)}\sin(\omega t)\cosh(\Omega t)\nonumber - 
\sin(\omega(t-s)) +  \sin(\omega t)\cosh(\Omega s)\Big)\Big] \nonumber \\
&+& 
\frac{\lambda^2\sigma}{32 \hbar} \int_0^t
ds\cosh(\Omega(t-s))\cos(\omega(t-s)) \Big\}.\label{Da}
\eeqa
Este caso es una generalizaci\'on del modelo de juguete
utilizado en la Ref.\cite{robin}, donde los autores
no consideran la interacci\'on entre el subsistema $B$
y el ba\~no t\'ermico. Es f\'acil obtener los resultados
de la Ref.\cite{robin} a partir de los nuestros al poner
$\gamma_0 = 0$. En los resultados num\'ericos 
a continuaci\'on,  representaremos a una part\'icula Browniana que pierde
coherencia debido a un entorno compuesto con un grado
de libertad inestable.

\item {\bf Caso ({\mbox b}): 
 Oscilador invertido + Oscilador arm\'onico + ${\cal E}$}.
En este caso, el subsistema $A$ est\'a representado por
un oscilador invertido que obedece la ecuaci\'on de 
movimiento cl\'asica $\ddot x(s) -
\omega^2 x(s)=0$ y el subsistema $B$, por un oscilador arm\'onico
que cumple con $\ddot q(s) + \Omega^2 q(s)=\frac{\lambda}
{M_{\rm B}} x(s)$. En particular, este modelo sencillo sirve para
estudiar el proceso de p\'erdida de coherencia en
sistemas ca\'oticos debido a entornos completamente arm\'onicos
\cite{PLA,fermazzidiana,ZP}.
Resulta muy directo obtener el coeficiente
de difusi\'on en este caso, realizando las sustituciones 
$\omega \rightarrow i \omega$ y $\Omega
\rightarrow i \Omega$ en el caso anterior. El coeficiente de difusi\'on es
\beqa {\cal D}_{\rm b}
&=&  \frac{\Omega^2}{(\omega^2+\Omega^2)^2}\Big\{\frac{2 \gamma_0 k_B
T}{\hbar \Omega^2} \lambda^2 \int_0^t ds 
\Big[\frac{
\sin(\Omega s)}{\sin(\Omega t)}(\cos(\Omega t)\cosh(\omega
t)-1)\nonumber \\
&-& \cos(\Omega s)\cosh(\omega t)
+\cosh(\omega(t-s))\Big]\nonumber \\
&\times & \Big[\Omega\Big(\frac{
\sin(\Omega s)}{\sin(\Omega t)^2}\cos(\Omega t) 
(1- \cos(\Omega t)\cosh(\omega t))+\sin(\Omega s)\cos(\omega t)\Big)
\nonumber \\
&+&  \omega \Big(-\frac{ \sin(\Omega s)}{\sin(\Omega
t)}\sinh(\omega t)\cos(\Omega t)\nonumber - 
\sinh(\omega(t-s)) +  \sinh(\omega t)\cos(\Omega s)\Big)\Big] 
\nonumber \\
&+& 
\frac{\lambda^2\sigma}{32 \hbar} \int_0^t
ds\cos(\Omega(t-s))\cosh(\omega(t-s)) \Big\}.\label{Db}
\eeqa

\item {\bf Caso ({\mbox c}): 
Oscilador arm\'onico + Oscilador arm\'onico + ${\cal E}$}.
Por completitud, tambi\'en consideraremos el caso de
dos osciladores arm\'onicos acoplados a trav\'es del
t\'ermino de interacci\'on de la Ec.(\ref{intAB}).
El procedimiento para obtener el coeficiente de difusi\'on
en este caso, es similar
al realizado para el caso (a). Para
obtener ${\cal D}_{\rm c}$, hay que reemplazar
$\Omega \rightarrow i \Omega$ en ${\cal D}_{\rm a}$.

\item {\bf Caso ({\mbox d}): 
Oscilador invertido + Oscilador invertido +
${\cal E}$}.
Finalmente, consideraremos dos osciladores invertidos acoplados
 y uno de ellos acoplado tambi\'en a un ba\~no t\'ermico
a alta temperatura. Este coeficiente de difusi\'on
ser\'a denominado ${\cal D}_{\rm d}$ y es f\'acil obtenerlo,
reemplazando $\Omega \rightarrow i \Omega$ en el 
coeficiente ${\cal D}_{\rm b}$. Veremos en \mbox{nuestros}
resultados num\'ericos, que este caso es el m\'as sensible
a las perturbaciones (ya que ambos subsistemas son inestables)
cuando no hay un entorno externo a ambos subsistemas ($\gamma_0=0$).
\end{itemize}

Una vez conocidos los coeficientes de difusi\'on, podemos
estudiar num\'ericamente sus comportamientos para distintos valores
de los par\'ametros del modelo. Con el fin de elegir 
situaciones representativas, en la columna izquierda  de la Fig.\ref{figure3-1},
 mostramos el coeficiente de difusi\'on normal cuando 
las frecuencias naturales de ambos subsistemas, $A$ y $B$, son iguales
($\omega \approx \Omega$) mientras que, en la columna derecha,  
cuando la frecuencia de $A$ es \mbox{considerablemente} mayor que 
la de $B$ ($\omega > \Omega$).
Cada situaci\'on ser\'a considerada, adem\'as, en el caso en
que los subsistemas est\'an aislados ($\gamma_0 = 0$), 
y para distintos valores del par\'ametro libre $\gamma_0 k_B
T$. No est\'a dem\'as aclarar que los tiempos analizados est\'an
limitados a aquellos donde la aproximaci\'on perturbativa 
es  v\'alida.

\begin{figure}[!ht]
\begin{center}
\includegraphics[width=13cm]{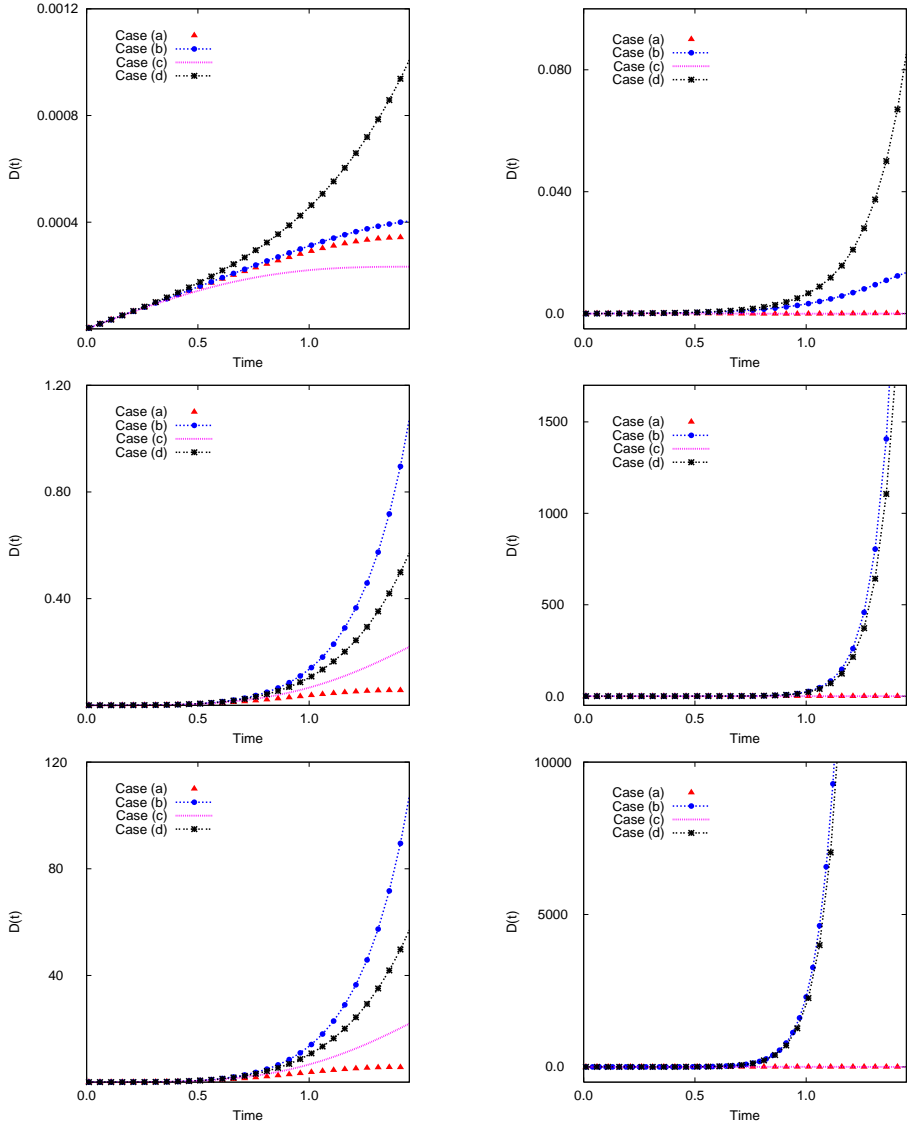}
\caption{Mostramos la comparaci\'on entre los distintos coeficientes
de difusi\'on correspondientes a cada caso. Los gr\'aficos que se
encuentran en el extremo superior de la figura, corresponden al
caso $\gamma_0=0$, es decir, los subsistemas $A$ y $B$ est\'an
aislados. Los gr\'aficos en la fila del medio corresponden al 
valor $\gamma_0k_BT =1$ y, los del extremo inferior, 
a $\gamma_0k_BT =100$. Los valores de los par\'ametros utilizados fueron
elegidos por conveniencia y son: $\omega = \Omega = 1$ y $\sigma =
0.01$ (en la columna
izquierda); y $\omega = 5 \Omega$, $\sigma = 0.01$ (en la columna
derecha). Los casos (b) y (d) muestran una tasa de crecimiento mayor
a tiempos cortos para ambos conjuntos de par\'ametros.}
\label{figure3-1}
\end{center}
\end{figure}

En la Fig.\ref{figure3-1}, podemos notar la diferencia 
de comportamiento entre los distintos \mbox{coefi-} cientes de difusi\'on,
particularmente, entre  el caso en que el subsistema $A$ es estable
y cuando no lo es. En particular, el comportamiento exponencial
en los casos (b) y (d) se debe, justamente, a que el subsistema
$A$ es inestable \cite{PLA,fermazzidiana} (es decir, un oscilador invertido).
Por el contrario, en los casos (a) y (c), los coeficientes muestran
un comportamiento oscilatorio ya que el subsistema $A$ es un oscilador
arm\'onico. La diferencia entre el comportamiento oscilatorio y el
exponencial de los coeficientes es evidente a tiempos 
tales que $\omega t\geq 1$. Para tiempos m\'as cortos, 
todos los coeficientes son, pr\'acticamente, equivalentes. 
La din\'amica del oscilador invertido se hace m\'as evidente
cuando $\Omega t \geq 1$ y, los casos (b) y (d), se empiezan 
a diferenciar entre ellos.
Los gr\'aficos en el extremo superior de la Fig.\ref{figure3-1}
son una generalizaci\'on de los resultados obtenidos en la 
Ref.\cite{robin}, donde los autores consideran un oscilador
acoplado a otro invertido sin entorno adicional 
(es decir, en nuestro caso $\gamma_0=0$).
Los gr\'aficos de la fila del medio de dicha figura son para
un valor chico de $\gamma_0k_B T$ 
(pensandolo como  una temperatura ``efectiva")
 mientras que, los del extremo
inferior, son para un valor m\'as grande de dicho par\'ametro.

En el caso particular de $\gamma_0=0$, es f\'acil ver que 
${\cal D}_{\rm d}$ crece m\'as r\'apido, y de forma exponencial,
mientras que los otros coeficientes se mantienen con una amplitud
m\'as chica y con una tasa de crecimiento menor. Esto resultar\'a importante
para evaluar los tiempos de p\'erdida de coherencia en la siguiente 
secci\'on. Las inestabilidades inherentes al subsistema $A$
aumentan exponencialmente la difusi\'on originada debido a 
la interacci\'on con el ba\~no. Cuando el subsistema $B$
tambi\'en es inestable, entonces se ve una mayor sensibilidad
(exponencial) a las perturbaciones que en cualquier otro caso.
Sin embargo, es importante remarcar que es un modelo muy
simplificado, ya que ambos osciladores no est\'an acotados 
por debajo y por lo tanto, para tiempos largos desarrollar\'an 
divergencias no f\'isicas \cite{robin} (y por eso resulta 
importante estudiar el modelo s\'olo dentro de los l\'imites
del desarrollo perturbativo). De esta forma, concluimos que
 para tener una idea m\'as precisa de las consecuencias f\'isicas
 de la existencia de grados de libertad inestables en el entorno,
no podemos prescindir de la interacci\'on con el
resto del Universo, modelado en nuestro trabajo como
un n\'umero muy grande de osciladores arm\'onicos. 

La difusi\'on que sufre el subsistema $A$ es el resultado
directo de la interacci\'on de $A$ con $B$, y  de $B$
con el ba\~no t\'ermico ${\cal E}$. El entorno reacciona a la interacci\'on
generando disipaci\'on y difusi\'on en el subsistema $B$ y $A$.
Tomemos como ejemplos los casos (b) y (d):  en ambos casos, 
el subsistema $A$ es inestable.
Por un lado,  en el caso (b), como $B$ es
un oscilador arm\'onico, el proceso de difusi\'on es
m\'as efectivo, ya que gracias a su comportamiento oscilatorio,
genera ruido en $A$ peri\'odicamente. Mientras, en el caso (d), $B$ es
un oscilador invertido que no est\'a acotado por debajo. Luego,
la propagaci\'on de sus estados es ilimitada (en la siguiente secci\'on 
se volver\'a sobre esto). Parte de la informaci\'on es entregada
al entorno; pero a tiempos largos, la din\'amica intr\'insica de
este subsistema implicar\'a una escasa provisi\'on de difusi\'on al
subsistema $A$ y, consecuentemente, menos efectiva. En la pr\'oxima
secci\'on se estimar\'an los tiempos de p\'erdida de coherencia
y se dar\'a una explicaci\'on m\'as cuantitativa del proceso.
Los casos (a) y (c) son substancialmente diferentes ya que el
subsistema en cuesti\'on ($A$) es un oscilador arm\'onico.
En estos casos, se puede ver en la Fig.\ref{figure3-1}, que
la respuesta del entorno efectivo (B
+ $\cal{E}$) es menos efectiva que en los dos casos mencionados
antes. Esto se debe fundamentalmente a la estabilidad
del subsistema $A$. En los graficos que est\'an en las filas del
medio de la Fig.\ref{figure3-1}, mostramos el comportamiento
de estos coeficientes cuando la temperatura es ``baja''. 
En este caso, cuando $\gamma_0 k_B T=1$, podemos
ver que ${\cal D}_{\rm b}$ crece apenas m\'as r\'apido que 
${\cal D}_{\rm d}$, y ambos, mucho m\'as r\'apido que 
${\cal D}_{\rm a}$ y ${\cal D}_{\rm c}$.
Esto puede ser explicado teniendo en cuenta las propiedades
din\'amicas del entorno compuesto acoplado al sistema como
hicimos con anterioridad. El susbistema $B$ no es un buen
proveedor de difusi\'on si es inestable. A tiempos largos,
es decir $\Omega t> 1$, los estados del oscilador invertido
se propagan mucho en el espacio de fases. La informaci\'on
debe ir desde ${\cal E}$ hasta $A$ a trav\'es de $B$, lo cual
no resulta un proceso efectivo.
El comportamiento din\'amico del caso (a) deber\'ia ser
similar al del caso (b). Sin embargo, la diferencia entre ambos coeficientes que se 
observa en la Fig.\ref{figure3-1},  muestra que no lo es.
La diferencia entre tener un oscilador invertido en el subsistema
$A$ o en el subsistema $B$ resulta crucial para distinguir
los comportamientos; la cual se manifiesta claramente en un
crecimiento exponencial (o no) del coeficiente de difusi\'on
y en los tiempos de p\'erdida de coherencia que estimaremos en 
la pr\'oxima secci\'on.

Como ya dijimos, en los casos (b) y (d), la propagaci\'on
del estado inicial en $A$ es exponencialmente sensible a las 
fluctuaciones que vienen del entorno compuesto (B + $\cal{E}$),
y \'este reacciona inmediatamente ante dicha interacci\'on perdiendo
informaci\'on de manera m\'as r\'apida que en cualquier otro caso.
La pregunta que podr\'ia surgir, ante la evidencia de los
gr\'aficos presentados, es por qu\'e ${\cal D}_{\rm b}$ crece m\'as 
r\'apido que ${\cal D}_{\rm d}$, si en este \'ultimo caso hay
grados inestables tanto en el subsistema como en el entorno.
La respuesta es que, en el caso (b) la difusi\'on es m\'as
efectiva y el sistema tiene mayor p\'erdida de coherencia porque
el subsistema $A$ inestable pierde  informaci\'on debido a un
entorno uniforme (no \'ohmico) compuesto por osciladores arm\'onicos
($B$ m\'as los infinitos del entorno $\cal E$). En el caso de temperatura ``baja'',
la mayor contribuci\'on al comportamiento difusivo proviene
de la din\'amica del sistema $A+B$ \'unicamente. A tiempos
cortos, $\omega t< 1$, los grados de libertad inestables de $A$
dominan el comportamiento temporal, y los casos (b) y (d)
se comportan de manera similar.
Sin embargo, cuando $\omega t \geq 1$, existe una notable diferencia
entre ${\cal D}_{\rm b}$ y ${\cal D}_{\rm d}$ si las frecuencias de
ambos subsistemas son del mismo orden (columna 
izquierda de la Fig. \ref{figure3-1}). Cuando $A$ tiene una frecuencia
mayor, entonces es este subsistema el que domina la din\'amica y
ambos coeficientes,  ${\cal D}_{\rm b}$ y ${\cal D}_{\rm d}$, son
pr\'acticamente iguales en una escala temporal m\'as larga (columna 
derecha de la Fig. \ref{figure3-1}).
Cuando la temperatura del ba\~no aumenta, no hay diferencia entre los
casos (b) y (d), ya que el t\'ermino que se deriva del entorno $\cal E$ 
(proporcional a la temperatura), domina en el 
coeficiente de difusi\'on. Para valores altos de $\gamma_0 k_B T$
obtenemos una clara jerarqu\'ia entre los comportamientos
de los distintos casos (extremo inferior de la
Fig. \ref{figure3-1}). De nuevo, se puede observar que los coeficientes
de difusi\'on son mayores cuando los grados inestables est\'an en
el subsistema $A$. De esta forma, podemos concluir que la presencia 
de inestabilidades  aumenta la p\'erdida de
coherencia. Este efecto es a\'un m\'as importante cuando el subsistema
inestable $A$ est\'a acoplado a un \'unico grado de libertad ca\'otico
(caso (d) en ausencia de ba\~no t\'ermico), o bien, cuando est\'a 
acoplado a un entorno compuesto, formado por $B$+ $\cal{E}$, a muy alta
temperatura del ba\~no, como los casos (b) y (d) de la fila inferior de la figura.
En la Fig.\ref{figure3cap4}, mostramos
el coeficiente de difusi\'on para un valor alto de $\gamma_0 k_B T$
para los casos (a) y (c), en los cuales el subsistema $A$ es un oscilador
arm\'onico. Esta figura resulta \'util para comparar  los coeficientes
que tienen un comportamiento oscilatorio (que se desarrolla a tiempos
m\'as grandes y de amplitudes m\'as chicas) con los otros
dos que tienen un comportamiento exponencial en el tiempo.
\begin{figure}[!ht]
\begin{center}
\includegraphics[width=10cm]{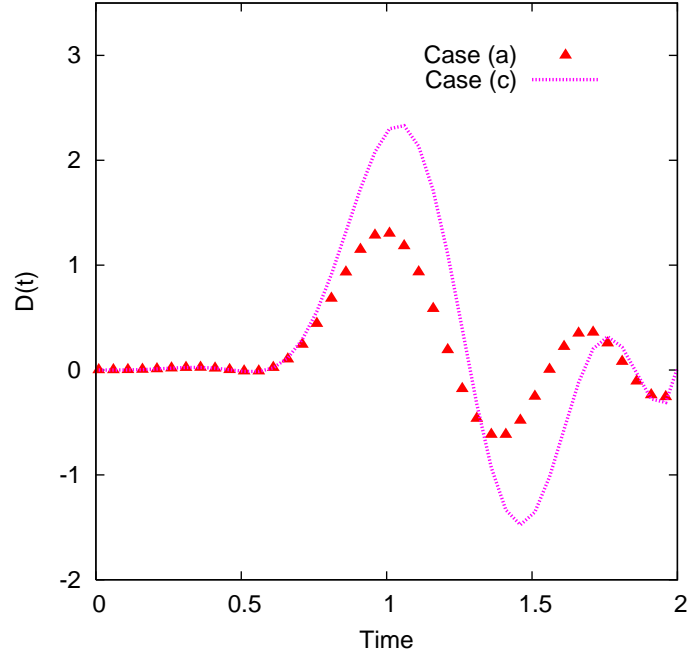}
\caption{Coeficientes de difusi\'on para los casos (a) y (c) con
$\omega =5 \Omega $, $\sigma = 0.01$ y $\gamma_0 k_B T = 100$. 
Podemos ver la evoluci\'on de estos coeficientes en una escala
distinta a la de la Figura \ref{figure3-1}.} 
\label{figure3cap4}
\end{center}
\end{figure}

\section{P\'erdida de coherencia en $A$}
\label{decoherencecomp}

Despu\'es de haber integrado los grados de libertad del entorno $\cal E$,
y, posteriormente, las coordenadas $q,q'$ correspondientes al subsistema
$B$, obtuvimos el coeficiente de difusi\'on responsable del proceso de 
p\'erdida de coherencia en el subsistema $A$. Definiremos nuevamente
el factor de p\'erdida de coherencia, como hemos hecho en otros
cap\'itulos de esta Tesis, 
\begin{equation}
\Gamma (t) = \exp\left\{- \int_0^t {\cal D}(s) ~ ds\right\}.
\end{equation}
Como se deduce de su definici\'on, $\Gamma (t)$ es inicialmente uno 
ya que no hay interacci\'on entre el subsistema y el 
entorno. A medida que evoluciona, esta cantidad va disminuyendo
hasta hacerse cero en el caso que la p\'erdida de coherencia sea total
(o tiempo infinito).
Para analizar este factor, presentaremos los resultados num\'ericos obtenidos
en los mismos casos de la secci\'on anterior y, para los mismos conjuntos de
valores de los par\'ametros del modelo.
\begin{figure}[!ht]
\begin{center}
\includegraphics[width=14cm]{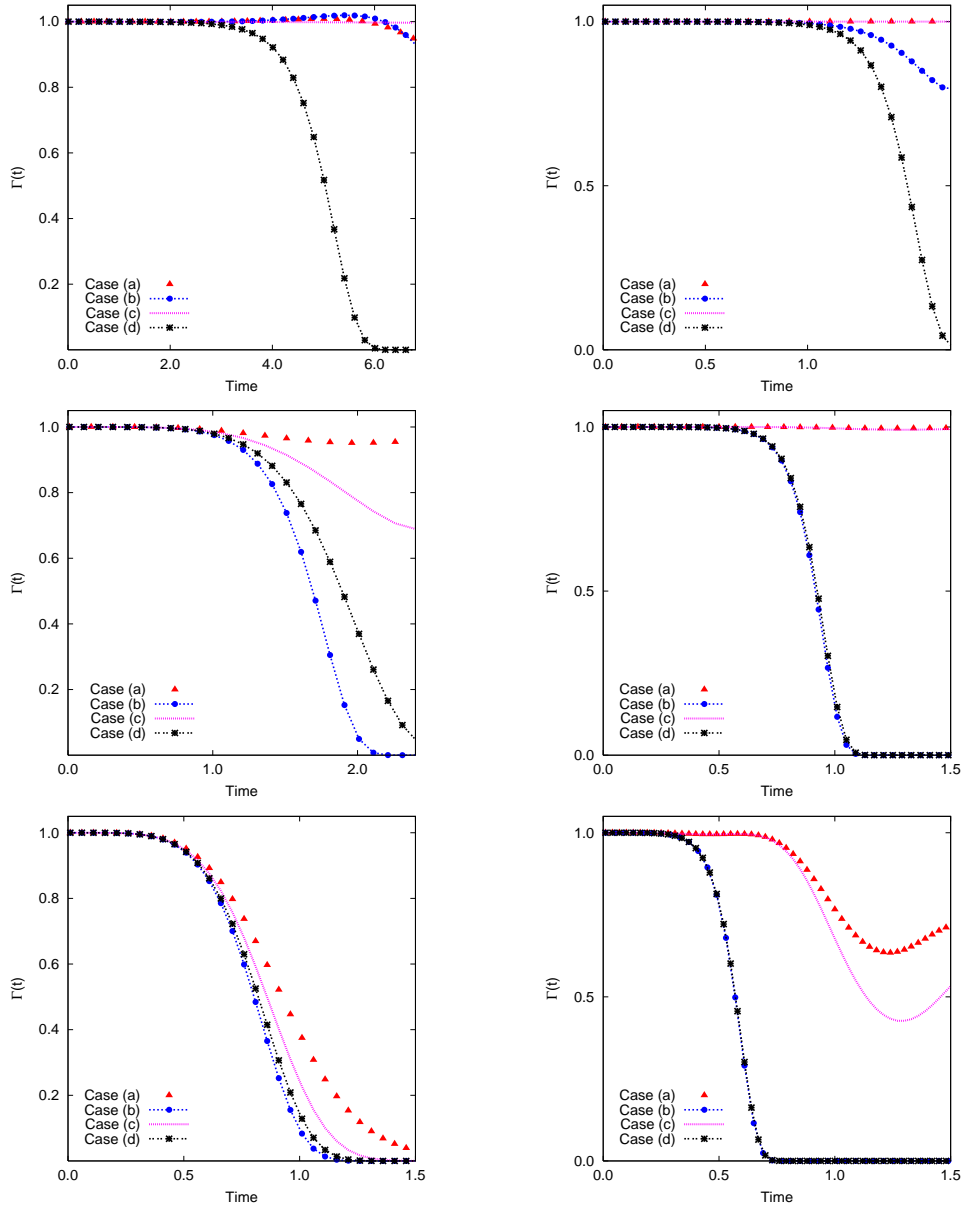}
\caption{Factores de p\'erdida de coherencia para los mismos
conjuntos de par\'ametros que en la Figura
\ref{figure3-1}. Cuando el sistema compuesto est\'a aislado, la p\'erdida
de coherencia es m\'as importante en el caso (d). 
Sin embargo, cuando  $\gamma_0\not= 0$, el caso (b) resulta con mayor
p\'erdida de coherencia .} \label{figure3-3}
\end{center}
\end{figure}
En la Fig.\ref{figure3-3}, mostramos los gr\'aficos para los 
factores de p\'erdida de coherencia, en los distintos casos: 
en la columna izquierda $\omega \approx
\Omega$ y en la columna derecha $\omega > \Omega$, en ambos casos
para $\gamma_0=0$, $\gamma_0 k_B T=1$ y $\gamma_0 k_B T=100$.
En los gr\'aficos del extremo superior, cuando el sistema $A+B$ est\'a aislado,
podemos notar que el tiempo de p\'erdida de coherencia 
es m\'as chico para el caso (d) que
para el caso (b), y en ambos casos, bastante m\'as r\'apido que para los
casos (a) y (c). Esto se debe fundamentalmente al 
hecho que el subsistema $A$,
que est\'a \'unicamente acoplado al subsistema $B$ genera ruido y disipaci\'on a 
grandes escalas. Estos fen\'omenos son m\'as intensos cuando $B$ es tambi\'en
inestable, es decir, el caso (d) es ``dos veces exponencial". En el caso (a), notamos
que el entorno, formado s\'olo por $B$, induce cambios menores en el comportamiento
oscilatorio de $A$.  El  estiramiento de los estados del entorno a lo largo de sus
direcciones inestables se refleja en el sistema como difusi\'on. Lo mismo ocurre
en el caso (b), con la diferencia  \'unica y esencial  que el subsistema que evoluciona
a lo largo de su direcci\'on inestable es el $A$, mientras que el entorno (s\'olo $B$)
oscila. Como este proceso de estiramiento genera difusi\'on, a mayor 
extensi\'on mayor difusi\'on en el sistema. Por lo tanto, resulta l\'ogico que el
caso (d) sea el mejor ejemplo de este proceso, ya que sus dos grados de 
libertad son inestables y este modelo resulta, entonces, doblemente difusivo.
El caso (c) se muestra por completitud pero la p\'erdida de coherencia
ocurre a escalas temporales m\'as largas 
 ya que, en este caso, no hay grados de libertad inestables.

Tan r\'apido como la interacci\'on es prendida y el sistema $A+B$ deja
de estar aislado, el oscilador $B$ disipa tanto en el ba\~no t\'ermico como
en el subsistema $A$. Esto se puede observar en los gr\'aficos del medio
y extremo inferior de la Figura  \ref{figure3-3}. A temperaturas muy altas,
no hay diferencia entre los casos (d) y (b), ya que ambos pierden
coherencia en la misma escala temporal. El ba\~no t\'ermico
domina el proceso de difusi\'on. Sin embargo, s\'i se observan diferencias
entre estos dos casos y aquellos donde el subsistema $A$ es un oscilador
arm\'onico, o sea casos (a) y (c). En la Figura \ref{figure3-4} mostramos
el comportamiento del factor de p\'erdida de coherencia para estos dos
\'ultimos casos, en una escala temporal m\'as larga donde la p\'erdida
de coherencia se hace importante aunque no total (caso (a)), a\'un
cuando la temperatura del entorno es muy alta. 

\begin{figure}[!ht]
\begin{center}
\includegraphics[width=10cm]{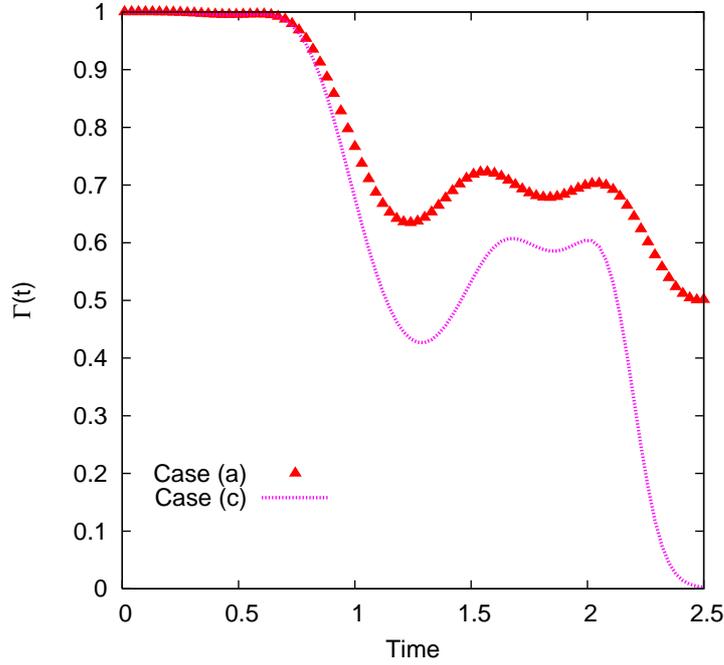}
\caption{Factor de p\'erdida de coherencia para los casos (a) y (c)
para   $\gamma_0k_BT = 100$, $\omega = 5
\Omega = 5$. Se puede observar que, a pesar que los comportamientos son
muy parecidos, la p\'erdida de coherencia en el caso (c) es m\'as importante
que el en caso (a), donde  el subsistema $B$ es un oscilador invertido.} 
\label{figure3-4}
\end{center}
\end{figure}

\subsection{Estimaci\'on del tiempo de p\'erdida de coherencia}

En esta secci\'on, nos dedicaremos a estimar anal\'iticamente los
tiempos de p\'erdida de coherencia. Como ya hemos dicho,
estos tiempos indican una escala temporal donde 
se pueden encontrar rasgos cl\'asicos en la evoluci\'on del sistema,
como por ejemplo, una funci\'on de Wigner definida positiva en el espacio de
fases.

Cuando el subsistema $A$ es un oscilador invertido, un punto inestable
se forma en el centro del espacio de fases, el cual tiene asociadas
direcciones estables e inestables \cite{ZP}. Estas direcciones
est\'an caracterizadas con los coeficientes de Lyapunov $\Lambda$
con parte negativa real e imaginaria positiva. Para tener una medida
cuantitativa de estos tiempos, debemos considerar que la
din\'amica del sistema tambi\'en da lugar a la posibilidad de que los
estados se contraigan a lo largo de la direcci\'on estable. Es decir,
la expansi\'on exponencial del paquete gaussiano a lo largo de 
una de las direcciones, debido al punto hiperb\'olico en el origen, es 
compensada con la contracci\'on exponencial en la otra direcci\'on.
El ancho del paquete gaussiano en la direcci\'on del momento 
depende del tiempo de la forma $\sigma_p(t) = \sigma_p(t_0)
 \exp{[\Lambda t]}$, donde $\sigma_p(t_{0})$ es el ancho del
paquete al tiempo inicial.  El coeficiente de Lyapunov
est\'a definido por $\Lambda = 2 \omega^2$.
Los efectos difusivos limitan el proceso de contracci\'on de la
funci\'on de Wigner. La cota sobre el ancho del paquete
est\'a dada por $\sigma_{\rm
c}=\sqrt{2{\cal D}_{\rm i}/\Lambda}$ \cite{leshouches,jpphabzurek} 
donde ${\rm i}$ es b o d). Existe otra escala m\'as, $t_{\rm max}$
relacionada al tiempo en que la p\'erdida de coherencia empieza a
ser efectiva, que es posterior al tiempo en que la contracci\'on 
lleg\'o a su valor l\'imite. Usaremos esta escala para estimar el
tiempo de p\'erdida de coherencia $t_D$.

La evoluci\'on del paquete gaussiano se lleva a cabo en
dos etapas diferenciadas. Durante la primera,  la evoluci\'on 
est\'a dominada por la parte unitaria de la ecuaci\'on maestra
y se mantiene dentro de un \'area pr\'acticamente constante.
Este proceso dura hasta que el estiramiento del estado
inicial es mayor que el ancho cr\'itico del estado. Durante
esta etapa, la difusi\'on no altera mucho la apariencia de 
la funci\'on de Wigner, la cual se ver\'a expandida o contra\'ida.
Cuando $\sigma \sim \sigma_c$, la difusi\'on empieza a cobrar 
importancia y comienza la segunda etapa de la evoluci\'on.
El estado ya no se contrae m\'as, pero s\'i es posible que siga
expandi\'endose a una velocidad fijada por el coeficiente 
de Lyapunov $\Lambda$. Como resultado, el \'area (o bien, el 
volumen) en el espacio de fases aumenta. Uno puede
estimar el tiempo correspondiente a la transici\'on de una evoluci\'on
reversible a una irreversible, como
\beq
t_{\rm c} = \frac{1}{\Lambda} \ln{\frac{\sigma_p(0)}{\sigma_{\rm c}}}.
\label{criticalt}\eeq
En nuestro modelo de juguete, podemos usar esta escala como
la escala t\'ipica de p\'erdida de coherencia, poniendo 
$t_D \approx \frac{1}{\Lambda} \ln{\frac{\sigma_p(t_{\rm max})}
{\sigma_{\rm c}}}$. De esta forma obtenemos
\beq
t_D =  t_{\rm max} + \frac{1}{\Lambda} \ln{\frac{\sigma_p(0)}{\sigma_{\rm c}}}.
\label{decotime}\eeq

Si usamos los mismos valores de los par\'ametros que en la 
Figura \ref{figure3-3},  podemos estimar num\'ericamente esta escala
como
$t_{D_{\rm b}} \sim 7.7$ y  $t_{D_{\rm d}} \sim 6.4$, para
el primer conjunto de par\'ametros (columna izquierda)
($\omega =\Omega = 1$) si $\gamma_0 = 0$; 
$t_{D_{\rm b}} \sim 2.4$ y
$t_{D_{\rm d}} \sim 2.7$ para $\gamma_0k_B T = 1$; y $t_{D_{\rm
b}} \sim 1.6$ y $t_{D_{\rm d}} \sim 1.7$, en el caso 
$\gamma_0k_B T = 100$. 
Para el conjunto de par\'ametros de la derecha de la Figura 
\ref{figure3-3} ($\omega= 5\Omega = 5$), los valores
temporales estimados son : $t_{D_{\rm b}}
\sim 3.0$ y $t_{D_{\rm d}} \sim 2.7$ si $\gamma_0 = 0$;
$t_{D_{\rm b,d}} \sim 0.1$, cuando $\gamma_0k_B T = 1$; y $t_{D_{\rm b,d}}
\sim 0.6$ en el caso que $\gamma_0k_B T = 100$. 
Todos estos resultados coinciden con los tiempos de p\'erdida
de coherencia que pueden leerse de los gr\'aficos, es decir, aquellos
tiempos donde el factor $\Gamma (t)$ se hace muy chico.

Podemos ver que el proceso de p\'erdida de coherencia es m\'as lento
en el caso (d) que en el (b) cuando el sistema $A+B$  no est\'a aislado,
ya que se verifica, a partir de la ecuaci\'on (\ref{criticalt}) para cada caso,
\beq t_{D_{\rm b}} - t_{D_{\rm d}} = \frac{1}{\Lambda}
\ln{\frac{\sigma_{\rm c}^{\rm d}} {\sigma_{\rm c}^{\rm b}}}=
\frac{1}{2\Lambda}\ln{\frac{{\cal D}_{\rm d}}{{\cal D}_{\rm b}}}.
\eeq Esto concuerda con los resultados num\'ericos presentados
para los coeficientes de difusi\'on en la Figura \ref{figure3-1},
ya que ${\cal D}_{\rm d} < {\cal D}_{\rm b}$, lo cual implica que
 $t_{D_{\rm b}} < t_{D_{\rm d}}$. En el caso en que el sistema $A+B$
est\'a aislado del ba\~no t\'ermico ($\gamma_0=0$), de los gr\'aficos 
es f\'acil ver que
${\cal D}_{\rm d} \geq {\cal D}_{\rm b}$,  que  implica  $t_{D_{\rm b}} \geq
t_{D_{\rm d}}$, de nuevo concordando con nuestros
c\'alculos num\'ericos.

Los tiempo de p\'erdida de coherencia para los casos (a) y (c)
ocurren como en los sistemas arm\'onicos usuales. Podemos
tener una estimaci\'on de estas escalas usando el
resultado del MBC a alta temperatura, es decir, el tiempo
de p\'erdida de coherencia estimado como aquel tiempo $t_D$
tal que $1 \approx L^2 \int_0^{t_D}{\cal D}(s) ds$  (hay que tomar
la distancia t\'ipica del MBC, o sea, por ejemplo,  $L$ como $2\sigma$, como
medida de la dispersi\'on en posici\'on del paquete gaussiano
inicial). \\

De esta forma, hemos analizado el proceso de p\'erdida de coherencia
inducido en un subistema debido a la presencia de un entorno compuesto.
El entorno compuesto fue \mbox{modelado} por un oscilador (o antioscilador) acoplado a
un conjunto de osciladores arm\'onicos a temperatura alta. El subsistema principal, 
pod\'ia ser un oscilador o antioscilador, seg\'un el caso analizado.
En este contexto, hemos mostrado que los osciladores arm\'onicos son
capaces de retener la informaci\'on por un per\'iodo de tiempo m\'as largo y, de esta forma,
generar difusi\'on en el subsistema $A$ m\'as eficientemente que un oscilador invertido.
Esta es la raz\'on principal por la cual, en general, el caso (b) pierde m\'as coherencia  que el
caso (d); y (c) que (a).
Es importante remarcar que, para tener un modelo m\'as realista del entorno compuesto, uno
deber\'ia considerar un potencial de pozo doble para el subsistema $B$ en los casos (a) y (d).
Este tipo de potencial tiene las mismas caracter\'isticas que el oscilador invertido, pero la ventaja
que su espacio de fases est\'a acotado, lo cual dar\'ia una mejor medida del efecto global inducido en
el subsistema $A$. En esa situaci\'on, el caso (d) resultar\'ia el m\'as ``decoherente" para cualquier
valor de $\gamma_0 k_B T$ \cite{dwPRE}. De todos modos, 
si mantenemos el an\'alisis a tiempos cortos de la evoluci\'on,
los potenciales son equivalentes.
\newpage
\thispagestyle{empty}
\cleardoublepage

\chapter{Fases geom\'etricas y p\'erdida de coherencia}
\label{c5}
\markboth{ Fases geom\'etricas y p\'erdida de coherencia}
{Cap\'itulo 5}

En este cap\'itulo, nos ocuparemos de las fases que adquieren
 los sistemas cu\'anticos abiertos.
Trataremos con fases de naturaleza geom\'etrica y nuestro 
principal inter\'es ser\'a cuantificar
el efecto del entorno sobre  las fases en s\'i mismas, y, en 
particular, en los experimentos
posibles para medirlas.

Un enfoque alternativo para estudiar la decoherencia, 
es analizar este problema  considerando
las  fases de las part\'iculas  como un proceso 
estad\'istico; es decir, estudiar c\'omo el entorno modifica la funci\'on de onda
de las part\'iculas. Este proceso se lo conoce como {\it dephasing}. Ambos enfoques son 
similares y dicha equivalencia fue demostrada en \cite{SternAhaImry}.

Para introducir el concepto de fase geom\'etrica, 
lo primero que buscamos son ejemplos
de ellas en la vida diaria; es decir ``cl\'asica".
 La pregunta que nos surge es si podemos
encontrar alguna relaci\'on entre un astronauta,
un gato y el p\'endulo de Foucault. Resulta raro, 
pero todos ellos utilizan el concepto
de fase geom\'etrica como principio fundamental 
de su movimiento b\'asico. 
Supongamos que un astronauta est\'a en el espacio libre, inicialmente 
de espaldas a su nave espacial. Si \'el quiere darse la vuelta para volver a 
su nave, \textquestiondown c\'omo lo logra si no hay nada con qu\'e 
empujarse ni  apoyarse para dar la vuelta? Su momento angular
es inicialmente cero, y para dar la vuelta, pareciera que necesita generar
cierto momento angular. Pero en ausencia de fuerzas, esto resulta imposible. 
Sin embargo, cualquiera que ha visto un gato caer, sabe que este razonamiento 
es falso. Los gatos se enfrentan con este problema casi todos los d\'ias, y, de alguna
forma, logran darse vuelta y caer sobre sus patas. 
La explicaci\'on es que no usan las leyes de la din\'amica para darse vuelta, 
sino la topolog\'ia del espacio. En la Fig.\ref{hombre}
mostramos c\'omo hace un astronauta para dar la vuelta. 
\begin{figure}[!ht]
\center
\includegraphics[width=15cm]{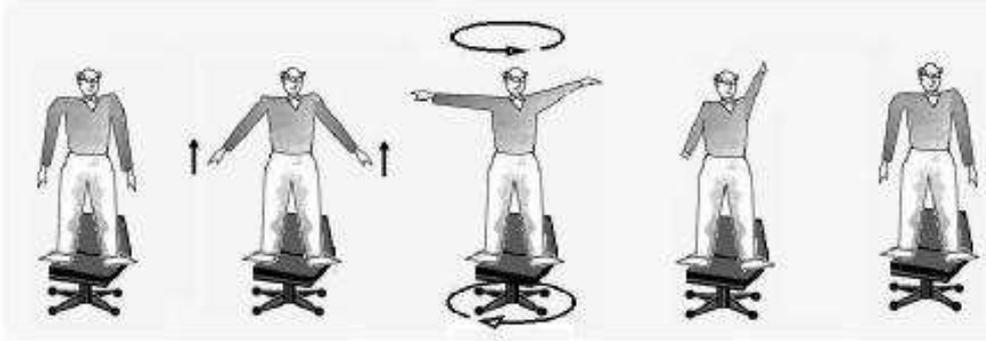}
\caption{Una persona parada en una silla giratoria, puede hacerla girar
usando simplemente un efecto geom\'etrico y manteniendo el momento
angular cero. En la posici\'on tres, la persona rota los brazos para un lado 
mientras el cuerpo lo rota para el otro. Repitiendo el ciclo
varias veces, la persona puede rotar el \'angulo deseado. Con este
efecto, el  gato logra caer sobre sus patas si inicialmente estaba de espalda
al piso..}
 \label{hombre}
\end{figure}

 El p\'endulo de Foucault puede ser explicado de la misma forma 
 \cite{Hammond}. Imaginemos que el p\'endulo est\'a 
suspendido a cierta latitud $\theta$, y observamos su movimiento
a medida que la Tierra rota sobre su propio eje. Se sabe que, despu\'es
de una rotaci\'on de la Tierra, el p\'endulo adquirir\'a
una fase respecto al plano original de movimiento. Para ver \'esto,
escribamos las ecuaciones de movimiento del p\'endulo.
El Lagrangiano del problema es
\begin{equation}
 L=\frac{m}{2} (\frac{dx ^2}{dt} + \frac{dy ^2}{dt}) -
\frac{m \omega^2}{2} (x^2+y^2) - m\Omega \cos\theta (x\frac{dy}{dt}
-y\frac{dx}{dt}), \nonumber 
\end{equation}
donde $\Omega$ es la frecuencia de oscilaci\'on de la Tierra 
($2 \pi$ por d\'ia), $m$ es la masa del p\'endulo y $\omega$ la frecuencia natural
de balanceo del p\'endulo. La soluci\'on a las ecuaciones de 
Euler-Lagrange, en la coordenada $z=x+i y$ es, en el l\'imite adiab\'atico es:
\begin{equation}
 z(t) \sim x_0 e^{-i \Omega \cos \theta t} e^{-i \omega t}. \nonumber
\end{equation}
Esta soluci\'on nos resulta conveniente para visualizar las contribuciones
que conforman la fase del movimiento del p\'endulo: una fase
din\'amica $\omega t$ y otra geom\'etrica $\Omega \cos \theta t$. 
Despu\'es
de una rotaci\'on completa de la Tierra, la fase geom\'etrica cl\'asica es 
$2 \pi (1-\cos \theta)$.

\section{Fases geom\'etricas cu\'anticas}

Supongamos que tenemos un conjunto de estados, y en particular,
un estado en el punto A y otro en el punto B, como indica la 
Fig.\ref{estados}. \textquestiondown Cu\'al es el \'angulo entre
estos dos vectores\footnote{ Como en  Mec\'anica Cu\'antica es posible
pensar las fases como vectores, esta pregunta resulta an\'aloga
a saber cu\'al es la fase relativa a ambos estados cu\'anticos.}?
Pero, \textquestiondown c\'omo podemos medir la fase entre dos
estados que est\'an en posiciones diferentes? Una alternativa,
es transportar uno de los estados hacia la posici\'on del otro y,
cuando est\'an pr\'oximos, medir el \'angulo entre ambos. El
\'unico inconveniente que podemos encontrar en este procedimiento
es asegurarnos que no estamos introduciendo ninguna fase adicional
al transportar el estado. 
\begin{figure}[!ht]
\center
\includegraphics[width=9cm]{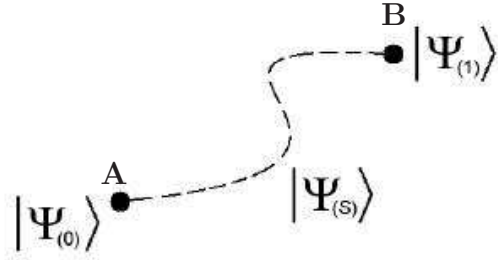}
\put(-185,46){\bf A}
\put(-80,106){\bf B}
\caption{\textquestiondown C\'omo comparamos las fases de dos estados
distintos cuando la probabilidad entre estados es la \'unica
cantidad bien definida en Mec\'anica Cu\'antica?}
 \label{estados}
\end{figure} El camino m\'as directo es una geod\'esica
y la correspondiente evoluci\'on en este camino es conocida como
{\it transporte paralelo}. Para definirlo, debemos mirar una evoluci\'on
infinitesimal, es decir de $\vert \psi(s)\rangle $ a $\vert \psi(s+ds) \rangle$.
Si no queremos fases adicionales, debemos pedir
\begin{equation}
 \rm Arg\{\langle \psi(s) \vert \psi(s+ds) \rangle\}=0.
\end{equation}
Esto equivale a pedir que $\langle \psi(s) \vert \psi(s+ds) \rangle$
sea una cantidad real, 
\begin{equation}
 \rm Im\{\langle \psi(s) \vert \psi(s+ds) \rangle\}=Im\{\langle 
\psi(s) \vert  d \vert \psi(s) \rangle\} = 0
\end{equation}
a segundo orden. Pero, como $\langle \psi(s) \vert  d \vert 
\psi(s) \rangle$ es imaginario puro, la condici\'on equivale a
 \begin{equation}
\langle \psi(s) \vert  d \vert \psi(s) \rangle = 0.
\end{equation}
Si la evoluci\'on satisface esta ecuaci\'on, entonces la fase es
transportada {\it paralelamente}. Esta definici\'on, sin embargo, no
es invariante de gauge (o de fase). Es decir, si en lugar de $\vert 
\psi(s)\rangle $ usamos otro estado equivalente, 
$\vert \tilde{\psi}(s)\rangle= e^{i \alpha(s)}\vert \psi(s)\rangle $, la
condici\'on de transporte paralelo cambia seg\'un,
\begin{equation}
\langle \tilde{\psi}(s)\vert {\rm d} \vert \tilde{\psi}(s) \rangle=
\langle {\psi}(s)\vert {\rm d} \vert {\psi}(s)\rangle + i \frac{\rm d \alpha}{ds} ds.
\end{equation}
Por tanto, debemos integrar la expresi\'on $\langle \psi(s) \vert  
d \vert \psi(s) \rangle$ en un lazo cerrado, de modo de obtener una cantidad
invariante de gauge. Esto nos da una expresi\'on para la fase
geom\'etrica la cual hay que exponenciar (la integral de $\rm 
d \alpha/ ds$ en una curva cerrada da $2 \pi$, y $e^{i 2 \pi}$ es uno). 
De esta forma,
la fase geom\'etrica que se deduce del transporte paralelo es 
\begin{equation}
 \gamma=\int_i^f \langle \psi(s) \vert \frac{d}{d s} \vert \psi(s) \rangle
ds.
\label{fasevladko}
\end{equation}
Ahora, \textquestiondown c\'omo es posible que,
si estamos pidiendo que en cada paso infinitesimal la diferencia de
fase sea cero, en un lazo cerrado, la fase total es distinta de cero?
La respuesta est\'a en la curvatura del espacio en cuesti\'on. Cuando
una cantidad se anula infinitesimalmente, pero en una regi\'on finita no, se 
la conoce como {\it no integrable}. De este modo,
podr\'iamos afirmar brevemente que las fases geom\'etricas son una
 manifestaci\'on de factores de fase no integrables en Mec\'anica Cu\'antica.

\subsection{Esfera de Bloch}
Los estados de dos niveles son ubicuos en la Naturaleza. Estos
estados pueden ser convenientemente representados en una esfera, de
forma tal que todos los estados puros est\'an ubicados en la
superficie, mientras que los estados mixtos se encuentran dentro de ella.
Existe una correspondencia uno a uno entre los puntos de la esfera
y los estados de un sistema de dos niveles, como por ejemplo, el
esp\'in $1/2$ de $\rm SU(2)$. 
Tratemos de aplicar el concepto de fase geom\'etrica a un estado en la 
esfera de Bloch, como muestra la Fig.\ref{esfera}.

\begin{figure}[!ht]
\center
\includegraphics[width=8cm]{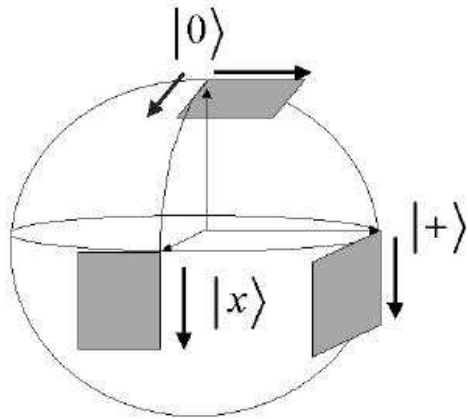}
\caption{Ejemplo de transporte paralelo en la esfera de Bloch. 
La evoluci\'on puede ser implementada de diferentes maneras. El estado
vector est\'a inicialmente en el Polo Norte de la esfera, en el plano
de la hoja. Luego, es transportado hacia el Ecuador, rotado $90^{\circ}$
y llevado de vuelta al Polo Norte. El estado vector final est\'a
rotado respecto del inicial en $90^{\circ}$; es decir, apunta fuera
de la hoja.}
 \label{esfera}
\end{figure}

Supongamos que hacemos evolucionar el estado $\vert 0 \rangle$ al estado
$\vert + \rangle = \vert 0 \rangle + \vert 1 \rangle$ y luego, al estado 
$\vert x \rangle = \vert 0\rangle + i \vert 1 \rangle$. Finalmente, 
llevamos el estado a $\vert 0 \rangle$ nuevamente. 
En la esfera de Bloch, esto equivale
a llevarlo del Polo Norte al Ecuador, rotarlo $90^{\circ}$ y 
moverlo  de vuelta hacia el Polo Norte. \textquestiondown Cu\'al es la
fase geom\'etrica correspondiente a esta situaci\'on? 
Para poder calcularla, comenzamos con un estado paralelo al Ecuador, 
en el plano de la hoja. 
Si lo transportamos paralelamente a lo largo del recorrido cerrado
indicado, obtenemos un estado rotado respecto del inicial, apuntando
hacia afuera del plano de la hoja. Esto sucede a pesar de que a cada paso
infinitesimal, el vector de estado  permanece parelelo a s\'i mismo.
El \'angulo entre el estado inicial y el final es $\pi/2$, que es
equivalente al \'area recorrida por el  vector de estado durante el transporte
(o bien, el correspondiente \'angulo s\'olido del transporte).
De esta forma, los estados ortogonales $\vert 0 \rangle$ y 
$\vert 1 \rangle$ evolucionan de la siguiente manera:
\begin{eqnarray}
 \vert 0 \rangle &=& e^{i \Omega/2} \vert 0 \rangle \nonumber \\
\vert 1 \rangle &=& e^{-i \Omega/2} \vert 1 \rangle, \nonumber
\end{eqnarray}
donde $\Omega$ es el \'angulo s\'olido (el factor $1/2$ es porque
los estados ortonormales est\'an separados una distancia $\pi$
en la esfera de Bloch). Esto demuestra que los estados ortogonales
adquieren fases opuestas de igual magnitud durante la evoluci\'on.
La fase puede ser calculada seg\'un la definici\'on de la 
Ec.(\ref{fasevladko}), o de manera ``discreta'':
\begin{equation}
 \rm Arg\{\langle 0 \vert + \rangle \langle + \vert x \rangle
\langle x \vert 0 \rangle\}.
\end{equation}
Esta \'ultima formulaci\'on fue originalmente realizada 
por Pancharatnam \cite{Panchat}. Este \mbox{formalismo} es muy poderoso e
 importante en varias teor\'ias f\'isicas \cite{Frankel}. Pero antes de ver un 
ejemplo concreto, deber\'iamos analizar c\'omo se implementa
 f\'isicamente el transporte paralelo.

\subsection{Implementaci\'on adiab\'atica del transporte paralelo}

La implementaci\'on del transporte paralelo fue 
realizada por Berry \cite{Berry}, quien descubri\'o la
fase geom\'etrica en Mec\'anica Cu\'antica en 1984.

La funci\'on de onda del sistema es una funci\'on de los par\'ametros
del mismo y del tiempo, $\vert \psi(s(t),t) \rangle$. Supongamos que
el Hamiltoniano del sistema es una funci\'on solamente de $s$, $H=H(s(t))$.
Adem\'as, supongamos que estos par\'ametros var\'ian muy lentamente, 
de modo que el sistema, que inicialmente se encuentra en un autoestado
del Hamiltoniano, se mantiene en un autoestado del Hamiltoniano
instante a instante, es decir
\begin{equation}
 H(s(t)) \vert \Psi_n(s(t),t) \rangle = E_n(s(t)) \vert \Psi_n(s(t),t) \rangle.
\end{equation}
La ecuaci\'on de Schr\"{o}dinger para este sistema es
\begin{equation}
 i \frac{d}{dt}\vert \Psi_n(s(t),t) \rangle = H(s(t),t) \vert \Psi_n(s(t),t) \rangle,
\end{equation}
donde hemos asumido $\hbar=1$. Nuestra intenci\'on es mostrar que 
la evoluci\'on adiab\'atica implementa naturalmente el transporte
paralelo de la fase del estado cu\'antico. Multiplicando
la ecuaci\'on de Schr\"{o}dinger por $\langle \Psi_n \vert$ y
considerando la ecuaci\'on de autovalores, obtenemos
\begin{equation}
 i \langle \Psi_n \vert \frac{d}{dt} \vert \Psi_n \rangle = E_n.
\end{equation}
Cada estado gana un fase din\'amica adem\'as de la geom\'etrica
a medida que evoluciona. Cuando se quiere medir la fase geom\'etrica, 
lo ideal es librarse de la fase din\'amica. Para eso, definimos una nueva 
funci\'on de onda considerando la fase din\'amica
\begin{equation}
 \vert \Phi(s(t),t) \rangle := e^{i E_n(s(t))} \vert \Psi_n(s(t),t),
\rangle
\end{equation}
la cual verifica una ecuaci\'on equivalente a la de transporte
paralelo
\begin{equation}
 \langle \Phi(s(t),t) \vert \frac{d}{dt} \vert \Phi(s(t),t) \rangle =0.
\end{equation}
De esta forma se ve que la parte geom\'etrica de la fase 
cu\'antica es transportada paralelamente cuando el estado 
evoluciona seg\'un la
ecuaci\'on de Schr\"{o}dinger en la aproximaci\'on adiab\'atica.

La forma de derivar una expresi\'on cerrada para la fase geom\'etrica
es la siguiente. Para eso, definimos
\begin{equation}
 \vert \Phi(s(t),t) \rangle = e^{i \gamma(t)} \vert \Psi_n(s(t))
\rangle,
\end{equation}
con 
\begin{equation}
\frac{d}{dt} \gamma(t) = -i \langle \Phi(s(t)) \vert \frac{d}{ds} \vert
\Phi (s(t)) \rangle \frac{ds}{dt}.
\end{equation}
De esta forma,
\begin{equation}
 \frac{d}{d s} \gamma= \beta, ~~~{\rm con} ~~~ \beta=-i \langle
 \Phi(s(t)) \vert \frac{d}{ds} \vert
\Phi (s(t)) \rangle.
\end{equation}
Integrando sobre una curva cerrada $\delta S$, obtenemos 
la fase geom\'etrica deseada
\begin{equation}
 \gamma=\oint_{\delta S} \beta.
\end{equation}
Resumiendo, la fase cu\'antica que adquiere un estado durante
su evoluci\'on unitaria consiste de dos contribuciones, $\phi=\delta
+ \gamma$:\begin{description}
 \item {\bf Din\'amica}: definida por $\delta=\int E(t) dt$ y,
\item {\bf Geom\'etrica}: definida por $\gamma= \int \langle \Psi (s)
\vert \frac{d}{ds} \vert \Psi(s) \rangle \frac{ds}{dt} dt$.
\end{description}

En este contexto, estudiaremos diferentes modelos de 
sistemas cu\'anticos 
donde es posible encontrar fases geom\'etricas. 
Sin embargo, veremos que,
cuando estos sistemas interact\'uan con entornos, las fases 
originalmente geom\'etricas,
pierden esta caracter\'istica, y  dependen de los par\'ametros del 
entorno. Adem\'as, estudiaremos c\'omo, en los distintos casos, 
acoplar estos sistemas cu\'anticos a entornos extensos 
cl\'asicos o cu\'anticos, induce p\'erdida de coherencia en 
el sistema original.

\subsection{Efecto Aharonov-Bohm y Aharonov-Casher}

El efecto Aharonov-Bohm \cite{ABohm} puede ser explicado en funci\'on del concepto de
fase geom\'etrica, ya que resulta la manisfestaci\'on del campo electromagn\'etico
en la fase relativa de dos part\'iculas que interfieren. El experimento comienza
con la preparaci\'on de dos paquetes de ondas, uno correspondiente a cada electr\'on.
La interferencia se analiza en presencia de un solenoide, por lo cual cada part\'icula
recorrer\'a un camino a trav\'es de dos lados opuestos de este solenoide. El campo
magn\'etico en la regi\'on exterior al solenoide es cero pero el potencial no. 
Cuando las part\'iculas alcanzan
el punto final de la evoluci\'on, se observa que el haz de electrones gana una fase
$\exp(\frac{-i e \Phi}{\hbar c})$, donde $\Phi$ es el flujo del campo. 
Esta fase ha sido
experimentalmente observada, y depende del camino, lo cual equivale 
a decir  que es no integrable. A pesar de \'esto,  si damos una vuelta alrededor del flujo, siempre
adquirimos la misma fase, sin importar el camino que tomemos. La fase depende del 
n\'umero de veces que el flujo ha sido rodeado. Esta es una caracter\'istica topol\'ogica
que resulta muy \'util en computaci\'on cu\'antica. Este mismo efecto, pero con part\'iculas
neutras de momento dipolar constante, se denomina Aharonov-Casher \cite{ACasher}.

La pregunta que nos surge es \textquestiondown c\'omo se modifica 
este resultado si el experimento no
se realiza en presencia de un campo est\'atico sino en un campo electromagn\'etico
dependiente del tiempo? Adem\'as, queremos saber 
cu\'anta injerencia tiene esta fase en la reducci\'on de la visibilidad 
del patr\'on de interferencia generado por las part\'iculas neutras, a veces muy masivas.

\section{P\'erdida de coherencia inducida por una fase de Aharonov-Casher fluctuante}
\label{pra72}

En muchos casos, la interacci\'on con el entorno no puede ser eliminada; por ejemplo,
en el caso de part\'iculas cargadas y \'atomos neutros con momento dipolar
permanente, la interacci\'on con el campo magn\'etico es crucial. Es esta  
interacci\'on, justamente, la que  induce una reducci\'on en la 
visibilidad del patr\'on de interferencia.  En este contexto, las interacciones
de las fluctuaciones de vac\'io del campo electromagn\'etico han sido
consideradas  un agente ``decoherente" en \cite{Alejandro}. All\'i, los
autores estudiaron el experimento de dos rendijas para electrones
en presencia de contornos conductores, los cuales modifican la estructura
del vac\'io y, por tanto, las predicciones de los efectos de p\'erdida de 
coherencia (respecto al caso donde no hay contornos). En \cite{Vourdas}
se analiz\'o el efecto del ruido cu\'antico en experimentos de
interferencia con electrones para distintos tipos de campos de  microondas. A pesar
de la naturaleza cu\'antica del ruido, por ejemplo fluctuaciones de vac\'io,
el ruido cl\'asico est\'a siempre presente, ya sea en un campo
externo dependiente del tiempo, o bien, en variables aleatorias
que parametrizan el entorno. Por esto, en \cite{Vourdas2}, los
autores estudiaron la destrucci\'on del patr\'on de interferencia
de electrones debido tanto al ruido cu\'antico como al cl\'asico.

El experimento de Aharonov-Bohm (AB) puede ser un buen ejemplo para poner
a prueba las predicciones de la decoherencia. Este experimento comienza con
la preparaci\'on  de los dos paquetes de onda de los electrones, $\varphi_1(\vec{x})$
y $\varphi_2(\vec{x})$, en una superposici\'on coherente. Se asume que cada
part\'icula sigue un camino cl\'asico bien definido, $C_1$ y $C_2$, respectivamente.
La funci\'on de onda total contempla la presencia del entorno,
\beq \psi
(t = 0)=  \left[\varphi_1(\vec x) + \varphi_2(\vec x)\right]
\otimes \chi_0(\vec y), \eeq
donde $\chi_0(\vec y)$ representa el estado cu\'antico inicial del entorno,
cuyo conjunto de coordenadas es denotado por  $\vec y$. A medida que
el tiempo transcurre, los estados del electr\'on se entrelazan con el entorno,
y la funci\'on total a un tiempo dado $t$ es
\beq \psi (t) =  \varphi_1(\vec x,t) \otimes
\chi_1 (\vec y,t) +
             \varphi_2(\vec x,t) \otimes \chi_2 (\vec y,t).
\eeq
De esta forma, los dos estados $\varphi_1$
and $\varphi_2$ del electr\'on se correlacionan con dos estados
diferentes del entorno. La probabilidad de encontrar a la part\'icula
en una posici\'on dada a un tiempo $t$ (por ejemplo, cuando se
observa el patr\'on de interferencia) es
\beq \mbox{Prob}(\vec x,t) = \vert \varphi_1(\vec
x,t)\vert^2 + \vert \varphi_2(\vec x,t)\vert^2 + 2 {\rm Re}\left(
\varphi_1(\vec x,t) \varphi^*_2(\vec x,t) \int d^3y ~\chi_1^*(\vec
y,t)\chi_2(\vec y,t)\right). \label{Fcap5} \eeq
El factor de solapamiento $F = \int d^3y ~\chi_1^*(\vec y,t)\chi_2(\vec
y,t)$ es responsable de dos efectos. Por un lado, origina un corrimiento
en las franjas de interferencia del patr\'on; y por el otro, su valor absoluto
da lugar a una disminuci\'on del contraste en dichas franjas. En ausencia 
de entorno, $F=1$. Cuando los dos estados
del entorno se vuelven ortogonales, el estado final del mismo identifica
el camino que sigui\'o el electr\'on. La p\'erdida de coherencia se 
manifiesta tan pronto como las ondas parciales del electr\'on logran
ubicar al entorno en estados ortogonales de modo que $F=0$.

La p\'erdida de coherencia cu\'antica tambi\'en puede ser explicada,
alternativamente, como el efecto del entorno sobre las ondas parciales
del electr\'on. Cuando un potencial est\'atico act\'ua sobre una de estas
ondas parciales, \'esta adquiere una fase,
\beq \phi = - \int V[x(t)] dt, \eeq
y de esta forma, el t\'ermino de interferencia est\'a multiplicado por un 
factor $e^{i\phi}$. Este factor resulta un posible agente  ``decoherente".
El efecto que origina est\'a \'intimamente relacionado al c\'aracter estad\'istico de
$\phi$, particularmente, cuando el potencial no es est\'atico. De esta manera,
cualquier fuente de ruido estoc\'astico dar\'a lugar a un t\'ermino decreciente.
Para un caso general, $\phi$ no est\'a definida; es decir, se debe describir
 mediante una funci\'on distribuci\'on $P(\phi)$. Desde este punto de vista
estad\'istico, la fase se escribe como
\beq \langle e^{i\phi}\rangle = \int e^{i\phi} P(\phi
) d\phi.\label{IF} \eeq

De esta manera, la incerteza en la fase produce un t\'ermino decreciente que 
tiende a eliminar el patr\'on de interferencia. Este {\it dephasing} se debe a
la presencia de un entorno ruidoso acoplado al sistema y puede ser 
tambi\'en representado por la funcional de influencia de 
Feynman y Vernon \cite{Feynman} (m\'etodo que hemos usado
en cap\'itulos anteriores y da lugar a la ecuaci\'on maestra para el caso del
MBC (Ec.(\ref{master}))). En la Ref.\cite{SternAhaImry}, los autores demostraron 
la equivalencia formal entre ambos enfoques de modo que
\beq \langle e^{i\phi}\rangle = F =  \int d^3y  ~ \chi_1^*(\vec
y,t)\chi_2(\vec y,t). \label{overlap}\eeq
El factor $F$ contiene informaci\'on acerca de la naturaleza estad\'istica
del ruido. De esta forma, el ruido, ya sea cl\'asico o cu\'antico, hace que
$F$ sea menor que 1. Nuestra intenci\'on es cuantificar c\'omo destruye 
el patr\'on de interferencia para un experimento con  part\'iculas.

En \cite{Ford}, el factor de solapamiento $F$ fue evaluado de un punto de
vista alternativo a los trabajos existentes del tema. Los autores estudiaron
el efecto  de un campo electromagn\'etico dependiente del
tiempo en la coherencia de los electrones. Para \'esto, incluyeron el origen
estad\'istico de la fase $\phi$ de AB. Sin embargo, no consideraron que 
la fase se originaba en las fluctuaciones de vac\'io ni debido al campo
electromagn\'etico dependiente del tiempo.  Ellos consideraron la
existencia de una variable aleatoria $t_0$,  definida como el tiempo
de emisi\'on del electr\'on. Esta variable produc\'ia una fase $\phi$
fluctuante, por tanto, era necesario calcular un promedio temporal
para obtener un resultado concreto acerca de su influencia en el
patr\'on de interferencia. En este ejemplo sencillo, el rol del entorno
cu\'antico es reemplazado por un campo externo cl\'asico dependiente
del tiempo. Sin embargo, el efecto es similar. Los autores consideraron
una onda monocrom\'atica linealmente polarizada, que se propaga
en la direcci\'on perpendicular al plano que contiene a los electrones
(o haz de electrones). La reducci\'on del contraste de las franjas
result\'o ser suficientemente grande como para ser observada.

En esta secci\'on seguiremos dicha idea. Evaluaremos el factor 
de solapamiento $F$ para part\'iculas coherentes neutras con momento
dipolar permanente (el\'ectrico y magn\'etico) en presencia de un
campo electromagn\'etico cl\'asico dependiente del tiempo. Consideraremos
dos casos diferentes que podr\'ian ser interesantes a nivel experimental.
En primer lugar, los dipolos interact\'uan con una onda electromagn\'etica 
plana linealmente polarizada. Este caso es la generalizaci\'on de los
resultados obtenidos en \cite{Ford}  para la fase de 
Aharonov-Casher. En segundo lugar, estudiaremos el caso de dipolos
coherentes que interfieren dentro de una gu\'ia de ondas de 
secci\'on rectangular. Veremos que, el efecto de p\'erdida de
coherencia no afecta visiblemente
 el patr\'on de interferencia en todos los casos.

\subsection{Dipolos coherentes y una onda plana}
\label{dipolosplanewave}

La interacci\'on cl\'asica y cu\'antica de un dipolo con un
campo electromagn\'etico arbitrario fue estudiada en detalle 
por J.Anandan en \cite{Anandan}. El Lagrangiano cl\'asico de interacci\'on
 es $\frac{1}{2}P_{\mu\nu}F^{\mu\nu}$ donde $P_{\mu\nu}$
es el tensor antisim\'etrico del dipolo \cite{Alejandro, Anandan}.
En el sistema de coordenadas de la part\'icula, el dipolo
el\'ectrico ($\mbf{d}$) y magn\'etico ($\mbf{m}$) pueden ser
obtenidos a partir de $P_{0i}= d_i$ y
$P_{ij}=\epsilon_{ijk}m_k$, respectivamente.

En el caso cu\'antico, la fase que adquieren  dos 
part\'iculas neutras con momento dipolar el\'ectrico y magn\'etico
 debido a la presencia del campo electromagn\'etico
dependiente del tiempo se llama fase Aharonov-Casher (AC)
\cite{ACasher} y se define
\beq \phi=-
\oint_{\delta \Omega} a_{\nu}(x) dx^{\nu}, \label{faseAC} \eeq
donde $a_{\nu}(x)=(-\mbf{m} \cdot \mbf{B} -\mbf{d} \cdot \mbf{E},
\mbf{d} \times \mbf{B} - \mbf{m} \times \mbf{E})$ juega el mismo
rol que el potencial cuadrivector $A_\nu$ en la fase de 
Aharonov-Bohm (AB) \cite{Ford}, $\delta \Omega=C_1-C_2$ 
es un camino espacio temporal 
cerrado, y $C_1$, $C_2$ son los caminos recorridos por los dipolos
que interfieren.

Para evaluar la fase AC de la Ec.(\ref{faseAC}), consideraremos
una onda monocrom\'atica \mbox{linealmente} polarizada, de 
frecuencia $\omega$, que se propaga en la direcci\'on $\hat{y}$.
Los momentos dipolar el\'ectrico y magn\'etico est\'an en las
direcciones $\hat{z}$ y $\hat{x}$, respectivamente. Adem\'as,
asumiremos que el recorrido de las part\'iculas neutras est\'a
acotado \'unicamente al plano $\itm{{\hat x}-{\hat z}}$, como
muestra la Fig.\ref{figtraydip1}. Podemos escribir la onda
como $\mbf{E}(x)=E_0 \sin(wt-ky)~\hat{z}$,
$\mbf{B}(x)=E_0 \sin(wt-ky)~\hat{x}$ y calcular $a_\nu$,
\beqa
a_{\nu}(x)&=&(-d_z E_z-
m_x B_x, m_y E_z,d_z B_x-m_x E_z,-d_y B_x)\nonumber \\
&=& E_0 (-d_z - m_x , m_y ,d_z - m_x,-d_y) \sin (\omega t - k y)
\nonumber\\ &\equiv &{\tilde a}_{\nu}\sin (\omega t - k y).
\label{anu}
\eeqa

Asumiremos que la fase AC depende de una variable 
aleatoria $\xi = \omega t_0$ determinada por el tiempo 
de emisi\'on $t_0$ de las part\'iculas. Este
tiempo se define en el momento que el centro del paquete 
de ondas es emitido. Cuando el tiempo de medici\'on del 
experimento es mayor que el
tiempo de vuelo de las part\'iculas, el resultado obtenido es el promedio
temporal sobre $t_0$. De esta forma podemos escribir la fase AC como
\beq \phi (t_0) =-\oint_{\delta\Omega} {\tilde a}_\nu \sin
(\omega t - k y + \omega t_0) ~ dx^\nu, \eeq
la cual, antes de promediar en $t_0$, puede ser re-escrita como
\beq \phi (t_0) = A \cos
(\omega t_0) + B \sin (\omega t_0), \eeq donde \beqa A &=&
-\oint_{\delta\Omega} {\tilde a}_\nu(x) \sin (\omega t - k
y)~dx^\nu,
\nonumber \\
B &=& -\oint_{\delta\Omega} {\tilde a}_\nu \cos (\omega t - k y)~dx^\nu .
\label{AyB}\eeqa
El promedio sobre la fase fluctuante $\phi$, la cual genera ruido
cl\'asico, da lugar a un factor de solapamiento o de ``decoherencia''
\beq
F = \langle e^{i\phi}\rangle = \lim_{T\rightarrow \infty}\frac{1}{2T}
\int_{-T}^{T}dt_0 \exp\left\{i\left[A \cos (\omega t_0) + B \sin (\omega t_0)
\right]\right\}=
J_0(\vert C\vert ),\label{promediophi}
\eeq
donde $J_0$ es la funci\'on de Bessel de orden 0. El
m\'odulo del n\'umero complejo $C = A + i B$ mide el grado
de p\'erdida de coherencia en el sistema. El factor de decoherencia $F$ 
decrece de uno a cero a medida que $\vert C\vert$ var\'ia de
cero al primer cero de la funci\'on de Bessel $J_0$. Para valores
 m\'as grandes de $\vert C\vert$, el factor $F$ oscila con
una amplitud cada vez m\'as chica. Para una distribuci\'on
gaussiana $P(\phi)$ con $\langle \phi^2 \rangle \ll 1$,
obtenemos, en el l\'imite $\vert C\vert \ll 1$, $\langle 
e^{i\phi}\rangle \approx 1
- \langle \phi^2\rangle = 1 - \vert C\vert^2/2$.

Una caracter\'istica de las fases geom\'etricas usuales AB y AC es
que no dependen de la velocidad de las part\'iculas \cite {Berry}, 
y adem\'as, no hay fuerzas aplicadas sobre ellas \cite{APV}. Es decir,
las fases mencionadas s\'olo dependen de la topolog\'ia del camino
cerrado $\delta \Omega$. Resulta evidente que estas propiedades ya no
son v\'alidas cuando el campo externo depende expl\'icitamente del
tiempo. En nuestro caso, la part\'icula tiene una fuerza neta sobre
ella. Luego, para analizar la dependencia de la fase AC con la trayectoria,
evaluaremos dicha fase para distintos recorridos. Veremos 
que la dependencia de la fase con la velocidad est\'a fuertemente
relacionada con  el camino elegido.

\subsubsection{Trayectorias sim\'etricas}

En esta secci\'on, estudiaremos la fase AC que adquieren dos part\'iculas
neutras con momento dipolar el\'ectrico y magn\'etico permanente
cuando siguen trayectorias sim\'etricas como las mostradas en la 
Fig.\ref{figtraydip1}.
\begin{figure}[!ht]
\center
\includegraphics[width=12cm]{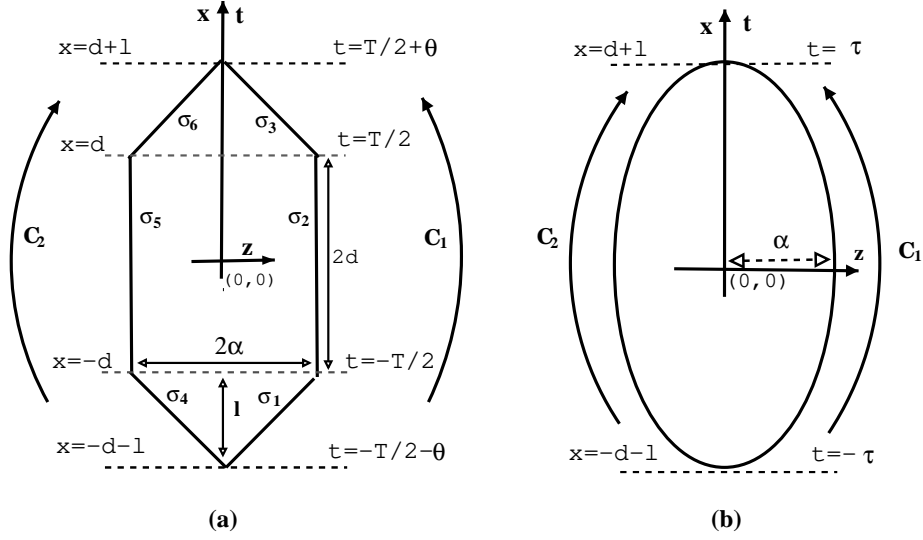}
\caption{Recorridos ${\cal C}_1$ y ${\cal C}_2$ para la trayectoria: 
(a) usada en \cite{ford} y  (b) el\'iptica.}
\label{figtraydip1}
\end{figure}
En primer lugar, utilizaremos la misma trayectoria propuesta
 en \cite{Ford} para hacer m\'as f\'acil la comparaci\'on
de los efectos de p\'erdida de coherencia entre part\'iculas cargadas
y neutras con momento dipolar permanente. Para evaluar la
Ec.(\ref{faseAC}) en el caso de la Fig.\ref{figtraydip1}(a), debemos
calcular \beq \phi=\oint_{\delta \Omega} a_{\nu}(x) dx^{\nu}= \sum_{i=1}^6
\int_0^1 a_{\nu} (\sigma_i^{\nu}(u))\cdot \frac{d\sigma_i ^{\nu}}{d u}
\label{parametrize}
du, \eeq 
donde $\sigma_i$,  con $i=1...6$, parametriza los distintos
segmentos de la trayectoria. El camino ${\cal C}_1$ es:
\begin{eqnarray}
\sigma_{1}(u) & = & (-T/2-\theta + \theta u,-d-l+u l,0,u \alpha)  \nonumber \\
\sigma_{2}(u) & = & (-T/2 + Tu,-d+2du,0,\alpha) \nonumber \\
\sigma_{3}(u) & = & (T/2+u\theta, d+ul,0,\alpha -u \alpha),
\quad \quad \mathrm{for} \quad
0 \leq u\leq 1;
\label{param-sym}
\end{eqnarray}
mientras que el camino ${\cal C}_2$ se parametriza:
\begin{eqnarray}
\sigma_{4}(u) & = & (-T/2-\theta + \theta u,-d-l+u l,0,-u \alpha) \nonumber \\
\sigma_{5}(u) & = & (-T/2+T u,-d+2 d u,0,- \alpha) \nonumber \\
\sigma_{6}(u) & = & (T/2+u \theta, d+ul,0,-\alpha +u \alpha), \quad \mathrm{for} \quad
0 \leq u\leq 1.
\end{eqnarray}
Realizando la integral en la Ec.(\ref{parametrize}) y usando
las definiciones de la Ec.(\ref{AyB}), obtenemos $B_d=0$ y
\beq
\vert C_d\vert\equiv
\vert A_d \vert = 4 E_0 d_y \big(\frac{ 2 \alpha}{\omega \theta} \big)
\sin \big(\frac{\omega \theta}{2}\big) \sin \big(\frac{\omega (T +
\theta)}{2} \big),
\label{Cd}
\eeq
donde $2\alpha$ es la m\'axima separaci\'on entre las part\'iculas,
$d_y$ es el momento dipolar el\'ectrico en la direcci\'on $\hat y$,
y $T,\theta$ son tiempos caracter\'isticos de la trayectoria. 
El sub\'indice $d$ indica que los valores encontrados para $A,B$ y $C$
corresponden al caso de dipolos.

Para part\'iculas no relativistas, esperamos que 
$\omega\theta,\,\,\omega T\gg 1$. Por tanto, para tener una estimaci\'on de $C_d$, 
podemos reemplazar las funciones arm\'onicas  por el valor t\'ipico
$1/\sqrt 2$. De manera de hacer expl\'icita la dependencia 
con la velocidad, podemos escribir $\theta= s/ v$, donde $s$ es la longitud
del primer y tercer segmento del camino, definido por 
$s =\sqrt{\alpha^2+ l^2}$. As\'i, la Ec.(\ref{Cd}) queda
\beq \vert C_d\vert \approx
\frac{2}{\pi}E_0
d_y \big( \frac{\alpha}{s} \big)\lambda  v \approx
\frac{2}{\pi} e E_0
\big( \frac{\alpha}{s} \big)\lambda L  v
. \label{Cdbis} \eeq
Ac\'a $L$ es el largo caracter\'istico de un \'atomo con momento
dipolar el\'ectrico $d= e L$ ($L \approx 10^{-9} m$), y $\lambda$
la longitud de onda del campo  electromagn\'etico. El resultado
an\'alogo para electrones calculado fue 
$|C_e|\approx \frac{1}{\pi^2} e E_0 \lambda^2 (\frac{\alpha}{s}) v$ 
\cite{Ford}. Suponiendo, de manera
ingenua, que las part\'iculas cargadas y las neutras tuvieran la misma
velocidad durante el recorrido de la trayectoria, la relaci\'on
entre ambos efectos ser\'ia
\beq |C_d|
\approx |C_e| \bigg(\frac{L}{\lambda}\bigg). \label{comparacioncap5} \eeq
A pesar que el resultado de la Ec.(\ref{comparacioncap5}) es bastante
desalentador (si uno buscaba obtener un efecto observable o, 
al menos, del mismo orden que en el caso de part\'iculas cargadas), 
hay que tener en cuenta que la secci\'on eficaz de scattering de 
una part\'icula neutra
es mucho menor que la de los electrones \cite{schwinger}
\footnote{Para conocer  en detalle
la cuenta ver Ref.\cite{Casher}.}. Este hecho da lugar a que, aumentando 
la intensidad del campo electromagn\'etico externo, aumente el 
valor del factor $F$ algunos \'ordenes de magnitud. De esta manera, aumentan
 las posibilidades concretas de que el efecto
 sea observable en part\'iculas neutras con
momento dipolar permanente.

Por otro lado, si las part\'iculas neutras siguen una trayectoria
el\'iptica, como la mostrada en la Fig.\ref{figtraydip1}(b), los
c\'alculos para estimar $C$ son similares. La trayectoria en este caso
se parametriza seg\'un
\beqa
\sigma_1(u) = (\tau \sin(u),(d+l) \sin(u),0,~~\alpha \cos(u))
 \quad \quad \quad \mathrm{for} -\pi/2 \leq u \leq \pi/2 \nonumber \\
\sigma_2(u) = (\tau \sin(u),(d+l) \sin(u),0,-\alpha \cos(u))
 \quad \quad \quad \mathrm{for} -\pi/2 \leq u \leq \pi/2,
\label{param-ellip}
\eeqa
donde $\tau$ es el tiempo de vuelo de los dipolos y $(d +l)$ es la 
longitud total del recorrido. En este caso,
\beq |C^d_{\rm ellip}|= 2 \pi \alpha E_0 d_y
\mathrm{J}_1 [\omega \tau], \label{Cel}\eeq
con $\mathrm{J}_1$ la funci\'on de Bessel de orden 1. Usando la
aproximaci\'on asint\'otica de esta funci\'on para $\omega \tau \gg 1$,
obtenemos
\begin{equation}
|C^d_{\rm ellip}| \approx \frac{\sqrt 2\pi\alpha E_0d_y}{(\omega \tau )^{1/2}}
= \sqrt \pi\alpha e E_0 L \left(\frac{v \lambda}{s'}\right)^{1/2}
\label{Cd_ellip} \end{equation}
donde $s'$ es la distancia recorrida por cada una de las part\'iculas
neutras a una velocidad $v$, en un tiempo $\tau$. Es importante
destacar que, mientras $\vert C_d\vert$ en la Ec.(\ref{Cdbis}) depende
linealmente con la velocidad, para la trayectoria el\'iptica 
$|C^d_{\rm ellip}|$ lo hace como $\sqrt v$.

Resulta interesante comparar este resultado con \'aquel obtenido para
electrones siguiendo la trayectoria el\'iptica, el cual no fue estimado
en \cite{Ford}. Por tanto, luego de estimarlo usando la definici\'on 
de la fase AB, obtenemos
\beq |C^e_{\rm
ellip}|=  2 \pi \alpha e E_0 \lambda \mathrm{J}_1[\omega \tau]. \eeq
Comparando esta expresi\'on con la Ec.(\ref{Cel}), resulta 
evidente que la dependencia con la velocidad es similar en ambos casos.

Finalmente, podemos resaltar que las  trayectorias estudiadas hasta 
ahora son sim\'etricas respecto del eje $\hat x $ (o bien, el eje  $\hat t$). En estos
casos, s\'olo contribuye el t\'ermino $A$ de la Ec.(\ref{AyB}) a la
fase AC. Esto se debe a la paridad de los integrandos en los ejes
mencionados.

\subsubsection{Trayectorias asim\'etricas}

En esta secci\'on, nos ocuparemos del efecto de p\'erdida de coherencia
en el patr\'on de interferencia de dos part\'iculas neutras cuando
recorren una trayectoria asim\'etrica, como la mostrada en la Fig.\ref{figtraydip2}.
\begin{figure}[!ht]
\center
\includegraphics[width=7cm]{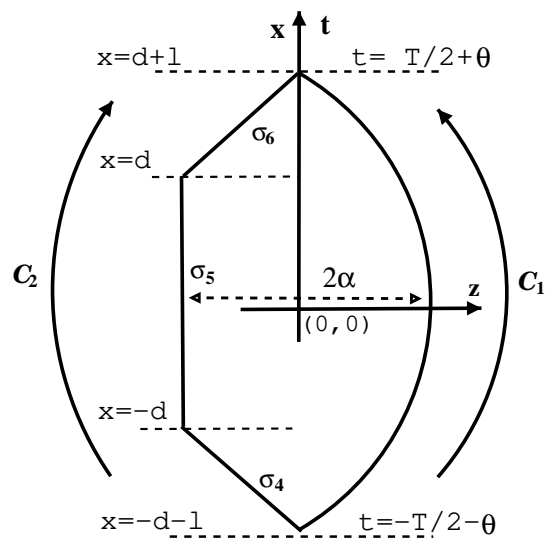}
\caption{Recorridos ${\cal C}_1$ y ${\cal C}_2$ de una trayectoria asim\'etrica.}
 \label{figtraydip2}
\end{figure}
La parametrizaci\'on de la trayectoria, en este caso, puede ser 
le\'ida de las Ecs.(\ref{param-sym}) y (\ref{param-ellip}).
Luego de realizar las correspondientes integrales, obtenemos
los coeficientes $C^d_{\rm asym}= A^d_{\rm asym} + i B^d_{\rm asym}$,
\beqa A^d_{\rm
asym} &=&\pi d_y E_0 \alpha \mathrm{J}_1[\omega \tau] +
\frac{4E_0d_y \alpha}{\omega \theta} \sin [\frac{\omega\theta}{2}]
\sin [\frac{\omega}{2}(T + \theta)],\nonumber \\
B^d_{\rm asym} &= & (d_z + m_x)\frac{2E_0}{\omega} \left[
\sin [\frac{\omega \theta}{2}] \cos[\frac{\omega}{2}(T + \theta)] +
 \sin[\frac{\omega T}{2}] \right] \nonumber \\
&+&  \frac{2 m_y E_0 l}{\omega \theta} \sin
[\frac{\omega \theta}{2}] \cos[\frac{\omega}{2}(T+\theta)] +
\frac{4 d E_0 m_y}{\omega T} \sin[\frac{\omega T}{2}]
.\eeqa
El primer t\'ermino en el coeficiente $B$ es el m\'as importante
en el r\'egimen de bajas velocidades, ya que los otros t\'erminos
son de \'orden ${\cal
O}(1/\sqrt{\omega \tau})$, o bien, ${\cal O}(1/\omega \theta)$,
ambos mucho menor que uno. Por tanto, la cantidad $|C^d_{\rm asym}|$, 
est\'a dominada por dicho t\'ermino y puede ser aproximada por,
\beq |C^d_{\rm asym}| \approx \frac{e}{\pi}  E_0  L
\lambda\,\, , \label{Cd_asym} \eeq
independiente de la velocidad.

Para part\'iculas cargadas, la situaci\'on es bastante diferente. El factor
de decoherencia $F$, usando la trayectoria asim\'etrica, al ser estimado
da $C^e_{\rm asym}= A^e_{\rm asym} +i B^e_{\rm asym}$, con
\beqa A^e_{\rm asym} &=& 2\pi e E_0 \frac{\alpha}{\omega}
\mathrm{J}_1[\omega \tau] + \frac{4 e E_0 \alpha}{\omega^2 \theta}
\sin [\frac{\omega\theta}{2}]
\sin [\frac{\omega}{2}(T + \theta)],\nonumber \\
B^e_{\rm asym} &= & 0
.\eeqa
Por tanto, podemos aproximar esta cantidad por
\beq
|C^e_{\rm asym}| \approx \frac{e}{ \sqrt{2\pi}}  E_0 \alpha 
\bigg(\frac{v\lambda^3}{s'}\bigg)^{1/2},\label{Ce_asym} \eeq
que depende de la velocidad, de la misma manera que 
para el caso de la trayectoria el\'iptica.

\subsection{Dipolos coherentes en una gu\'ia de ondas}
\label{dipolosguia}

En esta secci\'on, consideraremos el campo generado dentro de una gu\'ia
de ondas (en la direcci\'on del eje $\hat{y}$), de secci\'on
rectangular, como aquel campo que interact\'ua con nuestros dipolos.
Para el modo TE, los campos en la gu\'ia, tomando parte real, son
\begin{eqnarray}
B_y & = & B_0 \cos(k_x x) \cos(k_z z) \exp(i(k_y y - \omega t)), \nonumber \\
B_x & = & \frac{-i k_y k_x}{\gamma^2} B_0 \sin(k_x x) \cos( k_z z)
 \exp(i(k_y y - \omega t)), \nonumber \\
B_z & = & \frac{-i k_y k_z}{\gamma^2} B_0 \cos(k_x x) \sin(k_z z)
\exp(i(k_y y - \omega t)), \nonumber \\
E_x & = & \frac{ i \omega k_z}{\gamma ^2} B_0 \cos(k_x x) \sin(k_z z)
\exp(i(k_y y - \omega t)), \nonumber \\
E_z & = & \frac{ i \omega k_x}{\gamma ^2} B_0 \sin(k_x x) \cos(k_z z)
\exp(i(k_y y - \omega t)),
\end{eqnarray}
donde $k_x = \frac{ m \pi }{b}$, $k_z = \frac{ l \pi}{a}$ (con
$m$, $l$ enteros y $a,b$ las dimensiones de la gu\'ia), 
$\gamma= \sqrt{ (\frac{ l \pi}{a})^2 + (\frac{m \pi}{b})^2}$,
$\omega \equiv k = \sqrt{\gamma^2+k_y^2}$ y $B_0$ un 
n\'umero complejo. Por tanto, escribiremos 
$B_0= |B_0| \exp (i \omega t_0)$, donde $t_0$ es el tiempo
de emisi\'on de la part\'icula, como fue definido en la 
Secci\'on \ref{dipolosplanewave}.

Consideremos ahora la trayectoria de la Fig.\ref{figtraydip3}. Las
part\'iculas son dispuestas de un lado ($x_{\rm i}$) de la gu\'ia, 
a un tiempo dado $t_0$. Se deja evolucionar el sistema y
 se observa el patr\'on de interferencia en alg\'un punto  $x_{\rm f}$. 
Como los campos
son nulos fuera de la gu\'ia de onda, las part\'iculas s\'olo
interact\'uan con el campo durante los trayectos rectos del recorrido.
\begin{figure}[!ht]
\center
\includegraphics[width=5cm]{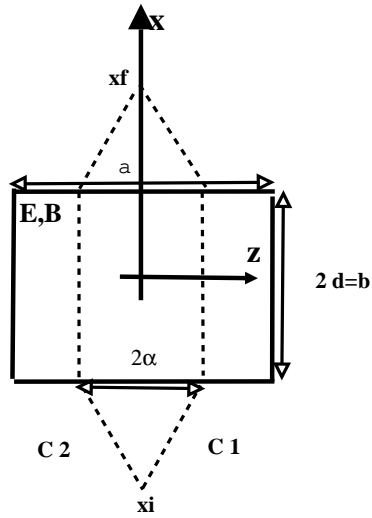}
\caption{ Las part\'iculas neutras con momento dipolar 
permanente interfieren en presencia de un campo electromagn\'etico dependiente
del tiempo. Las part\'iculas recorren los caminos 
$C_1$ y $C_2$ en el plano $\itm{x-z}$ dentro de una gu\'ia de ondas.}
\label{figtraydip3}
\end{figure}
Considerando $C^d_{\rm guia} = A^d_{\rm guia}
+ i B^d_{\rm guia}$, obtenemos para este caso
\beqa
A^d_{\rm guia} & = & 0 \nonumber \\
B^d_{\rm guia} & = & -\frac{ 4
}{( \omega T)^2-( m \pi)^2} \vert B_0\vert \frac{k_z}{\gamma^2}
\sin (\frac{l \pi \alpha}{a}) \bigg[ \omega T \cos(\frac{ m
\pi}{2}) \sin(\frac{\omega T}{2})
\nonumber \\
& - & m \pi \cos(\frac{\omega T}{2}) \sin(\frac{ m \pi}{2}) \bigg]
\bigg( d_x \omega T - m_z k_z T- 2 d_y k_y \bigg),
\label{te general}
 \eeqa
donde $2d=b$ es la distancia total del experimento dentro de la gu\'ia,
$2 \alpha$ es, nuevamente, la separaci\'on m\'axima de las part\'iculas,
$m,l$ son los modos de la gu\'ia y $T$ el tiempo de vuelo dentro
de la misma.

Para el primer modo de la gu\'ia, es decir $l=1,m=0$, esta expresi\'on
tiene un aspecto m\'as simple
 \beq C^d_{\rm {TE_{10}}}= -\frac{4 }{\omega T} \vert
B_0\vert \frac{a}{\pi} \sin(\frac{\pi \alpha}{a})
\sin(\frac{\omega T}{2})
 (d_x \omega T-m_z k_y T-b d_y k_y).
\eeq
Podemos estimar la cantidad $|C^d_{\rm {TE_{1,0}}}|$ 
de la misma manera
que lo hicimos en la secci\'on precedente. As\'i,
\beq |C^d_{\rm {TE_{10}}}| \approx
\frac{2\sqrt 2}{\pi} B_0~a~\sin(\frac{\pi \alpha}{a})
d_x,\label{Cd_guia10} \eeq
donde asumimos que $|\mbf{d}| >> |\mbf{m}|$. Este resultado
es independiente de la velocidad de los dipolos nuevamente. Si
realizamos la misma estimaci\'on pero para part\'iculas cargadas, obtenemos
\beq
|C^e_{\rm {TE_{10}}}|\approx \frac{1}{(2 \pi^3)^{1/2}} e B_0 a \lambda
v \sin \big(\frac{\alpha \pi}{a} \big)
.\eeq
Comparando ambos resultados, notamos que en el 
caso de las part\'iculas cargadas, la fase s\'i depende 
de la velocidad de las mismas.
Este hecho reduce la posibilidad de observar el efecto 
experimentalmente
ya que la velocidad de las part\'iculas es no relativista, por tanto
el efecto resulta de muy poca magnitud.

Volviendo al caso de los dipolos y la gu\'ia de onda, podemos estimar 
el factor $C^d_{\rm {TE}}$ para un modo arbitrario de la cavidad, es decir
$m \ne 0$ and $l \ne 0$. El resultado que obtenemos es
\beqa C^d_{\rm TE} &\approx& - \frac{4 B_0}{(\omega
T)^2-(m \pi)^2} \frac{k_z}{\gamma^2} \sin(\frac{l \pi \alpha}{a})
\bigg[ \omega T \cos(\frac{m \pi}{2}) \sin(\frac{ \omega T}{2}) -
m \pi \cos(\frac{\omega T}{2}) \sin(\frac{m \pi}{2})\bigg]
\nonumber\\
&&\bigg(d_x \omega T -m_z k_z T- b d_y k_y \bigg) \sin(\omega t_0).
 \eeqa
Es f\'acil ver que el resultado es despreciable en el caso de
$m$ impar y $m\pi \ll \omega T$. En cualquier otro caso,
obtenemos
\beq
|C_{\rm TE}^d| \approx B_0 a \vert\sin (\frac{\pi\alpha}{a})\vert \bigg[
d_x^2 + (\frac{b}{\omega T}d_y k_z)^2 - \frac{2 b}{\omega T} d_x d_y
k_y\bigg]^{1/2}, \label{TEgeneral} \eeq
que tiene la misma magnitud que el modo m\'as bajo de la cavidad $\rm {TE}_{10}$ de la Ec.(\ref{Cd_guia10}).

Por otro lado, para el modo TM, los campos son
\begin{eqnarray}
B_x & = & \frac{i\omega k_z}{\gamma^2} E_0 \cos(k_x x) \cos(k_z z)
\exp(i(k_y y - \omega t)), \nonumber \\
B_z & = & \frac{-i \omega k_x}{\gamma^2} E_0 \sin(k_z z) \cos( k_x x)
 \exp(i(k_y y - \omega t)), \nonumber \\
E_x & = & \frac{i k_y k_x}{\gamma^2} E_0 \cos(k_x x) \sin(k_z z)
\exp(i(k_y y - \omega t)), \nonumber \\
E_y & = & E_0 \sin(k_x x) \sin(k_z z)
\exp(i(k_y y - \omega t)), \nonumber \\
E_z & = & \frac{i k_y k_z}{\gamma^2} E_0 \sin(k_x x) \cos(k_z z)
\exp(i(k_y y - \omega t)).
\end{eqnarray}
donde $E_0= |E_0| \exp (i \omega t_0)$. Despu\'es de un poco de 
algebra, es posible mostrar que
\beqa C^d_{\rm TM}&=& \frac{ -4}{( \omega T)^2-( m
\pi)^2} \vert E_0\vert \sin (\frac{l \pi \alpha}{a}) \bigg[ k_x
\bigg( \omega T \cos(\frac{ m \pi}{2})
\sin(\frac{\omega T}{2}) 
- m \pi \cos(\frac{\omega T}{2}) \sin(\frac{ m \pi}{2}) \bigg)
\nonumber \\
&\times&\bigg( -d_x \frac{k_y}{\gamma^2} T + m_z \frac{\omega}{\gamma^2}T+
b d_y \frac{\omega}{\gamma^2} \bigg)+ 
\bigg( \omega T \cos(\frac{\omega T}{2}) \sin(\frac{m \pi}{2})
- m
\pi \cos(\frac{m \pi}{2}) \sin(\frac{\omega T}{2}) \bigg)
\nonumber \\
&\times& \bigg(-d_y
T-bm_z \bigg) \bigg] \sin(\omega t_0). \eeqa
Usando un razonamiento an\'alogo al modo TE, 
obtenemos una contribuci\'on de la fase, nuevamente independiente de la velocidad,
pero proporcional a $m_z$. Como hemos estado asumiendo que 
$|\mathbf{m}|<<|\mathbf{d}|$, esta contribuci\'on resulta 
$|C_{\rm TM}^d| \ll |C_{\rm TE}^d|$.
Finalmente, podemos mencionar que en el modo TM, no hay componente del
campo magn\'etico a lo largo del eje de la gu\'ia, y consecuentemente,
la fase AB es nula. Por tanto, la fase AC es la \'unica que esperamos encontrar. 

\subsection{Estimaciones num\'ericas}

Cuando comparamos los factores de p\'erdida de coherencia para 
part\'iculas cargadas y neutras, por ejemplo en la Ec.(\ref{comparacioncap5}),
asumimos que los par\'ametros relevantes del experimento 
(velocidad, separaci\'on m\'axima) se manten\'ian
iguales en ambos casos. Aunque te\'oricamente es una
 l\'inea de pensamiento v\'alida, resulta poco realista. 
Por tanto, en esta secci\'on,
consideraremos los valores reales utilizados en experimentos 
de interferometr\'ia con part\'iculas cargadas y neutras. En todos los
casos, asumiremos las dimensiones de la gu\'ia 
aproximadamente $a\sim\rm{cm}$.

Por un lado, en experimentos de interferometr\'ia con 
haces de electrones, los paquetes
 de onda pueden ser separados hasta una distancia 
de $100 \mu\rm{m}$ \cite{Hasselbach}. Una velocidad t\'ipica, no relativista es
$v_e \sim 0.1$, lo cual implica una relaci\'on $\omega T \sim 10$
para un campo de longitud de onda $\lambda \sim 100 \mu\rm{m}$.

Por otro lado, en interferometr\'ia con \'atomos neutros, la m\'axima
distancia de separaci\'on posible es $1~\rm{mm}$
\cite{Keith,Pfau}. Las velocidades de los \'atomos son 
aproximadamente $v_d \sim 10^{-5}$ \cite{Greenberger}.

La densidad de energ\'ia del campo electromagn\'etico se calcula seg\'un
$\rho_{\rm onda}= E_0^2/2$, para una onda plana. La densidad de energ\'ia
para el campo dentro de la gu\'ia de onda es $\rho_{\rm guia}= 
(a B_0)^2/\lambda^2$. Considerando los valores t\'ipicos de $\lambda$ y $a$
mencionados, podemos ver que $a/\lambda \sim 1$ para dipolos, mientras que 
para electrones se verifica $a/\lambda \sim
10$. Como estamos usando las unidades de Lorentz Heaviside 
$\hbar=c=1$, $\rho$ tambi\'en es el flujo de energ\'ia del campo electromagn\'etico.
Por tanto, asumiremos un flujo aproximado de $\rm{Watts}/\rm{cm}^2$.

Con todos estos valores experimentales, podemos estimar el factor
 $F=J_0(\vert C \vert)$, con los correspondientes $C$ 
calculados en la secci\'on precedente. Los resultados los presentamos
en la Tabla \ref{Tabla}.
\begin{table}[h!]
\center
\begin{tabular}{|c|c|c|}
  \hline
\textbf{Trayectorias} & \textbf{Electrones}& \textbf{Dipolos}\\
  \hline \hline
   $~~|C|~~~ $ & $1$  &  $ 10^{-6} $\\ \hline
  $|C_{\rm ellip}|~$& $ 10$  & $
  10^{-3}$ \\
  \hline
   $|C_{\rm asym}|~$& $10$ &
    $10^{-1}$\\
  \hline
  $|C_{\rm {TE_{1,0}}}|$ & $1$ &
  $10^{-1}$ \\
  \hline
\end{tabular}
\caption{ \'Orden de magnitud del valor absoluto del factor $C$
para todas las trayectorias estudiadas, tanto para electrones como
dipolos.} \label{Tabla}
\end{table}
Como se puede observar, los resultados para los electrones son 
de orden uno o mayor, lo cual significa que el efecto es 
experimentalmente observable. Por el contrario, los resultados 
para los dipolos son mucho m\'as chicos en algunos casos, 
mientras que en otros son s\'olo un orden menor.
En el caso de electrones, hemos demostrado que es posible
obtener una destrucci\'on completa del patr\'on de interferencia, fijando
el valor de $C$ en un cero de la funci\'on de Bessel. Por otro lado,
en un experimento de interferencia con \'atomos neutros, el mejor
montaje experimental resulta la trayectoria asim\'etrica, ya que el
factor $C$ no depende de la velocidad de la part\'icula y podr\'ia ser
no despreciable.
Sin embargo, podr\'iamos considerar otro tipo de part\'iculas neutras. 
Las mol\'eculas  $C_{60}$ y $C_{70}$ (fulerenos) han sido 
utilizadas para experimentos de interferometr\'ia. Estos 
sistemas mesosc\'opicos \cite{Facchi,Brezger} est\'an 
compuestos de un n\'umero grande de \'atomos,
pero a\'un as\'i, pueden ser descriptos por una funci\'on de onda.
En este contexto, podemos estimar el factor de decoherencia $F$
utilizando los valores experimentales de dichos sistemas. A pesar de que
puede ser considerado un modelo muy sencillo para estas mol\'eculas, nos
puede dar una estimaci\'on cuantitativa de la magnitud de este efecto
en sistemas neutros muy masivos. Estas mol\'eculas tienen una velocidad similar a los
\'atomos neutros que hemos considerado, pero viajan 
distancias m\'as grandes. Por tanto, debemos considerar 
una longitud de onda m\'as grande,
por lo menos del mismo orden de la distancia total que recorren los 
fulerenos. Adem\'as, debemos resaltar que en el 
experimento de Hornberger
{\it et al.} \cite{Hornberger:2003}, el laser que se utiliz\'o era 
de potencia $26\rm{Watts}/\rm{cm}^2$ (del orden de nuestra 
estimaci\'on). Con estos valores, y considerando la trayectoria 
asim\'etrica, el
factor $C$ para los fulerenos es $|C^{\rm fulerenos}_{\rm
asim}| \approx 1$, indicando que la atenuaci\'on o destrucci\'on 
 del patr\'on
de interferencia puede ser observable. De hecho, 
la atenuaci\'on de dicho
patr\'on fue observada en \cite{Hornberger:2003}.

\subsection{Aplicaci\'on: p\'erdida de visibilidad en un experimento
de dos rendijas}

En esta secci\'on estudiaremos la disminuci\'on de la visibilidad
de las franjas de interferencia en experimentos hom\'onimos
con part\'iculas, ya sea \'atomos neutros fr\'ios y fulerenos.
La pregunta que nos gustar\'ia responder es \textquestiondown
c\'omo afecta la decoherencia o el {\it dephasing} 
al patr\'on de interferencia 
en estos experimentos ? Es decir, \textquestiondown c\'omo
se reduce la visibilidad en estos experimentos y en qu\'e escala
caracter\'istica?

Por lo general, el principal problema que tienen los experimentos
de interferencia con part\'iculas es determinar las verdaderas
razones por las cuales hay p\'erdida de coherencia espacial. 
Este hecho se manifiesta en una reducci\'on de la visibilidad de las franjas de interferencia;
es decir, una disminuci\'on del contraste entre m\'aximos y m\'inimos.
En principio, estos efectos podr\'ian explicarse debido a la 
interacci\'on de las part\'iculas con las rendijas, generando vibraciones 
o interacciones de Van der Waals \cite{Grisent}; o bien, en 
la diferencia de tama\~no de las rendijas \cite{Bozic}.
En este trabajo no consideraremos dichos efectos porque, en general, y
bajo condiciones experimentales apropiadas, resultan despreciables. 
Por tanto, asumiremos que los haces de part\'iculas
son coherentes y su di\'ametro mucho menor que el ancho de las rendijas. 
Es posible encontrar factores que afectan la coherencia espacial de 
las part\'iculas durante su trayecto hacia la pantalla.
Estos efectos din\'amicos {\it decoherentes} se los puede atribuir a
las colisiones  con las mol\'eculas de aire y fotones t\'ermicos.
La no monocromacidad del haz y la aleatoridad en su emisi\'on 
(debido a la dificultad experimental de producir el mismo estado 
inicial para todos los haces de part\'iculas)
tambi\'en pueden originar una p\'erdida de coherencia espacial a 
trav\'es del c\'aracter fluctuante de la fase, es decir el {\it dephasing} 
estuadiado anteriormente en este cap\'itulo.

En lo que sigue estudiaremos num\'ericamente el patr\'on de 
intererencia entre 
part\'iculas de masa M, que pasan por dos rendijas separadas 
una distancia $2 L_0$ en la direcci\'on $\hat x$. El patr\'on es 
observado
en una pantalla distante, ubicada en la direcci\'on $\hat y$. 
La part\'iculas pasan a trav\'es de las rendijas
y viajan una distancia $L$  hasta alcanzar la pantalla en un tiempo
de vuelo $t_L=ML/p_0$, donde $p_0$ es el momento inicial en la
direcci\'on $\hat y$. Asumiremos que las part\'iculas son coherentes
en la direcci\'on $\hat x$; mientras que la din\'amica en la direcci\'on 
$\hat y$ ser\'a la de una part\'icula libre viajando
una distancia $L$ en un tiempo de vuelo $t_L$.

El experimento comienza con la preparaci\'on de un estado inicial, el
cual corresponde a una superposici\'on coherente de dos paquetes de 
ondas, cada uno centrado en cada una de las rendijas, factorizado
de la siguiente manera \cite{Venugopalan,Viale}
$\psi_S(\vec{x},0)= (\phi_1(x,0) + \phi_2(x,0)) \otimes \chi(y,0)$,
donde $|\phi_1|^2$ y $|\phi_2|^2$ corresponden a la amplitud 
de probabilidad de la part\'icula de pasar por la rendija 1 y por la
rendija 2, respectivamente; mientras que $\chi(y,t)$ es una
funci\'on de onda gaussiana en la direcci\'on $\hat y$. Estamos
suponiendo, adem\'as, invariancia traslacional en $\hat z$.

Cuando el sistema es abierto, la evoluci\'on est\'a plagada de 
caracter\'isticas no unitarias, como fluctuaci\'on y disipaci\'on;
independientemente de cu\'an chico sea el acoplamiento de \'este a su
entorno. Asumiremos que, inicialmente, el estado total del sistema, 
part\'icula + entorno, se factoriza de la siguiente manera
\begin{equation}
\Psi(\vec{x},\vec{X},0)= [\phi_1(x,0)
+\phi_2(x,0)]\otimes \chi(y,0) \otimes
\zeta(\vec{X},0),
\end{equation}
donde hemos introducido la funci\'on $\zeta(\vec{X},t)$ 
para describir el estado  del entorno.

El patr\'on de interferencia observado en la pantalla a un 
tiempo dado $t$, corresponde a la distribuci\'on de probabilidad 
asociada a la funci\'on de onda a dicho tiempo, definida por
\beq P(\vec{x},t)= \rho_r(x,x,t)|\chi(y,t)|^2
=\bigg(|\phi_1(x,t)|^2+ |\phi_2(x,t)|^2 
+ 2 { \Gamma}(t){\rm Re} (\phi_1^*(x,t) \phi_2(x,t))\bigg)
|\chi(y,t)|^2
\label{gamma} \eeq
donde hemos escrito expl\'icitamente la relaci\'on entre el patr\'on
de interferencia y la matriz densidad reducida del sistema. 
En esta expresi\'on podemos
notar la aparici\'on del factor de solapamiento o de decoherencia, 
mencionado en varias secciones de esta Tesis. $\Gamma(t)$ contiene
la informaci\'on del entorno, independientemente de la forma que se haya obtenido.

Para especificar el efecto del entorno debemos decidir qu\'e 
camino seguir. Podr\'iamos suponer que el entorno est\'a 
modelado por un
n\'umero grande de osciladores arm\'onicos y resolver la 
ecuaci\'on maestra Ec.(\ref{master}) con los coeficientes 
correspondientes a la temperatura del entorno. 
De esta forma, obtendr\'iamos la $\rho_r(\vec{x},t)$ y de ah\'i 
podr\'iamos conocer el patr\'on de interferencia a cualquier
 tiempo (Ec.(\ref{gamma})). As\'i estar\'iamos estudiando, 
el efecto {\it decoherente} din\'amico que produce la 
interacci\'on de las part\'iculas con las paredes de las
 rendijas al pasar a trav\'es de ellas; $\Gamma(t)$ ser\'ia 
simplemente $\Gamma(t)=\exp(-{\cal D}t)$ con ${\cal D}= 2M \gamma_0 k_B T$ 
\footnote{Los experimentos realizados a temperatura ambiente 
con fulerenos y \'atomos fr\'ios satisfacen la condici\'on de  
temperatura alta.}. Otra alternativa, ser\'ia pensar que la 
reducci\'on de la visibilidad del patr\'on de interferencia 
se debe a las colisiones de las part\'iculas con las mol\'eculas 
de aire durante su tiempo de vuelo. En ese caso, la 
ecuaci\'on din\'amica que habr\'ia que resolver ser\'ia una 
ecuaci\'on maestra markoviana \cite{Walls85,Diosi95} del tipo 
\beq i
\frac{\partial \rho_r}{\partial t} = [H,\rho_r] - i \Lambda
[x,[x,\rho_r]].\label{scatering} \eeq
El efecto del entorno est\'a resumido en el factor $\Lambda$
adicionado a la evoluci\'on libre del sistema. Este modelo considera
los efectos difusivos generados por $\Lambda$ pero desprecia 
la disipaci\'on \cite{Vacchini00,Vacchini01}. Por ejemplo, 
en \cite{Viale}, los autores consideraron $\Lambda=\Lambda_{\rm aire} +
\Lambda_{\rm fotones}$. De esta forma, 
el efecto del entorno est\'a contenido en $\Gamma(t)=\exp(-\Lambda t)$
 y, por lo general, $\Lambda$ es estimada fenomenol\'ogicamente
 a partir de la longitud de onda de las part\'iculas y 
la secci\'on eficaz de las mismas. En este contexto, 
otros modelos de decoherencia, como por ejemplo 
el de Hornberger, Sipe y Arndt \cite{Horn}, se pueden aplicar.
Finalmente, otra alternativa es considerar un modelo de 
{\it dephasing} como el desarrollado en este cap\'itulo. 
La interacci\'on entre las part\'iculas y el campo 
electromagn\'etico cl\'asico dependiente del tiempo 
induce una fase de Aharonov fluctuante
\cite{Casher,CasherJPA}. En este caso, $\Gamma=F=
\langle e^{i\phi}\rangle$ con $F$ y $\phi$ definidas en la Ec.(\ref{promediophi}).

En lo que sigue, s\'olo nos limitaremos a desarrollar 
el modelo de {\it dephasing} y analizar num\'ericamente 
la visibilidad de su 
patr\'on de interferencia en funci\'on del tiempo. Los otros 
casos pueden ser le\'idos en detalle en \cite{fringeModern}. Por \'ultimo,
compararemos nuestros resultados num\'ericos 
con los datos de un experimento de interferencia con \'atomos fr\'ios y fulerenos.

\subsubsection{An\'alisis num\'erico del patr\'on de interferencia}

Para reproducir el patr\'on de interferencia de un experimento
verdadero, deberemos ser un poco m\'as precisos al 
escribir la funci\'on de onda del sistema. En particular, 
a un tiempo inicial, 
tendremos, para el sistema cu\'antico
\beq
\psi(\vec{x},0)= N \bigg( \exp(-\frac{(x-L_0)^2}{4 \sigma_{x0}^2})
+ \exp(-\frac{(x+L_0)^2}{4 \sigma_{x0}^2}) \bigg)
   \exp(-\frac{y^2} {4 \sigma_{y0}^2}-i k_y y) \label{2gauscap5}
\eeq
donde $2 L_0$ es la separaci\'on inicial de los centros de los
paquetes gaussianos, $\sigma_{x0}$, $\sigma_{y0}$ y $k_y$
son par\'ametros libres del modelo que deben ser ajustados 
seg\'un los
datos experimentales. Adem\'as, asumiremos que 
$\Delta p_y << p_y$, de modo que el momento en esa 
direcci\'on est\'a bien definido y
el paquete de onda tiene una longitud de onda asociada 
$\lambda_{dB}$ tal que $\lambda_{dB}
\sim \hbar/p_y << \Delta y$.

La matriz densidad correspondiente a la funci\'on de onda 
de la Ec.(\ref{2gauscap5}). es
\beqa \rho_r(\vec{x},\vec{x}',0)&=&
\rho_r(x,x',0) \otimes  \rho_r(y,y',0) \nonumber \\
&=&  2 N^2 \bigg(\cosh(2 L_0 (x+x')) + 
 \cosh(2 L_0 (x-x')) \bigg)\chi(y,0)^*\chi(y',0). \nonumber \eeqa
La evoluci\'on din\'amica de esta matriz puede ser 
reproducida utilizando una matriz densidad gaussiana en la
 aproximaci\'on de Born \cite{Viale,Joos} (que no implica que estemos mirando
la interferencia de campo lejano o Fraunhofer).
De esta forma, la intensidad que se registra en la 
pantalla a un tiempo posterior $t$, es
\beq P(x,t)= e^{-\tilde{N}(t)} e^{-4C(t)
(x^2-L_0^2)}\bigg(\cosh(8 C(t)
L_0 x) 
+ \Gamma(t) \cos(4 B(t) L_0 x) \bigg), \label{prob} \eeq
donde hemos absorbido el t\'ermino gaussiano $|\chi(y,t)|^2$
de la direcci\'on $\hat y$ en la normalizaci\'on $e^{-\tilde{N}(t)}$.

Simularemos el experimento de interferometr\'ia 
de la siguiente manera. Haremos evolucionar la matriz densidad inicial 
 seg\'un el m\'etodo num\'erico detallado 
en el Ap\'endice A. La \'unica diferencia es que fijaremos 
$\gamma_0=0$, de modo de eliminar la contribuci\'on del 
entorno, ya que \'esta est\'a contenida en el factor $F$ que 
calculamos con anterioridad para
neutrones (en la Tabla \ref{Tabla}) y fulerenos. De esta forma,
 tendremos acceso a la matriz densidad del sistema, o mejor dicho, 
a los par\'ametros $A(t),~B(t),~C(t)$, a todo tiempo.
Cuantificaremos la p\'erdida de contraste del patr\'on de interferencia 
definiendo una funci\'on llamada visibilidad $\nu$,
de particular importancia a nuestros fines,
\[ \nu=\frac{ I_{\rm max}-I_{\rm min}}
{I_{\rm max}+I_{\rm min}},\]
donde $ I_{\rm max}$ y $I_{\rm min}$ representan los m\'aximos
y m\'inimos vecinos de interferencia,  respectivamente. Es f\'acil 
ver que, en funci\'on de la matriz densidad, podemos escribir esta
 cantidad aproximando
\[ \nu(t) 
\sim \frac{|\rho_{\rm int}(x,x,t)|}{\rho_{11}(x,x,t)
+ \rho_{22}(x,x,t)}, \]
donde $\rho_{ii}=|\phi_i(x,t)|^2$, con
$i=1,2$ y $\rho_{\rm{int}}$ contiene los t\'erminos de interferencia.
El valor de esta funci\'on var\'ia desde 0 (no hay franjas) hasta 1
(visibilidad total de las franjas). En nuestro modelo de {\it dephasing},
 la funci\'on visibilidad obtenida num\'ericamente es:
\[ \nu_C(t) \approx \frac{J_0(\vert C \vert)}{\cosh(8 L_0 C(t) x)}.
\]
Recordemos que el factor $\Gamma=F$ es constante en el tiempo a 
diferencia del factor de decoherencia que se hubiera obtenido al 
considerar otro modelo, por ejemplo $\Gamma(t)=\exp(-{\cal D} t)$ o
$\Gamma(t)=\exp(-\Lambda t)$. En la Fig.\ref{visibilityc} mostramos
nuestros resultados.
\begin{figure}[!ht]
\center
\includegraphics[width=17cm]{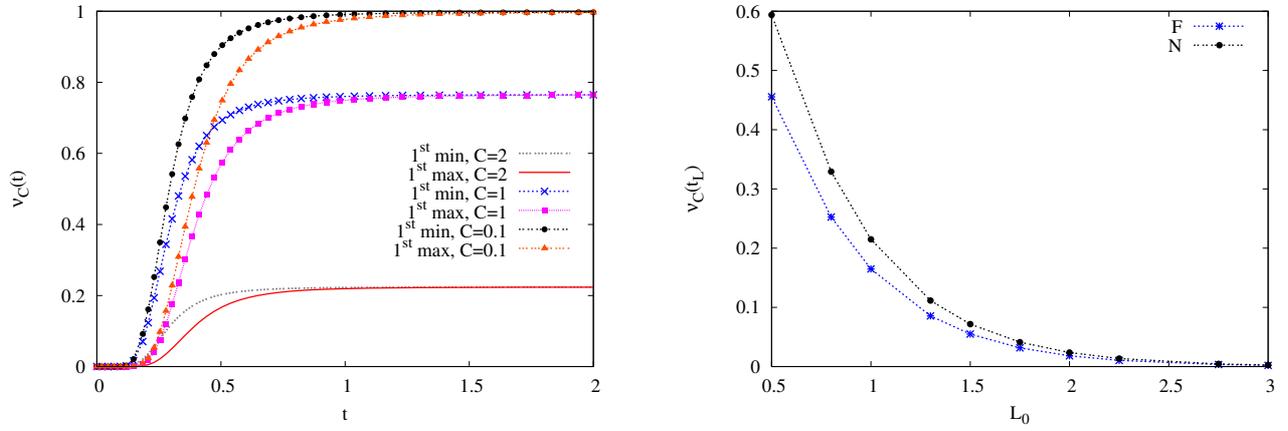}
\caption{{\it Izquierda}: Evoluci\'on temporal de la funci\'on visibilidad 
$\nu_C(t)$ para neutrones ($C_{\rm neutrons}=0.1$) y
fulerenos ($C_{\rm
fullerenes}=1$ y $C_{\rm
fullerenes}=2$) en presencia de un campo electromagn\'etico cl\'asico 
dependiente del tiempo. Las curvas son para el primer m\'aximo y
 primer m\'inimo del patr\'on. Vemos que todas las curvas alcanzan 
un valor asint\'otico distinto de cero. {\it Derecha}: La funci\'on 
visibilidad $\nu_C(t)$ como funci\'on de la separaci\'on de las rendijas 
$L_0$ para un tiempo fijo $t_L=0.2$ s con $\sigma_{x0}=0.5~s^{-1}$. 
Las curvas son para el primer m\'aximo de interferencia  en el caso de
 neutrones y fulerenos. $L_0$ est\'a medido en unidades de $s^{-1}$.}
 \label{visibilityc}
\end{figure}
En el modelo que aqu\'i estamos estudiando, la interacci\'on entre
las part\'iculas neutras con momento dipolar permanente y el campo
electromagn\'etico cl\'asico est\'a siempre presente y no se puede
``apagar''. Sin embargo, como hemos demostrado con anterioridad 
en el caso de los neutrones, el efecto es muy peque\~no, totalmente 
despreciable  de realizar un experimento. Pero, resulta
 inesperadamente relevante a la hora para experimentos 
part\'iculas masivas como, por ejemplo, los fulerenos $C_{60}$ o 
$C_{70}$, ya que $C_F \sim {\cal O}(1)$.
\begin{figure}[!ht]
\center
\includegraphics[width=10cm]{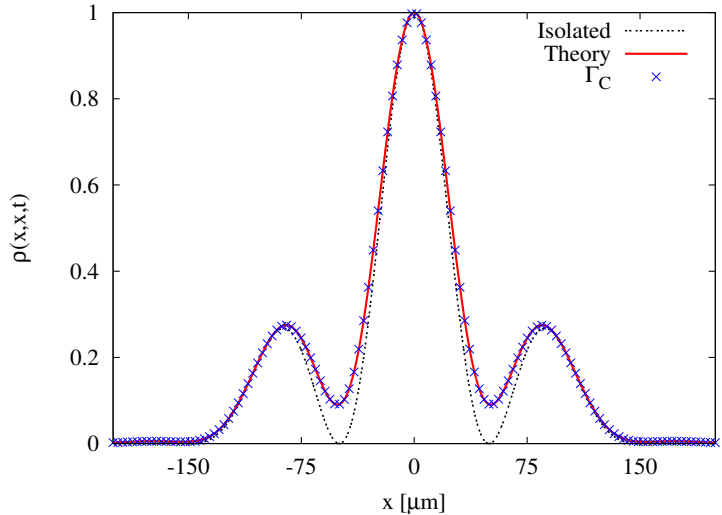}
\caption{El patr\'on de interferencia, con una visibilidad 
$\nu_C \sim 0.62$, que se observar\'ia en una pantalla, 
tras considerar el modelo de {\it dephasing} para un 
experimento de interferencia entre mol\'eculas masivas. 
Las curvas graficadas representan el caso aislado, la 
predicci\'on te\'orica utilizando los datos experimentales 
reportados en \cite{Hornberger:2003} y los resultados 
nuestros utilizando el modelo de {\it dephasing} con $C=1$.}
 \label{patronvisc}
\end{figure}

Por \'ultimo, aplicaremos nuestro modelo de {\it dephasing} a los
datos experimentales reportados en \cite{Hornberger:2003} durante
la observaci\'on del patr\'on de interferencia de fulerenos $C_{70}$.
El resultado se presenta en la Fig.\ref{patronvisc}. Para estas 
part\'iculas masivas, podemos observar que el patr\'on de interferencia 
se aten\'ua cuando el sistema es abierto. Los efectos incoherentes 
contenidos en el factor de solapamiento $\Gamma_C$ son suficientes 
para reproducir los efectos del entorno en el patr\'on de 
interferencia de un experimento verdadero con fulerenos. 
La \mbox{atenuaci\'on} de dicho patr\'on fue observada en \cite{Hornberger:2003}, 
en un experimento donde el campo cl\'asico de nuestro modelo 
puede representar el laser utlizado en dicho experimento.

\section{Fases geom\'etricas en el modelo esp\'in-bos\'on}

Como hemos mencionado antes, los sistemas cu\'anticos 
tienen la capacidad de retener la 
informaci\'on de sus movimientos en el espacio de Hilbert a 
trav\'es de la adquisici\'on de una fase geom\'etrica. Estos 
factores de fase dependen s\'olo de la geometr\'ia del camino 
elegido por el sistema durante  la evoluci\'on y no del camino mismo.
La fase de Berry es un fen\'omeno cu\'antico relacionado a 
la evoluci\'on adiab\'atica de un sistema como hemos 
mencionado al principio de este cap\'itulo. Recientemente, 
se ha sugerido que debiera ser posible observar la fase de 
Berry (FB) en una estructura superconductora \cite{FalciNature}, 
y posiblemente, usarla para controlar la evoluci\'on de un 
estado cu\'antico \cite{Jones}. Sin embargo, esta afirmaci\'on 
 no tiene en cuenta el
acoplamiento del sistema cu\'antico al entorno, el cual, en 
dichas estructuras, no resultan despreciables \cite{Nakamura}.  
La presencia del entorno y su
injerencia en la FB tambi\'en resulta de gran importancia  en 
experimentos que
intentan manipular sistemas de dos niveles, o {\it qubits} \cite{Jones}.

En este contexto, estudiaremos la fase de Berry de un sistema
que no est\'a completamente aislado. Los sistemas reales, 
est\'an acoplados a un entorno y esta interacci\'on se hace 
evidente cuando se estudia el 
espectro de energ\'ias del sistema, ya que se vuelve 
un espectro cont\'inuo.  Originalmente, la FB se defini\'o
 para sistemas cuyos estados estaban separados por un intervalo 
finito de energ\'ia. Al pasar de un espectro discreto a uno cont\'inuo,
 uno esperar\'ia que los par\'ametros del Hamiltoniano no pudieran
 ser variados de forma suficientemente lenta como para ser 
considerada una evoluci\'on adiab\'atica. Esto significa que no 
ser\'ia posible observar una FB en ese sistema, o de forma m\'as 
general, en ning\'un sistema de la Naturaleza.
Este argumento resulta demasiado simplista ya que se han realizado 
experimentos que observaron la FB, ya sea de forma directa o 
indirecta \cite{Ao, Zhu}. Consideraremos un sistema cu\'antico sencillo, 
de dos niveles, que exhibe FB y lo acoplaremos a un entorno, de 
manera bilineal, el cual est\'a compuesto por un n\'umero muy grande 
de osciladores arm\'onicos. Intentaremos responder dos preguntas. En primer 
lugar,  \textquestiondown bajo qu\'e condiciones la FB puede ser 
observada? Y en segundo lugar,  \textquestiondown si la fase que se 
observa es igual a la FB del sistema aislado? Si no lo es, 
\textquestiondown es \'esta de naturaleza
geom\'etrica?

Consideraremos sistema a aquellos grados de libertad de 
los cuales podemos tener \mbox{control}. De esta manera, podemos
 preparar el sistema en el estado que deseamos y elegir los
 par\'ametros del Hamiltoniano bajo los cuales queremos que 
evolucione. El entorno consiste de  todos los grados de libertad 
a los cuales  no tenemos acceso experimental. 
Generalmente, lo m\'as eficiente que se puede lograr 
es asegurar que el universo, sistema + entorno, est\'e en 
equilibrio t\'ermico a temperatura T. Si, adem\'as, se logra 
bajar la temperatura T a cero,  entonces podr\'iamos preparar 
al universo en su estado
fundamental. 

La idea original de Berry, que consideraba una evoluci\'on 
adiab\'atica y c\'iclica, fue \mbox{generalizada} en varios aspectos. 
Mientras que la mayor\'ia de ellas han sido para estados 
puros, algunas pocas se han concentrado en analizar esta 
fase para el caso de estados mixtos, probablemente motivadas 
en la implementaci\'on de {\it qubits}. En este contexto, 
en \cite{Sjoqvist} los autores
presentaron una definici\'on alternativa de las fases geom\'etricas
 (FG) para operadores densidad no degenerados, basados en
 interferometr\'ia cu\'antica.
Por otro lado, en \cite{Singh} se defini\'o la FG para un estado
 mixto, pero desde un punto de vista cinem\'atico. Adem\'as, 
se extendi\'o el estudio al caso de operadores densidad 
degenerados. Ambos enfoques, el cinem\'atico y el interferom\'etrico, 
son equivalentes, lo cual fue demostrado recientemente 
en \cite{Zanardi}. 
Finalmente, una buena generalizaci\'on de las FGs al 
caso de una evoluci\'on no unitaria para estados mixtos 
fue realizada  en \cite{Tong}, lo cual puede ser
 fundamental  para la evaluar la solidez de la llamada 
Computaci\'on Cu\'antica Geom\'etrica. All\'i, se
propuso una representaci\'on funcional de la fase 
geom\'etrica, la cual obtuvieron tras la sustracci\'on de la
 fase din\'amica de la fase total adquirida por el sistema 
a trav\'es de una transformaci\'on de gauge.

En este contexto, la FG de un estado mixto que 
evoluciona de manera no unitaria fue definida 
como \cite{Tong} :
\begin{equation} \Phi =
{\rm arg}\{\sum_k \sqrt{ \varepsilon_k (0) \varepsilon_k (\tau)}
\langle\Psi_k(0)|\Psi_k(\tau)\rangle  e^{-\int_0^{\tau} dt \langle\Psi_k|
\frac{\partial}{\partial t}| {\Psi_k}\rangle}\}, \label{fasegeo}
\end{equation}
donde  $\varepsilon_k(t)$ son los autovalores y 
 $|\Psi_k\rangle$ los autovectores de la matriz densidad reducida del sistema
$\rho_{\rm r}$. En la expresi\'on Ec.(\ref{fasegeo}), $\tau$ denota el tiempo
en el cual el sistema aislado completa un ciclo de la evoluci\'on. 
Cuando se considera el efecto del entorno sobre el sistema 
cu\'antico, \'este ya no sigue una evoluci\'on c\'iclica. 
No obstante, en adelante consideraremos una evoluci\'on 
cuasi-c\'iclica, es decir ${\cal P}:t ~\epsilon~[0,\tau]$  con 
$\tau=2 \pi/\Omega$ ($\Omega$ es la frecuencia del sistema) 
\cite{Tong}. Hay que destacar que la fase de la Ec.(\ref{fasegeo})
 es invariante de gauge y que corresponde a la fase geom\'etrica
 unitaria en el caso que el estado es puro y el sistema cerrado  
\cite{Sjoqvist, Singh}.

El conocimiento de las FGs nos hace pensar que \'estas pueden
 ser observadas en experimentos que se llevan a cabo en una 
escala caracter\'istica suficientemente ``lenta" donde las 
correcciones no adiab\'aticas de la evoluci\'on pueden ser ignoradas, 
pero, a su vez, lo suficientemente ``r\'apida" para evitar que el entorno
 destruya completamente el patr\'on de interferencia \cite{Gefen}. 
Hasta ahora, no ha habido evidencia experimental de la existencia 
de FGs para estados mixtos durante evoluciones no unitarias.
 Lo que haremos en esta secci\'on es analizar c\'omo se ven 
afectadas las FGs debido al proceso de p\'erdida de coherencia. 
\mbox{Veremos} cu\'an robustas 
resultan frente a los efectos difusivos y tambi\'en, bajo qu\'e 
condiciones pueden ser observadas.
En este contexto, presentaremos un modelo esp\'in-bos\'on sencillo 
y calcularemos las correcciones a la fase geom\'etrica unitaria 
debido a diferentes tipos de entornos. Finalmente, estimaremos
 los tiempos de p\'erdida de coherencia a partir de los cuales las 
FGs ya no resultan observables
porque, literalmente, desaparecen.

En lo que sigue, estudiaremos un sistema de dos niveles, 
o {\it qubit}, acoplado
a un ba\~no t\'ermico.  La principal caracter\'istica de este modelo, 
es que resulta resoluble de manera \mbox{exacta} y la p\'erdida de 
coherencia es el \'unico efecto difusivo inducido en el sistema, 
es decir, no hay disipaci\'on. Este caso, es una situaci\'on particular 
del esp\'in-bos\'on de Legget \cite{legget} 
(donde el t\'ermino relacionado al efecto t\'unel cu\'antico es
$\Delta=0$) y ha sido ampliamente usado para estudiar 
la p\'erdida de coherencia en Computaci\'on Cu\'antica \cite{Eckert}. 
Particularmente, este modelo resulta  relevante 
en las propuestas de observaci\'on FGs en nanocircuitos 
superconductores \cite{Falci}.
Nos concentraremos en entornos \'ohmicos y supra\'ohmicos, 
a temperatura alta, baja y cero.

El Hamiltoniano que describe la acci\'on completa del sistema 
de dos niveles  con el entorno es
\begin{equation}
H_{\rm SB}=  \frac{1}{2} \hbar \Omega
 \sigma_z +\frac{1}{2} \sigma_z
\sum_{k} \lambda_{k} (g_k a^{\dagger}_{k}+g_k^*a_{k}) + 
\sum_{k} \hbar \omega_{k}
a_{k}^{\dag} a_{k}, \label{HSB}
\end{equation}
donde el entorno est\'a descripto en funci\'on de un conjunto 
de osciladores arm\'onicos acoplados linealmente al sistema 
de dos niveles, a trav\'es de la coordenada de posici\'on de 
estos \'ultimos. El Hamiltoniano de interacci\'on, proporcional a
 $\sigma_z$, indica que el estado del entorno es sensible s\'olo 
a los valores de $\sigma_z$; es decir, observa \'unicamente si 
el sistema est\'a en $|\uparrow>$ o
$|\downarrow>$. La raz\'on para considerar este acoplamiento 
es que los efectos de un acoplamiento proporcional a 
$\sigma_x$ y/o $\sigma_y$ resultan importantes para determinar 
la renormalizaci\'on de la frecuencia natural del sistema.
Sin embargo, en este caso $[\sigma_z,H_{\rm int}]=0$ y por tanto,
 la ecuaci\'on maestra correspondiente se simplifica bastante, ya 
que no hay efectos ni de renormalizaci\'on ni disipaci\'on. Adem\'as, 
si $[\sigma_z,H_{\rm int}]=0$, las poblaciones de los t\'erminos de 
la diagonal se mantienen constante, es decir
\begin{eqnarray}
 \rho_{11}=\rm Tr_{\cal E} {\vert 1 \rangle \langle 1 \vert \rho(t)}= 
\langle 1 \vert \rho_r(t) \vert 1 \rangle; \nonumber \\
\rho_{00}=\rm Tr_{\cal E} {\vert 0 \rangle \langle 0 \vert \rho(t)}= 
\langle 0 \vert \rho_r(t) \vert 0 \rangle, \nonumber 
\end{eqnarray}
donde $\rho(t)$ es la matriz densidad del sistema total. Antes de 
determinar la ecuaci\'on maestra de este sistema, podemos notar
 que el Hamiltoniano de interacci\'on en la hom\'onima representaci\'on es
\begin{equation}
 H_I(t)=e^{i H_0 t} H_I e^{-i H_0 t}= \sum_k
\sigma_z \bigg(g_k a^{\dagger}_{k}e^{i \omega_k t}+
g_k^*a_{k}e^{-i \omega_k t}\bigg),
\end{equation}
y el operador de evoluci\'on temporal unitario en la misma 
representaci\'on
\begin{equation}
 U(t)=T_{\leftarrow} \exp\bigg(-i \int_0^t ds H_I(t) \bigg).
\end{equation}
Como el conmutador del Hamiltoniano de interacci\'on a 
dos tiempos distintos es una funci\'on del tipo,
\begin{equation}
 [H_I(t),H_I(t')]=-2 i \sum_k \vert g_k \vert^2 
\sin(\omega_k(t-t')) \equiv -2 i \varphi (t-t'),
\end{equation}
obtenemos
\begin{eqnarray}
 U(t)&=&\exp\bigg( -\frac{1}{2} \int_0^t ds \int_0^t ds' [H_I(s),H_I(s')] 
\theta(s-s') \bigg) \exp \bigg(-i \int_0^t ds H_I(s) \bigg) \nonumber \\
& & \exp \bigg(i \int_0^t ds \int_0^t ds' \varphi(s-s') \theta(s-s') \bigg) V(t),
\end{eqnarray}
donde hemos definido al operador unitario $V(t)$ como
\begin{equation}
 V(t)= \exp \bigg(\frac{1}{2} \sigma_z \sum_k  (\alpha_k a^{\dagger}_{k}
+\alpha_k^*a_{k}) \bigg), \label{1cap5}
\end{equation}
cuyas amplitudes $\alpha_k= 2 g_k\frac{1-e^{i \omega_k t}}{\omega_k}$.
Con \'esto, podemos conocer el comportamiento en el tiempo 
de las coherencias de la matriz densidad del sistema asumiendo 
que a tiempo cero el sistema y el entorno no est\'an 
correlacionados, es decir $\rho(0)=\rho_r(0) \otimes \rho_B$,
 con $\rho_B$ un estado t\'ermico de equilibrio,
\begin{equation}
 \rho_{r_{ij}}(t)=\langle i \vert \rho_r(t) \vert j \rangle =
\langle i \vert \rm Tr_{\cal E}{V(t) \rho(0) V^{-1}(t)}\vert j \rangle,
\label{2cap5}
\end{equation}
es decir,
\begin{equation}
 \rho_{r_{10}}(t)=\rho_{r_{01}}^*(t)=\rho_{r_{10}}(0) e^{{\cal A}(t)},
\end{equation}
donde hemos elegido el nombre $\cal A$ en analog\'ia al coeficiente
de visibilidad $A_{\rm int}$ del Cap\'itulo \ref{c2}.
El coeficiente ${\cal A}(t)$ lo podemos conocer con la ayuda 
de las Ecs.(\ref{1cap5}) y (\ref{2cap5}),
\begin{equation}
 {\cal A}(t)=\ln {\rm Tr_B} \bigg\{\exp\bigg(\sum_k  (\alpha_k 
a^{\dagger}_{k}+\alpha_k^*a_{k}) \rho_B \bigg) \bigg\}= \sum_k
 \ln \bigg\langle \exp(\alpha_k a^{\dagger}_{k}+\alpha_k^*a_{k})
\bigg \rangle .
\end{equation}
Los corchetes angulares denotan el valor de expectaci\'on con
 respecto al estado inicial del entorno. Esta expresi\'on puede 
ser f\'acilmente escrita como
\begin{equation}
 \chi(\alpha_k,\alpha_k^*) \equiv \bigg\langle \exp(\alpha_k
a^{\dagger}_{k}+\alpha_k^*a_{k})
\bigg \rangle \equiv \exp \bigg\{-\frac{1}{2} \vert \alpha_k\vert ^2 
\bigg\langle \{a_k,a_k^{\dagger}\} \bigg \rangle\bigg\}
\end{equation}
Si, ahora, hacemos el pasaje a una densidad espectral cont\'inua $I(\omega)$,
 obtenemos la funci\'on de decoherencia,
definida seg\'un:
\begin{equation}
{\cal A}(t)=-\int_0^{\infty} d\omega I(\omega) \coth\bigg(\frac{\omega}{2 k_B T}\bigg)
\frac{(1-\cos(\omega t))}{\omega^2}.\label{Gcap5}
\end{equation}
De esta manera, hemos obtenido la expresi\'on expl\'icita de la funci\'on
de decoherencia en este modelo. Resulta evidente que ${\cal A}(t)$,
depende tanto de la temperatura del entorno, como de la forma de la
densidad espectral $I(\omega)$. Este procedimiento, nos 
da el mismo resultado que si hubieramos realizado el m\'etodo perturbativo
de la serie de Dyson, mencionado en el Cap\'itulo \ref{c1}, para obtener
la ecuaci\'on maestra del sistema reducido. En ese caso,
 asumiendo  que (i) el sistema y el entorno 
no est\'an inicialmente correlacionados y (ii) el entorno 
est\'a inicialmente en equilibrio,
 la  ecuaci\'on maestra exacta para la matriz densidad reducida es,
\begin{equation}
\dot{\rho_{\rm r}} = -i \Omega [\sigma_z,\rho_{\rm r}] - {\cal D}(t)
[\sigma_z,[\sigma_z,\rho_{\rm r}]], \label{mastercap5}
\end{equation}
donde 
\begin{equation}
{\cal D}(s)=\int_0^s ds' \int_0^{\infty} d\omega I(\omega)
\coth\bigg(\frac{\omega}{2 k_B T}\bigg) \cos(\omega(s-s')).
\nonumber
\end{equation}
El modelo describe un mecanismo puramente ``decoherente'', 
conteniendo \'unicamente el t\'ermino de difusi\'on ${\cal D}(t)$.  
Los m\'etodos se relacionan a partir de 
\mbox{${\cal A}(t)= \int_0^t ds {\cal D}(t)$}.

\subsection{C\'alculo de la fase geom\'etrica}

En esta secci\'on, calcularemos la fase geom\'etrica para el 
modelo esp\'in-bos\'on. Como ya mencionamos,
las poblaciones de los t\'erminos diagonales de la matriz densidad
 se mantienen constante; es decir, $\rho_{\rm r_{ii}}(t)=
\rho_{\rm r_{ii}}(0)$, para $i=0,1$.  Luego, la soluci\'on para 
los t\'erminos fuera de la diagonal ser\'a 
\begin{equation}
\rho_{\rm r_{01}}(t)= e^{-i \Omega t -
{\cal A}(t)}\rho_{\rm r_{01}}(0)
\end{equation}
donde $\rho_{\rm r_{01}}(0)$ es
 una constante determinada a partir de las condiciones iniciales. De 
esta forma, definimos el coeficiente de decoherencia 
como $\Gamma(t)=\exp(-{\cal A}(t))$, de manera an\'aloga a los otros
modelos estudiados en esta Tesis.
Como ${\cal A}(t)$ es una cantidad real, la matriz densidad reducida
 se completa pidiendo que $\rho_{\rm r_{01}}(0)=\rho_{\rm r_{10}}(0)$ 
ya que $\rho_{\rm r_{10}}={\rho_{\rm r_{01}}}^{*}$. 
Para un sistema de dos niveles, el espacio de
 Hilbert puede ser representado por los estados sobre y dentro de la 
esfera de Bloch (Fig.\ref{esfera}). Luego, podemos asumir que,
 inicialmente, nuestro estado est\'a en la superficie de 
dicha esfera (estado puro)
\begin{equation} 
|\Psi (0) \rangle=\cos(\theta_0/2) |e \rangle + \sin(\theta_0/2) |g \rangle .
\nonumber\end{equation}
As\'i, la constante $\rho_{\rm r_{01}}(0)$ se determina a partir de 
$\rho_{\rm r}(0)=|\Psi(0)\rangle \langle \Psi(0)|$.  Para tiempos $t>0$, 
la matriz densidad reducida es
\begin{equation}
\rho_{\rm r}(t)=\bigg(\begin{array}{cc} \cos(\theta_0/2)^2 ~~~~~~~~~~~~
 1/2 \sin(\theta_0) e^{i \Omega t - {\cal A}(t)}\\
1/2 \sin(\theta_0)e^{-i \Omega t - {\cal A}(t)}~~~~~~~~~~~~
 \sin(\theta_0/2)^2
\end{array} \bigg). \nonumber
\end{equation}

Para calcular la fase geom\'etrica definida en la Ec.(\ref{fasegeo}), 
s\'olo necesitamos los autovalores y autovectores de la matriz 
densidad reducida. Los autovalores de esta matriz se calculan
 f\'acilmente, obteniendo
\begin{equation}
\label{varep+}
\varepsilon_{\pm}(t) = \frac{1}{2} \pm \frac{1}{2}
\sqrt{\cos(\theta_0)^2
+ \exp(-2 {\cal A}(t)) \sin(\theta_0)^2}. 
\end{equation}
En el caso de los autovectores, s\'olo necesitaremos $|\Psi_{+}(t)
 \rangle$ ya que $\varepsilon_{-}(0)=0$, y por tanto, la \'unica 
contribuci\'on a la fase geom\'etrica saldr\'a del autovector 
$|\Psi_{+}(t) \rangle$ y su correspondiente autovalor. 
El autovector buscado se escribe
\begin{equation}
\label{Phi+}
|\Psi_{+} (t) \rangle = e^{-i \Omega t} \sin(\theta_t/2)
|e\rangle + \cos(\theta_t/2) |g\rangle, 
\end{equation}
con $\tan (\theta_t/2) = \exp(- {\cal A}(t) ) \cot (\theta_0/2)$.
Es f\'acil  ver que, cuando ${\cal A}=0$, obtenemos los 
resultados de la evoluci\'on unitaria. En particular, lo que 
resulta importante calcular es el factor $\langle \Psi_k| \dot{\Psi}_{k} \rangle$ de la
Ec.(\ref{fasegeo}). Haciendo la derivada temporal del autovector, obtenemos
$\langle \Psi_k| \dot{\Psi}_{k} \rangle= -i \Omega
\sin(\theta_t/2)^2$.
De esta forma, la fase geom\'etrica de este sistema abierto es
\begin{equation}
\Phi={\mathrm {arg}}\bigg\{\sqrt{\varepsilon_+(\tau)
\varepsilon_+(0)} \langle \Psi_+(\tau)| \Psi_+(0)\rangle e^{i
\Omega \int_0^{\tau}dt  \sin(\frac{\theta_t}{2})^2}\bigg\}.
\nonumber
\end{equation}
Este resultado es v\'alido para cualquier matriz densidad que
 tenga estos autovalores y autovectores, independientemente
 de la expresi\'on exacta de ${\cal A}(t)$. Reemplazando las 
Ecs. (\ref{varep+}) y (\ref{Phi+}) en la expresi\'on de la fase, 
la fase geom\'etrica asociada a una evoluci\'on semic\'iclica 
${\cal P}:t ~\epsilon~[0,\tau]$ (con
$\tau=2 \pi/\Omega$  \cite{Tong}), asumiendo adem\'as 
$\cos(\theta_0/2) \geq 0$,  es
\begin{equation}
\Phi= \Omega \int_0^{\tau}~dt~ \sin
\bigg(\frac{\theta_t}{2}\bigg)^2.
\end{equation}

Para evaluar esta integral, haremos una expansi\'on en 
potencias de la constante disipativa ($\gamma_0$). De 
esta forma, a orden cero en dicha constante, obtendremos 
la fase geom\'etrica unitaria $\Phi^{U}$ y, a primer order, la 
correcci\'on no unitaria a dicha fase. As\'i,
\begin{eqnarray}
\Phi= \Phi^{U} + \delta \Phi
\approx  \pi (1-\cos(\theta_0)) 
+ \gamma_0 \frac{\Omega}{2} \sin(\theta_0)^2 \cos(\theta_0)
\int_0^{\tau}~dt \bigg[\frac{\partial
{\cal A}(t)}{\partial \gamma_0} \bigg]\bigg|_{\gamma_0=0}, \nonumber
\end{eqnarray}
donde $\Phi^{U}=\pi (1-\cos(\theta_0))$ es el resultado de la fase
 para un sistema de dos niveles cu\'antico cerrado.
Para evaluar la correcci\'on a la fase unitaria, debemos especificar
 qu\'e tipo de entorno tenemos; es decir, su densidad espectral
 y su temperatura. En lo que sigue analizaremos diferentes casos:
\begin{description}
 \item {\bf Entorno \'ohmico a temperatura alta}.
En este caso, la densidad espectral es de la forma $I(\omega)= 
\gamma_0/4~ \omega e^{-\omega/\Lambda}$ y, adem\'as 
vale $\hbar \Lambda \ll k_B T$. As\'i, podemos aproximar 
${\mathrm {coth}}(\beta \hbar \omega /2)$ en la Ec.(\ref{Gcap5}) 
por $2 k_B T/(\hbar \omega)$ y el coeficiente de decoherencia
 que se obtiene es 
${\cal A}_{\rm n=1,HT}(t)= (\gamma_0 \pi k_B T) t/\hbar$. De esta forma, 
podemos estimar la correcci\'on a la fase unitaria  como
\begin{equation}
\delta \Phi_{n=1,\rm HT} = \pi^2 \gamma_0 \bigg(\frac{\pi k_B T}{\hbar
\Omega} \bigg) \sin(\theta_0)^2 \cos(\theta_0). \label{ohHTcap5}
\end{equation}
Esta correcci\'on es proporcional a $\gamma_0 k_B T/\hbar 
\Omega$, por lo cual resulta grande para entornos calientes 
y no deber\'ia ser despreciada (aunque $\gamma_0 \ll 1$). 
Nuestro resultado es formalmente similar a aquel encontrado
  en la Ref. \cite{Yi}. Sin embargo, en ese trabajo, no 
existe un modelo espec\'ifico sobre el entorno considerado ni una 
formulaci\'on sobre la din\'amica de los sistemas abiertos.

\item {\bf Entorno \'ohmico a temperatura cero}.
En este caso, hay una escala temporal menos en 
comparaci\'on al caso de temperatura alta; por lo cual 
es esperable que la fase se corrija de una manera 
significativamente distinta. El factor ${\mathrm {coth}}(\beta\hbar\omega
/2)$, en la definici\'on de ${\cal A}(t)$ de la Ec. (\ref{Gcap5}) 
puede ser reemplazado por $1$ y ${\cal A}_{n=1,\rm T=0}(t)=
\gamma_0/2 \log(1+\Lambda^2 t^2)$. As\'i, la correcci\'on a la fase es
\begin{equation}
\delta \Phi_{n=1,\rm T=0}= \pi \gamma_0 \bigg(-1+\log(\frac{2 \pi
 \Lambda}{\Omega})\bigg)
\sin(\theta_0)^2 \cos(\theta_0). \label{ohT0cap5}
\end{equation}
Lo importante de este caso, es que la correcci\'on a la fase 
unitaria puede asociarse a  las fluctuaciones cu\'anticas de
 vac\'io del entorno.
\end{description}

Podemos adem\'as considerar entornos no \'ohmicos. En el 
caso de entornos supra\'ohmicos, se manifiestan efectos no
 lineales en la din\'amica de tiempos cortos, debido a las 
correlaciones del ba\~no generadas por $\omega^n$ con $n>1$. Lo 
interesante de esta situaci\'on es que el campo electromagn\'etico 
puede ser modelado como un entorno supra\'ohmico, lo cual
 puede ser muy importante en \'optica cu\'antica y en el c\'alculo 
de las fases geom\'etricas. Por tanto, evaluaremos la correci\'on 
a la fase geom\'etrica unitaria en presencia de dicho tipo de entorno.

\begin{description}
\item {\bf Entorno supra\'ohmico a temperatura alta}.
Haciendo la misma aproximaci\'on para la 
${\mathrm {coth}}(\beta \hbar \omega /2)$ de la Ec.(\ref{Gcap5}) 
que en el caso \'ohmico, obtenemos el coeficiente de 
decoherencia en este caso ${\cal A}_{n=3,\rm HT}=\frac{2 k_B T}{\hbar} \gamma_0
\Lambda (\frac{\Lambda^2 t^2}{1+\Lambda^2 t^2})$. 
De esta forma, es f\'acil estimar la correcci\'on a la fase unitaria como
 \begin{equation}
\delta \Phi_{n=3,\rm HT}= \pi \gamma_0 \bigg( \frac{2 k_B T}{\hbar
\Lambda} \bigg) \sin(\theta_0)^2 \cos(\theta_0).\label{nohHTcap5}
\end{equation}

\item {\bf Entorno supra\'ohmico a temperatura cero}.
De la misma forma que en los otros casos, podemos 
estimar la correcci\'on a primer orden en $\gamma_0$ 
de la fase del sistema abierto como
\begin{equation}
\delta \Phi_{n=3,\rm T=0} = \pi \gamma_0
 \sin(\theta_0)^2 \cos(\theta_0).
 \label{nohT0cap5}
\end{equation}
\end{description}
Como $\Lambda $ es la m\'axima frecuencia presente 
en el entorno, resulta v\'alido asumir que $\Omega \leq \Lambda$. 
Si el par\'entesis de la Ec.(\ref{ohT0cap5}) 
es de orden uno y, por tanto, las correcciones a temperatura cero 
(Ecs.(\ref{ohT0cap5}) y (\ref{nohT0cap5})), tanto del entorno \'ohmico 
como del supra\'ohmico, resultan similares  en valor absoluto.
 Lo mismo ocurre en el caso de temperatura alta con las correcciones
 a la fase de las Ecs.(\ref{ohHTcap5}) y Ec.(\ref{nohHTcap5}).
 
Estos resultados son llamativamente interesantes y hacen 
evidente la solidez del \mbox{modelo}, y de la fase en s\'i misma.
Si bien la fase
unitaria del sistema no depende del camino elegido, la correcci\'on
no unitaria depende de los par\'ametros del entorno. A\'un as\'i,
estas fases podr\'ian ser observables bajo condiciones apropiadas.
A continuaci\'on analizaremos dichas condiciones.

\subsection{Tiempos de p\'erdida de coherencia}

En adelante, estimaremos los  tiempos de p\'erdida de coherencia en
 los diferentes casos mencionados anteriormente. Esta escala
 temporal corresponde al tiempo en el cual desaparecen los
 t\'erminos no diagonales de la matriz densidad, y resulta importante
 a la hora de planificar un experimento para medir las fases 
geom\'etricas.  Cuando ${\cal A}(t)$ es grande, las 
coherencias de la matriz desaparecen en una escala temporal 
$t_{\cal D}$ asociada a ${\cal A}(t)$, o bien, al factor de decoherencia
$\Gamma(t)$. De esta forma,  las 
interferencias cu\'anticas son eliminadas y desaparece el
patr\'on de interferencia. Por tanto, 
 estimaremos esta escala temporal $t_{\cal D}$ pidiendo la 
condici\'on ${\cal A} (t_D) \approx 1$ en cada caso.
\begin{figure}[!ht]
\centering
\includegraphics[width=12cm]{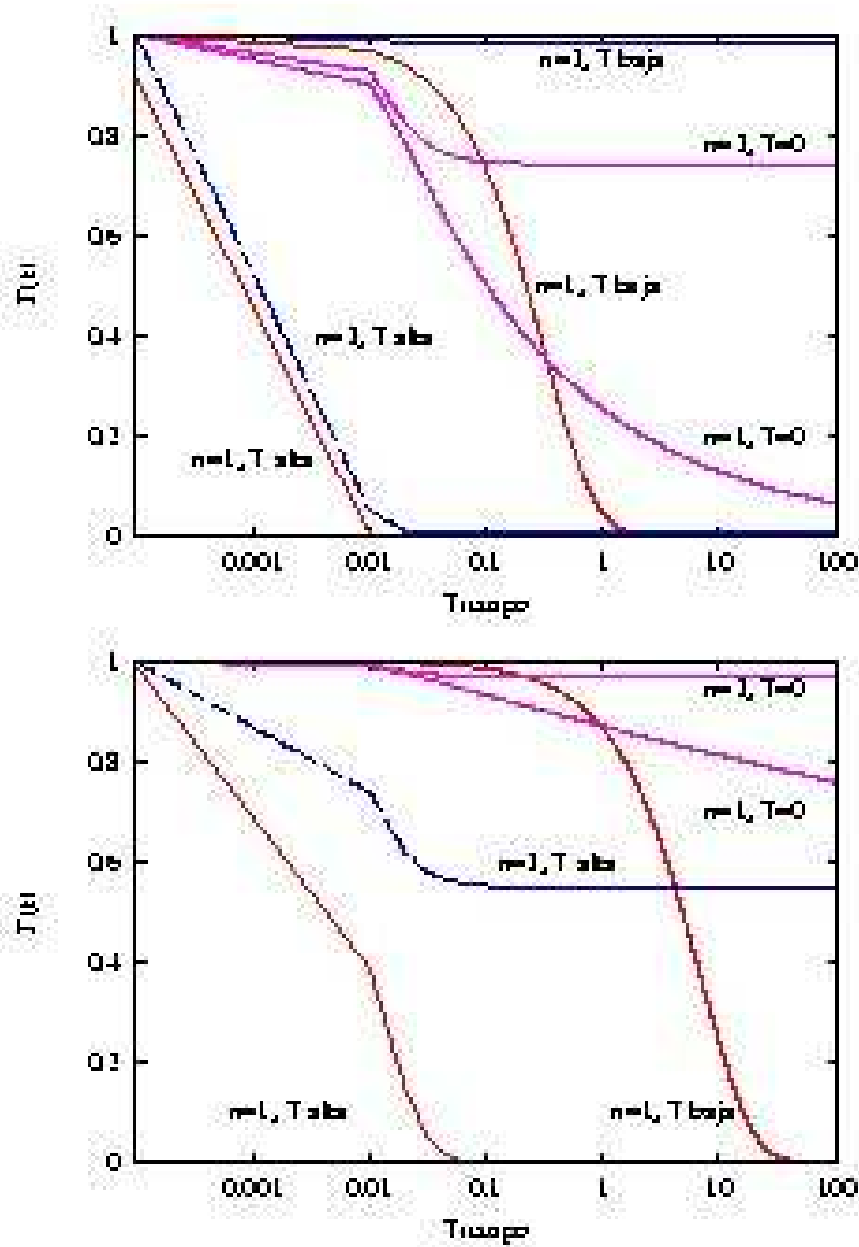}
\caption{Comparaci\'on entre los coeficientes $\Gamma(t)$ 
para el entorno \'ohmico  ($n=1$) y el supra\'ohmico 
($n=3$) para distintas temperaturas del entorno: 
alta ($T=1000$), baja ($T=1.55$) y cero ($\hbar=1=k_B)$. 
Los valores de los par\'ametros usados son $\Lambda=100\Omega$ y
$\gamma_0=0.3$ en el gr\'afico superior mientras que 
$\Lambda=100\Omega$ y $\gamma_0=0.03$ en el inferior. 
El tiempo est\'a medido en unidades de  $\Omega$. Se 
puede notar que el l\'imite de baja temperatura no es igual 
al caso de temperatura estrictamente cero.}
\label{fases1}
\end{figure}

Para el entorno \'ohmico a temperatura alta  el coeficiente de 
decoherencia calculado es
${\cal A}_{n=1, \rm HT}(t)=(k_B T \pi \gamma_0/\hbar) t$. 
En este caso, el tiempo de p\'erdida de coherencia es 
extremadamente corto, ya que 
$t_D^{n=1,\rm HT}=\hbar/(k_B T \pi \gamma_0)$, como se 
puede observar en la Fig.\ref{fases1}. Este es el \'unico 
caso que no depende expl\'icitamente de la frecuencia de 
corte del entorno $\Lambda$.
Por otro lado, cuando el entorno \'ohmico est\'a a temperatura
 cero, el tiempo de decoherencia que se obtiene es 
$t_D^{n=1, \rm T0}= e^{1/\gamma_0}/\Lambda$ para 
$\Lambda t \geq 1$. El proceso de p\'erdida de coherencia 
se hace presente en este entorno, pero en una escala 
temporal m\'as larga que la del caso de temperatura alta.
 Esto concuerda con las estimaciones encontradas en el 
Cap\'itulo \ref{c2} de esta Tesis para estos entornos.
 Como se puede observar en el gr\'afico inferior de la 
Fig.\ref{fases1}, la p\'erdida de coherencia se ve retrasada 
cuanto m\'as chico es el valor de $\gamma_0$.

En cuanto al entorno supra\'ohmico, veremos que el 
proceso de decoherencia tiene caracter\'isticas bastante 
diferentes. En el caso de temperatura alta, el coeficiente
 de decoherencia estimado es 
${\cal A}_{n=3,\rm HT}(t)=(2 k_B T \gamma_0) \Lambda t^2/\hbar$ si
$\Lambda t \ll 1$.  As\'i, la estimaci\'on del tiempo de 
decoherencia es directa y se obtiene $t_D^{n=3,\rm HT}=1/\Lambda 
\sqrt{\hbar \Lambda/(2 k_B T \gamma_0)}$. Sin embargo, si 
$\Lambda t \geq 1$, el coeficiente de decoherencia es 
cualitativamente diferente: ${\cal A}_{n=3,\rm HT}= 2 k_B T 
\gamma_0/(\hbar \Lambda)$, ya que es constante en el tiempo.
 En este caso, la p\'erdida de
 coherencia en el sistema s\'olo se llevar\'a a cabo si 
$2 k_B T \gamma_0 \gg \hbar \Lambda$ y eso debe 
ocurrir en un tiempo
$t < 1/\Lambda$, como se puede observar en el gr\'afico
 superior de la Fig.\ref{fases1}. Por el contrario, en el 
gr\'afico inferior de dicha figura, el valor de $\gamma_0$ 
es bastante m\'as chico y $k_B T \gamma_0 < \hbar \Lambda$,
 lo cual implica que ${\cal A}_{n=3,\rm HT}$ nunca ser\'a de 
orden uno y la decoherencia no ser\'a efectiva en dicho entorno. 
Finalmente, si el entorno supra\'ohmico est\'a a temperatura cero,  
el coeficiente de decoherencia estimado es 
${\cal A}_{n=3,\rm T0}= \gamma_0 \Lambda^4 t^4/(1+\Lambda^2 t^2)^2$. 
En la Fig.\ref{fases1}, podemos notar que
 no hay p\'erdida de coherencia en el sistema para este caso. 
Esto se debe a que ${\cal A}_{n=3,\rm T0}= \gamma_0$
 para $\Lambda t \geq 1$. El factor de decoherencia
 entonces ser\'a constante $e^{-\gamma_0}$. 
Como $\gamma_0 < 1$, este factor nunca ser\'a cero.

En s\'intesis podemos notar que, a pesar de ser un modelo 
muy sencillo de decoherencia, reobtenemos las estimaciones 
de los tiempos de decoherencia realizados en entornos 
\'ohmicos y no \'ohmicos del Cap\'itulo \ref{c2}. Para medir
 la fase geom\'etrica del sistema, la fase din\'amica debe 
ser eliminada, ya sea utilizando la t\'ecnica de {\it spin echo} 
para espines en campos magn\'eticos \cite{Ekert2}, o bien, 
utilizando el transporte paralelo y asegurarse, as\'i, que la 
fase din\'amica es cero a todo tiempo. A pesar que el 
estado no adquiere una fase localmente, s\'i la adquiere 
globalmente, la cual es igual a la fase geom\'etrica en el 
caso del sistema cu\'antico aislado. En el caso de sistemas 
abiertos hemos demostrado c\'omo se modifican estas fases
 en funci\'on de los par\'ametros que caracterizan al entorno. 
Consideramos que 
 estas estimaciones resultan importantes a la hora de 
planificar un experimento de medici\'on de fases geom\'etricas 
para estados mixtos en evoluciones no unitarias \cite{Du}, 
particularmente interesante en el marco de la  
Computaci\'on Cu\'antica. Seg\'un nuestro an\'alisis, 
podemos afirmar que el mejor montaje experimental 
para lograr la medici\'on de esta fase es el entorno 
supra\'ohmico en el caso de temperatura cero y r\'egimen 
subamortiguado.  Es decir, podr\'ia implementarse en
una cavidad electromagn\'etica superconductora.            
\newpage
\thispagestyle{empty}
\cleardoublepage

\chapter{Decoherencia de dominios y defectos durante una transici\'on de fase }
\label{c6}
\markboth{Decoherencia en Teor\'ia de Campos }
{Cap\'itulo 6}


En este cap\'itulo final, presentaremos un an\'alisis completo acerca del
proceso de p\'erdida de coherencia para un campo escalar
que sufre una transici\'on de fase. Las transiciones de fase con
ruptura de simetr\'ia traen como consecuencia la formaci\'on de 
configuraciones de campo  macrosc\'opicas estables, llamadas
dominios y defectos topol\'ogicos. El estudio del proceso de
decoherencia en este contexto es crucial para entender la
aparici\'on de estas configuraciones cl\'asicas a partir de
un campo cu\'antico con ruptura de la simetr\'ia.
Este cap\'itulo resulta una continuaci\'on a una serie de
trabajos  realizados por Lombardo y colaboradores \cite{ferdiego,
fermazziray, fermazziray-npb}.

Por un lado, el ordenamiento que
sufre el campo luego de una transici\'on de fase de este tipo, se debe al crecimiento
de la amplitud de las longitudes de onda inestables del mismo, las cuales surgen 
autom\'aticamente a partir de los m\'aximos inestables del potencial. A partir de los 
trabajos de Guth y Pi \cite{guthpi}, resulta conocido el hecho que los modos inestables
generan correlaciones a trav\'es del proceso de ``squeezing" (igual que lo
que ocurre en \'Optica Cu\'antica). Por otro lado, 
 la diagonalizaci\'on de la funcional de decoherencia
resulta inevitable al considerar el granulado grueso del campo. 

Existen varias escalas temporales relevantes para ser tenidas en cuenta
en la descripci\'on de una transici\'on de fase. Si \'esta es r\'apida, la condici\'on
inicial del campo escalar puede ser descripta por un campo libre con
un potencial invertido, es decir (masa)$^2 <0$. \'Esto es v\'alido hasta que la
funci\'on de onda del campo explora los m\'inimos del potencial en un 
tiempo, conocido como el tiempo espinodal $t_{\rm esp}$. 
En particular,  el ordenamiento del campo
luego de una transici\'on de fase se debe al crecimiento de la amplitud de
los modos inestables (longitudes de onda largas). Por tanto, los modos de
longitudes de onda cortas del campo, junto con los otros campos
externos con los cuales el campo sistema interact\'ua, forman un entorno
efectivo que, al efectuarle un proceso de 
 granulado grueso,  induce decoherencia en el campo 
sistema \cite{ferdiego}. Es por esto, que existe otra escala temporal asociada
al entorno, llamada tiempo de decoherencia $t_{\cal D}$. Cuando 
$t > t_{\cal D}$, el par\'ametro de orden se vuelve una entidad cl\'asica \cite{fermazzidiana}.

Como es sabido las escalas temporales involucradas en el proceso de decoherencia
son dif\'{\i}ciles de obtener en teor\'{\i}a de campos y para estimarlas es necesario realizar
aproximaciones. En el marco de la teor\'{\i}a cu\'antica de campos, F.C. Lombardo y F.D.
Mazzitelli \cite{ferdiego} desarrollaron un m\'etodo para estudiar en forma cuantitativa y de manera
aproximada los efectos difusivos que producen decoherencia. Para ello consideraron un
campo escalar $\varphi$ con autointeracci\'on $\lambda \varphi^4$  y usaron 
una longitud de onda cr\'{\i}tica para
separar a los modos del campo en dos sectores: aquellos con longitudes de onda mayores
a la cr\'{\i}tica y los modos restantes. En este modelo estudiaron la decoherencia para los
modos de longitud de onda ``larga" como funci\'on de la longitud cr\'{\i}tica. En particular,
realizaron dicho estudio para un campo $\varphi$ acoplado conformemente a la curvatura de
un espacio-tiempo de de Sitter y mostraron que la p\'erdida de coherencia es efectiva si
la longitud cr\'{\i}tica es mayor al radio de Hubble en el contexto de modelos Inflacionarios
de Cosmolog\'ia.

 Continuando con el desarrollo del m\'etodo, estos autores junto con R.J. Rivers
\cite{fermazziray,fermazziray-npb}, consideraron el problema de la validez 
de la aproximaci\'on cl\'asica para el
par\'ametro de orden de una transici\'on de fase de segundo orden, producida por la
ruptura de una simetr\'{\i}a. Trabajando en el espacio-tiempo de Minkowski, estudiaron la
p\'erdida de coherencia para este par\'ametro de orden, el cual estaba constitu\'ido b\'asicamente
por el modo homog\'eneo del campo $\varphi$. En este contexto, adem\'as de la autointeracci\'on
$\lambda \varphi^4$, los autores propusieron un conjunto de campos con temperatura, a
los cuales el campo $\varphi$ pod\'ia estar acoplado de distintas maneras. Luego estimaron un
tiempo de decoherencia asociado a cada uno de los distintos t\'erminos de la acci\'on de
interacci\'on del par\'ametro de orden. Finalmente, mostraron que dichos tiempos, para
acoplamientos d\'ebiles, son t\'ipicamente m\'as cortos que el tiempo para el cual
las no  linealidades del problema comienzan a ser relevantes, concluyendo que es posible realizar un
an\'alisis cl\'asico en tal r\'egimen. 

En el contexto de la Teor\'ia Cu\'antica de Campos (TCC), 
podemos analizar la decoherencia de dos maneras,
en algunos casos equivalentes.  En trabajos realizados
por Lombardo, Mazzitelli y Rivers \cite{fermazziray,fermazziray-npb,rayfer},
se estudi\'o la diagonalizaci\'on de $\rho$ a partir de una ecuaci\'on maestra.
 Otra posibilidad, es que la diagonalizaci\'on se analice
a partir  de la funcional de
influencia directamente. 
En cualquier enfoque que tomemos, lo importante es definir  en qu\'e
base del campo se intenta lograr dicha diagonalizaci\'on, la cual, de todas
formas, se obtiene de manera aproximada. Si consideramos un campo sistema
de infinitos grados de libertad, el camino m\'as aconsejable es el funcional.

Este cap\'itulo resulta una ``secuela" a los trabajos mencionados,
ya que aqu\'i  exploraremos de qu\'e forma dichas transiciones, naturalmente, convierten
la descripci\'on cu\'antica del Universo en cl\'asica.
En adelante, como en los trabajos previos,
usaremos la formaci\'on de dominios luego de una transici\'on de fase
para caracterizar la aparici\'on del comportamiento cl\'asico.
En particular, recrearemos este resultado pero utilizando
un formalismo m\'as sencillo.

\section{El modelo}

Consideraremos el caso de un campo real cu\'antico $\phi$ con un 
potencial de pozo doble. Como ya hemos mencionado, el ordenamiento
del campo luego de una transici\'on de fase, se debe al crecimiento
de los modos de longitud de onda larga. Para estos modos, el entorno
est\'a compuesto de los modos de longitud de onda corta m\'as los
campos externos con los cuales el campo sistema interact\'ua en ausencia
de reglas de selecci\'on \cite{huzhang,boya}. La inclusi\'on de campos
externos se debe, por un lado, al hecho que un campo escalar aislado
es f\'isicamente poco realista; y por el otro, al hecho que nos provee
de un marco para realizar  aproximaciones que simplifican algunos
c\'alculos \cite{fermazziray-npb}. Para ser espec\'ificos, la acci\'on m\'as
sencilla de un campo escalar y varios campos entorno $\chi_{\rm a}$ es
\begin{equation}
S[\phi , \chi ] = S_{\rm sistema}[\phi ] + S_{\rm entorno}[\chi ] +
S_{\rm interaccion}[\phi ,\chi ], \label{action0}
\end{equation}
donde (con $\mu^2$, $m^2 >0$ )
\begin{equation}
S_{\rm sistema}[\phi ] = \int d^4x\left\{ \frac{1}{2}\partial_{\mu}
\phi\partial^{\mu} \phi + {\frac{1}{2}}\mu^2 \phi^2 -
\frac{\lambda}{4}\phi^4\right\}, \nonumber \end{equation}
\begin{equation}
S_{\rm entorno}[\chi ] = \sum_{\rm a=1}^N\int d^4x\left\{
\frac{1}{2}\partial_{\mu}\chi_{\rm a}
\partial^{\mu}
\chi_{\rm a} - \frac{1}{2} m_{\rm a}^2 \chi^2_{\rm a}\right\},
\nonumber
\end{equation}
y las acciones m\'as relevantes entre sistema y entorno son de la
forma bicuadr\'atica:
\begin{equation}
S_{\rm interaccion}[\phi ,\chi ] = - \sum_{\rm a=1}^N\frac{g_{\rm a}}{8}
\int d^4x ~ \phi^{\rm 2} (x) \chi^{\rm 2}_{\rm a} (x).
\label{Sint}
\end{equation}
Es v\'alido resaltar que, a\'un si no se consideraran los campos externos
con una interacci\'on cuadr\'atica con el campo sistema 
del tipo de la Ec.(\ref{Sint}), la interacci\'on entre los modos de \'este de longitud
de onda larga y corta se podr\'ia escribir en funci\'on de t\'erminos
de esta forma (adem\'as de varios otros), y, as\'i,  mostrar que un t\'ermino 
de esta forma es obligatorio.

A pesar que el campo sistema $\phi$ sufre una p\'erdida de coherencia
debido a la presencia de sus propios modos de longitud de onda corta 
\cite{ferdiego},  primero consideraremos el caso en que el entorno
est\'a compuesto \'unicamente por los campos externos $\chi_{\rm a}$.
Como los entornos poseen un efecto acumulativo sobre la aparici\'on del
comportamiento cl\'asico, la inclusi\'on de un componente adicional en
el entorno reduce el tiempo que le lleva al sistema comportarse de manera
cl\'asica. Por \'esto, resulta l\'ogico incluir de a una 
las partes del entorno, ya que de esa forma estamos obteniendo 
una cota superior del
tiempo de decoherencia en cada paso \cite{fermazziray-npb}.

En \cite{fermazziray,fermazziray-npb,rayfer}, se ha demostrado
que para que el an\'alisis sea m\'as robusto es necesario
 que el entorno tenga un fuerte impacto en el campo sistema (y
no al rev\'es). La forma m\'as sencilla de implementar eso es pedir
que $N \gg 1$, o sea el n\'umero de campos externos $\chi_{\rm a}$
presentes en el entorno con masas comparables $m_{\rm a} \backsimeq
\mu$, d\'ebilmente acoplados al campo escalar $\phi$, con 
$\lambda$, $g_{\rm a} \ll 1$ (para m\'as detallas ver \cite{ferdiego}).
De esta forma,  existen $N$ campos
externos interactuando con el campo sistema, pero s\'olo un potencial
propio del sistema incapaz de reaccionar contra ese entorno expl\'icito.

Para la consistencia de los c\'alculos a un lazo, alcanza pedir que
$g_{\rm a}=g/\sqrt{N}$. Con nuestra elecci\'on de sistema y entornos, no hay
interacciones directas del tipo $\chi^4$. Y aquellas que son indirectas, mediadas
por lazos en $\phi$, son despreciables en un factor $g/\sqrt{N}$.
Consideraremos que tanto el estado inicial del sistema como el del entorno, es 
un estado t\'ermico de equilibrio, a una temperatura alta $T_0$, mayor que 
la temperatura cr\'itica $T_c$. Supondremos, luego, un cambio global
en el entorno (por ejemplo, la expansi\'on del Universo Temprano),
que se puede caracterizar por el cambio en la temperatura, 
de modo que $T_f < T_c$. Es decir, la transici\'on de fase no se debe
a los efectos inducidos por los campos entorno. 

La condici\'on inicial asumida no corresponde al caso en 
que la matriz densidad total est\'a no correlacionada en
los campos $\phi$ y $\chi$, ya que es la interacci\'on entre
\'estos la que conduce a la restituci\'on de la simetr\'ia 
perdida en el caso de temperatura alta. As\'i, la condici\'on
inicial conduce a una acci\'on efectiva para las cuasiparticulas
$\phi$,
\begin{equation}
S^{\rm eff}_{\rm sistema}[\phi ] = \int d^4x\left\{
\frac{1}{2}\partial_{\mu} \phi\partial^{\mu} \phi - \frac{1}{2}
m_{\phi}^2(T_0) \phi^2 - \frac{\lambda}{4}\phi^4\right\}
\label{Stherm}
\end{equation}
donde, seg\'un la aproximaci\'on de campo medio, 
$m_{\phi}^2(T_0)= -\mu^2(1-T_0^2/T_{\rm c}^2)$ para 
$T\approx T_{\rm c}$. Como resultado, podemos
tomar una matriz densidad a temperatura $T_0$, factorizada
de la forma ${\hat\rho}[T_0] = {\hat\rho}_{\phi}[T_0] \times
{\hat\rho}_{\chi}[T_0]$, donde ${\hat\rho}_{\phi}[T_0]$
est\'a determinado por $S^{\rm eff}_{\rm sistema}[\phi
]$ y ${\hat\rho}_{\chi}[T_0]$ por $S_{\rm entorno}[\chi_{\rm a} ]$.
Esto significa los campos $\chi_{\rm a}$ tienen un gran efecto
en el campo $\phi$, mientras que \'este tiene un efecto 
despreciable en los campos entorno.

Si asumimos que el cambio en la temperatura no es muy lento, 
las inestabilidades exponenciales del campo sistema $\phi$ crecen
tan r\'apido que el sistema logra poblar los m\'inimos de potencial
mucho antes que la temperatura sea bastante m\'as chica que la
temperatura cr\'itica \cite{moro}. Como la temperatura $T_c$ no tiene
una importancia particular para los campos entorno, para 
tiempos chicos en la evoluci\'on podemos considerar la
temperatura del entorno es fija  $T_{\chi}\approx T_c$. Adem\'as,
por simplicidad, las masas de los campos $\chi_{\rm a}$ ser\'an
consideradas $m_{\rm a} \simeq \mu$.

En este contexto, nos alcanza con pedir que la transici\'on sea
instant\'anea. Transiciones m\'as lentas hacen que los c\'alculos
anal\'iticos sean demasiado complicados, sin cambiar la naturaleza
cualitativa de los resultados \cite{ferdiego}.

\section{M\'etodo de la funcional de Decoherencia}

La idea de historias consistentes nos provee un marco alternativo
para estudiar la clasicalidad en sistemas cu\'anticos abiertos.
La evoluci\'on cu\'antica de un sistema puede ser considerada
como una superposici\'on lineal de un granulado fino de
 historias consistentes.
Nosotros intentaremos distinguir entre soluciones
cl\'asicas emergentes luego de una transici\'on de fase cu\'antica.
Por \'esto, trabajaremos en la base de configuraciones del campo.
Si consideramos que el campo $\phi(x)$ representa un historia, entonces
la amplitud cu\'antica de \'esta es $\Psi [\phi] \sim e^{iS[\phi]}$
en unidades de $\hbar=1$. En la aproximaci\'on que estudiaremos
aqu\'i, la cantidad que nos interesar\'a es un granulado grueso
de historias, es decir
\begin{equation}
\Psi [\alpha] = \int {\cal D}\phi ~ e^{iS[\phi]}\alpha [\phi],
\end{equation}
donde $\alpha [\phi]$ es una funci\'on filtro que define el
granulado grueso, o traza sobre los grados de libertad de
los campos $\chi_{\rm a}$.
As\'i, la funcional de decoherencia para dos historias
del campo es:
\begin{equation}
 {\cal D}[\alpha^+,\alpha^-] = \int {\cal D}\phi^+{\cal
 D}\phi^-~e^{i(S[\phi^+]-S[\phi^-])}\alpha^+ [\phi^+]\alpha^-
 [\phi^-].
\end{equation}
${\cal D}[\alpha^+,\alpha^-]$ no se factoriza porque las
historias $\phi^{\pm}$ no son independientes. En este contexto,
decoherencia significa f\'isicamente  que dos historias del
granulado grueso adquieren realidades individuales y por tanto,
se les debe asignar probabilidades definidas en el sentido cl\'asico.

Una condici\'on necesaria y suficiente para la validez de las
reglas de suma de probabilidades (es decir, sin interferencias
cu\'anticas) \cite{Gri} es que se cumpla
\begin{equation}
 {\rm Re}{\cal D}~[\alpha^+,\alpha^-]\approx 0,
\end{equation}
cuando $\alpha^+\neq\alpha^-$, aunque en muchas situaciones, la
condici\'on a\'un m\'as restrictiva ${\cal D}[\alpha^+,\alpha^-]\approx 0$
se cumple \cite{Omn}. En ese caso, las historias son {\it consistentes}.

Para nuestra aplicaci\'on particular, obtendremos 
como historia de grano grueso a aquella obtenida a partir de un
granulado grueso sobre todas las configuraciones 
de grano fino donde el campo $\phi$ permanece cercano
a su configuraci\'on cl\'asica $\phi_{\rm cl}$.
La funci\'on filtro la definimos como
\begin{equation}
 \alpha_{\rm cl}[\phi ] = \int {\cal D}J~
 e^{i\int J(\phi - \phi_{\rm cl})}\alpha_{\rm cl}[J].
\end{equation}
En un caso general, $\alpha[\phi]$ es una funci\'on suave
(excluyendo el caso donde $\alpha[\phi]=\rm const$, donde
no hay granulado grueso). Usando, entonces, 
\begin{equation}
J\phi \equiv \int d^4x J(x) \phi (x),
\end{equation}
podemos escribir la funcional de decoherencia entre dos
historias cl\'asicas como
\begin{eqnarray}
{\cal D}[\alpha^+,\alpha^-] = \int {\cal D}J^+{\cal
 D}J^-~e^{i W[J^+,J^-] - (J^+ \phi_{\rm cl}^+ -
 J^- \phi_{\rm cl}^-)} \alpha^+[J^+]\alpha^{-*}[J^-],
\end{eqnarray}
donde 
\begin{equation}
e^{i W[J^+,J^-]} = \int  {\cal D}\phi^+ {\cal D}\phi^- ~
e^{i(S[\phi^+] - S[\phi^-] + J^+\phi^+ - J^-\phi^-)}
,\end{equation}
es la funcional generatriz  de camino temporal cerrado (CPT) 
\cite{calhu}.
En principio, podr\'iamos examinar soluciones cl\'asicas
generales  para ver la consistencia del modelo, pero,
en la pr\'actica, es conveniente limitarse a soluciones
particulares $\phi^{\pm}$, teniendo en cuenta la
naturaleza del proceso de p\'erdida de coherencia que estamos
estudiando. Inicialmente, como ya hemos mencionado, realizamos
la separaci\'on del par\'ametro de orden del campo $\phi$
y los campos expl\'icitos del entorno $\chi_{\rm a}$, donde,
seg\'un la aproximaci\'on de fase estacionaria sobre $J$,
\begin{equation}
 {\cal D}[\phi^+_{\rm cl},\phi^-_{\rm cl}] \sim
 F[\phi^+_{\rm cl},\phi^-_{\rm cl}].
\end{equation}
 $F[\phi^+, \phi^-]$ es la funcional de influencia 
de Feynman y Vernon \cite{Feynman}. La funcional de influencia
se escribe en funci\'on de la acci\'on de influencia
$A[\phi^+_{\rm cl},\phi^-_{\rm cl}]$  como
\begin{equation}
  F[\phi^+_{\rm cl},\phi^-_{\rm cl}]=
 \exp \{iA[\phi^+_{\rm cl},\phi^-_{\rm cl}]\}.
\end{equation}
De esta forma, la funcional de decoherencia que resulta es
\begin{equation}
|{\cal D}[\phi^+_{\rm cl},\phi^-_{\rm cl}]| \sim
 \exp \{-{\rm Im}\delta A[\phi^+_{\rm cl},\phi^-_{\rm cl}]\},
 \end{equation}
donde $\delta A$ es la contribuci\'on a la acci\'on debido
a la presencia del entorno.

En este contexto, una vez que hemos elegido las soluciones
de inter\'es, las historias adyacentes se vuelven consistentes
a un tiempo de decoherencia $t_{\cal D}$, para el cual
\begin{equation} 
1\approx {\rm Im~\delta A}\vert_{t = t_{\cal D}}. \label{tD2}
\end{equation}
Como estamos considerando un acoplamiento d\'ebil con los
campos externos, podemos expandir la funcional de influencia
$F[\phi^+, \phi^-]$ a segundo orden no trivial en la constante
de acoplamiento para $N$ grande. Los t\'erminos de orden superior
son despreciados ya que decaen r\'apidamente con las potencias
de $N$. La forma general de la acci\'on de influencia es
\cite{ferdiego,calhumaz}
\begin{eqnarray}
\delta A[\phi^+,\phi^-] &= &\{\langle S_{\rm
int}[\phi^+,\chi^+_{\rm a}]\rangle_0 - \langle S_{\rm
int}[\phi^-,\chi^-_{\rm a}]\rangle_0\} 
+\frac{i}{2}\{\langle S_{\rm int}^2[\phi^+,\chi^+_{\rm a}]\rangle_0 - \big[\langle
S_{\rm int}[\phi^+,\chi^+_{\rm a}]\rangle_0\big]^2\}\nonumber
\\ &-& i\{\langle S_{\rm int}[\phi^+,\chi^+_{\rm a}] S_{\rm int}[\phi^-,\chi^- _{\rm a}]
\rangle_0 - \langle S_{\rm int}[\phi^+,\chi^+_{\rm
a}]\rangle_0\langle S_{\rm
int}[\phi^-,\chi^-_{\rm a}]\rangle_0\} \label{inflac} \nonumber \\
&+& \frac{i}{2}\{\langle S^2_{\rm int}[\phi^-,\chi^-_{\rm
a}]\rangle_0 - \big[\langle S_{\rm int}[\phi^-,\chi^-_{\rm
a}]\rangle_0\big]^2\}.\nonumber
\end{eqnarray}
Un paso adicional que se deber\'ia realizar a esta altura, 
 es la separaci\'on del
campo $\phi$ entre sus modos de longitud de onda corta y larga,
es decir $\phi=\phi_{< \Lambda} + \phi_{> \Lambda}$, con $\Lambda$
una frecuencia de corte. En el caso de una transici\'on de
fase, podemos tomar $\Lambda \sim \mu$. Los modos de longitud de onda
larga, $\vert k \vert < \mu$, determinan los dominios, mientras
que los de longitud de onda corta $\vert k \vert > \mu$ act\'uan
como un entorno impl\'icito de los primeros. Sin embargo, ese paso
no ser\'a inclu\'ido en esta Tesis, ya que \'unicamente sirve para 
acortar los tiempos de decoherencia, los cuales como se ver\'a,
ya son suficientemente cortos.

Cabe destacar que, a\'un si ignoramos este paso, ya que lo
\'unico que buscamos es acotar el tiempo de decoherencia 
$t_{\cal D}$, \'este
no resulta \'unico pues depende de la soluci\'on cl\'asica
que se considere para el campo sistema. Sin embargo, en 
la pr\'actica, todas las soluciones razonables que se eligen,
dan lugar a una cota superior para $t_{\cal D}$, 
cualitativamente similar en todos los casos.  Esto se debe,
en primer lugar, a que para los modos de longitud de onda
largas del campo sistema, el perfil espacial de
la soluci\'on cl\'asica del mismo no  resulta relevante, ya que
es su crecimiento exponencial en el tiempo el que fija la escala
para la aparici\'on del comportamiento cl\'asico.  Por \'esto mismo, es
que $t_{\cal D}$ es insensible a estos modos, y adem\'as, depende
de manera logar\'itmica, \'unicamente, de los par\'ametros de la teor\'ia.
Asimismo, con la escala de masa que estamos asumiendo en este trabajo,
la escala de la inversa de la longitud de Compton, ya queda determinada
la tasa del crecimiento exponencial y la distancia natural sobre la
cual no vamos a discrimar entre soluciones cl\'asicas. Distintas elecciones
de estas \'ultimas, entonces, implican diferencias en $\mu t_{\cal D}$ del
order de la unidad, valores much\'isimos m\'as peque\~nos que 
$\mu t_{\cal D}$, que, por tanto, ignoraremos. 
Para acoplamientos
d\'ebiles como el que nos concierne en este caso, el
tiempo de decoherencia es relativamente sencillo de calcular,
principalmente debido a la forma de la interacci\'on
$S_{\rm int}[\phi,\chi]$ de la Ec.(\ref{Sint}).
 En lo que sigue veremos que, en nuestro modelo en particular,
este tiempo resulta menor que el tiempo \'unico $t^*$, tiempo al
cual la transici\'on se completa, definido en funci\'on del campo
sistema seg\'un la siguiente relaci\'on:
\begin{equation}
\langle \phi^2\rangle_{t^*}\sim \eta^2 = 6\mu^2/\lambda\,\,\,,
\label{deftsp}
\end{equation}
donde el promedio es sobre el campo sistema y $\eta$
es  el valor final del campo.
En la Ec.(\ref{deftsp}), se desprecian \footnote{Estamos suponiendo
que la din\'amica del sistema est\'a dominada por el modo
homog\'eneo, es decir $\phi(\vec{x},t) \sim \phi(t)=e^{\mu t}$} 
los modos de longitud de onda corta ($|k|
>\mu$) de $\phi$. Esto da $\langle \phi^2\rangle_{t^*}$ 
(de manera m\'as expl\'icita, y en sinton\'ia con los otros
cap\'itulos de esta Tesis, $ \langle
\phi^2\rangle_{t^*} = Tr\{\rho_{\rm r}\phi^2\}$ donde 
$\rho_{\rm r}$ es la matriz densidad reducida del campo 
sistema). De nuevo, cualquier ambig\"uedad que se presente
en $\mu t_{\cal D}$, es del orden de la unidad, equivalente
a la escala temporal en la cual los efectos no lineales
son importantes, en comparaci\'on al comportamiento
exponencial del campo libre \cite{karra}.
Es sabido que el efecto del entorno es inducir el
proceso de p\'erdida de coherencia en el sistema, y la
posterior clasicalizaci\'on del mismo. Por tanto, el comportamiento
cl\'asico del campo sistema se puede describir mediante las 
ecuaciones estoc\'asticas cl\'asicas de Langevin que \'este
satisface \cite{nunoray,nunobett}.  A la vez, es necesario remarcar, 
que estamos asumiendo un acoplamiento tal
 que el sistema no vuelva a tener coherencia (o ``recohera") 
hasta despu\'es del tiempo $t^*$ \cite{nuno}.

As\'i, para el acoplamiento bicuadr\'atico de la Ec.(\ref{Sint}), la
funcional de influencia est\'a dada por:
\[
{\rm Re} \delta A = \frac{g^2}{8} \int d^4 x\int d^4y ~ \Delta (x)
K(x-y) \Sigma (y), \]
\[
{\rm Im} \delta A = - \frac{g^2}{16} \int d^4x\int d^4y ~ \Delta
(x) N (x,y) \Delta (y), \]
donde $K(x-y) = {\rm Im} G_{++}^2(x,y) \theta (y^0-x^0)$  es
el n\'ucleo de disipaci\'on y $ N(x-y) = {\rm Re} G_{++}^2(x,y)$ 
el de ruido (difusi\'on). $G_{++}$ es la funci\'on de Green de
camino temporal cerrado del campo $\chi$ a temperatura 
$T_0$ \cite{fermazziray-npb}.
Hemos definido $\Delta =\frac{1}{2}(\phi^{+2} -
\phi^{-2})$ and $\Sigma =\frac{1}{2} (\phi^{+2} + \phi^{-2})$.
Buscamos soluciones cl\'asicas de la forma
\[\phi_{\rm cl}(\vec x, s) =  f(s,t)\Phi(\vec x),\]
donde, en principio, $f(s,t)$ verifique $f(0,t)= \phi_{\rm i}$ y
 $f(t,t) =\phi_{\rm f}$, mientras que  $\Phi(\vec x)$ indica
la configuraci\'on espacio del campo.

Comenzaremos mostrando que, a partir del uso de la Ec.(\ref{tD2}),
podemos obtener los resultados previamente obtenidos por los
autores en la Refs.\cite{fermazziray,fermazziray-npb}, pero 
de manera m\'as sencilla (evitando la ecuaci\'on maestra que es bastante
complicada en Teor\'ia de Campos). Ya que consideramos que $t_{\cal D}
< t^{*}$, nos resulta suficiente restringirnos a una condici\'on inicial definida
por un campo gaussiano libre de $(\rm masa)^2$ negativa. De hecho, 
para ecuaciones de Langevin ideales, esta suposici\'on resulta una buena
aproximaci\'on para la formaci\'on de dominios en el r\'egimen no 
lineal   \cite{moro}. Buscamos campos cl\'asicos que, luego de un cambio
s\'ubito de la temperatura global, tengan la forma \cite{fermazziray}
\begin{equation}
\phi_{\rm cl}^{\pm}(s,\vec x)= e^{\mu s}\phi_{\rm f}^{\pm}
\cos(k_0\, x)\cos(k_0\, y)\cos(k_0\, z),\label{cboard}
\end{equation}
donde $\phi_{\rm f}^\pm$ es la configuraci\'on final del campo. Esta
es una aproximaci\'on de un modo solo para un estructura regular
de dominio tipo {\it chequer-board}. Las longitudes de ondas cortas
pueden ser introducidas sin modificar significativamente los resultados
(para m\'as detalles ver \cite{fermazziray-npb}). Para un 
cambio brusco como el mencionado anteriormente, usaremos el
comportamiento a tiempos largos del modo homog\'eneo
 presente en el campo sistema ($k_0=0$), 
$\phi_{\rm cl}^{\pm}(s,\vec x)\sim e^{\mu s}\phi_{\rm f}^{\pm}$.
El factor exponencial, como es usual, se deriva del crecimiento
de los modos inestables de longitudes de ondas largas.

Por todo esto, ${\rm Im}\delta A[\phi^+_{\rm cl},\phi^-_{\rm cl}]$
toma la siguiente forma:
\begin{eqnarray}
{\rm Im}~\delta A = \frac{g^2 V T_c^2\pi}{64} \Delta_{\rm f}^2
\int_0^\infty \frac{dk}{(k^2 + \mu^2)^2} 
\times \frac{1 + e^{4\mu t} - e^{2\mu t} \cos{(2\sqrt{k^2 +
\mu^2})}}{(k^2 + 2 \mu^2)}, \label{ImS}
\end{eqnarray}
donde $\Delta_{\rm f} = 1/2 (\phi_{\rm f}^{+2} - \phi_{\rm
f}^{-2})$, y $T_c$ es la temperatura cr\'itica del entorno.
La presencia del volumen $V$, se debe a que estamos
considerando configuraciones de campo distrubu\'idas en
un volumen $V$ (ondas planas en cada direcci\'on).
 Este volumen se interpreta como el
volumen m\'inimo necesario en el cual no hay superposiciones
coherentes de estados macrosc\'opicamente distinguibles
del campo sistema. M\'as adelante, consideraremos 
configuraciones donde este factor no est\'a presente.

Luego de pedir que $\mu t \gg 1$, la integral en momento
puede ser resuelta anal\'iticamente, de manera de obtener
\begin{equation}
{\rm Im}~\delta A \sim \frac{g^2 V T_c^2\pi^2}{256}\frac{(3 -
2\sqrt{2})}{\mu^3} \Delta_{\rm f}^2 e^{4\mu t}. \label{ImS2}
\end{equation}
Una vez obtenida esta expresi\'on, estamos en condiciones
de evaluar el tiempo de decoherencia $t_{\cal D}$ para
una variaci\'on en la amplitud del campo, como
\begin{equation}
\mu t_D \sim \frac{1}{2} \ln\left\{\frac{16 \mu^{\frac{3}{2}}}{g
T_c \Delta_{\rm f} V^{\frac{1}{2}}\pi\sqrt{(3 -
2\sqrt{2})}}\right\}.
\end{equation}
Usando un valor conservativo para el volumen, $V = {\cal
O}(\mu^{-3})$, obtenemos
\begin{equation}
\mu t_D \sim \frac{1}{2} \ln\left\{\frac{16 \mu^{3}}{g T_c
\Delta_{\rm f}}\right\}\label{mutD}.
\end{equation}
Podemos re-escribir esta expresi\'on en t\'ermino de $\Delta_{\rm f} = \bar\phi
\bar\Delta /2$, con $\bar\phi = \phi_{\rm f}^+ + \phi_{\rm f}^-$
y, $\bar\Delta = \phi_{\rm f}^+ - \phi_{\rm f}^-$. Cuando la transici\'on
se termina, ${\bar\phi}^2 \simeq \eta^2 \sim
\lambda^{-1}$. A un tiempo $t_{\cal D}$, tomaremos el valor
${\bar\phi}^2 \sim {\cal O}(\mu^2\alpha/\lambda )$.
El valor de $\alpha$ debe ser determinado, considerando
que $\lambda < \alpha < 1$ y que a tiempo $t_{\cal D}$, 
$\langle\phi^2\rangle_t \sim \alpha\eta^2$. En un trabajo previo
\cite{fermazziray}, los autores mostraron que el
valor  de $\alpha$ se determina seg\'un $\alpha \approx \sqrt{\mu/T_c}$.
Asimismo, fijamos $\bar\Delta \sim 2 \mu$, es decir, no discriminamos
entre amplitudes de campo que difieren en ${\cal O}(\mu)$. 
$\mu^{-1}$ indica el grosor de las paredes de los
contornos de los dominios a medida que el campo
se establece en sus valores fundamentales. Finalmente,
por simplicidad, fijamos los acoplamientos $g \sim \lambda$.
Con esto, obtenemos la cota superior para el tiempo $t_{\cal D}$,
\begin{equation}
\mu t_D \sim \frac{1}{2}
\ln\left\{\frac{\eta}{T_c\sqrt{\alpha}}\right\},
\end{equation}
valor que coincide exactamente con \'aquel obtenido en
la Ref.\cite{fermazziray}, a partir del uso de la ecuaci\'on
maestra.

Si ahora, adem\'as, trazamos sobre los modos de longitud de
onda corta, como en \cite{ferdiego}, 
obtendr\'iamos un t\'ermino adicional en ${\rm Im} \delta A$,
cualitativamente similar a Ec.(\ref{ImS2}), que servir\'a
para preservar la condici\'on $t_{\cal D} < t^*$, haciendo que 
el tiempo $t_{\cal D}$ sea a\'un m\'as chico. En lo que sigue,
no consideraremos este entorno impl\'icito.

A modo comparativo, buscamos $t^*$ para el cual 
$\langle\phi^2\rangle_t \sim \eta^2$, definido en las 
Refs.\cite{fermazziray,fermazziray-npb}
\begin{equation}
\mu t^* \sim \frac{1}{2} \ln\left\{\frac{\eta}{\sqrt{\mu
T_c}}\right\},
\end{equation}
donde $\mu^{-1} < t_D < t^*$, con 
\begin{equation}
\mu t^* - \mu t_D \simeq \frac{1}{4}
\ln\left\{\frac{T_c}{\mu}\right\} > 1, \label{tD3}
\end{equation}
para un acoplamiento d\'ebil, \'o temperaturas iniciales
suficientemente elevadas.

A esta altura, hemos demostrado que el m\'etodo de 
la funcional de decoherencia nos da las mismas 
conclusiones que si hubi\'esemos resuelto la
ecuaci\'on maestra.  Por esto, nos dedicaremos a resolver
configuraciones de campo que resultar\'ian de un 
alto costo anal\'itico al ser encaradas desde el punto de
vista de la ecuaci\'on maestra.  En este contexto, 
nos ocuparemos de un dominio localizado en 
vez de configuraciones de  campo distribu\'idas 
en todo el  volumen $V$.

La orientaci\'on de esos dominios es irrelevante. De manera
m\'as sencilla, consideramos las soluciones de un dominio
cl\'asico, para el modo $k_0=0$, de la forma:
\begin{equation}
\phi^\pm (s,\vec x) = \phi_{\rm f}^\pm e^{\mu s}\tanh(\mu
x),\label{kink}
 \end{equation}
que relaciona dominios adyacentes.
Nuestro inter\'es primario es determinar el tiempo de 
decoherencia inducido por un peque\~no apartamiento
en las paredes del dominio, a trav\'es de la evaluaci\'on
de la acci\'on de influencia para las configuraciones
cl\'asicas del campo 
\begin{equation}
\phi^\pm (x,s) = \eta f(s,t) \Phi^\pm (x), ~~~~~{\rm donde} 
~~~~~\Phi^\pm (x) = \Phi (x \pm \delta/2)
\end{equation}
As\'i, tomamos $\delta$ como un peque\~no apartamiento
en la posici\'on de la pared, y consecuentemente, expandimos
la soluci\'on cl\'asica (o, equivalentemen $\Delta(x)$)
 en potencias de $\mu \delta$, a primer orden:
\begin{equation}
\Delta(s,\vec x) \approx \mu ~\delta ~\eta^2 ~e^{2\mu s}
~\tanh(\mu x) ~{\rm sech}^2(\mu x).
\end{equation}

De esta forma, la parte imaginaria de la acci\'on de influencia,
despu\'es de haber integrado en el tiempo y asumido que 
$\mu t \gg 1$, se puede escribir como:
\begin{eqnarray}
&&{\rm Im}\delta A \approx
\frac{g^2T_c^2\eta^4\delta^2\mu^2}{64(2\pi)^6}~ e^{4\mu t}~\int
d^3x
\int d^3y \int d^3k \int d^3p  ~~\frac{e^{-i(\vec p + \vec k)(\vec
x - \vec
y)}}{(k^2 + \mu^2)(p^2 + \mu^2)}  \nonumber \\
&\times &\frac{\tanh(\mu x)~{\rm sech}^2(\mu x)~\tanh(\mu y)~{\rm
sech}^2(\mu y)}{\left[(\sqrt{k^2 + \mu^2}+\sqrt{p^2 + \mu^2})^2 +
4 \mu^2\right]}\nonumber .\end{eqnarray}

Estas integrales pueden ser calculadas de manera exacta, aunque
en parte de manera anal\'itica y num\'erica, para obtener como
resultado
\begin{equation}
{\rm Im}\delta A \approx 0.2 \frac{g^2T_c^2\eta^4
L^2\delta^2}{1024} \frac{e^{4\mu t}}{\mu^2}, \label{imkinks}
\end{equation}
donde $L^2$ es un t\'ermino de superficie, an\'alogo al volumen
de la configuraci\'on anterior. Este factor se debe a que estamos
considerando una soluci\'on de un {\it kink} en una dimensi\'on,
pero encuadrada en un espacio de tres dimensiones. El
factor $L^2$ representa el espacio libre de dos dimensiones
del correspondiente {\it kink} en tres dimensiones. De manera
conservativa, podemos fijar $L ={\cal
O}(\mu^{-1})$, como la m\'inima escala espacial necesaria
en donde no hay superposiciones coherentes de estados
macrosc\'opicos del campo.

Para conocer el tiempo de decoherencia debido al apartamiento 
$\delta$, el cual llamaremos ${\bar t}_D$, asumimos $\mu {\bar t}_D
> 1$. As\'i,  ${\bar t}_D$ puede ser estimado de la \'ultima
ecuaci\'on, y por tanto, su  magnitud se define seg\'un,
\begin{equation}
\mu {\bar t}_D \sim \frac{1}{2} \ln\left\{\frac{74\mu}{g T_c
\eta^2 L \delta}\right\}.
\end{equation}
A simple vista, \'este resulta parece muy similar a todos aquellos
encontrados por los autores en la serie de trabajos mencionados.
La diferencia entre las amplitudes de campo est\'a reemplazada
por la distancia $\delta$ del desplazamiento del contorno del
dominio. Sin embargo, la diferencia principal est\'a en la
potencia de $\eta$ dentro del logaritmo. Si tomamos
$\delta = \gamma \mu^{-1}$, podemos acotar ${\bar t}_D$
de la siguiente manera
\begin{equation}
\mu {\bar t}_D \sim \frac{1}{2} \ln\left\{\frac{12\mu}{\gamma
T_c}\right\}.
\end{equation}

Como $T_c \gg \mu$ \footnote{De hecho, est\'a demostrado que $T_c^2/\mu^2 \sim
24/\lambda$ en 
\cite{fermazziray-npb}.}, este resultado restringe los posibles valores
de $\gamma$, de modo que se verifique $\mu {\bar t}_D > 1$,
es decir, $\gamma \leq \mu/T_c\ll 1$, y, por tanto, $\delta\ll \mu^{-1}$.
Si $\gamma$ es mayor, a tiempos muy cortos de nuestro modelo
($t \sim \mu^{-1}$), obtenemos r\'apidamente que ${\rm Im}~\delta A
> 1$, de la Ec.(\ref{imkinks}).  De esta manera, el sistema se
comporta cl\'asicamente 
desde un tiempo muy temprano de la evoluci\'on, 
ya que $\mu {\bar t}_D = {\cal O}(1)$.

Como el tama\~no estimado del  ``n\'ucleo" ({\it core}) del
dominio es del orden de 
${\cal O}(\mu^{-1})$, asumimos que $\delta \sim {\cal O}(\mu^{-1})$
 (es decir, $\gamma = {\cal O}(1)$) como la longitud
m\'inima donde no hay superposiciones coherentes de
estados macrosc\'opicos del campo. As\'i,
la decoherencia ocurre a partir del tiempo que satisface
\begin{equation}
\mu t^* - \mu {\bar t}_D > 0.
\end{equation}

A\'un, si pedimos que la decoherencia ocurra en escala
de una fracci\'on del ancho de la pared del dominio (es decir, 
$\gamma \ll 1$), podemos asegurar que
\begin{equation}
\mu t^* - \mu {\bar t}_D \sim \frac{1}{4}
\ln\left\{\frac{\gamma^2T_c}{\lambda \mu}\right\}  > 0.
\end{equation}
Este resultado implica que la decoherencia que ocurre
en un sistema debido al desplazamiento de los contornos
es muy anterior a las otras estudiadas (${\bar t}_D < t_D$).
Esto sugiere que, para el estudio de la decoherencia, resultan
de menor importancia las configuraciones de campo desplazadas
una peque\~na distancia que aquellas configuraciones extendidas, en las
cuales el campo fluct\'ua y se observan configuraciones de distinta
amplitud.

Lo que resulta m\'as interesante, es el hecho que si uno
realiza los c\'alculos en dimensi\'on $1+1$, el coeficiente $L$
aparece s\'olo en el primer ejemplo (es decir, una onda plana).
Por lo tanto, en lugar de considerar un dominio tipo {\it tablero
de ajedrez} como en los trabajos 
\cite{fermazziray,fermazziray-npb}, podemos
restringirnos \'unicamente a dos dominios adyacentes con
contornos dados por  la Ec.(\ref{kink}). Si consideramos el caso
sencillo de dos paredes separadas una distancia,
tal que $\phi^+_{\rm f}=\phi^-_{\rm f}-\epsilon$, donde $\epsilon$
representa una fluctuaci\'on peque\~na alrededor del valor final
de la configuraci\'on de campo. Suponiendo que el valor
final de la configuraci\'on es $\phi^+_{\rm f}=\sqrt{\alpha}
\eta$, con $\alpha$ el coeficiente autoconsistente mencionado
con anterioridad y $\epsilon= {\cal O}(\mu)$, podemos reobtener
el resultado de la Ec.(\ref{tD3}) de manera exacta. Nuevamente,
observamos que hay decoherencia antes que la transici\'on de
fase se complete. Adem\'as,  el campo tiene correlaciones cl\'asicas
a tiempos tempranos de la evoluci\'on (debido a la naturaleza
casi gaussiana del r\'egimen
 \cite{fermazziray,diana}) que completan el escenario
cl\'asico a tiempo $t^*$.

Por \'ultimo, la pregunta que surge es:
\textquestiondown qu\'e informaci\'on nos aporta este 
 an\'alisis \mbox{acerca} del comportamiento cl\'asico
de defectos topol\'ogicos, como por ejemplo v\'ortices,
cuyas \mbox{separaciones}  son una medida del tama\~no de los
dominios en circunstancias normales \cite{kibble2}? Si consideramos
que los v\'ortices tienen sus n\'ucleos en l\'ineas de ceros 
(de manera an\'aloga que los n\'ucleos de las paredes de los dominios), 
se ha demostrado en la Ref.\cite{rayfermazzi-plb} que el \mbox{mecanismo} de
producci\'on de v\'ortices cl\'asicos tiene varias partes. El entorno
elige los modos de longitud de onda largos del campo sistema
a tiempos cortos, anteriores o del orden del tiempo en el cual la
transici\'on de fase se completa. En particular, aquellos modos
que son del orden de la separaci\'on de la l\'inea de ceros (que caracterizan
los dominios cl\'asicos), perder\'an coherencia antes que la transici\'on
finalice, mientras que aquellos modos que son del orden del grosor
del v\'ortice cl\'asico no pierden coherencia nunca. Para que la
l\'inea de ceros se desarrolle en n\'ucleos de  v\'ortices, el campo
debe tener una energ\'ia proporcional a las soluciones de los
v\'ortices de las ecuaciones cl\'asicas de Euler-Lagrange.
Esto implica que las fluctuaciones del campo sean funciones
``picudas" alrededor de los modos de longitud de onda largas,
para evitar as\'i fluctuaciones que originen movimientos en los n\'ucleos
y la creaci\'on de peque\~nos bucles de v\'ortices. La densidad resultante
de la l\'inea de ceros puede ser deducida en el r\'egimen lineal.
El an\'alisis realizado en este cap\'itulo nos induce a pensar que
la decoherencia debida a desplazamiento de los v\'ortices es
irrelevante en comparaci\'on a la decoherencia debido a fluctuaciones
del campo.            
\cleardoublepage
\chapter*{Conclusiones}
\label{conclusiones}
\addcontentsline{toc}{chapter}{Conclusiones}

\markboth{Sistemas cu\'anticos bajo la influencia de condiciones
externas.}
{Conclusiones}


Esta Tesis tuvo
como objetivo principal estudiar el proceso de decoherencia en distintos sistemas
f\'isicos, con el prop\'osito de contribuir al estudio del rol de las fluctuaciones de vac\'io
en los procesos difusivos en sistemas cu\'anticos abiertos.

Inicialmente, hemos estudiado el Movimiento Browniano Cu\'antico (MBC) a
temperatura estrictamente cero. En particular, estudiamos los
efectos de las fluctuaciones puramente cu\'anticas del entorno sobre
el sistema, siendo \'estas la \'unica fuente de ruido presente.
Hemos calculado  los
coeficientes difusivos de la ecuaci\'on maestra de manera perturbativa 
y estimado los  tiempos de decoherencia, de manera
anal\'itica y num\'erica.  En este contexto, hemos 
mostrado que el
proceso de decoherencia a temperatura cero depende fundamentalmente
de los coeficientes de difusi\'on normal ${\cal D}(t)$ y an\'omalo $f(t)$ \cite{PLA}.
Tambi\'en, hemos extendido dicho an\'alisis al caso de entornos generales,
es decir supra\'ohmicos y sub\'ohmicos.  Hemos comparado
los distintos factores de ``decoherencia" para los casos
de temperatura alta y cero. En ese caso,
hemos mostrado que, 
 mientras el entorno supra\'ohmico 
a temperatura alta resulta muy efectivo induciendo p\'erdida
de coherencia, a  temperatura cero
no resulta un entorno difusivo, y por tanto, no hay p\'erdida de coherencia
(al menos significativa) \cite{PLA2}. 
Por el contrario, el entorno sub\'ohmico es muy efectivo induciendo
p\'erdida de coherencia en el sistema, tanto a temperatura alta como en el
caso que su temperatura sea cero.

Como paso siguiente, nos ocupamos del estudio de la activaci\'on
energ\'etica en entornos generales a temperatura arbitraria. En el
caso de  temperatura alta, relacionamos dicho fen\'omeno a la activaci\'on
t\'ermica. Sin embargo, hemos demostrado que este proceso ocurre
a\'un en ausencia de temperatura del entorno. 
El sistema se ``activa" energ\'eticamente, ya no inducido por la temperatura
del entorno sino por sus fluctuaciones cu\'anticas.  Mas a\'un,  hemos mostrado
que existe una relaci\'on muy estrecha entre la p\'erdida de coherencia
y la activaci\'on de la energ\'ia, ambos procesos inducidos por la interacci\'on
entre el sistema y el entorno. Los sistemas que sufren una mayor o total
p\'erdida de coherencia, son aquellos cuya energ\'ia tiene un 
crecimiento m\'as evidente.

Las fluctuaciones cu\'anticas del entorno, presentes en el estado inicial
del mismo, claramente deben jugar alg\'un papel relevante en la activaci\'on.
Estas fluctuaciones ya no son fluctuaciones de vac\'io del sistema total.
Igualmente,  resulta sorprendente que tengan tanta injerencia en la evoluci\'on del
sistema. La naturaleza puramente cu\'antica del entorno, la cual puede
ser correctamente despreciada en el caso de temperatura alta, da lugar
a efectos importantes en el caso del entorno a temperatura cero.
En t\'erminos de la ecuaci\'on maestra, las fluctuaciones cu\'anticas del
entorno generan t\'erminos difusivos no nulos, tanto normales ${\cal D}(t)$
como an\'omalos $f(t)$. Esto es particularmente cierto, en el caso del
coeficiente  $f(t)$. En el caso \'ohmico, hemos demostrado  en la
Secci\'on \ref{ohmcap2} que dicho 
coeficiente depende logar\'itmicamente de la frecuencia de corte $\Lambda$,
con lo cual puede ser de gran magnitud. Los efectos difusivos,
inducidos por las fluctuaciones cu\'anticas del entorno, son la mayor
diferencia con el caso de alta temperatura, y resultan responsables, 
a $T=0$ de la excitaci\'on energ\'etica de la part\'icula. A pesar que
este proceso es muy diferente a la ``activaci\'on t\'ermica", creemos
que a\'un as\'i puede ser interpretado en funci\'on de un escenario
cl\'asico modificado. La mayor dificultad para encontrarlo, es lograr 
que un ba\~no t\'ermico
cl\'asico simule las propiedades cu\'anticas de un entorno a $T=0$.

Posteriormente, aplicamos nuestros resultados acerca del proceso de 
decoherencia y excitaci\'on energ\'etica en el MBC, a un caso de 
mayor complejidad. Por tanto, hemos considerado un oscilador arm\'onico en un
pozo de potencial doble acoplado a un entorno de infinitos osciladores
arm\'onicos. Adem\'as de las escalas temporales conocidas, como
el tiempo de decoherencia $t_{\cal D}$ y el tiempo de excitaci\'on $t_{\rm act}$, 
aparece una nueva
escala temporal relevante: el tiempo de tuneleo $\tau$.
En s\'intesis, hemos descripto la din\'amica del sistema abierto en funci\'on
de tres fen\'omenos principales: p\'erdida de coherencia, efecto t\'unel
y excitaci\'on  energ\'etica. Tanto el primero como el \'ultimo proceso son
inducidos por la presencia del entorno, mientras que el segundo es de
naturaleza puramente cu\'antica. Hemos demostrado que el entorno 
puede ser modelado de modo de que la decoherencia inhiba el
efecto t\'unel ( $t_{\cal D} \ll \tau\ll t_{\rm act}$), 
aunque, de hecho, 
cualquier otra permutaci\'on es en principio posible entre los tres procesos que describen
la din\'amica del sistema \cite{dwPRE}. En este Cap\'\i tulo nos hemos concentrado 
en la jerarqu\'\i a que se mencion\'o antes. 
Para el caso del sistema cerrado, las simulaciones num\'ericas reprodujeron
todas las propiedades conocidas del efecto t\'unel. El an\'alisis entre el caso
abierto y el cerrado nos result\'o fundamental; por un lado, para
 entender que, cuando el
entorno est\'a a temperatura cero, las fluctuaciones cu\'anticas del entorno
tienen un rol importante en la evoluci\'on y, por el otro, 
 confirmar nuestro estudio de la Secci\'on \ref{ET0}: el sistema se excita energ\'eticamente para tiempos
posteriores al tiempo de p\'erdida de coherencia, a\'un en contacto con un 
entorno a $T=0$.

Por otra parte, resulta muy interesente la extensi\'on del estudio de la din\'amica
del sistema  al
caso de superposiciones macrosc\'opicas de estados cu\'anticos, como por ejemplo
condensados de Bose-Einstein en un pozo doble de potencial. Hemos extendido
nuestras consideraciones acerca  del efecto t\'unel, la decoherencia y la activaci\'on
al caso en el que la condici\'on
inicial en el sistema es una superposici\'on coherente de dos paquetes gaussianos
\cite{dwJCS}.
En los experimentos de BECs de gases at\'omicos dilu\'idos alkalinos-met\'alicos, los
\'atomos atrapados son enfr\'iados por evaporaci\'on e intercambian part\'iculas 
con el entorno (parte no condensada del sistema). La coherencia de BECs macrosc\'opicos 
cu\'anticos resulta en el
efecto t\'unel de \'atomos entre  modos, de manera an\'aloga al efecto t\'unel
en pares de Cooper de una juntura Josephson. \'Esta es un \'area en la que seguiremos trabajando.

En el Cap\'itulo \ref{c4} de esta Tesis, hemos estudiado el proceso de decoherencia
inducido por entornos compuestos  utilizando
el formalismo de integrales funcionales de Feynman y Vernon.
El entorno compuesto fue modelado por un oscilador (o antioscilador) acoplado a
un conjunto de osciladores arm\'onicos a temperatura alta. El subsistema principal, 
fue considerado un oscilador o antioscilador, seg\'un el caso analizado.
En este contexto, hemos mostrado que los osciladores arm\'onicos son
capaces de retener la informaci\'on por un per\'iodo de tiempo m\'as largo que los osciladores 
invertidos y, de esta forma,
generar difusi\'on en el subsistema $A$ m\'as eficientemente. 
Es importante remarcar que, para tener un modelo m\'as realista del entorno compuesto, uno
podr\'ia considerar un potencial de pozo doble para el subsistema $B$ en los casos donde este subsistema sea un oscilador invertido \cite{compositePRA, compositeModern}. 
Este tipo de potencial tiene las mismas caracter\'isticas que el oscilador invertido, pero la ventaja
es que su espacio de fases est\'a acotado, lo cual dar\'ia una mejor medida del efecto global inducido en
el subsistema $A$.  De todos modos, 
si mantenemos el an\'alisis a tiempos cortos de la evoluci\'on,
los potenciales son equivalentes.

En el Cap\'itulo \ref{c5} de esta Tesis, hemos estudiado distintos sistemas
f\'isicos donde aparecen fases ge\'ometricas. Cuando el sistema interact\'ua con un entorno, las fases 
originalmente geom\'etricas, pierden esta caracter\'istica, 
y  dependen de los par\'ametros del 
entorno.  Hemos abarcado sistemas muy diversos.
En particular, estudiamos el ``dephasing" inducido
por la aleatoridad en la emisi\'on de las part\'iculas en un experimento
de interferencia con part\'iculas neutras con momento dipolar permanente.
Esta incerteza experimental, se traduce en una fase exponencial decreciente
que tiende a eliminar las coherencias cu\'anticas del sistema, y por tanto,
genera una reducci\'on en la visibilidad del patr\'on de interferencia.
Hemos utilizado datos experimentales para estimar este factor de
decoherencia en el caso de part\'iculas cargadas, neutras y fulerenos.
Los resultados obtenidos, resultaron particularmente interesantes en el
caso de los fulerenos.  Estos 
sistemas mesosc\'opicos est\'an 
compuestos de un n\'umero grande de \'atomos,
pero a\'un as\'i, pueden ser descriptos por una funci\'on de onda.
En este contexto, hemos  estimado el factor de decoherencia 
utilizando los valores experimentales de dichos sistemas. A pesar de que
puede ser considerado un modelo muy sencillo para estas mol\'eculas, nos
 di\'o una estimaci\'on cuantitativa de la magnitud de este efecto
en sistemas neutros muy masivos. Con estos valores, y considerando la trayectoria 
asim\'etrica propuesta en \cite{Casher}, el
factor de decoherencia estimado para los fulerenos,  la 
destrucci\'on completa del patr\'on
de interferencia puede ser observable \cite{CasherJPA}. De hecho, 
la atenuaci\'on de dicho
patr\'on fue observada experimentalmente en el a\~no 2003.
De esta forma, hemos aplicado nuestro modelo de {\it dephasing} a los
datos experimentales reportados en dicho experimento durante
la observaci\'on del patr\'on de interferencia de fulerenos $C_{70}$
\cite{fringeModern, fringeJCS}.
As\'i, hemos podido observar que el patr\'on de interferencia 
se aten\'ua cuando el sistema es abierto. Los efectos incoherentes 
contenidos en el factor de decoherencia son suficientes 
para reproducir los efectos del entorno en el patr\'on de 
interferencia de un experimento verdadero con fulerenos. 

En cuanto a la fase geom\'etrica cu\'antica que adquiere el
modelo esp\'in-bos\'on estudiado en el Cap\'itulo \ref{c5},
hemos demostrado c\'omo se ve afectada dicha fase por la presencia 
del entorno.  Para medir
 la fase geom\'etrica del sistema, la fase din\'amica debe 
ser eliminada, ya sea utilizando la t\'ecnica de {\it spin echo} 
para espines en campos magn\'eticos, o bien, 
utilizando el transporte paralelo y asegurarse, as\'i, que la 
fase din\'amica es cero a todo tiempo. A pesar que el 
estado no adquiere una fase localmente, s\'i la adquiere 
globalmente, la cual es igual a la fase geom\'etrica en el 
caso del sistema cu\'antico aislado. En el caso de sistemas 
abiertos hemos demostrado c\'omo se modifican estas fases
 en funci\'on de los par\'ametros que caracterizan al entorno. 
Consideramos que 
 estas estimaciones resultan importantes a la hora de 
planificar un experimento de medici\'on de fases geom\'etricas 
para estados mixtos en evoluciones no unitarias,
particularmente interesante en el marco de la  
Computaci\'on Cu\'antica. Seg\'un nuestro an\'alisis, 
podemos afirmar que el mejor montaje experimental 
para lograr la medici\'on de esta fase es el entorno 
supra\'ohmico en el caso de temperatura cero y r\'egimen 
subamortiguado \cite{fasesPRA}.  En la actualidad continuamos 
trabajando sobre generalizaciones de estos modelos para determinar 
c\'omo se altera el car\'acter geom\'etrico de la fase modificada por 
distintos tipos de entornos.

Finalmente, en el Cap\'itulo 6 de esta Tesis, hemos estudiado el 
proceso de decoherencia en Teor\'ia Cu\'antica de Campos, 
haciendo uso del m\'etodo de la funcional de decoherencia.
Este m\'etodo  nos permiti\'o reobtener resultados conocidos
de manera m\'as sencilla \cite{PLB} ya que 
hemos demostrado que el m\'etodo de 
la funcional de decoherencia nos da las mismas 
conclusiones que si hubi\'esemos resuelto la
ecuaci\'on maestra. 

Calculamos el tiempo de decoherencia para configuraciones de 
campo con ruptura de la simetr\'ia. El aporte m\'as significativo 
de este Cap\'itulo radica en el c\'alculo y comparaci\'on entre los 
tiempos de decoherencia obtenidos para diferentes configuraciones cl\'asicas de 
campo (defectos). Por ejemplo, mostramos que el tiempo de decoherencia asociado 
a configuraciones de campo de kinks de igual amplitud, pero desplazadas 
en posici\'on, es notablemente menor que el tiempo de decoherencia asociado 
a configuraciones que difieren s\'olo en fluctuaciones sobre la 
amplitud del campo. Esto es de relevancia para el an\'alisis del n\'umero 
de defectos topol\'ogicos y su evoluci\'on durante transiciones de fase 
cu\'anticas, y el rol que en el contexto del mecanismo de Kibble-Zurek, 
tienen las fluctuaciones cu\'anticas del entorno.

Con todo esto, podemos concluir que hemos estudiado el
proceso de decoherencia en diversos sistemas f\'isicos.
Para ello, hemos usado tanto t\'ecnicas funcionales como
estad\'isticas, resolviendo la din\'amica de los sistemas 
ya sea, de manera perturbativa o exacta, ana\'\i tica y num\'erica.  Incluso, hemos acoplado
nuestros sistemas a entornos tanto cl\'asicos como puramente cu\'anticos.
A\'un quedan
muchos temas abiertos y preguntas sin contestar, que forman 
parte de nuestros planes de trabajo futuros.             
 \cleardoublepage
 \newpage
\thispagestyle{empty}
 \cleardoublepage
  \newpage
\appendix 
\label{AA}
\addcontentsline{toc}{chapter}{Ap\'{e}ndice}


\chapter{Representaci\'on matem\'atica de dos paquetes gaussianos}
\label{dospaquetes}

Para estudiar el proceso de p\'erdida de coherencia, usaremos
una soluci\'on gaussiana para resolver la ecuaci\'on maestra 
Ec.(\ref{master}). La matriz de densidad  que propondremos es
 \cite{Joos}
\begin{eqnarray}
\rho_{\rm r}(x,x',t)&=& e^{-N(t)} \exp \bigg\{- \bigg[ A(t) (x-x')^2 
+ i B(t)(x+x')+ C(t) (x+x')^2 \nonumber \\
&+& i K(t)(x-x') + E(t)(x+x') \bigg] \bigg\}, \label{1gaus}
\end{eqnarray}
donde $e^{-N(t)}$ asegura la conservaci\'on de la traza, $A(t)$ 
describe el rango de coherencia y $C(t)$  especifica la extensi\'on
del sistema en el espacio de coordenadas. Todos los coeficientes
$A(t),~B(t),$...,$E(t)$, dependen del tiempo y son funciones
reales (para que la matriz sea herm\'itica). En nuestro caso,
estudiaremos la evoluci\'on din\'amica de dos paquetes gaussianos,
inicialmente localizados en $x=\pm L_0$. Por lo tanto, debemos
reemplazar $x\rightarrow x+L_0$ y $x\rightarrow x-L_0$  
en la Ec.(\ref{1gaus}) y superponer ambas soluciones para representar
as\'i la din\'amica de una superposici\'on lineal de
dos paquetes de ondas inicialmente deslocalizados, como anal\'iticamente
estudiamos en la Secciones \ref{decoohm} y  \ref{deconoohm}.

La soluciones que usaremos entonces es,
\begin{eqnarray}
\rho_{\rm r}(x,x',t)&=& 2 e^{-N(t)} e^{-4 L_0^2 C(t)}
\exp \bigg\{ - \bigg[ A(t) (x-x')^2 
+ i B(t)(x^2-x'^2)+ C(t) (x+x')^2 \nonumber \\
&+& i K(t)(x-x') + E(t)(x+x') \bigg] \bigg\}
  \bigg(  \cosh [4 L_0 C(t) (x+x')
- i 2 L_0 B(t) (x-x')  \nonumber \\
&-& 2 L_0 E(t)]
+ e^{-4 L_0^2 (A(t)-C(t))} \cosh [4 L_0 A(t) (x-x') 
+ i 2 L_0 B(t) (x+x') \nonumber \\
&-&  i 2 L_0 K(t)] \bigg).
\label{2gaus}
\end{eqnarray}

Esta soluci\'on, al ser reemplazada en la ecuaci\'on maestra
(Ec.(\ref{master})), conduce a un conjunto de ecuaciones
diferenciales acopladas para todos los coeficientes
dependientes del tiempo $A(t),~B(t),$..., $E(t)$,
\begin{eqnarray}
\dot{A}(t) &=& \frac{4}{M} + {\cal D}(t) - 2 A(t) \gamma(t) +
 2 B(t) f(t) \nonumber \\
\dot{B}(t) &=& \frac{2}{M} (B(t)^2- 4 A(t) C(t))
+ \frac{M}{2} (\Omega^2 +
\delta \Omega^2(t)) 
-B(t) \gamma(t) - 4 C(t) f(t) \nonumber \\
\dot{C}(t) &=& \frac{4}{M} B(t) K(t) \nonumber \\
\dot{K}(t) &=& \frac{2}{M} (B(t) K(t) - 2 A(t) E(t))
- 2 \gamma(t) K(t)
- 4 f(t) E(t) \nonumber \\
\dot{E}(t) &=& \frac{2}{M} (2 C(t) K(t) + B(t) E(t)),
\label{sistema}
\end{eqnarray}
donde el punto representa la derivada parcial $\partial_t$
y $\delta \Omega^2(t)$,
$\gamma(t)$, ${\cal D}(t)$ y $f(t)$ son los coeficientes de
la ecuaci\'on maestra definidos en el conjunto de ecuaciones de
la Ec.(\ref{coefdef}), para cada tipo de entorno considerado 
 ($n=1/2$, $n=3$ y $n=1$). Pidiendo que 
$\rm{Tr}~{\rho_r}=1$ para todo tiempo, obtenemos el valor
de la normalizaci\'on $N(t)$ como funci\'on del tiempo y de los
otros coeficientes de la soluci\'on (\ref{2gaus}),
\begin{equation}
e^{-N} =
\frac{\sqrt{C/\pi} ~ e^{-{\frac{E^2}{4 C}}}}{
\cosh(4 L_0 E) + e^{- L_0^2 \frac{B^2}{C}-4 L_0^2 A}
\cos(L_0 E \frac{B}{C}+ 2L_0 K)}, \nonumber
\end{equation}
donde hemos omitido escribir la dependencia expl\'icita del tiempo
de los coeficientes $A(t)$, $B(t)$,..., $E(t)$ por simplicidad. Es importante
destacar que esta expresi\'on se convierte en el valor 
de la normalizaci\'on calculado por los autores en \cite{Joos}, en el caso
de un \'unico paquete gaussiano, es decir $x_0=0$ y $L_0=0$
 ($e^{-N}=1/2\sqrt{C/\pi}$).

En el Cap\'itulo 2 de esta Tesis, resolvimos num\'ericamente
este conjunto de ecuaciones Ec.(\ref{sistema})  para los distintos
tipos de entornos presentados, es decir \'ohmicos y no \'ohmicos,
 en el l\'imite de temperatura alta y tambi\'en
cuando la temperatura de los mismos es cero. Para ello, utilizamos
un m\'etodo standard de Runge-Kutta, de paso variable, de orden 5
para distintas condiciones iniciales de los coeficientes 
$A(0)$, $B(0)$, $C(0)$, $K(0)$, $E(0)$ (inicialmente 
$A(0)=C(0)=1/\sqrt{\sigma_{x0}^2}$). Con esta  programa calculamos
las cantidades necesarias para el an\'alisis num\'erico del proceso
de p\'erdida de coherencia y excitaci\'on energ\'etica del sistema de
prueba (como por ejemplo, la Entrop\'ia Lineal y la Energ\'ia media
del  sistema). Todos los resultados presentados son robustos frente
a cambios en los par\'ametros de integraci\'on del m\'etodo num\'erico
utilizado.

\section{P\'erdida de coherencia}

Observando la soluci\'on propuesta de la Ec.(\ref{2gaus}), podemos
ver que el t\'ermino que mide las interferencias, es decir que
proviene de $\rho_1(x,x') \rho_2(x',x) + \rho_2(x,x') \rho_1(x',x)$,
es el segundo t\'ermino de dicha expresi\'on y 
tiene una exponencial decreciente multiplicando. Por lo tanto, si
la cantidad $L_0^2(A(t)-C(t))$ resulta positiva, entonces, las interferencias
tender\'an a desaparecer y la cantidad $\Gamma(t)=\exp(-4 L_0^2 (A(t)-C(t)))$
puede ser usada como medida de p\'erdida de coherencia.


\section{Entrop\'ia Lineal y Energ\'ia media del sistema}
\label{Aentropia}

La expresi\'on de la Entrop\'ia Lineal se define como
\begin{equation}
Sl(t) = \int_{-\infty}^{\infty} dx \int_{-\infty}^{\infty} dx' \rho_{\rm r}(x,x',t)
\rho_{\rm r}(x',x,t)
\end{equation}
Usando como matriz densidad reducida la soluci\'on de la
Ec.(\ref{2gaus}), la Entrop\'ia Lineal queda, como funci\'on 
de los coeficientes $A(t)$, $B(t)$,..., $E(t)$,
\beqa 
Sl(t)&=&
\bigg(\exp(- N(t))\bigg)^2  \frac{\pi}{\sqrt{A(t) C(t)}} \times
 \bigg\{ \frac{1}{8} \bigg(2 + \exp(-8 L_0^2 \frac{B(t)^2}{A(t)})
-8 L_0^2 C(t)) \nonumber \\
&+& \exp(-8 L_0^2 A(t)) 
+ \frac{1}{2}  \exp(-6 L_0^2 (A(t)+C(t))) \cos(4 L_0^2 B(t)) \bigg) \bigg\}.
 \nonumber
\eeqa

La expresi\'on para la Energ\'ia media del sistema es
\[ \langle E(t) \rangle=\frac{1}{2M} \langle p^2(t) \rangle
+ \frac{1}{2}M \bigg(\omega^2+ \delta \Omega^2(t)\bigg) \langle
x^2(t) \rangle \]  Recordando la definici\'on del valor medio
usando la matriz densidad ($<x^2>=\int_{-\infty}^{\infty} dx \rho(x,x) x^2$),
obtenemos para $<x^2>$  y $<p^2>$,
\beqa 
\langle x^2(t) \rangle &=&
\frac{e^{-N(t)}}{16 C(t)^2} \sqrt{\frac{\pi}{C(t)}}
\times \bigg\{ C(t) + 8 L_0^2 C(t)^2 
+ \bigg[ C(t) + 2 B(t)^2 L_0^2 \bigg] \times \nonumber \\
& & \exp \bigg(-L_0^2 [4 A(t) +\frac{B(t)^2}
{C(t)}] \bigg) \bigg\} \nonumber
\eeqa
y
\beqa
\langle p^2(t) \rangle &=& \frac{e^{-N(t)}}{2 C(t)} \sqrt{\frac{\pi}{C(t)}}
\times
 \bigg\{ \bigg[ 4 A(t) C(t) + B(t)^2 + 32 B(t)^2 C(t) L_0^2 \bigg] \nonumber \\
&+& \exp\bigg(-L_0^2 (\frac{4 A(t) C(t)+B(t)^2}{C(t)})\bigg)
\bigg(\frac{-2 L_0^2}{C(t)} [B(t)
+ 4 A(t) C(t)]^2 \nonumber \\
&+& B(t)^2+4C(t) A(t) \bigg) \bigg\}. \nonumber
\eeqa
Con estas expresiones es posible recuperar aquellas soluciones
para una gaussiana cuando $x_0=0$ y $L_0=0$ \cite{Joos}.

\newpage
\chapter{Coeficientes de difusi\'on 
y tiempos de p\'erdida de coherencia}

\section{Supra\'ohmico en el l\'imite de  temperatura alta}

Para poder calcular los coeficientes de la ecuaci\'on maestra Ec.(\ref{master})
cuando la part\'icula de prueba est\'a acoplada a un entorno supra\'ohmico,
suponiendo que este \'ultimo se encuentra a temperatura alta, 
es decir $\hbar \Lambda \ll k_B T$,
usaremos la misma densidad espectral definida en la Ec.(\ref{densidadnoohm}).
Usando las definiciones de la Ec.(\ref{coefdef}), obtenemos

\beqa
{\cal D}(t)_{n=3}&=&  \frac{2M  \gamma_0 k_B T}{2 \pi (\Lambda
+ \Lambda^3 t^2)^2} \bigg\{ 4 \Lambda^3 t
\cos(\Omega t)
+ (1+\Lambda^2 t^2) \Omega \bigg[ -2 \Lambda \sin(\Omega t) \nonumber \\
&+& (1+\Lambda^2 t^2) \bigg(-i \cosh(\frac{\Omega}{\Lambda})
 [{\rm Ci}(\Omega(t-\frac{i}{\Lambda}))
- {\rm Ci}(\Omega(t+\frac{i}{\Lambda}))-i\pi ] \nonumber \\
&-& \sinh(\frac{\Omega}{\Lambda})[{\rm Shi}
(\Omega(t-\frac{i}{\Lambda})) +{\rm Shi}
(\Omega(t+\frac{i}{\Lambda}))] \bigg) \bigg]\bigg\} 
\label{DsupHT} \eeqa
y \beqa f(t)_{\rm n=3} &=& 2M k_B T  \frac{\gamma_0}{\pi}
\bigg\{-\frac{\Omega \cos(\Omega t)}{\Lambda + \Lambda^3 t^2} -
\frac{2 \Lambda t \sin(\Omega t)}{ (1+\Lambda^2 t^2) }
+ \frac{1}{2 \Lambda^2} \bigg[ \Omega \bigg(2 \Lambda  \label{fsupHT}\\
&-&\Omega \bigg({\rm Ci}
(\Omega(t-\frac{i}{\Lambda}))
+{\rm Ci}(\Omega(t+\frac{i}{\Lambda})) - {\rm Ci}(\frac{-i \Omega}{\Lambda})
-{\rm Ci}(\frac{i \Omega}{\Lambda})\bigg) \sinh(\frac{\Omega}{\Lambda}) \nonumber\\
&-& i \Omega \cosh(\frac{\Omega}{\Lambda}) \bigg(-2 i {\rm Shi}
(\frac{ \Omega}{\Lambda}) -  {\rm Shi}(\Omega(t-\frac{i}{\Lambda}))
+ {\rm Shi}(\Omega(t+\frac{i}{\Lambda})) \bigg) \bigg) \bigg] \bigg\}.\nonumber 
\eeqa

Los dos coeficientes que se derivan del n\'ucleo de disipaci\'on,
$\delta \Omega^2_{n=3} (t)$ and $\gamma(t)_{n=3}$, son similares
a los que se calcularon en la Secci\'on \ref{nonohmcap2}, para un entorno
supra\'ohmico a temperatura cero, ya que este n\'ucleo no depende de la
temperatura del entorno (a diferencia del n\'ucleo de ruido).
Con los coeficientes difusivos se puede estimar
el coeficiente de visibilidad $A_{\rm int}$ (Ec.(\ref{aintcap2})) como
hemos realizado  para los
entornos a temperatura cero. Con los coeficientes de las Ecs.(\ref{DsupHT})
y (\ref{fsupHT}), se puede estimar  dicho
coeficiente en el caso que $\Lambda t \ll 1$, seg\'un
$A_{\rm int}\approx (2 M k_B T L_0^2\gamma_0) \Lambda t^2$.
El tiempo de p\'erdida de coherencia $t_D$ se calcula f\'acilmente,
pidiendo $A_{\rm int}(t_D)\sim 1$. De esta forma, se obtiene
\begin{equation}
t_D\sim \frac{1}{\Lambda}
\sqrt{\frac{\Lambda}{(2 M k_B T L_0^2\gamma_0)}}.
\label{tdsupHT}
\end{equation}
Sin embargo, si se cumple que $\Lambda t \geq
1$,  el coeficiente de visibilidad puede ser estimado como
$A_{\rm int}\approx 2M k_B T L_0^2\gamma_0/\Lambda$.
Sorprendentemente,  los efectos difusivos  s\'olo ser\'an importantes
en este caso si $2M k_B T L_0^2\gamma_0 \gg \Lambda$,
es decir si $A_{\rm int} \geq 1$, lo cual deber\'ia ocurrir en un tiempo
$t \leq 1/\Lambda$ (Ec.(\ref{tdsupHT})). Adem\'as, si $\gamma_0$ 
es suficientemente chica, tal que $k_B T L_0^2\gamma_0 < \Lambda$, 
entonces
el coeficiente de visibilidad $A_{\rm int}$ nunca ser\'a de orden uno
y los efectos de p\'erdida de coherencia en este caso no ser\'an 
importantes. 
\begin{figure}[!h]
\includegraphics[width=12cm]{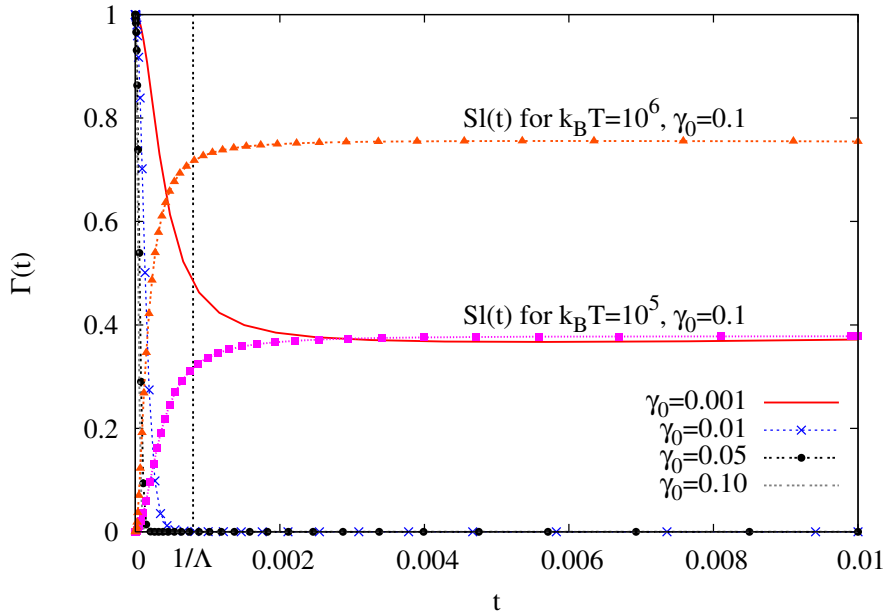}
\caption{Factor de p\'erdida de coherencia $\Gamma(t)$ del sistema
Browniano acoplado a un entorno supra\'ohmico en el l\'imite
de temperatura alta. Los par\'ametros est\'an medidos en unidades de 
la frecuencia natural $\Omega$. Consideramos los casos
$k_BT=10^5$, $\Lambda=2000$, $\Omega=0.1$,
$L_0=2$ para distintos valores de $\gamma_0$. Adem\'as, se presenta
el comportamiento de la Entrop\'ia lineal $Sl(t)$ para el valor m\'as
grande de $\gamma_0$, tanto para $k_BT=10^5$ como $k_BT=10^6$.
Este entorno es particularmente dependiente del valor de los par\'ametros
del modelo y no resulta tan efectivo induciendo efectos difusivos en 
el sistema como el entorno \'ohmico.}
\label{figsupapb}
\end{figure}
Podemos verificar nuestras estimaciones anal\'iticas
con la ayuda del la Figura \ref{figsupapb}.
Las interferencias son efectivamente destru\'idas s\'olo
para valores relativamente ``grandes" de $\gamma_0$, si recordamos
que trabajamos en el r\'egimen subamortiguado. Por ejemplo, en la
Fig.\ref{figsupapb}, se puede observar que, si $\gamma_0=0.001$,
el coeficiente $\Gamma(t)$ alcanza un valor final de $0.4$, mientras que,
si $\gamma_0=0.1$, este mismo coeficiente se hace finalmente cero
en una escala temporal $t_D \sim 1/(\sqrt{ \Lambda M\gamma_0k_B
T})$, implicando que la p\'erdida de coherencia es total en este \'ultimo
caso. Como mencionamos en la Secci\'on \ref{nonohmcap2}, el
proceso de p\'erdida de coherencia se debe \'unicamente a la ``patada"
inicial, despu\'es de la cual, el sistema permanece constante,
como sugiere la Figura. All\'i mismo, las curvas que muestran
un comportamiento creciente, simulan la Entrop\'ia Lineal del
sistema  para los valores $\gamma_0=0.1$ cuando 
$k_BT=10^5$ y $k_B T=10^6$, respectivamente. 
En estos casos, el coeficiente $\Gamma(t) \approx 0$ para
tiempos del orden $t <1/\Lambda$. 
 Si observ\'asemos el patr\'on de interferencias
para los valores de la Fig.\ref{figsupapb} y $\gamma_0=0.1$,
no observar\'iamos franjas de interferencia. Sin embargo, si hici\'eramos
lo mismo para el caso en que $\gamma_0=0.001$ (manteniendo fijos
los otros par\'ametros), entonces, las interferencias estar\'ian apenas
atenuadas, pero seguir\'ian existiendo. Como, por lo que ya demostramos,
el sistema luego permanece constante, las interferencias nunca desaparecer\'ian
para este caso.

\section{Sub\'ohmico en el l\'imite de  temperatura alta}

Para el caso sub\'omico los coeficientes difusivos, en el l\'imite de 
alta temperatura son y $\Lambda/\Omega > 1$, se obtiene

\beqa {\cal D}(t)_{n=1/2}&\simeq& 4M k_B T  \frac{\gamma_0}{\pi}
\bigg({\rm Si}((\Lambda -\Omega)t) +{\rm Si}((\Lambda+\Omega)t)
\bigg)  \label{Dsubcap2} \eeqa y \beqa f(t)_{\rm n=1/2} &\simeq& 2M k_B T
\frac{\gamma_0}{\pi} \bigg( {\rm Ci}((\Lambda
-\Omega)t) -{\rm Ci}((\Lambda+\Omega)t) \nonumber \\
&-&\log(\Lambda-\Omega) + \log(\Lambda + \Omega) \bigg). 
\label{fsubcap2}
\eeqa

El factor de visibilidad, en el caso $\Lambda \gg \Omega$, puede ser
f\'acilmente estimado a partir de los coeficientes y la Ec.(\ref{aintcap2})
$A_{\rm int} \approx 32/\pi ~M L_0^2
k_BT\gamma_0 t ~{\rm Si}(\Lambda t)$. De esta forma, el tiempo
de p\'erdida de coherencia que se obtiene es
\begin{equation}
t_D \sim 1/(16M L_0^2 K_B T)
\end{equation}

\begin{figure}[!ht]
\includegraphics[width=12cm]{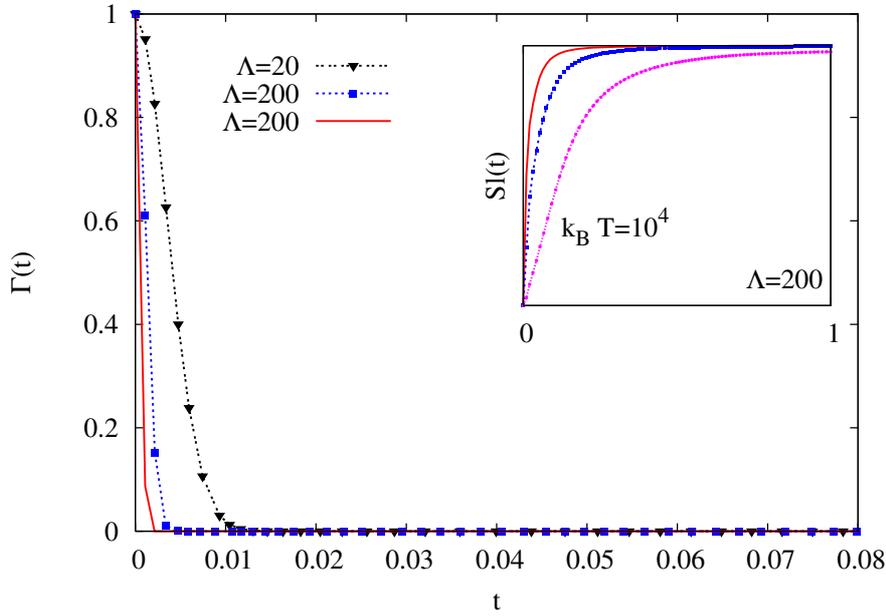}
\caption{Evoluci\'on temporal de la exponencial decreciente $\Gamma(t)$
del sistema acoplado a un entorno sub\'ohmico a temperatura alta.
Todos los par\'ametros est\'an medidos en unidades de la frecuencia
natural del sistema $\Omega$. Consideramos los casos 
$\Omega=0.1$, $L_0=2$, $\gamma_0=0.001$ (y $\gamma_0=0.005$
en el caso de la l\'inea s\'olida del gr\'afico), $k_B T=10^5$ 
(en el caso indicado $k_B T=10^4$) para distintos valores de 
la frecuencia de corte $\Lambda$. En el recuadro, graficamos la 
evoluci\'on temporal de la Entrop\'ia lineal $Sl(t)$ para un valor fijo de 
la frecuencia de corte $\Lambda=200$ y distintos valores de $k_B T$
y $\gamma_0$.} 
\label{figsubapb}
\end{figure}

En la  Fig.\ref{figsubapb} presentamos la evoluci\'on temporal
del factor de p\'erdida de coherencia $\Gamma(t)$ para diferentes
valores de $\gamma_0$ y $\Lambda$. La dependencia con estos
valores es similar a la registrada en todos los casos anteriores: cuanto 
m\'as grande es el acoplamiento con el entorno y m\'as grande
la frecuencia de corte, m\'as r\'apido se hace cero el coeficiente 
$\Gamma(t)$ (l\'inea s\'olida
para $\Lambda=200$ y $\gamma_0=0.005$). En el recuadro
de la Figura, mostramos la evoluci\'on temporal de la Entrop\'ia lineal
$Sl(t)$ para dos valores del gr\'afico principal 
($\gamma_0=0.005$ y $\gamma_0=0.001$, siempre a $k_B
T=10^5$) y una curva adicional para el caso $\gamma_0=0.001$,
$k_B T=10^4$. Es f\'acil notar en dicha Figura, que la Entrop\'ia
lineal alcanza un estado de saturaci\'on m\'as temprano cuando
la temperatura del ba\~no y el acoplamiento con el mismo
son mayores.

 \cleardoublepage
  \newpage
\thispagestyle{empty}
 \cleardoublepage
  \newpage
\thispagestyle{empty}
\cleardoublepage

\cleardoublepage
\newpage
%


\addcontentsline{toc}{chapter}{Bibliograf\'ia}
\begin{thebibliography}{99}
\bibitem{Zurek1} W. H. Zurek, Phys. Rev. {\bf D 24}, 1516 (1981).
\bibitem{Zurek2} W. H. Zurek, Phys. Rev. {\bf D 26}, 1862 (1982).
\bibitem{Zurek3} W. H. Zurek, Physics Today {\bf 44}, 36 (1991)
\bibitem{Feynman} R. Feynman y F. Vernon, Ann. Phys. (N. Y.) 
{\bf 24}, 118 (1963).
\bibitem{SternAhaImry} A. Stern, Y. Aharonov y Y. Imry,
Phys. Rev. {\bf A 41}, 3436 (1990).
\bibitem{GraIng} H. Grabert, P. Schram y G.L. Ingold, Phys. Rep. 
{\bf 168}, 115 (1988).
\bibitem{HuPazZhangI} B.L. Hu, J.P. Paz y Y. Zhang, Phys. Rev. 
{\bf D 45}, 2843 (1993).
\bibitem{Buttiker} K. E. Nagaev y  M. Buttiker, Europhys. Lett. {\bf 58} (4), 475 (2002).
\bibitem{Jordan} A. N. Jordan y M. Buttiker, Phys. Rev. Lett {\bf 92}, 247901
(2004).
\bibitem{PLA} Fernando C. Lombardo y Paula I. Villar, Phys. Lett.
{\bf A  336}, 16 (2005).
\bibitem{PLA2} Fernando C. Lombardo y Paula I. Villar, Phys. Lett.
{\bf A  371}, 190 (2007).
\bibitem{Dounas} D. R. Dounas-Frazer, A. M. Hermundstad y  L. D. Carr,
Phys. Rev. Lett. {\bf 11}, 200402 (2007).
D. R. Dounas-Frazer, A. M. Hermundstad  y L. D. Carr, {\it Preprint} quant-ph/0610166
(2006).
\bibitem{Shin2} Y. Shin, G. B. Jo, M. Saba, T. A. Pasquini, W. Ketterle y  
D. E. Pritchard, Phys. Rev. Lett. {\bf  95}, 170402 (2005).
\bibitem{dwPRE} Nuno D. Antunes, Fernando C.
Lombardo,  Diana Monteoliva y Paula I. Villar, Phys. Rev.
{\bf  E 73}, 066105 (2006).
\bibitem{dwJCS} Fernando C. Lombardo,  Diana Monteoliva
y  Paula I. Villar, Journal of Physics: Conference Series {\bf 67}, 012067 (2007)
\bibitem{ChouYuHu} C. H. Chou, T. Yu y B. L .Hu, {\it Preprint}: quant-ph/
0703088v1 (2007).
\bibitem{compositePRA} Fernando C. Lombardo y Paula I. Villar, Phys.Rev {\bf A 72},
034103 (2005).
\bibitem{compositeModern}Fernando C. Lombardo y Paula I. Villar, 
Int.J.Mod.Phys. {\bf B 20},  2951(2006).
\bibitem{Casher} Fernando C. Lombardo, Franciso D. Mazzitelli y
Paula I. Villar, Phys. Rev. {\bf A 72}, 042111 (2005).
\bibitem{CasherJPA} Fernando C. Lombardo y
Paula I. Villar, J. Phys. {\bf A 39}, 1-8  (2006).
 \bibitem{fasesPRA} Fernando C. Lombardo y
Paula I. Villar, Phys. Rev.  {\bf A 74},
 042311 (2006).
\bibitem{fringeModern} Fernando C. Lombardo y
Paula I. Villar, International Journal of Modern Physics B, en prensa (2007).
\bibitem{fringeJCS} Paula I. Villar y Fernando C. Lombardo, Journal
of Phys.: Conference Series {\bf 67}, 012041 (2007).
\bibitem{ferdiego} F.  Lombardo y F. D.
Mazzitelli, Phys.Rev. D {\bf 53},  2001 (1996). 
\bibitem{fer97} Fernando C. Lombardo, Francisco D. Mazzitelli,  
Phys. Rev. {\bf D 55},  3889-3892 (1997).
\bibitem{fer98} Fernando C. Lombardo y Francisco D. Mazzitelli, 
Phys. Rev. {\bf D 58}, 024009 (1998).
\bibitem{ferrusso} Fernando C. Lombardo, Francisco D. Mazzitelli y Jorge G. Russo,
 Phys. Rev. {\bf D 59},  064007 (1999).
\bibitem{ferPhD} Fernando C. Lombardo, ``Transici\'on
Cu\'antico-Cl\'asica en Teor\'ia de Campos'', Tesis Doctoral,
FCEyN, UBA, 1998.
\bibitem{fermazziray} F. C. Lombardo, F. D. Mazzitelli y R. J. Rivers,
Phys. Lett. {\bf B 523}, 317 (2001).
\bibitem{rayfermazzi-plb} R. J. Rivers, F. C. Lombardo y F. D. Mazzitelli,
Phys. Lett {\bf B 539}, 1 (2002).
\bibitem{fermazziray-npb} F. C. Lombardo, F. D. Mazzitelli y R. J. Rivers, 
Nucl. Phys. {\bf B 672}, 462 (2003).
\bibitem{PLB}F. C. Lombardo,  R. Rivers y P. I. Villar, Phys. Lett 
{\bf B 648}, 64  (2007).

\bibitem{Facchi} P. Facchi, S. Pascazio y T. Yoneda, Open Systems and Information
Dynamics {\bf 14}, 139 (2007).
\bibitem{Brezger} B. Brezger {\it et al.}, Phys. Rev. Lett. {\bf 88}, 100404 (2002).
\bibitem{Hornberger:2003} K. Hornberger, S. Uttenthaler, B. Brezger, L. Hackermuller,
M. Arndt y A. Zeilinger, Phys. Rev. Lett. \textbf{90}, 160401 (2003).
\bibitem{Breuer} H.P.Breuer y F. Petruccione,  {\it The theory
of open quantum systems}. Oxford University Press (2002).
 \bibitem{Lindenberg} K. Lindenberg y B.J. West, Phys. Rev. Lett. {\bf51}, 1370 (1983).
\bibitem{Carmichel} H. Carmichel, {\it An open system approach to quantum optics}.
Springer Verlag, (1993).
\bibitem{UnruhZurek} W. G. Unruh y W. H. Zurek, Phys. Rev. 
{\bf D40}, 1071 (1989).
\bibitem{Caldeira} A. O. Caldeira y A. J. Leggett, Physica {\bf A 121},
587-616, (1983).  A. O. Caldeira y A. J. Leggett,
Ann. Phys. (N.Y.) {\bf 149}, 374 (1983).
\bibitem{Hakim}V. Hakim y V. Ambegoakar, Phys. Rev. 
{\bf A32}, 423 (1985).
\bibitem{Haake} F. Haake y R. Reibold, Phys. Rev. {\bf A 32}, 2462 (1985).
\bibitem{Paz} J. P. Paz, {\it The Physical Origin of Time Asymmetry}, 
ed. by J. Halliwell, J. Perez Mercader y W.
Zurek (Cambridge University Press, Cambridge, 1994)
\bibitem{HuPazZhangII}B. L. Hu, J. P. Paz y Y. Zhang, Phys. Rev. 
{\bf D47}, 1576 (1993).
\bibitem{jppdavila} L. D\'avila Romero y J. P. Paz, 
Phys. Rev. {\bf A 55}, 4070 (1997).
\bibitem{leshouches} J. P. Paz y W. H. Zurek, {\it
Environmet-induced decoherence and the transition from quantum to
classical}, lectures at the 72nd Les Houches Summer School on
"Coherent Matter Waves"; (1999). arXiv: quant-ph/0010011


\bibitem{ford} G. W. Ford y R. F. O'Connell, J. Optics {\bf B5}, S349 (2003).
\bibitem{imry} Y. Imry, arXiv: cond-mat/0202044.
\bibitem{sinha} S. Sinha, Phys. Lett. {\bf A 228}, 1 (1997).
 \bibitem{jpphabzurek} J.P. Paz, S. Habib y W.H. Zurek, Phys. Rev.
  {\bf D 47}, 488 (1993).
\bibitem{Hu2} C. H. Fleming, B. L. Hu y A. Roura, 
{\it Solutions to Master Equations of 
Quantum Brownian Motion is a General Environment with 
External Force}; e-print: arXiv:0705.2766v1 [quant-ph].
\bibitem{legget} A. J. Leggett, S. Chakravarty, A. T. Dorsey, M. P. A. Fisher,
 A. Garg y W. Zwerger, Rev. Mod. Phys. {\bf 59}, 1 (1987).
\bibitem{sonnentag} P. Sonnentag y F. Hasselbach, 
Braz. J. of Phys. {\bf 35}, 385 (2005); 
Phys. Rev. Lett. {\bf 98}, 200402 (2007).
\bibitem{tongvojta} Ning-Hua Tong y Matthias Vojta, Phys. Rev. Lett. {\bf 97},
016802 (2006).
\bibitem{caleg}  A. O. Caldeira y A. J. Leggett, Phys. Rev. Lett. 
{\bf 46}, 211 (1981); A.O. Caldeira y A.J. Leggett, Ann. Phys.
\textbf{149}, 374 (1983).
\bibitem{gral} U. Weiss, {\it Quantum Dissipative Systems}; World Scientific, Singapore 
(1993); J.M. Martinis, M.H. Devoret y J. Clarke, Phys. Rev. {\bf B 35}, 4682 (1987); A. 
Wallraff, T. Duty, A. Lukashenko y A.V. Ustinov, Phys. Rev. Lett. {\bf 90}, 
037003 (2003).
\bibitem{hanggi} P. H\"anggi, P. Talkner y M. Borkovek, Rev. Mod. Phys. {\bf 62}, 
251 (1990).
\bibitem{Coleman} S. Coleman, {\it Aspects of Symmetry}; Cambridge University Press, 
NY (1985).
\bibitem{gral2} J. Ankerhold y H. Grabert, Phys. Rev. Lett. {\bf 91}, 
016803 (2003); E. Calzetta, A. Roura y 
E. Verdaguer, Phys. Rev. Lett. {\bf 88}, 010403 (2002).
\bibitem{tunel} W. A. Lin y L. E. Ballentine, Phys. Rev. {\bf A 45}, 3637
(1992); R. Uttermann, T. Dittrich y P. H\"anggi, Phys.Rev. {\bf E49}, 273 (1994)
; T. Dittrich, B. Oelschlaegel y P. H\"anggi, Europhys. Lett. {\bf 22}, 5
(1993); S. Kohler, R. Utermann, P. H\"anggi y T. Dittrich; Phys.Rev. {\bf E 58},
 7219 (1998).
\bibitem{diana_jpp} D. Monteoliva y J. P. Paz,
Phys. Rev. Lett, {\bf 85}, 3375 (2000); Phys. Rev. {\bf E 64}, 056238 (2001).
\bibitem{Shin1} Y. Shin, M. Saba, T.A. Pasquini, W. Ketterle, 
D. E. Pritchard y A. E. Leanhardt, Phys. Rev. Lett. {\bf 92}, 050405 (2004).
\bibitem{Anderson} E. Anderson, T. Calarco, R. Folman, M. Anderson, B. Hessmo
y Schmiedmayer, Phys. Rev. Lett {\bf 88}, 100401 (2002).
\bibitem{Menotti} C. Menotti, J. R. Anglin,  J. I. Cirac y P. Zoller, Phys.
Rev. {\bf A 63}, 023601 (2001).
\bibitem{Pitaevskii} L. Pitaevskii y S. Stringari, Phys. Rev. Lett. {\bf 87},
180402 (2001).
\bibitem{prezhdo}O. V. Prezhdo y P. Rossky, Phys. Rev. Lett. {\bf 81}, 5294 (1998).
\bibitem{robin} R. Blume-Kohout y W.H. Zurek, Phys. Rev. {\bf A 68}, 032104 (2003).
\bibitem{fermazzidiana} F.C. Lombardo, F.D. Mazzitelli y D. Monteoliva,
Phys. Rev. {\bf D 62}, 045016 (2000).
\bibitem{nunoferdiana} N.D. Antunes, F.C. Lombardo y D. Monteoliva, Phys.
Rev. {\bf E64}, 066118 (2001).
\bibitem{elze} Hans-Thomas Elze, {\it Vacuum-Induced Quantum Decoherence and the Entropy Puzzle}; hep-ph/9407377
\bibitem{kapral1} K. Shiokawa y R. Kapral, J. Chem. Phys. {\bf 117}, 7852 (2002)
\bibitem{anu} A. Venugopalan, Phys. Rev. {\bf A 61}, 012102 (1999)
\bibitem{ZP} W.H. Zurek y J.P. Paz, Phys. Rev. Lett. {\bf 72}, 2508 (1994

\bibitem{Hammond} R. T. Hammond, Contemp. Phys. {\bf 36}, 103 (1995).
\bibitem{Panchat} S. Pancharatnam, Proc. Indian Acad. Sci., Sect. {\bf A 44}, 247 (1956).
\bibitem{Frankel} T. Frankel, {\it The geometry of Physics},
Cambrigde University Press, Cambridge, 2000.
\bibitem{Berry} M. V. Berry, Proc. Roy. Soc. {\bf A 329}, 45 (1984).
\bibitem{ABohm} Y. Aharonov y D. Bohm, Phys. Rev. {\bf 115}, 485 (1959).
\bibitem{ACasher}  Y. Aharonov y A. Casher, Phys. Rev. Lett. {\bf 53}, 319
(1984).
\bibitem{Alejandro} F. D. Mazzitelli, J. P. Paz y A. Villanueva, Phys. Rev. {\bf A 68},
062106 (2003).
\bibitem{Vourdas}  C. C. Chong, D. I. Tsomokos y A. Vourdas, Phys. Rev. {\bf A 66},
033813 (2002).
\bibitem{Vourdas2} A. Vourdas, Phys. Rev. {\bf A 64}, 053814 (2001).
\bibitem{Ford} Jen-Tsung Hsiang y L. H. Ford, Phys. Rev. Lett. {\bf 92},
250402 (2004).
\bibitem{Anandan} J. Anandan, Phys. Rev. Lett. {\bf 85}, 1354 (2000).
\bibitem{APV} Y. Aharonov, S. A. Pearle y L. Vaidman, Phys. Rev. \textbf{A 37},4052 (1988).
\bibitem{schwinger} J. Schwinger, L. L. DeRaad, K. Milton y W. Tsai, {\it Classical
Electrodynamics}, Advanced Book Program, Westview Press, 1998.
\bibitem{Hasselbach} F. Hasselbach, Z. Phys. B: Condens. Matter {\bf 71},
443 (1988).
\bibitem{Keith} D. W. Keith, C.R.Ekstrom, Q.A.Turchette y D.E.Pritchard,
Phys.Rev.Lett. {\bf 66}, 2693 (1991).
\bibitem{Pfau} T. Pfau, C. Kurtsiefer, C. S. Adams, M. Sigel y J. Mlynek,
Phys. Rev. Lett. {\bf 71}, 3427 (1993).
\bibitem{Greenberger} D. M. Greenberger, D. K.A twood, J. Arthur, C. G. Shull
y M. Schlenker, Phys. Rev. Lett. {\bf 47}, 751 (1981).
\bibitem{Grisent} R. E. Grisent, W. Sch\"{o}llkopf, J. P. Toennies, G. C. Hegerfeldt y  T. K\"{o}hler, Phys. Rev. Lett. {\bf 83}, 1755 (1999).
\bibitem{Bozic} M. Bozic, D. Arsenovic y L. Vuskovic, Phys. Rev.{\bf A 69}, 053618 (2004).
\bibitem{Venugopalan}T. Qureshi y A. Venugopalan, quant-ph/0602052.
\bibitem{Viale} A. Viale, M. Vicari y N. Zanghi, Phys. Rev. A {\bf 68},063610 (2003).
\bibitem{Walls85} D. F. Walls y G. J. Milburn, Phys. Rev. {\bf A 31}, 2403 (1985).
\bibitem{Diosi95} L. Diosi, Europhys. Lett. {\bf 30}, 63 (1995).
\bibitem{Vacchini00} B. Vacchini, Phys. Rev. Lett. {\bf 84}, 1374 (2000).
\bibitem{Vacchini01} B. Vacchini, Phys. Rev. {\bf E 63}, 066115 (2001).
\bibitem{Horn} K. Hornberger, J. E. Sipe y M. Arndt, Phys. Rev. A {\bf 70},053608 (2004).
\bibitem{Joos} D.Guilini, et. al., {\it Decoherence and the Appearance of
 a Classical World in Quantum Theory}, Springer, Berlin, 1996.
\bibitem{FalciNature} G. Falci {\it et al.}, Nature {\bf 407}, 355 (2000).
\bibitem{Jones} J. A. Jones, V.Vedral, A. Eckert y G. Castagnoli, Nature {\bf 403}, 869 (2000).
\bibitem{Nakamura} Y. Nakamura, Yu. A. Pashkin y J. S.Tsai, Nature
{\bf 398}, 786 (1999).
\bibitem{Ao} P. Ao y D. J. Thouless, Phys. Rev. Lett. {\bf 70}, 2158 (1993).
\bibitem{Zhu} X. -M. Zhu, E.Br\"andstr\"om y A. Sundqvist, Phys. Rev. Lett. {\bf 78}, 122 (1997).
\bibitem{Zanardi} P. Zanardi y M. Rasetti, Phys. Lett. {\bf A 264}, 94 (1999).
\bibitem{Sjoqvist} E. S\"{o}qvist, A. K. Pati, A. Ekert, J. S. Anandan, M. Ericsson, D. K. L. Oi y V. Vedral,  Phys. Rev. Lett. {\bf 85}, 2845 (2000). 
\bibitem{Singh} K. Singh, D. M. Dong, K. Basu, J. L. Chen, 
y J. F. Du, Phys. Rev. A, {\bf 67}, 032106 (2003). 
\bibitem{Tong}  D. M. Tong, E. Sjöqvist, L. C. Kwek,
 y C. H. Oh, Phys. Rev. Lett. {\bf 93}, 080405 (2004).
\bibitem{Gefen} R. S. Whitney, Y. Makhlin, A. Shnirman,
 y Y. Gefen, Phys. Rev. Lett. {\bf 94},  070407 (2005).
\bibitem{Eckert} G. M. Palma, K. Suominen y A. Ekert,  
Proc. R. Soc. London, Ser.A 452, 567 (1996);
L. Viola  y S. Lloyd, Phys. Rev. A {\bf 58},
 2733 (1998).
\bibitem{Falci} G. Falci, R. Fazio, G. M. Palma, J. Siewert 
y V. Vedral,  Nature (London) {\bf 407},
355 (2000).
\bibitem{Yi} X. X. Yi, D. M. Tong, L. C. Wang, L. C. Kwek y 
C. H. Oh, Phys. Rev. A 73, 052103 (2006).
\bibitem{Ekert2} A. Ekert {\it et al.}, J. Mod. Opt. {\bf 47},
2501 (2000).
\bibitem {Du} J. Du, P. Zou,M. Shi, L. C. Kwek,J. Pan, C. H. Oh, A. Ekert, D. K.L. Oi, 
y M. Ericsson, Phys. Rev. Lett. {\bf 91}, 
100403 (2003); I. Fuentes-Guridi, E. R. Livine, Phys. Rev. Lett. {\bf 94}, 020503 (2005).
\bibitem{guthpi} A. Guth y S. Y. Pi, Phys. Rev {\bf D}32, 1899
(1985).
\bibitem{rayfer} R. J. Rivers y F. C. Lombardo, Int. J.Theor. Phys. {\bf 44},
1855 (2005); R. J. Rivers y F. C. Lombardo, Brazilian Journal of Physics 35, 397 (2005).
\bibitem{huzhang} B. L. Hu, in {\it Relativity and Gravitation: Classical and Quantum}, J.C.D.
Olivo et al (Eds.), World Scientific, Singapore (1991)
\bibitem{boya}  D. Boyanovsky, H. J. de Vega y R. Holman, Phys.
Rev. {\bf D49}, 2769 (1994); D. Boyanovsky, H. J. de Vega, R.
Holman, D. -S. Lee y A. Singh, Phys. Rev. {\bf D 51}, 4419
(1995); S. A. Ramsey y B. L. Hu, Phys. Rev. {\bf D 56}, 661 (1997)
\bibitem{moro} E. Moro y G. Lythe, Phys.Rev. {\bf E 59} R1303 (1999).
\bibitem{Gri} R.B. Griffiths, J. Stat. Phys. {\bf 36}, 219 (1984)
\bibitem{Omn} R. Omnes, J. Stat. Phys. {\bf 53}, 893 (1988); Ann. Phys. {\bf 201}, 354 (1990); Rev. Mod. Phys. {\bf 64}, 339 (1992)
\bibitem{calhu} E. Calzetta y B. L. Hu, Phys. Rev. {\bf D 35}, 495 (1987)
\bibitem{calhumaz}E. A. Calzetta, B.L. Hu y F. D. Mazzitelli, Phys. Rep. {\bf 352}, 459 (2001).
\bibitem{karra} G. Karra y R. J. Rivers, Phys.Lett. B {\bf 414}, 28 (1997).
\bibitem{nunoray} N. D. Antunes, P. Gandra, R. J. Rivers y A. Swarup, Phys. Rev. {\bf D 73}, 085012 (2006);
 N. D. Antunes, P. Gandra, and R. J. Rivers, Phys. Rev. {\bf D 71}, 105006 (2005)
\bibitem{nunobett} N. D. Antunes, L. M. A. Bettencourt y  W. H. Zurek, Phys. Rev. Lett. {\bf 82}, 2824 (1999)
\bibitem{nuno} N. D. Antunes, F. C. Lombardo y D.  Monteoliva, Phys. Rev.
{\bf E 64}, 066118 (2001);  Nuno .D. Antunes, Fernando.C. Lombardo, D.
Monteoliva y Paula I. Villar, Phys. Rev. {\bf E 73}, 066105 (2006)
\bibitem{diana} F. C. Lombardo, F. D. Mazzitelli y  D. Monteoliva, Phys. Rev.
{\bf D62}, 045016 (2000)
\bibitem{kibble2} T. W. B. Kibble, Phys. Rep. {\bf 67}, 183 (1980)



\end{thebibliography}
\end{document}